\documentstyle[sprocl,epsf,twoside]{article}

\def\beq{\begin{equation}}   
\def\eeq{\end{equation}}

\def\lsim{\mathrel{\rlap{\lower3pt\hbox{\hskip0pt$\sim$}}
    \raise1pt\hbox{$<$}}}         
\def\gsim{\mathrel{\rlap{\lower4pt\hbox{\hskip1pt$\sim$}}
    \raise1pt\hbox{$>$}}}         

\begin{document}

\begin{flushright}
 JLAB-THY-00-44\\
\end{flushright}
\vskip 1cm

\title{HIGH-ENERGY QCD AND WILSON LINES\footnote{To be published 
in the Boris Ioffe Festschrift ``At the Frontier of Particle
Physics/Handbook of QCD'', edited by M. Shifman (World Scientific, 
Singapore, 2001)}}

\author{I.~BALITSKY}

\address{Phys. Dept., Old Dominion Univ., Hampton Blvd., 
Norfolk, VA 23529, USA\\
and\\
Theory Group, Jefferson Lab, 12000 Jefferson Ave., \\  
Newport News, VA 23606, USA}

\maketitle\abstracts{
At high energies the particles move very fast so their 
trajectories can
be approximated by straight lines collinear to their velocities. 
The proper degrees of freedom for the
fast gluons moving along the straight lines are the Wilson-line 
operators -- infinite  gauge factors  ordered along the straight 
line. I review the study of the high-energy scattering in terms 
of Wilson-line degrees of freedom.
}


\tableofcontents
\newpage

\section{Introduction}

Traditionally, high-energy scattering in perturbative QCD 
(pQCD) is 
studied by direct summation of Feynman diagrams. 
In the leading logarithmic approximation (LLA)
\begin{equation}
\alpha_s \ll 1,\qquad \alpha_s \ln {s\over m^2}\simeq 1~,
\label{flob1}
\end{equation}
the amplitudes at high energy 
are determined by the Balitsky-Fadin-Kuraev-Lipatov (BFKL) 
pomeron~\cite{bfkl} (for a review, see Ref.~2),
\begin{equation}
A(s)\sim \left({s\over m^2}\right)^{12{\alpha_s\over\pi} \ln 2}\,.
\label{flob2}
\end{equation} 
Here $m$ is the characteristic mass or virtuality of scattered particles 
(for example, for the small-$x$ deep inelastic scattering $m^2=Q^2$).
In order for 
perturbative QCD (pQCD) 
to be applicable, $m$ must be sufficiently large so that
$\alpha_s(m)\ll 1$. 

The power behavior of BFKL cross
section  (\ref{flob2}) violates  the Froissart bound and, therefore, the BFKL
pomeron describes only  the pre-asymptotic behavior at intermediate energies
when the cross sections are small in comparison to the geometric cross section 
$2\pi R^2$. In order to  find the true high-energy asymptotics 
by analysis of  Feynman diagrams we should sum up not only
the leading logarithms $(\alpha_s \ln s)^n$ but also the sub-leading
ones $\alpha_s(\alpha_s \ln s)^n$, then the sub-sub-leading terms
$\alpha_s^2(\alpha_s \ln s)^n$,  etc.  
This is almost equivalent to finding an exact
answer to arbitrary QCD amplitude in all orders in perturbation theory. A more
realistic approach is to unitarize the BFKL pomeron, i.e. to sum up the subset
of sub-leading logarithms which restores the unitarity in $s$ channel. Still,
it is a difficult problem which has been in a need of a solution for more
than 20 years.  One of the most popular ideas on solving this problem is
reducing   QCD at high energies to some sort of low-dimensional effective
theory which will be simpler than original QCD, maybe even to the extent of
exact solvability. The first step on this road is to identify proper degrees of
freedom for this effective theory. One of the possible choices is to formulate
high-energy scattering in terms of ``reggeized gluons.''\cite{lobzor}  An 
alternative and related approach~\cite{ing} is based on 
so-called Wilson lines -- infinite gauge links corresponding  to fast gluons
moving along the straight-line classical trajectories.

An important aspect of the Wilson-line approach to high-energy scattering is
the fact that it serves as a bridge between pQCD calculations 
and the semiclassical approach to high-energy scattering based on the solution 
of the classical equations 
for the fast-moving sources.\cite{lrmodel} The semiclassical QCD (sQCD) is 
applicable when the
coupling constant is small but the characteristic fields produced 
by colliding
particles are large, $\sim {1\over g}$. As advocated in 
Ref.~4, 
sQCD may be relevant for the
heavy-ion collisions because the coupling
constant can be relatively small due to high density of partons in the center
of the collision. The relevant ``saturation scale'' was estimated to be
$\sim 1$ GeV at RHIC and $\sim 2-3$~GeV at
LHC.\cite{jklw,finn,mu00}

\begin{figure}[htb]
\centerline{
\epsfysize=2.2cm
\epsffile{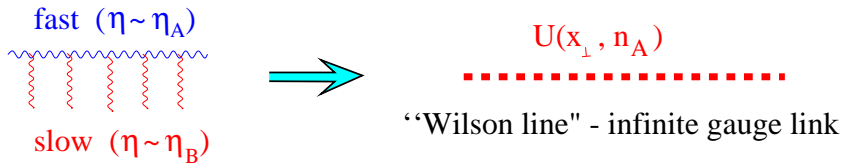}}
\caption{Propagator of a fast ``A'' gluon in the slow ``B'' background.}
\label{ofig1}
\end{figure}

\begin{figure}[htb]
\centerline{
\epsfysize=2.2cm 
\epsffile{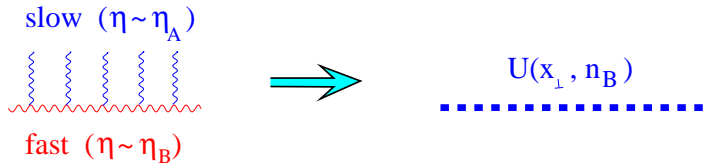}}
\caption{Gluon of ``B'' type viewed  from the rest frame of ``A'' gluons.} 
\label{ofig2}
\end{figure}

Let us demonstrate that the relevant degrees of freedom for the high-energy
scattering are Wilson lines.\cite{nacht}
As a
result of the high-energy  collision, we have a shower of produced particles in
the range of rapidity between those of the colliding particles. 
Consider two clusters of particles with 
different rapidities: ``A'' particles with rapidities close to $\eta_A$ and
``B'' particles with rapidities $\simeq\eta_B$. From the viewpoint of the
``B'' particles   the ``A'' gluon moves very fast, so its trajectory can be
approximated  by a straight line collinear to the gluon momentum, see 
Fig.~\ref{ofig1}.  
The propagator of such 
gluon reduces to the free propagator multiplied by the infinite 
 gauge factor (made from ``B'' gluons) ordered along the straight 
 line parallel to
$n_A$, the direction corresponding to the rapidity  $\eta_A$: 
\begin{equation}
U(x,n_A)=[\infty n_A+x,-\infty n_A+x] .
\label{flob3}
\end{equation}
Hereafter we use the notation
\begin{equation}
[x,y]\equiv P\exp ig\int_{0}^{1}du (x-y)^{\mu} A_{\mu}(ux+(1-u)y)
\label{defpexp}
\end{equation}
for the straight-line gauge link connecting the points $x$ and $y$.
 Therefore, the $B$ particles can interact with 
 $A$ fields only via the Wilson lines (\ref{flob3}).
Similarly, if we sit in the rest frame of the ``A'' gluons
the ``B'' particles are moving fast along the direction collinear 
to the vector
$n_B$ corresponding to rapidity $\eta_B$, see Fig.~\ref{ofig2}.
The propagator of these gluons
reduces to the Wilson line (made from ``A'' gluons) collinear to $n_B$  
\begin{equation}
U(x,n_B)=[\infty n_B+x,-\infty n_B+x] .
\label{flob4}
\end{equation}
Again, the relevant degree of freedom is the non-local Wilson line 
(\ref{flob4}) rather than the local field $A(x)$.
We see that the particles with different rapidities perceive each other as 
Wilson lines. The formal proof of this statement in terms of Feynman diagrams 
is given in the Appendix (see also Ref.~9). 

In this review I give a pedagogical introduction to the Wilson-line-based
approach to high energy scattering. After a short overview of the 
traditional approach, I shall present the operator expansion for 
high-energy scattering
which provides the operator language for the BFKL equation in the same way
as the usual light-cone expansion gives the operator description of the DGLAP 
equation. Unlike the latter, there is a symmetry between the coefficient 
functions and matrix elements in the high-energy operator expansion which 
can be 
summarized by the factorization formula for 
high-energy scattering. This factorization formula gives us the rigorous 
definition of the
effective action for a given interval of rapidity. In the last section  
we discuss the semiclassical approach to effective action related to the
problem of scattering of two shock waves in QCD.

\section{The hard pomeron in pQCD}

Since there are many excellent reviews of the traditional,
Feynman diagrams-based, approach to high-energy scattering
(see e.g. Refs.~2, 10), 
I will 
present here the short introduction to the subject so as to set up the stage 
for the subsequent analysis of the high-energy scattering in terms of
Wilson-line operators.
\subsection{High-energy $\gamma^\ast\gamma^\ast$ scattering}
For simplicity, we
consider the classical example of high-energy scattering of 
virtual photons with virtualities $\sim -~m^2$ 
\begin{eqnarray}
A(s,t)=
-i\int d^4xd^4yd^4z
e^{-ip_Ax-ip_By+ip'_Az}{\mbox{$\langle 0|$}} 
T\{j_A(x)j_B(y)j'_A(z)j'_B(0)\}{\mbox{$|0\rangle $}}.
\label{2.1.1}
\end{eqnarray}
Here $j_A(x)$ is 
electromagnetic current $j^{\mu}(x)$ 
multiplied by the polarization vector $e^A_{\mu}(p)$.
In the Regge limit ($s\gg m^2,t$) it is
convenient to use the Sudakov decomposition:   
\begin{equation}
p^{\mu}~=~\alpha_pp_1^{\mu}+\beta_pp_2^{\mu}+p_{\perp}^{\mu},
\label{2.1.2}
\end{equation}
where
$p_1^{\mu}$ and
$p_2^{\mu}$ are the
light-like vectors close to $p_A$ and 
$p_B$, respectively: 
\begin{equation}
p_A=p_1+{p_A^2\over s}p_2,\quad p_B=p_2+{p_B^2\over
s}p_1,\quad r\equiv p_B-p'_B=\alpha_r p_1+\beta_r p_2+r_\perp .
\label{2.1.3} 
\end{equation}
The momentum transfer $r=p'_A-p_A=\alpha_r p_1+\beta_r p_2+r_\perp$
 has components $\alpha_r\sim\beta_r\sim {m^2\over s}$ so 
$t\simeq -{\vec r}^2$. The typical diagram 
for the high-energy $\gamma^\ast\gamma^\ast$ amplitude 
is shown in Fig.~3 (recall that 
the diagrams with gluon exchanges dominate at high energies).

\begin{figure}[htb]
\centerline{
\epsfysize=5cm
\epsffile{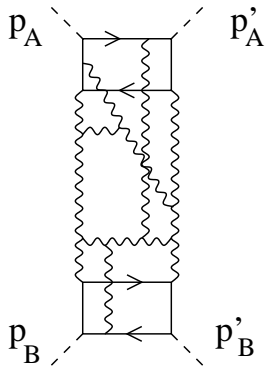}}
\caption{ A typical Feynman diagram for the high-energy 
$\gamma^*\gamma^*$ scattering.} \label{typdm}
\end{figure}
We will calculate the imaginary part of the 
amplitude $A(s,t)$
\begin{equation}
W={1\over \pi}{\rm Im} A.
\label{2.1.4}
\end{equation}
The real part of $A(s,t)$ can be 
restored using the dispersion relations. (It turns out that in the leading 
logarithmic approximation (LLA)
 the amplitude at high energy is purely imaginary, see e.g. the review
in Ref.~2). 

\begin{figure}[htb]
\centerline{
\epsfysize=3.3cm
\epsffile{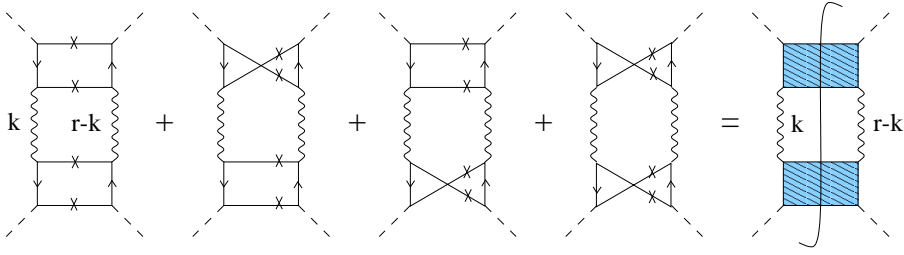}}
\caption{ Lowest-order diagrams for the high-energy scattering 
of virtual photons.} \label{fig2}
\end{figure}

\begin{figure}[htb]
\centerline{
\epsfysize=2.7cm
\epsffile{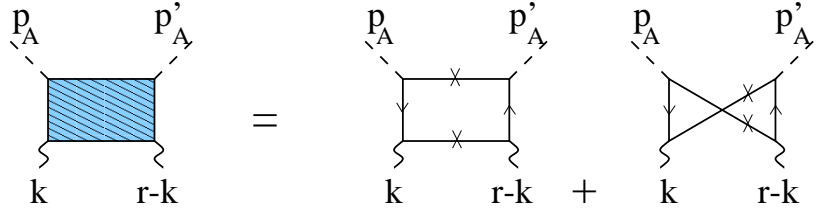}}
\caption{ Photon impact factor.} \label{ofig5}
\end{figure}

Let us start with the lowest-order diagrams shown in Fig.~4.
The integral over gluon momentum $k=\alpha_kp_1+
\beta_kp_2+k_{\perp}$ has the form
\begin{equation}
W^0={2\over\pi}g^4\int {d^4k\over 16\pi^4} {1\over k^2}
{1\over (r-k)^2}{\rm Im} (\Phi_A)^{ab}_{\xi\eta}(k,+r -k) 
{\rm Im}\Phi_B^{\xi\eta ab}(-k,k-r)
\label{2.1.5}
\end{equation}
where $(\Phi_A)^{ab}_{\xi\eta}(k, r-k)$ and 
$(\Phi_B)^{ab}_{\xi\eta}(-k,k-r)$
are  the upper and the lower blocks 
of the diagram in Fig.~\ref{fig2} (stripped of the strong 
coupling constant $g$). 
Here $a,b$ and $\xi,\eta$ are the color and
Lorentz indices, respectively.
In the Regge kinematics ($\equiv~s\gg$ everything else) 
$\alpha_k\sim {m^2\over s}$ 
and $\beta_k\sim x$ so $k^2\simeq -\vec{k}_{\perp}^2$. Moreover, 
alpha's in the upper block are $\sim 1$ 
so one can drop $\alpha_k$ in the upper block. Similarly, beta's in the 
lower block are $\sim 1$ 
hence one can neglect $\beta_k$ in the lower block. We get 
($\Phi^{ab}={\delta_{ab}\over N_c^2-1}\Phi^{cc}$)
\begin{eqnarray}
W^0&=&\hspace{-2mm}{2g^4\over (N_c^2-1)\pi}
\label{2a6}\\
&\times&\hspace{-2mm}\int {d^4k\over 16\pi^4} {1\over
\vec{k}_{\perp}^2} {1\over (\vec{r}-\vec{k})_{\perp}^2}{\rm Im} 
\left.\Phi^{aa}_{\xi\eta}(k,r -k)\right|_{\alpha_k=0} 
{\rm Im}\left.\Phi^{\xi\eta bb}(-k,k-r)\right|_{\beta_k=0},
\nonumber
\end{eqnarray}
where $N_c=3$ is the number of colors. At high energies, the metric tensor 
$g^{\mu\nu}$ in
the numerator of the  Feynman-gauge  gluon propagator reduces to 
 ${2\over s}p_2^{\mu}p_1^{\nu}$,
so the integral (\ref{2a6}) for 
the imaginary part factorizes into a product of two ``impact factors"
integrated with two-dimensional propagators
%
\begin{equation}
W^0={s\over \pi}g^4{N_c^2-1\over 4}\left(\sum e_q^2\right)^2
\int {d^2k_{\perp}\over 4\pi^2} {1\over \vec{k}_{\perp}^2}
{1\over (\vec{r}-\vec{k})_{\perp}^2}I^A(k_{\perp},r_{\perp}) 
I^B(-k_{\perp},-r_{\perp}),
\label{2a7}
\end{equation}
where
\begin{eqnarray}
I^A(k_{\perp},r_{\perp})\!\!\!\!\! &=&
\!\!\!\!\! {p_2^{\xi}p_2^{\eta}\over s(N_c^2-1)}
\left(\sum e_q^2\right)^{-1}
\left.\int {d\beta_k\over 2\pi}
{\rm Im} \Phi^{aa}_{\xi\eta}(k,r -k)\right|_{\alpha_k=0} ,
\label{fla8}\\
I^B(-k_{\perp},-r_{\perp})\!\!\!\!\! &=&\!\!\!\!\! 
{p_1^{\xi}p_1^{\eta}\over s(N_c^2-1)}\left(\sum
e_q^2\right)^{-1} \left.\!\!\!\! \int \!\! {d\alpha_k\over 2\pi}
{\rm Im} \Phi^{aa}_{N\xi\eta}(-k,k-r)\right|_{\beta_k=0} ,
\label{fla9}
\end{eqnarray}
and $\left(\sum e_q^2\right)$ is the sum of squared charges of active flavors.
The photon impact factor is given by the two one-loop diagrams shown in  
Fig.~\ref{ofig5}.

The standard calculation of these diagrams yields~\cite{mes}
\begin{equation}
{I}^{A}(k_{\perp},r_{\perp})=\bar{I}^{A}(k_{\perp},r_{\perp})- 
\bar{I}^{A}(0,r_{\perp}) ,
\label{fla10d}
\end{equation}
where
%
\begin{eqnarray}   
\bar{I}^{A}(k_{\perp},r_{\perp}) &=&
{1\over 2}\int _{0}^{1} \frac {d\alpha}{2\pi }
   \int _{0}^{1} \frac {d\alpha'}{2\pi }
   \left\{
   \vec{P}_{\perp}^2\alpha'\bar{\alpha'}-
[p_A^2\bar{\alpha}'+(p'_A)^2\alpha']\alpha\bar{\alpha}
   \right\}^{-1}\label{fla11}\\ 
&\times&
   \Bigg\{
      (1-2\alpha\bar{\alpha})(1-2\alpha'\bar{\alpha}')
      \vec{P}_{\perp}^2 (\vec{e}_A,\vec{e}'_A)_{\perp} 
+4\alpha\bar{\alpha}\alpha'\bar{\alpha'}[\vec{P}_{\perp}^2 
(\vec{e}_A,\vec{e}'_A)_{\perp}
\nonumber\\
&-&
2(\vec{P},\vec{e}_A)_{\perp}(\vec{P},\vec{e}'_A)_{\perp}]+
(e_A,e'_A)_{\perp}(p_A^2-(p'_A)^2)\alpha\bar{\alpha}
(1-2\alpha\bar{\alpha})
\nonumber\\
&\times&(1-2\alpha')+4\alpha\bar{\alpha}
(1-2\alpha)\alpha'(\vec{P},\vec{e}_A)_{\perp}
(\vec{r},\vec{e}'_A)_{\perp}    \Bigg\} 
\nonumber
\end{eqnarray}
for the transverse polarizations $A,A'=1,2$. 
Here $P_{\perp}\equiv k_{\perp}-r_{\perp}\alpha$ and $(a,b)_{\perp}$ 
denotes the (positive) scalar
product of transverse components of vectors $a$ and $b$. 

\subsection{The BFKL kernel}

In the next order in perturbation theory there are two types of diagrams for
the $\gamma^*\gamma^*$ amplitude: diagrams with 5-particle cut describing the
emission of an extra gluon and diagrams with 4-particle cut as in Fig.~4
but with an extra gluon  loop.

Let us at first consider the diagrams with the 5-particle cut 
shown in  Fig.~6. 
\begin{figure}[htb]
\centerline{
\epsfysize=7.7cm
\epsffile{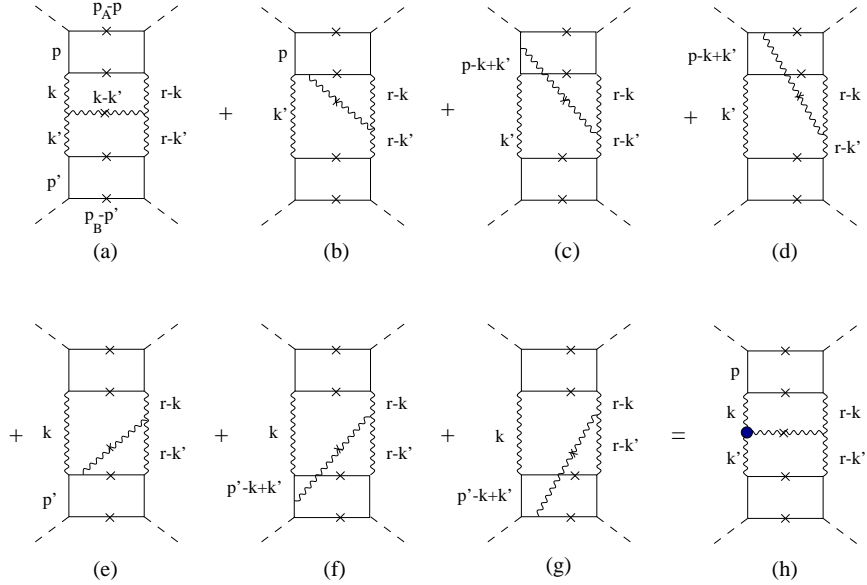}}
\caption{The effective vertex in LLA.}\label{lvertex}
\end{figure}
The contribution of the diagram shown
in Fig.~6a has the form   
\begin{eqnarray}
W_{(a)}^{(5)}&=&
{2\over\pi}g^6\int {d^4k\over 16\pi^4} \int {d^4k'\over 16\pi^4} 
{{\rm Im}\Phi_A^{\mu\nu ab}(k,r-k) \over k^2(r-k)^2}
{{\rm Im}\Phi_B^{\xi\eta mn}(-k',k'-r)\over (k')^2(r-k')^2}
\nonumber\\ 
&\times& f^{amc}f^{bnc}
\Gamma_{\mu\xi}^{~~~\sigma}(k,k')2\pi \delta((k-k')^2)
\theta(\alpha_k)
\Gamma_{\nu\eta\sigma}(r-k,r-k')\nonumber\\ 
&\times& {{\rm Im}\Phi_B^{\xi\eta mn}(-k',k'-r)\over (k')^2(r-k')^2},
\label{2.2.1}
\end{eqnarray}
where
\begin{equation}
\Gamma_{\mu\nu\lambda}(k,k')=
(k+k')_{\lambda}g_{\mu\nu}+(k'-2k)_{\nu}g_{\lambda\mu}+
(k-2k')_{\mu}g_{\nu\lambda}
\label{2.2.2}
\end{equation}
is the three-gluon vertex divided by g. (Strictly speaking, in order to
obtain $\Phi^A$ and $\Phi^B$ we must add the 
diagrams with permutations of the quark lines, as in Fig.~4).
As mentioned above, it is convenient to use Sudakov variables (\ref{2.1.2}): 
$k=\alpha_k p_1+\beta_k p_2+k_{\perp}$,  $k'=\alpha'_k p_1+\beta'_k
p_2+k'_{\perp}$. We will see that the logarithmic  contribution comes from the
region 
\begin{equation}
1\gg \alpha\gg \alpha'\sim {m^2\over s},~~~
{m^2\over s}\sim \beta\ll\beta'\ll 1, ~~~
\vec{k}_{\perp}^2\sim (k'_{\perp})^2\sim m^2.
\label{2.2.3}
\end{equation}
In this region 
$k^2=\alpha_k\beta_k s-\vec{k}_{\perp}^2\simeq -\vec{k}_{\perp}^2$. 
In the same way, $(k')^2=-(\vec{k}'_{\perp})^2$, 
$(r-k)^2=-(\vec{r}-\vec{k})_{\perp}^2$, and
$(r-k')^2=-(\vec{r}-\vec{k}')_{\perp}^2$. 
As we mentioned above, at high energies we can replace $g^{\mu\nu}$ 
in gluon propagators connecting the clusters with different rapidities 
by $2{p_2^\mu p_1^\nu\over s}$. With these approximations, the 
integral (\ref{2.2.1}) reduces to
\begin{eqnarray}
W_{(a)}^{(5)}&=&\!{2\over\pi}
g^6\Big({2\over s}\Big)^4\int {d\alpha_k d\beta_k\over
4\pi^2}{d^2k_{\perp}\over 4\pi^2}  \int{d\alpha'_k d\beta'_k\over 4\pi^2}
{d^2k'_{\perp}\over 4\pi^2}  {1\over \vec{k}_{\perp}^2}
{1\over (\vec{r}-\vec{k})_{\perp}^2}
\nonumber\\
&\times&\!{\rm Im} \Phi_A^{\ast\ast ab}(k,r-k)f^{amc}f^{bnc}
\Gamma_{\bullet\ast}^{~~~\sigma}(k,k')
2\pi \delta(\alpha_k\beta'_k s+(\vec{k}-\vec{k}')_{\perp}^2)\theta(\alpha_k)
\nonumber\\
&\times&\!\Gamma_{\bullet\ast\sigma}(r-k,r-k'){1\over (\vec{k}')_{\perp}^2}
{1\over (\vec{r}-\vec{k}')_{\perp}^2}
{\rm Im}\Phi_B^{\bullet\bullet mn}(-k',k'-r).
\label{2.2.4}
\end{eqnarray}
Since $\alpha$ in the upper block is $\sim 1$, one can neglect 
$\alpha_k$-dependence in $\Phi_A$ which leads to the replacement of 
$\int d\beta_k\Phi_A$ by the impact factor $I^A(k_{\perp},r_{\perp})$, 
see Eq.~(\ref{fla11}). Likewise, 
$\int d\alpha'_k\Phi_B\rightarrow I^B(k'_{\perp},r_{\perp})$ so we get
\begin{eqnarray}
W_{(a)}^{(5)}&=&{2g^6\over\pi}{N_c(N_c^2-1)\over 4}\left(\sum e_q^2\right)^2
\label{2.2.5}\\
\times\!\!\!\!\!&&\hspace{-0.6cm}\int{d^2k_{\perp}\over 4\pi^2}
{d^2k'_{\perp}\over 4\pi^2}I^A(k_{\perp},r_{\perp})  
{1\over \vec{k}_{\perp}^2(\vec{r}-\vec{k})_{\perp}^2}
{1\over (\vec{k}')_{\perp}^2(\vec{r}-\vec{k}')_{\perp}^2}
I^B(k'_{\perp},r_{\perp})\nonumber\\  
\times\!\!\!\!\!&&\hspace{-0.6cm}
\int {d\alpha_k d\beta'_k\over 4\pi^2}\Gamma_{\bullet\ast}^{~~~\sigma}(k,k')
2\pi \delta(\alpha_k\beta'_k s+(\vec{k}-\vec{k}')_{\perp}^2)\theta(\alpha_k)
\Gamma_{\bullet\ast;\sigma}(r-k,r-k').
\nonumber
\end{eqnarray}
Let us now turn to the diagram shown in Fig.~6b. Since the gluon with momentum
$k-k'$ now connects parts of the diagrams with different rapidities,  
we can replace $g^{\mu\nu}$ 
in this propagator
by $2{p_2^\mu p_1^\nu\over s}$. 
After that, the quark 
propagator with the momentum $p+k'$ in the upper block reduces to
\begin{equation}
t^a{\not\! p}_2{(\alpha_p+\alpha'_k){\not\! p}_1+
{\not\! p}_{\perp}+{\not\! k'}_{\perp}\over 
-(\alpha_p+\alpha'_k)(\beta_p+\beta'_k)s+(\vec{p}-\vec{k}')_{\perp}^2-
i\epsilon} 
{\not\! p}_2t^c\rightarrow t^a{\not\! p}_2{1\over -\beta'_k-i\epsilon}t^c ,
\label{2.2.6}
\end{equation}
(recall that $\alpha_p\sim 1, \beta_p\sim {m^2\over s}$). We see that in
the transverse space this propagator shrinks to a point so the answer for the 
upper block is again $I^A$ multiplied by 
${1\over\beta'_k+i\epsilon}$.  (The eikonal factor ${1\over\beta'_k+i\epsilon}$ 
is the Fourier transform of the first term of the expansion of 
Wilson-line propagator (\ref{flob3}) in powers of ``external slow field" 
represented by gluon with momentum $k'$). The right part of the diagram
in Fig.~6b is identical to that in Fig.~6a  so we obtain 
\begin{eqnarray}
W_{(b)}^{(5)}&=&\label{2.2.7}
i{2\over\pi}g^6{\rm Tr}\{t^at^ct^b\}\int{d^2k_{\perp}\over 4\pi^2} 
\int {d\alpha_k d\beta'_k\over 4\pi^2}
{d^2k'_{\perp}\over 4\pi^2}I^A(k_{\perp},r_{\perp}) f^{bac}\\
&\times&{1\over \beta'_k}{1\over (\vec{r}-\vec{k})_{\perp}^2}
2\pi \delta(\alpha_k\beta'_k s+(\vec{k}-\vec{k}')_{\perp}^2)\theta(\alpha_k)
\Gamma_{\bullet\ast\bullet}(r-k,r-k')\nonumber\\
&\times&{1\over (\vec{k}')_{\perp}^2}
{1\over (\vec{r}-\vec{k}')_{\perp}^2}I^B(k'_{\perp},r_{\perp}).
\nonumber
\end{eqnarray}
The contribution of the diagram in Fig.~6c is calculated in a similar 
way. One can replace
\begin{equation}
{t^c{\not\! p}_2\big[(\alpha_p-\alpha_k+\alpha'_k){\not\! p}_1+
{\not\! p}_{\perp}-{\not\! k}_{\perp}+{\not\! k'}_{\perp}
{\not\! p}_2t^a\big]
\over 
(\alpha_p-\alpha_k+\alpha'_k)(\beta_p-\beta_k+\beta'_k)s-
(\vec{p}-\vec{k}+\vec{k}')_{\perp}^2+i\epsilon}
\rightarrow t^c{\not\! p}_2{1\over \beta'_k-i\epsilon}t^a ,
\label{2.2.8}
\end{equation}
and, therefore,
\begin{eqnarray}
W_{(c)}^{(5)}&=&\label{2.2.9}
-i{2\over\pi}g^6{\rm Tr}\{t^bt^at^c\}\int{d^2k_{\perp}\over 4\pi^2} 
\int {d\alpha_k d\beta'_k\over 4\pi^2}
{d^2k'_{\perp}\over 4\pi^2}I^A(k_{\perp},r_{\perp}) f^{abc}\\ 
&\times&{1\over \beta'_k}{1\over (\vec{r}-\vec{k})_{\perp}^2}
2\pi \delta(\alpha_k\beta'_k s+(\vec{k}-\vec{k}')_{\perp}^2)\theta(\alpha_k)
\Gamma_{\bullet\ast\bullet}(r-k,r-k')\nonumber\\
&\times&{1\over (\vec{k}')_{\perp}^2}
{1\over (\vec{r}-\vec{k}')_{\perp}^2}I^B(k'_{\perp},r_{\perp}).
\nonumber\
\end{eqnarray}
Note that the sum of the results (\ref{2.2.5}), (\ref{2.2.7}), and (\ref{2.2.9}) 
may be 
obtained from the  contribution (\ref{2.2.5}) of the diagram in Fig.~6a. 
by the replacement 
\begin{equation} 
\Gamma_{\bullet\ast}^{~~\sigma}(k,k')\rightarrow
\Gamma_{\bullet\ast}^{~~\sigma}(k,k')-
{\vec{k}_{\perp}^2\over \beta'_k}p_2^{\sigma}.
\label{2.3.2.13}
\end{equation}

Now consider now the the diagram in Fig.~6d. The two quark propagators
carrying the momentum $k'$ give 
\begin{eqnarray}
&&{\not\! p}_2{(1-\alpha_p+\alpha'_k){\not\! p}_1-
{\not\! p}_{\perp}+{\not\! k'}_{\perp}\over 
(1-\alpha_p+\alpha'_k)({m^2\over s}-\beta_p+\beta'_k)s-
(\vec{p}-\vec{k}')_{\perp}^2+i\epsilon}\nonumber\\
&\times&
\not\!e^A_{\perp} {(\alpha_p+\alpha'_k){\not\! p}_1+{\not\! p}_{\perp}+
{\not\! k'}_{\perp}\over 
(\alpha_p+\alpha'_k)(\beta_p+\beta'_k)s-(\vec{p}-\vec{k}')_{\perp}^2+i\epsilon}
{\not\! p}_2\nonumber\\
&\rightarrow& {\not\! p}_2
{(1-\alpha_p){\not\! p}_1-
{\not\! p}_{\perp}+{\not\! k'}_{\perp}\over 
(1-\alpha_p)\beta'_k s}e^A_{\perp}
{(\alpha_p+\alpha'_k){\not\! p}_1+{\not\! p}_{\perp}+
{\not\! k'}_{\perp}\over 
\alpha_p\beta'_k s}.
\label{2.2.11}
\end{eqnarray}
Since we cannot keep both large terms $(1-\alpha_p)p_1$ and $\alpha_p p_1$ 
in the numerators this expression is ${m^2\over \beta'_ks}$ times smaller 
than the contribution (\ref{2.2.7}) of the diagram in Fig.~6b  so it vanishes 
in the LLA.
 
The diagrams in Fig 6e,f  are calculated in the same way as the diagrams in 
Fig 6b,c. Similarly, the result may be obtained from Eq.~(\ref{2.2.5}) by the 
replacement
\begin{equation} 
\Gamma_{\bullet\ast}^{~~\sigma}(k,k')\rightarrow
-{(\vec{k}')_{\perp}^2\over \alpha_k s}p_1^{\sigma}.
\label{2.2.12}
\end{equation}
In conclusion, the diagram in Fig.~6g vanishes in the LLA for the same reasons
as the Fig.~6c diagram.

Thus, the contribution of the diagrams in Fig.~6a--6e can be represented by one
diagram shown in Fig.~6h: 
\begin{eqnarray}
&&\hspace{-0.5cm}W_{(a+...g)}^{(5)}~=~{sg^6\over\pi}{N_c(N_c^2-1)\over 4}
\left(\sum e_q^2\right)^2\label{2.2.13}\\
&&\times~
\int{d^2k_{\perp}\over
4\pi^2}  {d^2k'_{\perp}\over 4\pi^2}I^A(k_{\perp},r_{\perp}) 
{1\over \vec{k}_{\perp}^2(\vec{r}-\vec{k})_{\perp}^2}
{1\over (\vec{k}')_{\perp}^2(\vec{r}-\vec{k}')_{\perp}^2}
I^B(k'_{\perp},r_{\perp})\nonumber\\ 
&&\times~
\int {d\alpha_k d\beta'_k\over 4\pi^2}L^{\sigma}(k,k')
2\pi \delta(\alpha_k\beta'_k s+(\vec{k}-\vec{k}')_{\perp}^2)\theta(\alpha_k)
\Gamma_{\bullet\ast\sigma}(r-k,r-k'),
\nonumber
\end{eqnarray}
where 
\begin{eqnarray}
L^{\sigma}(k,k')&=&
{2\over s}\Gamma_{\bullet\ast}^{\sigma}(k,k')-2
{(\vec{k}_{\perp})^2\over \beta'_k s}p_2^{\sigma} 
-2{(\vec{k}'_{\perp})^2\over\alpha_k s}p_1^{\sigma} 
\nonumber\\
&=&
(k+k')_{\perp}^{\sigma}
-(\alpha_k+2{\vec{k}_{\perp}^2\over \beta'_k s})p_1^{\sigma}-
(\beta'_k+2{(\vec{k}')_{\perp}^2\over \alpha_k s})p_2^{\sigma} 
\label{2.2.14}
\end{eqnarray}
is the Lipatov effective vertex for the gluon emission 
shown in Fig.~6h by a shaded circle. 
Note that unlike the usual three-gluon vertex, the effective vertex is 
gauge-invariant,
\begin{equation}
(k-k')_{\sigma}L^{\sigma}(k,k')=0.
\label{2.2.15}
\end{equation}

We have demonstrated that if we take the diagram in Fig.~6a and attach 
the left end of the $k-k'$ gluon line in all possible ways, 
the left 
three-gluon 
vertex in Fig.~6a is replaced by the effective vertex (\ref{2.2.14}). 
Likewise, the
sum of all possible attachments of the right end of this $k-k'$ 
gluon line
converts the right three-gluon vertex 
$\Gamma_{\bullet\ast\sigma}(r-k,r-k')$ 
into the effective vertex $L_{\sigma}(r-k,r-k')$.
Hence the  sum of all the diagrams 
with 5-particle cut takes
the form (see Fig.~7)
\begin{figure}[htb]
\centerline{
\epsfysize=4cm
\epsffile{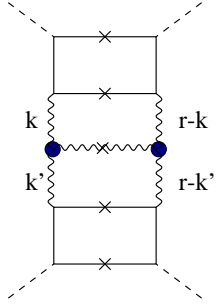}}
\caption{ Sum of the diagrams with gluon emission in LLA. Shaded
circle denotes the effective vertex.}\label{sum5}
\end{figure}
%
\begin{eqnarray}
W^{(5)}&=&\label{2.2.16}
{g^6\over\pi}{N_c(N_c^2-1)\over 4}\left(\sum e_q^2\right)^2
{s^2\over 2}\\
&\times&\int{d^2k_{\perp}\over 4\pi^2} 
{d^2k'_{\perp}\over 4\pi^2}{I^A(k_{\perp},r_{\perp})I^B(k'_{\perp},r_{\perp}) 
\over \vec{k}_{\perp}^2(\vec{r}-\vec{k})_{\perp}^2(\vec{k}')_{\perp}^2
(\vec{r}-\vec{k}')_{\perp}^2}
\nonumber\\ 
&\times&
\int {d\alpha_k d\beta'_k\over 4\pi^2}L^{\sigma}(k,k')
2\pi \delta(\alpha_k\beta'_k s+(\vec{k}-\vec{k}')_{\perp}^2)\theta(\alpha_k)
L_{\sigma}(r-k,r-k').
\nonumber
\end{eqnarray}
Since $\alpha_k\beta'_k s=-(\vec{k}-\vec{k}')_{\perp}^2$ 
due  to the
$\delta$-function, the product of two
Lipatov's vertices gives 
%
\begin{equation}
{1\over 2}L^{\sigma}(k,k')L_{\sigma}(r-k,r-k')=
-\vec{r}_{\perp}^2+
{\vec{k}_{\perp}^2(\vec{r}-\vec{k}')_{\perp}^2\over 
(\vec{k}-\vec{k}')_{\perp}^2}+
{(\vec{k}')_{\perp}^2(\vec{r}-\vec{k})_{\perp}^2\over 
(\vec{k}-\vec{k}')_{\perp}^2},
\label{2.2.17}
\end{equation}
which is proportional to the ``emission'' part of the BFKL kernel, see 
the Eq.~(\ref{2.2.19a}) below.
Now one can easily perform the remaining integrations over
$\alpha_k$ and $\beta'_k$ in the LLA  
\begin{equation}
s\!\int\! d\alpha_k d\beta'_k 
\delta(\alpha_k\beta'_k s+(\vec{k}-\vec{k}')_{\perp}^2)\theta(\alpha_k)=
\int^1_{m^2\over s} d\alpha_k{1\over \alpha_k}=
\ln{s\over m^2} ,
\label{2.2.18}
\end{equation}
and, therefore, the final result (for the diagrams with 5-particle cut) is
%
\begin{eqnarray}
W^{(5)}&=&{s\over\pi}g^4{N_c^2-1\over 4}{g^2\over 2\pi}N_c
\ln {s\over m^2}\label{2.2.19}\\
&\times&\int{d^2k\over 4\pi^2}
{d^2k'\over 4\pi^2}I^A(k_{\perp},r_{\perp}) 
{1\over \vec{k}_{\perp}^2 (\vec{r}-\vec{k})_{\perp}^2}
K_1(k_{\perp},k_{\perp}',r)
I^B(k'_{\perp},r_{\perp}) 
\nonumber
\end{eqnarray}
where
\begin{equation}
K_{(1)}(k_{\perp},k_{\perp}',r)=
-{\vec{r}_{\perp}^2\over (\vec{k}')_{\perp}^2(\vec{r}-\vec{k}'_{\perp})^2}
+{\vec{k}_{\perp}^2\over (\vec{k}')_{\perp}^2(\vec{k}-\vec{k}'_{\perp})^2}
+{(\vec{k}-\vec{r})_{\perp}^2\over
(\vec{k}'-\vec{r})_{\perp}^2(\vec{k}-\vec{k}'_{\perp})^2}
\label{2.2.19a}
\end{equation}
is the first part of the BFKL kernel coming from the diagrams with gluon
emission.

Apart from the diagrams with 5-particle cut shown in Fig.~6, there are
also diagrams with four-particle cut (``virtual corrections") 
of the type shown in Fig.~8. 
\begin{figure}[htb]
\centerline{
\epsfysize=7.7cm
\epsffile{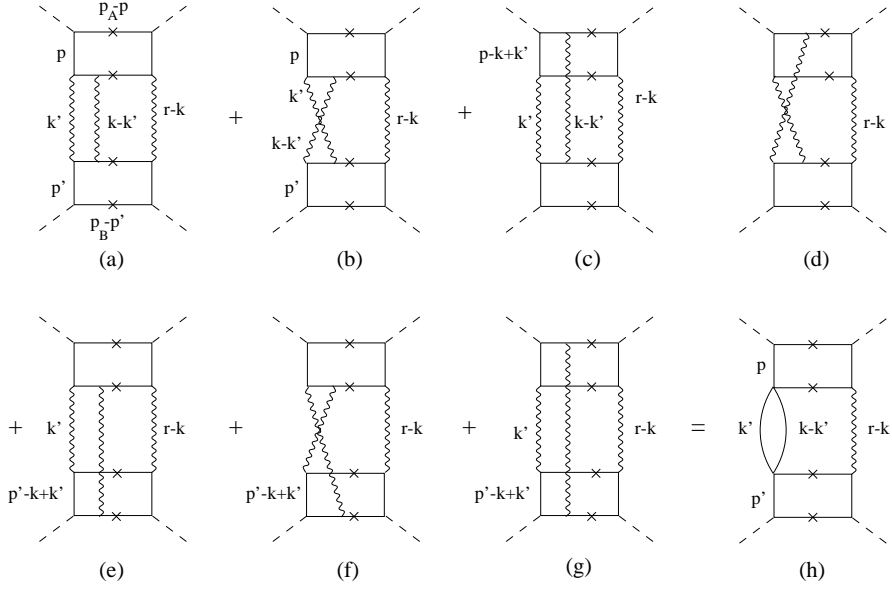}}
\caption{ Virtual corrections.}\label{vcorr}
\end{figure}
Let us consider the diagram shown in Fig.~8a. The integrals over 
$\alpha_k$ and $\beta_k$ are similar to the same integrals in the 
first-order diagram in Fig.~4  and therefore 
$\alpha_k\sim\beta_k\sim{m^2\over s}$. The logarithmic contribution 
comes from the region $1\sim\alpha_p\gg\alpha'_k\gg\alpha_k$. In this region
we can replace the quark propagator
with momentum $p-k'$ by the eikonal propagator (see Appendix 7.1),
\begin{equation}
\not\! p_2({\not\! p+\not\! k'})\not\! p_2\rightarrow 
\not\! p_2{1\over -\beta'_k+i\epsilon}.
\label{2.2.20}
\end{equation}
In addition, one can neglect $\beta'_k$ in comparison 
to $\beta'_p\sim 1$ in the lower block . The loop 
integral over $k'$ turns into
\begin{eqnarray}
&&\int{d\alpha'_k d\beta'_k\over 4\pi^2} 
{d^2k'\over 4\pi^2}\left[t^a{\not\!p_2\over -\beta'_k+i\epsilon}t^b\right] 
{1\over \alpha'_k\beta'_k s-\vec{k}_{\perp}^2}
{1\over \alpha'_k(\beta'_k-\beta_k)-
(\vec{k}-\vec{k}')_{\perp}^2}\nonumber\\
&\times&
\left[t^a{\not\!p_1(\beta'_p\not\!p_2+\not\!p'_{\perp}+
\not\!k'_{\perp})
\not\!p_1\over \alpha'_k\beta'_p s-
(\vec{p}'+\vec{k}')^2_{\perp}}t^b\right].
\label{2.2.21}
\end{eqnarray}
The integral over $\beta'_k$ is determined by the residue at 
$\beta'_k=0$ 
so we obtain
\begin{eqnarray}
&&\int_{m^2\over s}^1{d\alpha'_k \over 2\pi\alpha'_k} 
\int{d^2k'_{\perp}\over 4\pi^2}\Big[t^a\not\!p_2t^b\Big]
{1\over \vec{k}_{\perp}^2}
{1\over (\vec{k}-\vec{k}')_{\perp}^2}
\Big[t^a{\not\!p_1}t^b\Big]\nonumber\\
&=&\Big[t^a\not\!p_2t^b\Big]\Big[t^a{\not\!p_1}t^b\Big]\times
{g^2\over 4\pi^2}\ln {s\over m^2}\int{d^2k'_{\perp}\over 4\pi^2}
{1\over \vec{k}_{\perp}^2}
{1\over (\vec{k}-\vec{k}')_{\perp}^2}.
\label{2.2.22}
\end{eqnarray}
Let us add now the contribution of the diagram in Fig.~8b.  Like
the Fig.~8a case, we get the loop integral over $k'$ in the form
\begin{eqnarray}
&&\int{d\alpha'_k d\beta'_k\over 4\pi^2} 
{d^2k'\over 4\pi^2}\left[t^a{\not\!p_2\over -\beta'_k+
i\epsilon}t^b
\right] 
{1\over \alpha'_k\beta'_k s-\vec{k}_{\perp}^2}
{1\over \alpha'_k(\beta'_k-\beta_k)-
(\vec{k}-\vec{k}')_{\perp}^2}\nonumber\\
&\times&
\left[t^a{\not\!p_1(\beta'_p\not\!p_2+\not\!p'_{\perp}+
\not\!k'_{\perp})
\not\!p_1\over \alpha'_k\beta'_p s- 
(\vec{p}'+\vec{k}')^2_{\perp}}t^b\right]
\nonumber\\
&=&
\int_{m^2\over s}^1{d\alpha'_k \over 2\pi\alpha'_k} 
\int{d^2k'_{\perp}\over 4\pi^2}\Big[t^a\not\!p_2t^b\Big]
{1\over \vec{k}_{\perp}^2}
{1\over (\vec{k}-\vec{k}')_{\perp}^2}
\Big[t^a{\not\!p_1}t^b\Big]\nonumber\\
&=&\Big[t^a\not\!p_2t^b\Big]\Big[t^a{\not\!p_1}t^b\Big]\times
{g^2\over 4\pi^2}\ln {s\over m^2}\int{d^2k'_{\perp}\over 4\pi^2}
{1\over (\vec{k}')_{\perp}^2(\vec{k}-\vec{k}')_{\perp}^2}.
\label{2.2.23}
\end{eqnarray}
The diagrams shown in Fig.~8c--g do not give the logarithmic contribution for
the same reason as the diagram in Fig.~6d. 

We see that the sum of diagrams in Fig.~8a--g  reduces to the first-order
diagram in Fig.~\ref{fig2}a with the left gluon propagator 
${1\over-\vec{k}_{\perp}^2}$
 replaced by the factor
\begin{equation}
{1\over -\vec{k}_{\perp}^2}\rightarrow {g^2\over 4\pi}N_c\ln {s\over
m^2}\int{d^2k'_{\perp}\over 4\pi^2} {1\over
(\vec{k}')_{\perp}^2(\vec{k}-\vec{k}')_{\perp}^2}
\label{2.2.24}
\end{equation}
shown schematically in Fig.~8h. We get
\begin{eqnarray}
W_{(a+...g)}^{(4)}&=&-{s\over \pi}g^4{N_c^2-1\over 4}\left(\sum
e_q^2\right)^2 {g^2\over 4\pi}N_c\ln {s\over m^2}
\label{2.2.25}\\
&\times&
\int {d^2k_{\perp}\over 4\pi^2}
{I^A(k_{\perp},r_{\perp})I_B(k_{\perp},r_{\perp})\over {\vec
k}_{\perp}^2(\vec{r}-\vec{k})_{\perp}^2} \nonumber 
\left\{\int{d^2k'_{\perp}\over 4\pi^2}
{\vec{k}^2\over
(\vec{k}')_{\perp}^2(\vec{k}-\vec{k}')_{\perp}^2}\right\}.
\nonumber 
\end{eqnarray}
The diagrams with the gluon loop to the right of the cut lead to similar
replacement of the right gluon propagator 
${1\over -(\vec{k}-\vec{r})_{\perp}^2}$
by 
\begin{equation}
{1\over -(\vec{k}-\vec{r})_{\perp}^2}\rightarrow 
{g^2N_c\over 4\pi}\ln {s\over m^2}
\int{d^2k'_{\perp}\over 4\pi^2} {1\over
(\vec{k}'_{\perp}-\vec{r}_{\perp})^2(\vec{k}-\vec{k}')_{\perp}^2}.
\label{2.2.26}
\end{equation}
Thus we obtain the result
%
\begin{eqnarray}
W^{(4)}&=&-{s\over \pi}g^4{N_c^2-1\over 4}\left(\sum e_q^2\right)^2
{g^2\over 4\pi}N_c\ln {s\over m^2}\nonumber\\
&\times&\int {d^2k_{\perp}\over 4\pi^2} 
{I^A(k_{\perp},r_{\perp}) I_B(k_{\perp},r_{\perp})
\over {\vec k}_{\perp}^2(\vec{r}-\vec{k})_{\perp}^2}\nonumber\\
&\times&\int{d^2k'_{\perp}\over 4\pi^2}
\left\{{\vec{k}^2\over
(\vec{k}')_{\perp}^2(\vec{k}-\vec{k}')_{\perp}^2}
+
{(\vec{k}-\vec{r})^2\over
(\vec{k}'-\vec{r})_{\perp}^2(\vec{k}-\vec{k}')_{\perp}^2}\right\}
\label{2.2.27}
\end{eqnarray}
for the contribution of the diagrams with 4-particle cut. 

Adding the sum of the diagrams with real gluon emission  $W^5$ 
we obtain the
final result for the $\gamma^\ast\gamma^\ast$ scattering amplitude 
in the first
order in LLA. It can be represented in the form
\begin{eqnarray}
W^1&=&{s\over \pi}g^4{N_c^2-1\over 4}\left(\sum e_q^2\right)^2
{g^2\over 2\pi}N_c\ln {s\over m^2}\int{d^2k\over 4\pi^2}{d^2k'\over 4\pi^2}
\label{2.2.28}\\
&\times& 
I^A(k_{\perp},r_{\perp}) 
{1\over \vec{k}_{\perp}^2 (\vec{r}-\vec{k})_{\perp}^2}
K(k_{\perp},k_{\perp}',r)
I^B(k'_{\perp},r_{\perp}),
\nonumber
\end{eqnarray}
where
\begin{eqnarray}
&&K(k_{\perp},k_{\perp}',r)=K_{(1)}(k_{\perp},k_{\perp}',r)-
{1\over 2}\delta^{(2)}(k-k')\label{2.2.29}\\
&&\times\left\{\int{d^2k"_{\perp}\over 4\pi^2}
{\vec{k}^2\over
(\vec{k}")_{\perp}^2(\vec{k}-\vec{k}")_{\perp}^2}
+\int{d^2k"_{\perp}\over 4\pi^2}
{(\vec{k}-\vec{r})^2\over
(\vec{k}"-\vec{r})_{\perp}^2(\vec{k}-\vec{k}")_{\perp}^2}\right\}
\nonumber
\end{eqnarray}
is the BFKL kernel.\cite{bfkl} The explicit form of $K$ is
\begin{eqnarray}
&&\hspace{-8mm}K(k_{\perp},k_{\perp}',r)~=~
-{\vec{r}_{\perp}^2\over \vec{k'}_{\perp}^2(\vec{r}-\vec{k}')_{\perp}^2}+
{\vec{k}_{\perp}^2\over \vec{k'}_{\perp}^2
(\vec{k}-\vec{k}')_{\perp}^2}+
{(\vec{r}-\vec{k})_{\perp}^2\over (\vec{r}-\vec{k}')_{\perp}^2
(\vec{k}-\vec{k}')_{\perp}^2}\nonumber\\
&&\hspace{-8mm}-{1\over 2}\delta^{(2)}(k-k')\int{d^2k"_{\perp}\over 4\pi^2}
\left\{{\vec{k}^2\over
(\vec{k}")_{\perp}^2(\vec{k}-\vec{k}")_{\perp}^2}
+
{(\vec{k}-\vec{r})^2\over
(\vec{k}"-\vec{r})_{\perp}^2(\vec{k}-\vec{k}")_{\perp}^2}\right\} .
\label{2.2.30}
\end{eqnarray}
Note that both $W^{(5)}$ and $W^{(4)}$ are IR divergent but their sum $W^1$
given by Eq.~(\ref{2.2.28}) is IR finite. This is the usual Bloch-Nordsieck 
cancellation between th emission of real gluon in diagrams in Fig.~6 and
virtual gluon in Fig.~8.

\subsection{Bare pomeron in the LLA} 

The $\gamma^\ast\gamma^\ast$ amplitude in the first two orders in perturbation
theory may be represented in the operator form as
\begin{equation}
W^{(0+1)}=s{\cal C}\int{d^2k\over 4\pi^2}
I^A(k_{\perp},r_{\perp}){1\over \vec{k}_{\perp}^2
(\vec{r}-\vec{k})_{\perp}^2} \Big(1+ 
{g^2\over 8\pi^3}N_c\ln {s\over m^2} \hat{K}\Big)
I^B(k_{\perp},r_{\perp}),
\label{2.3.1}
\end{equation}
where ${\cal C}\equiv\alpha_s(N_c^2-1)(\sum e_q^2)^2$ and the operator 
$\hat{K}_r$ is defined by its kernel $K(k,k',r)$,
\begin{equation}
(\hat{K}_rf)(\vec{k}_{\perp})=\int{d^2k'\over 4\pi^2}
K(k_{\perp},k_{\perp}',r)f(\vec{k}'_{\perp}).
\label{2.3.2}
\end{equation}
We can demonstrate (and we will do this using the evolution equations for
the Wilson-line operators) that in the next orders in LLA the operator $K$ 
exponentiates:
\begin{equation}
W^{\rm LLA}=s{\cal C}\int{d^2k\over 4\pi^2}
I^A(k_{\perp},r_{\perp}){1\over \vec{k}_{\perp}^2 (\vec{r}-\vec{k})_{\perp}^2} 
\left({s\over m^2}\right)^{{g^2N_c\over 8\pi^3}\hat{K}_r}
I^B(k_{\perp},r_{\perp}) .
\label{2.3.3}
\end{equation}
It is convenient to represent the amplitude as an integral over the complex
momenta:
\begin{eqnarray}
W(s,t)&=&{s\over 2\pi i}\int_{\delta -i\infty}^{\delta +i\infty} 
d\omega \Big({s\over m^2}\Big)^{\omega}W(\omega,t),
\label{2.3.4}\\
W^{\rm LLA}(\omega,t)&=&{\cal C}\int{d^2k\over
4\pi^2} {I^A(k_{\perp},r_{\perp})\over \vec{k}_{\perp}^2
(\vec{r}-\vec{k})_{\perp}^2} 
{1\over \omega-{g^2\over 8\pi^3}N_c\hat{K}_r}
I^B(k_{\perp},r_{\perp}),
\nonumber
\end{eqnarray}
where $\omega=j-1$. The relation between the LLA and the 
power series for
$W(\omega,t)$ is
\begin{eqnarray}
W^{\rm LLA}(s,t)&=&s{\cal C}\sum_{n=1}^{\infty}{1\over n!}
\left(g^2\ln{s\over m^2}\right)^n f_n(t)~\Rightarrow\nonumber\\
W^{\rm LLA}(\omega,t)&=&{\cal C}\sum_{n=1}^{\infty}
{g^{2n}\over\omega^{n+1}} f_n(t)
\label{2.3.5}
\end{eqnarray}
where 
\begin{equation}
f_n(t)=\int{d^2k\over 4\pi^2}
I^A(k_{\perp},r_{\perp}){1\over \vec{k}_{\perp}^2 (\vec{r}-\vec{k})_{\perp}^2} 
\left({N_c\over 8\pi^3}\hat{K}_r\right)^n
I^B(k_{\perp},r_{\perp})
\label{2.3.6}
\end{equation}
are the coefficients of the LLA expansion.

 The asymptotics of the amplitude at $s\rightarrow
\infty$ is given by the rightmost singularity of the integrand in 
the right-hand side of Eq.~(\ref{2.3.4}) in the
$\omega$ plane.  The position of this singularity is given by
the maximal eigenvalue of the operator $\hat{K}_r$ determined by the
eigenfunction equation  
\begin{equation}
{\alpha_sN_c\over 2\pi^2}(\hat{K}_rf)(\vec{k}_{\perp})=
\omega f(\vec{k}_{\perp}).
\label{2.3.7}
\end{equation}
This equation is solved at arbitrary momentum transfer $r$~\cite{lip86}
yet it turns out that the maximal eigenvalue of Eq.~(\ref{2.3.3}) does not 
actually depend on $r$. For
simplicity, let us consider the case $r=0$ corresponding to total cross section
of $\gamma^\ast\gamma^\ast$ scattering. (In the next section we prove that
the position of singularity does not depend on $t=-\vec{r}_{\perp}^2$).

At $r=0$, the full and orthogonal set of eigenfunctions of the BFKL operator 
are simple powers  
\begin{equation}
f(\vec{k})=\big({\vec k}^2\big)^{-{1\over 2}+i\nu}e^{in\phi},
\label{2.3.8}
\end{equation}
with the eigenvalues 
%
\begin{equation}
\omega = 2N_c{\alpha_s\over\pi}\chi(\nu,n),~~~\chi(\nu,n)=
-{\rm Re} \Psi({|n|+1\over 2}+i\nu)-C .
\label{2.3.9}
\end{equation}
The maximal eigenvalue is 
$2N_c{\alpha_s\over\pi}\chi(0,0)=4{\alpha_s\over\pi}N_c\ln 2$, 
so the rightmost singularity (intercept of the ``hard pomeron") 
is located at
\begin{equation}
j=1+\omega_0,~~~~\omega_0=4{\alpha_s\over\pi}N_c\ln 2 ,
\label{2.3.10}
\end{equation}
so the
asymptotics at high energies in the LLA is  
\begin{equation}
\sigma\simeq \left({s\over m^2}\right)^{4{\alpha_s\over\pi}N_c\ln2} .
 \label{2.3.11}
\end{equation}
It is easy to see that the singularity at $\omega=\omega_0$ is the 
branch point ${1\over \sqrt{\omega-\omega_0}}$. 

As we mentioned in the introduction, the singularity at $j>1$ violates the
Froissart bound $\sigma \leq \ln^2s$. Recently, the next-to-leading correction
($\sim\alpha_s$) to the BFKL  kernel was found,\cite{nlobfkl} but the result
still violates the Froissart  bound, so the unitarization of the BFKL
pomeron is required. (Consequently, the BFKL pomeron (\ref{2.3.10}) is sometimes
called ``the bare pomeron in pQCD").

In the case of $\gamma^\ast\gamma^\ast$ scattering, it is possible to find the
explicit form of the cross section in the LLA. Expanding impact factors
$I(k,0)\equiv I(k)$ in a set of eigenfunctions (\ref{2.3.8}), we obtain 
%
\begin{eqnarray}
&&\hspace{-3mm}\sigma_{\rm tot}(p_{A},p_{B})~=~
g^{4}{1\over 2}(N_{c}^{2}-1)(\sum e_{i}^{2})^{2}\label{2.3.12}\\
&&\hspace{-3mm}\times~\int \! d\nu
({s\over
m^{2}})^{{2\alpha_{s}\over\pi }N_{c}\chi (\nu )}
\int \! \frac {dp_{\perp}}{4\pi ^{2}} I^{A}(p_{\perp})
(\vec{p}_{\perp}^2)^{-{3\over 2}+i\nu }
\int \!\frac {dp'_{\perp}}{4\pi ^{2}} I^{B}(p'_{\perp})
(p_{\perp}^{,2})^{-{3\over 2}-i\nu }.
\nonumber
\end{eqnarray}
Here we neglected the angle-dependent contributions coming from $n\not\!=0$
since they decrease with energy. At $s\rightarrow \infty$ the cross section 
(\ref{2.3.12}) is determined by the rightmost singularity in the $\nu $ plane
located at $\nu =0$ (in terms of $j$-plane it corresponds to 
Eq.~(\ref{2.3.10})) and the result is 
\begin{eqnarray}
\sigma_{\rm tot}(p_{A},p_{B})&=&
{1\over 2}g^4
{(N_c^2-1)\pi\over \sqrt{14\zeta
(3)N_c{\alpha_s\over\pi}\ln{s\over m^2}}}
(\sum
e_{i}^{2})^{2}\label{2.3.13}\\
&\times&({s\over m^{2}})^{{4\alpha_{s}\over\pi }N_{c}\ln2}
\int \! \frac {dp_{\perp}}{4\pi ^{2}} I^{A}(p_{\perp})
(\vec{p}_{\perp}^2)^{-{3\over 2}}
\int \!\frac {dp'_{\perp}}{4\pi ^{2}} I^{B}(p'_{\perp})
(\vec{p}_{\perp}^2)^{-{3\over 2}}
\nonumber
\end{eqnarray}
where $\zeta (3)\simeq 1.202$. 

\subsection{Diffusion in the transverse momentum and the BFKL equation 
with running coupling constant} 

\begin{figure}[htb]
\centerline{
\epsfysize=5.5cm
\epsffile{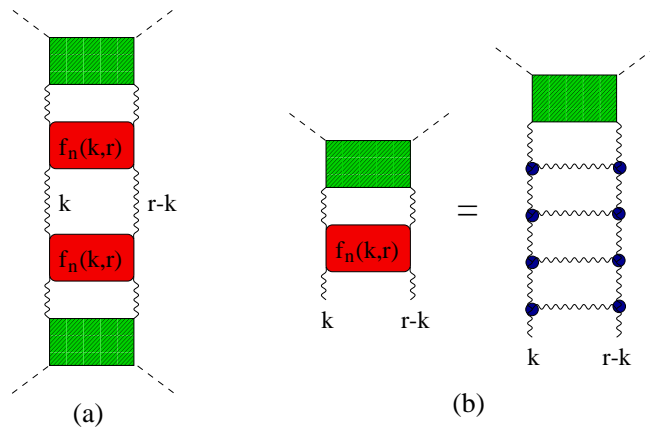}}
\caption{ Diffusion in $k_{\perp}$.} \label{fig8a}
\end{figure}

At first, let us demonstrate that the rightmost singularity of the  
BFKL equation is located at $\omega=\omega_0$ at $t\not\!=0$ as well 
(although its character changes from ${1\over\sqrt{\omega-\omega_0}}$ 
to $\sqrt{\omega-\omega_0}$). We shall see that
in higher orders in perturbation theory there is a ``diffusion''
in $k_{\perp}$ such that $\ln{\vec{k}_{\perp}^2\over m^2}\sim \sqrt{n}$ 
(where $n$ is the order of perturbation theory). 
To illustrate the diffusion, consider a rung of the BFKL ladder located
in the middle of the rapidity region (see Fig.~\ref{fig8a}).
Each of the upper or lower blocks in this diagram are  
``non-integrated gluon distribution''.
The $s\rightarrow\infty$ asymptotics is 
governed by the rightmost singularity of the function 
$W(\omega,t)$ (see Eq.~(\ref{2.3.4}))
which is determined by the asymptotics 
of the coefficients $f_n$
at $n\rightarrow\infty$.
For even $n$, these coefficients can be represented as
\begin{eqnarray}
&&f_{2n}(t)=\int{d^2k\over
4\pi^2} {1\over \vec{k}_{\perp}^2 (\vec{r}-\vec{k})_{\perp}^2} 
f_n^A(k_{\perp},r_{\perp})f_n^B(k_{\perp},r_{\perp}),
\label{2.4.1}
\end{eqnarray}
where 
\begin{equation}
f^A_{n}(k_{\perp},r_{\perp})=
\left({N_c\over 8\pi^3}\hat{K}_r\right)^n\! 
I^A(k_{\perp},r_{\perp}),~~~
f^B_{n}(k_{\perp},r_{\perp})=
\left({N_c\over8\pi^3}\hat{K}_r\right)^n\!  
I^B(k_{\perp},r_{\perp}).
\label{2.4.2}
\end{equation}
Let us demonstrate that the characteristic momenta $\vec{k}_{\perp}^2$ in 
the integral in Eq.~(\ref{2.4.1}) are $\sim m^2e^{\sqrt{n}}$.
At large transverse momenta $k_{\perp}$ the recursion formula 
$f^A_{n+1}(k_{\perp},r_{\perp})=
{N_c\over 8\pi^3}\hat{K}_rf^A_{n}(k_{\perp},r_{\perp})$ can be 
reduced to
\begin{eqnarray}
&&\hspace{-1cm}\omega \phi_{n+1}(\xi)=\label{2.4.3}\\
&&\hspace{-1cm}{g^2N_c\over 4\pi^2}\int d\xi'
\Bigg[{e^{(\xi-\xi')/2}\over 1-e^{\xi-\xi'}}\phi_n(\xi')-
\left({1\over 1-e^{\xi-\xi'}}-
{1\over \sqrt{1+4e^{2(\xi-\xi')}}}\right)\phi_n(\xi)\Bigg]
\nonumber
\end{eqnarray}
where $\xi=\ln{\vec{k}_{\perp}^2\over m^2}$ and 
$\phi_n(\xi)=\left({g^2\over \omega}\right)^n
{1\over |k_{\perp}|}f_n(\vec{k}_{\perp}^2)$. 
Next, we expand the function $\phi_n(\xi')$ 
in the integrand in Eq.~(\ref{2.4.3}) in Taylor series 
$\phi_n(\xi')=\phi_n(\xi)+(\xi'-\xi)\phi'_n(\xi)+
{1\over 2}(\xi'-\xi)^2\phi"_n(\xi)+...$. 
As we shall see below, at large $n$ and $k_\perp$ one can neglect higher terms 
in Taylor expansion, and then the recursion integral equation 
(\ref{2.4.3}) can
be approximated by the differential equation 
\begin{equation}
\omega {\partial\over\partial n}\phi(n,\xi)=
(\omega_0-\omega)\phi(n,\xi)+c{\partial^2\partial\xi^2}\phi(n,\xi) ,
\label{2.4.4}
\end{equation}
where $c={7\over\pi^2}g^2\zeta(3)$, $\zeta(3)\simeq$ 1.202. This equation 
describes the diffusion of the ``particle'' where $n$ serves
as a time and $\xi$ as a coordinate. It is well known that at large time $n$
the mean position $\xi$ of the ``particle'' is proportional to $\sqrt{n}$,
and therefore
our approximation of Eq.~(\ref{2.4.3}) by the diffusion equation (\ref{2.4.4})
is justified. 
 
Thus, we must find the solution of the diffusion equation (\ref{2.4.4})
with the ``wall-type'' boundary condition 
\begin{equation}
\left.\phi(n,\xi)\right|_{\xi=\xi_t}=0,\qquad
\xi_t\equiv \ln{\vec{r}_{\perp}^2\over m^2}
\label{2.4.5}
\end{equation}
which reflects the fact that our approximation is not valid at 
$\vec{k}_{\perp}^2<\vec{r}_{\perp}^2$. It is easy to check that the solution
of the Eq.~(\ref{2.4.4}) with the boundary condition (\ref{2.4.5}) 
behaves at large $\xi\sim\sqrt{n}$ as
\begin{equation}
\phi(n,\xi)\sim{(\xi-\xi_t)\over n^{3/2}}
e^{\left({\omega_0\over\omega}-1\right)n}
e^{-{\omega\over 4nc}(\xi-\xi_t)^2}
\label{2.4.6}
\end{equation}
where the coefficient of the proportionality may be determined 
by a more accurate analysis of the transition from the integral equation
(\ref{2.4.3}) to the diffusion equation (\ref{2.4.4}).

Substituting the estimate (\ref{2.4.6}) in the integral (\ref{2.4.1}), we
obtain
\begin{equation}
\left.\left({g^2\over \omega}\right)^nf_n\right|_{n\rightarrow\infty}\sim
{1\over n^{3/2}} e^{\left({\omega_0\over\omega}-1\right)n},
\label{2.4.7}
\end{equation}
which gives
\begin{equation}
W(\omega,t)\sim\sum \left({g^2\over \omega}\right)^n f_n=
\int_1^{\infty}dn {1\over n^{3/2}}e^{\left({\omega_0\over\omega}-1\right)n}=
\sqrt{\omega_0-\omega} .
\label{2.4.8}
\end{equation}
We see that the singularity is located at the same point $\omega=\omega_0$
as in the case of forward scattering, although its character is slightly
different: $\sqrt{\omega_0-\omega}$ instead of 
${1\over\sqrt{\omega_0-\omega}}$.\cite{bfkl}  

At $t=0$ there is no ``wall'' boundary condition (\ref{2.4.5}) which 
shows that the diffusion equation (\ref{2.4.4}) leads to 
$|\xi|\sim \sqrt{n}$. This means that the characteristic momenta $k_{\perp}$ 
are either very large, $\vec{k}_{\perp}^2\sim m^2e^{\sqrt{n}}$, or very 
small, $\vec{k}_{\perp}^2\sim m^2e^{-\sqrt{n}}$. The large contribution from 
the region of small $k_{\perp}$ region indicates the possibility of the
breakdown of perturbative QCD for high-energy scattering. 

We can safely apply pQCD to high-energy scattering if 
the characteristic transverse momenta of the gluons $k_{\perp}$
in the ladder are large. For the $\gamma^\ast\gamma^\ast$ with 
$p_A^2\sim p_A^2\sim m^2\gg \Lambda_{\rm QCD}^2$
one can check by explicit calculation that the characteristic
$k_{\perp}$ for the first few diagrams are $\sim m$. However, due to the
diffusion in $k_{\perp}$ , the leading contribution to the loop integrals
comes from the gluon momenta which are either very large, 
$\vec{k}_{\perp}^2\sim
m^2e^{\sqrt{n}}$, or very  small, 
$\vec{k}_{\perp}^2\sim m^2e^{-\sqrt{n}}$. Due to the
asymptotic freedom,  the fact that the $k_{\perp}$ may be very large at
$n\rightarrow\infty$ only strengthens the applicability of pQCD. On the
contrary,  the fact that $k_{\perp}$ may be small questions the applicability 
of pQCD to the high-energy $\gamma^\ast\gamma^\ast$ scattering. 

To take into account the asymptotic freedom, one may consider the BFKL
equation with the running coupling constant.
Each of the upper or lower blocks in the diagram in Fig.~\ref{fig8a} is a 
``non-integrated gluon distribution''
\begin{equation}
F^{A(B)}(k_{\perp},r_{\perp};s)=\sum {1\over n!}
\left(g^2\ln{s\over m^2}\right)^n 
f^{A(B)}_n(k_{\perp},r_{\perp})
\label{2.4.9}
\end{equation}
which satisfies the BFKL equation
\begin{eqnarray}
&&\omega F^{A(B)}(k_{\perp},r_{\perp};\omega)=
\label{2.4.10}\\
&&
I^{A(B)}(k_{\perp},r_{\perp})+ {g^2\over 8\pi^3}N_c\int
d^2k'_{\perp}K(k_{\perp},k'_{\perp},r_{\perp})
F^{A(B)}(k'_{\perp},r_{\perp};\omega) 
\nonumber
\end{eqnarray}
where $F(k_{\perp},r_{\perp};\omega)$ is a Mellin transform of 
Eq.~(\ref{2.4.9}):
$$
F(k_{\perp},r_{\perp};s)=
{1\over 2\pi i}\int d\omega\left({s\over m^2}\right)^{\omega}
F(k_{\perp},r_{\perp};\omega).
$$ 
In order to account for the 
asymptotic freedom, we can replace $g^2$ in the right-hand side of the 
Eq.~(\ref{2.4.10}) by $g^2(\vec{k}_{\perp}^2)$:\footnote{We 
have seen from the diffusion
equation that $(k'_{\perp})^2\sim \vec{k}_{\perp}^2$ in the 
adjacent rungs of the ladder so 
$g^2(\vec{k}_{\perp}^2)\equiv g^2((\vec{k}')_{\perp}^2)$.}
\begin{equation}
\omega F(k_{\perp},r_{\perp};\omega)= I(k_{\perp},r_{\perp})+
{g^2(k_{\perp})\over 8\pi^3}N_c\int d^2k'_{\perp}
K(k_{\perp},k'_{\perp},r_{\perp})
F(k'_{\perp},r_{\perp};\omega).
\label{2.4.11}
\end{equation}
This equation exceeds the LLA accuracy but it it can be demonstrated 
that in the case of large (or small) $\vec{k}_{\perp}^2$ the replacement
$g^2\rightarrow g^2(\vec{k}_{\perp}^2)$ agrees with the renormalization group
analysis~\cite{lip86}). Another arguments in favor of taking into account
these particular sub-leading logs follows from the analysis of the 
renormalon contributions.\cite{smallxren}

At large $k_{\perp}$ one can
replace the equation (\ref{2.4.11}) by the corresponding diffusion equation. 
It turns out that at large momentum transfer $|t|=\vec{r}_{\perp}^2$ the rightmost
singularity  of $F(k_{\perp},r_{\perp};\omega)$ is located simply at 
$t=12{\alpha_s(|t|)\over \pi}N_c\ln 2$. At $t=0$ the diffusion goes in both 
directions leading to the contributions coming from $k_{\perp}\sim
\Lambda_{\rm QCD}$. If one removes these contributions ``by hand'' 
(imposing the ``wall'' condition at $\vec{k}_{\perp}^2=\Lambda_{\rm QCD}$),
one
obtains a discrete set of Regge poles which condense from the  right to the
point $\omega=0$.\cite{lip86} A more satisfactory solution of the problem
of the diffusion to small $k_{\perp}$ would be to match the hard pomeron 
with the soft Landshoff-Donnachie pomeron (responsible for the
high-energy hadron-hadron scattering) which presumably comes from the
high-energy  exchanges by soft gluons 
(see, however, Ref.~15 
for an
alternative ``hard'' soft pomeron). Another possibility is that the 
diffusion to small $k_{\perp}$ disappears if one takes into account the
unitarization effects.\cite{mbraun1} 

The proper way to address the problem of running coupling constant in 
the BFKL equation is to use the NLO BFKL kernel in 
the renormalization-group analysis.\cite{kovmuller} 
The NLO correction to the anomalous 
dimension of the corresponding leading-twist gluon operator consists
of two parts: the conformal part and the running coupling part. 
The conformal part (see also Ref.~18) 
corrects the intercept
of the BFKL pomeron (\ref{2.3.10}), while the running coupling part, 
besides replacing $12{\alpha_s\over\pi}N_c\ln2$ by 
$12{\alpha_s(q^2)\over\pi}N_c\ln2$  in the leading order, 
leads to the non-Regge terms in the energy dependence of the cross section.
The numerical value of the correction to the
hard pomeron's intercept introduced by the conformal part of the NLO BFKL
kernel is large and negative. Its exact contribution is somewhat difficult
to estimate.\cite{barmbra,dross} There are hopes, however, that
collinear singularities causing this large NLO correction cancel each
other at higher orders in $\alpha_s$.\cite{cico}

\subsection{Reggeized gluons and unitarization of the pomeron}

As I mentioned above, the bare pomeron violates the Froissart bound so
we need to unitarize the BFKL pomeron. There are several approaches to 
the unitarization: effective reggeon field theory,\cite{many} the generalized
LLA~\cite{glla} equivalent to the quantum mechanics of reggeized 
gluons,\footnote{In the reggeon  
quantum mechanics, the unitarity is preserved only in the 
direct s-channel, 
while in a reggeon field theory the unitarity holds true in all the 
sub-channels
corresponding to different groups of particles in 
the final state.} and the 
dipole model.\cite{mu94,nnn}
We postpone the discussion of the dipole model until the next section and
turn the attention to reggeon-based schemes of the unitarization. 

The reggeized gluon can be defined as a ``hard pomeron'' for the 
quark-quark scattering. We have seen that the gluon propagator
 ${1\over \vec{k}_{\perp}^2}$ describing the exchange between two 
quarks to the left of the cut in Fig.~4 is replaced in the next
order by the factor (\ref{2.2.24}) coming from two diagrams in Fig.~8a,b. 
Thus, in the first two orders in perturbation theory the propagator 
describing the 
exchange between two quarks with gluon (color octet) quantum numbers 
in the $t$ channel has the form 
\begin{equation}
{1\over \vec{k}_{\perp}^2}\left(1-\alpha_s N_c\ln {s\over
m^2}\int{d^2k'_{\perp}\over 4\pi^2} {\vec{k}_{\perp}^2\over
(\vec{k}')_{\perp}^2(\vec{k}-\vec{k}')_{\perp}^2}\right).
\label{2.5.1}
\end{equation}
It can be demonstrated (either by direct summation of the Feynman 
diagrams~\cite{bfkl} or by evolution of the Wilson-line operators, 
see Sec.~3 below), that in the LLA the logarithmic factor in parenthesis  
exponentiates, therefore the exchange between two quarks is
described by the ``reggeized" gluon propagator
\begin{equation}
{1\over \vec{k}_{\perp}^2}
\left({s\over m^2}\right)^{\alpha_{\rm reg}(\vec{k}_{\perp}^2)},
\label{2.5.2}
\end{equation}
where
\begin{equation}
\alpha_{\rm reg}(t=-\vec{k}_{\perp}^2)=-\alpha_sN_c
\int{d^2k'_{\perp}\over 4\pi^2} {\vec{k}_{\perp}^2\over
(\vec{k}')_{\perp}^2(\vec{k}-\vec{k}')_{\perp}^2}
\label{2.5.3}
\end{equation}
is the trajectory of the reggeized gluon in the plane of complex momenta in
the leading order in $\alpha_s$.\footnote{This trajectory 
is IR divergent as it should be for the amplitude of the
scattering of the colored objects. For the scattering of white objects 
(like virtual photons discussed in the previous section) this divergence 
will cancel with the IR divergence for real gluon emissions. To avoid the 
infinities in the intermediate results, one can use the dimensional 
regularization (with $d=2+\epsilon$ transverse dimensions) or assume a small
gluon mass $\mu$.} Recently, this trajectory was computed
in the next-to-leading order in $\alpha_s$ by direct summation of Feynman 
diagrams~\cite{regg2} and by calculation of the two-loop anomalous
dimensions of the relevant Wilson-line operators.\cite{grishi} 

In terms of the reggeized gluons the BFKL ladder can be resummed as shown
in Fig.~9 where the dash-dotted line denotes reggeized gluon (\ref{2.5.2}) 
and the reggeon-reggeon-particle 
interaction is described by  Lipatov's vertex (\ref{2.2.14}). 
\begin{figure}[htb]
\centerline{
\epsfysize=6cm
\epsffile{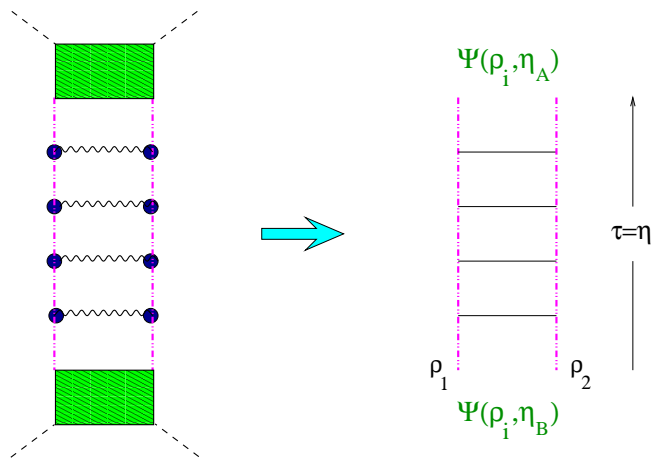}}
\caption{ BFKL ladder as a propagator of the 
two-reggeon state. Reggeized gluons are represented by dash-dot-dot lines.}
\label{2reg}
\end{figure}
(The expansion
of the reggeon trajectory in powers of $g^2$ reproduces the BFKL result
(\ref{2.3.3}) after combining the terms with like powers of $g^2$). 
This diagram can be interpreted as an evolution with respect to 
``time'' $\equiv$ rapidity of the two-particle 
state described by the wave function
$\Psi(\rho_1, \rho_2)$ 
in quantum mechanics with the Hamiltonian~\cite{lip86} 
\begin{eqnarray}
\hat{H}_{12}&=&\label{2.5.4}
{g^2N_c\over 16\pi^2}\Big\{\ln |\hat{p}_1|^2 + \ln |\hat{p}_2|^2 \\
&+& 
{1\over\hat{p}_1|^2\hat{p}_1|^2}
\big(\hat{p}_1^*\hat{p}_2\ln |\hat{\rho}_{12}|^2 \big(\hat{p}_1\hat{p}_2^*+
c.c.\big)+4C
\Big\}
\nonumber
\end{eqnarray}
where $\rho_j=x^{(j)}_{\perp 1}+ix^{(j)}_{\perp 2}$,  $\hat{p}_j=
i{\partial\over\partial \rho_j}$ (index $j=1,2$ numbers the particles), 
and $\hat{\rho}_{12}$ is the coordinate operator 
($\rho_{12}\equiv\rho_1-\rho_2$). The first two ``kinetic terms" 
correspond to the propagators of the reggeized gluons and the third 
term describes the interaction of reggeized gluons by exchange potential
coming from product of two Lipatov's vertices given by Eq.~(\ref{2.2.19a}).
The Hamiltonian (\ref{2.5.4}) has a property of holomorphic
 separability~\cite{lip9093}
\begin{equation}
\hat{H}_{12}=\hat{h}_{12}+\hat{h}_{12}^* ,
\label{2.5.5}
\end{equation}
where  
\begin{eqnarray}
\hat{h}_{12}=
{g^2N_c\over 16\pi^2}\left\{\ln \hat{p}_1\hat{p}_2+ 
{1\over\hat{p}_1}(\ln \rho_{12})\hat{p}_1+
{1\over\hat{p}_2}(\ln \rho_{12})\hat{p}_2
+2C
\right\} ,
\label{2.5.6}
\end{eqnarray}
and C=0.557 is Euler's constant.
The generalized LLA is the summation of the diagrams shown in Fig.~11 (see
the discussion in Ref.~29). 
\begin{figure}[htb]
\centerline{
\epsfxsize=11cm
\epsffile{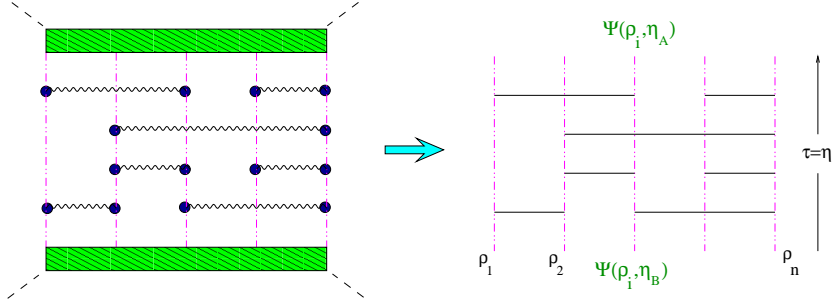}}
\caption{ Generalized LLA as quantum mechanics of the reggeized
gluons.} \label{GLLA}
\end{figure}
The number of reggeized gluons in $t$ channel 
is conserved, so the sum 
of the diagrams in Fig.~10a can be described by quantum mechanics of the 
reggeized gluons with pairwise interaction (\ref{2.5.4}),
\begin{equation}
\hat{H}=\sum_{i<k}T^a_iT^a_kH_{ik},
\label{2.5.7}
\end{equation}
where $H_{ik}$ is obtained from Eq.~(\ref{2.5.4}) by the trivial replacement
$1\rightarrow i,~ 2\rightarrow k$. 

The unitarity follows from the representation of the sum of these 
diagrams as a generalized eikonal~\cite{chengwubook} (see Fig.~12). 
\begin{figure}[htb]
\centerline{
\epsfxsize=8cm
\epsffile{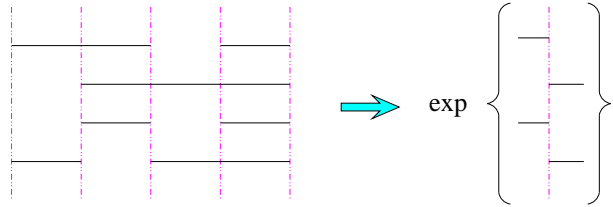}}
\caption{ Quantum mechanics of the reggeized gluons as 
a generalized eikonal.}\label{geik}
\end{figure}
In the multi-color limit ($N_c\rightarrow \infty,~ g^2N_c$-fixed), the
non-planar diagrams vanish hence only the interaction between the adjacent
reggeons survives (the unitarity still holds true).  The color structure
is then unique and the Hamiltonian reduces to~\cite{lip9093}
\begin{equation}
\hat{H}={1\over 2}\sum_{i=1}^{n}\hat{H}_{i,i+1},
\label{2.5.8}
\end{equation}
where ${1\over 2}$ comes from the fact that the adjacent gluons are in
the octet state. Using the property of the holomorphic separability 
(\ref{2.5.5}), it is possible to reduce the quantum mechanics
of the reggeons described by the Hamiltonian (\ref{2.5.8}) to the 
XXX Heisenberg model with spin $s=0$.\cite{lkf} Unfortunately, the 
explicit solution for the number of the magnets $k\geq 3$ ($\equiv$ 
number of the reggeons) has not yet been found. For the $k=3$ (the so-called
Odderon state of three reggeized gluons) the variational estimates give
the intercept at the value of $J$ slightly below 
1~\cite{janik,mbraun} (recently, another Odderon-type solution 
with intercept at $j=1$ was found in Ref.~34). 

In synopsis, we have found the subset of the non-LLA diagrams which
restores unitarity in the s-channel and in the large $N_c$ limit this
subset reduces
to the one-dimensional quantum mechanical model (XXX magnet with $s=0$).  
 
\section{Operator expansion for high-energy
scattering}

The expansion of the amplitudes at high energy in Wilson-line operators
is  very useful in a situation like small-$x$ DIS from the nucleon or
nucleus. As the usual light-cone expansion provides the operator language for 
the DGLAP
evolution, the high-energy OPE gives us the operator form of the BFKL 
equation. In the case of deep inelastic 
scattering there are two different
scales of transverse momentum $k_{\perp}$, and therefore it is 
natural to  factorize the amplitude in the product of contributions of 
hard and soft parts coming from
the regions of small and large
transverse  momenta, respectively. Technically we choose the factorization
scale $Q>\mu>m_N$, and the integrals over $\vec{k}_{\perp}^2>\mu^2$
give the coefficient functions in front of light-cone operators while the
contributions from $\vec{k}_{\perp}^2< \mu^2$ give matrix elements of
these operators normalized at the normalization point $\mu$. In the final
result for the 
structure functions the dependence on $\mu$ in the coefficient functions and
in the matrix elements cancels out yielding the $Q^2$ behavior
of structure functions of DIS.
 
In the case of the high-energy  (Regge ) limit, all the transverse momenta are
of the same order of  magnitude,  but colliding particles strongly differ in
rapidity, thus it is natural to factorize in the rapidity space.
Factorization in rapidity space means that a 
high-energy scattering amplitude can be represented as a convolution of 
contributions due to ``fast" and ``slow" fields. To be precise, we 
choose a certain rapidity $\eta_0$   to be a ``rapidity divide" 
and we call
fields with $\eta>\eta_0$ fast and fields with $\eta<\eta_0$ slow 
where $\eta_0$ lies in the region between spectator 
rapidity $\eta_A$ and target rapidity $\eta_B$. 
(The interpretation of these fields as
fast and slow is literally true only
in the rest frame of the target but we will use this 
terminology for any frame).
Similarly to the case of usual OPE, the integrals over fast fields give the 
coefficient functions in front of the relevant (Wilson-line)
operators while the integrals over slow fields form matrix elements of the
 operators.   For a 2$\Rightarrow$2 particle
scattering in Regge limit $s\gg m^2$  (where $m$ is
a common mass scale for all other momenta in the problem 
$t\sim p_A^2 
 \sim (p'_A)^2\sim p_B^2\sim (p'_B)^2\sim m^2$)
this operator expansion has the form~\cite{ing}
\begin{eqnarray}
A(p_A,p_B\Rightarrow p'_A,p'_B)&=&
\sum\int d^2x_1...d^2x_nC^{i_1...i_n}(x_1,...x_n)\nonumber\\
&\times&\langle p_B|{\rm Tr}\{U_{i_1}(x_1)...U_{i_n}(x_n)\}|p'_B\rangle .
\label{fla1.1}
\end{eqnarray}
(As usual, $s=(p_A+p_B)^2$ and $t=(p_A-p'_A)^2$).
Here  $x_i~(i=1,2)$ are the transverse coordinates 
(orthogonal to both $p_{A}$ and $p_{B}$) and 
$U_i(x)=U^{\dagger}(x){i\over g}{\partial\over\partial x_i}U(x)$ where 
the Wilson-line operator $U(x)$ is the 
gauge link ordered along the infinite
straight line corresponding to the ``rapidity divide'' $\eta_0$. Both 
coefficient functions and matrix elements in Eq.~(\ref{fla1.1}) depend 
 on the $\eta_0$ but 
this dependence is canceled in the physical amplitude just as the scale 
$\mu$ (separating coefficient functions and matrix elements) disappears 
from the final results for structure functions in case of usual 
factorization.
Typically, we have the factors $\sim (g^2\ln s/m^2-\eta_0)$ coming from
the ``fast" integral and the factors $\sim g^2\eta_0$ coming from
the ``slow" integral so they combine in a usual log factor
$g^2\ln s/m^2$. In the leading log approximation these factors
sum up into the BFKL pomeron.

Unlike usual factorization,
the expansion (\ref{fla1.1}) does not have 
the additional meaning of perturbative {\it versus} nonperturbative separation 
-- both the coefficient
functions and the matrix elements have perturbative and 
non-perturbative parts. This happens because the 
coupling constant in a
scattering process is determined by 
the scale of transverse momenta. When we perform 
the usual factorization 
in hard ($k_{\perp}>\mu$) and soft ($k_{\perp}<\mu$) momenta, 
we calculate the 
coefficient functions perturbatively (because 
$\alpha_s(k_{\perp}>\mu)$ is small) whereas
the matrix elements are non-perturbative. Conversely, when we factorize 
the amplitude in rapidity, both fast and slow parts have 
contributions coming from the regions of large and small 
$k_{\perp}$. In this 
sense, coefficient functions and matrix elements enter the expansion 
(\ref{fla1.1}) on equal footing. 

\subsection{High-energy OPE {\it vs} light-cone expansion}
   
 Let me remind the idea of the usual light-cone expansion for the
 deep inelastic scattering (DIS) at moderate $x$. First, we
  take formal limit $Q^2\rightarrow \infty$ and expand near the light cone 
  ($\equiv$in inverse powers of $Q^2$). The amplitude of DIS is then reduced
to the matrix elements of the 
 light-cone operators which are known as parton densities in the nucleon. 
At this step, the support lines for these operators are exactly light-like, 
leading to the logarithmical divergence in transverse
momenta.  The reason for this divergence is the following:  when we expand
T-product of electromagnetic currents near the light cone we assume that 
there are no hard quarks and gluons 
inside the proton. However, the  matrix elements of light-cone operators
contain formally unbounded integrations over $\vec{k}_{\perp}^2$, 
consequently there
are hard quarks and gluons in these matrix elements.  It is
well known how to proceed in this case: define the renormalized light-cone
operators with the integrations over the transverse momenta 
$\vec{k}_{\perp}^2>\mu^2$ cut off
and expand the T-product of electromagnetic currents in a set of these 
renormalized light-cone operators rather than in a set of the original 
unrenormalized ones (see e.g. Ref.~36). 
After that, the matrix elements of these
operators (parton densities) contain factors $\ln{\mu^2\over m^2}$ and
the corresponding coefficient functions contain $\ln{Q^2\over \mu^2}$.
When we calculate the amplitude we add these factors together, the 
dependence on the 
factorization scale $\mu$ cancels, and 
we get the usual DIS logarithmical factors $\ln{Q^2\over m^2}$.
An advantage of this method is that the dependence of structure
 functions on $Q^2$ 
is determined by the dependence of matrix elements of 
the light-cone operators on $\mu$ which is governed by
 the renormalization group.

To get the operator expansion for high-energy scattering, we will proceed 
in the same way. At first, we take the formal Regge limit 
$s\rightarrow\infty$ and 
demonstrate that the amplitude in this limit is reduced to 
matrix elements of the Wilson-line operators representing the two quarks 
moving with the speed of light in the gluon ``cloud.'' 
Formally, we obtain the operators $U$ ordered along  
light-like lines. Matrix elements of such operators contain divergent 
longitudinal integrations reflecting the fact that light-like gauge factor 
corresponds to a quark moving with speed of light (i.e., with infinite 
energy). 
The reason for this divergency is the same as in the case of usual 
light-cone expansion:
the fast-quark propagator in the gluon ``cloud''  
is replaced by the light-like
Wilson line assuming that there are no fast
gluons in the cloud.
However, when we calculate the matrix element of the 
Wilson-line operators with light-like support, the integration
 over the rapidities of the gluon  $\eta_p$ is unbounded so 
 our divergency
 comes from the fast part of the cloud 
  which does not really belong there. 
 Indeed, if the rapidity of the gluon 
 $\eta_p$ is of the order of the rapidity of the quark, this gluon 
 is a fast one. As a result, it 
 will contribute to the coefficient function (in front of the 
 operator constructed from the slow fields) rather than to 
 the matrix element of the operator. 
Similarly to the case of DIS, we need some regularization of the 
 Wilson-line operator which 
 cuts off the fast gluons. 
As demonstrated in Ref.~35, 
it can be done by changing the slope of the
supporting lines. If we wish the 
longitudinal integration stop at
$\eta=\eta_0$, we should order our gauge factors $U$ along a 
line parallel
to $n=\sigma p_1+ \tilde{\sigma}p_2$,
then  the coefficient 
functions in front of Wilson-line operators (impact factors) 
will contain logarithms 
$\sim g^2\ln 1/\sigma$.  Similarly to DIS, when we calculate the amplitude,
 we add the terms $\sim g^2\ln 1/\sigma$ coming from
the coefficient functions  to the terms $\sim g^2\ln
{\sigma\over m^2/s}$ coming from matrix elements  
so
that the dependence on the ``rapidity divide" $\sigma$ cancels and we get the
usual high-energy factors  $g^2\ln {s\over m^2}$ which are responsible for 
BFKL pomeron. Again, the advantage of this method is that the energy 
dependence
of the amplitude is determined by the renorm-group-like evolution equations
for the Wilson-line operators with respect to the slope of the line.

\subsection{High-energy asymptotics as a scattering from the shock-wave field.}

Consider again for simplicity the high-energy $\gamma^*\gamma^*$ scattering 
(\ref{2.1.1}). To put this amplitude in a form
symmetric with respect the top and bottom photons, we make a shift of the
coordinates in the currents by $(z_{\bullet},0,0_{\perp})$ and then reverse
the sign of $z_{\bullet}$. This gives:
\begin{eqnarray}
A(s,t)&=& -i{2\over s}\int 
d^{2}z_{\perp} dz_{\bullet} dz_{*} \int \! d^{4}x d^{4}y 
   e^{-ip_{A}\cdot x-ip_{B}\cdot y}
e^{-i\alpha_r z_{\bullet}+i\beta_r z_{\ast}-i(r,z)_{\perp}}
\nonumber\\
&\times&\Big\langle 0\Big|
T\{j_A(x_{\bullet},x_{*}+z_{*},x_{\perp}+z_{\perp})
             j'_A(0,z_{*},z_{\perp}) \nonumber\\
&\times&             j_B(y_{\bullet}+z_{\bullet},y_{*},y_{\perp})
             j'_B(z_{\bullet},0,0_{\perp})\} \Big|0\Big\rangle  .
\label{3.2.1}
\end{eqnarray}
As we discussed in Sec.~1, $\alpha_r\sim\beta_r\sim{m^2\over s}$ 
so it
can be neglected.

It is convenient to
start with the upper part of the diagram,
i.e., to study how fast quarks
move in an external gluonic field.
After that, functional integration
over the gluon fields will reproduce us the Feynman diagrams of the
type of Fig.~3:
\begin{eqnarray}
   A(s,t)&=& 
   -i{s\over 2}
   \int \! d^{2}z_{\perp}e^{-i(r,z)_{\perp}}
~~{\cal N}^{-1}
   \int {\cal D}A \, e^{iS(A)} {\rm det}(i\nabla )
\label{3.2.2}\\
&\times&\hspace{-2mm} 
   \left\{{2\over s}\int \! dz_{*} \int \!d^{4}x
           \, e^{-ip_{A}\cdot x}
           \left\langle T j_A(x_{\bullet},
x_{*}+z_{*},x_{\perp}+z_{\perp})
                j'_A(0,z_{*},z_{\perp})
           \right\rangle_{A}
    \right\}
\nonumber\\
&\times&\hspace{-2mm} 
   \left\{{2\over s}\int \! dz_{\bullet}\int \!
          d^{4}y \, e^{-ip_{B}\cdot y}
          \left\langle
              T j_B(y_{\bullet}+z_{\bullet},y_{*},y_{\perp})
j'_B(z_{\bullet}, 0, 0_{\perp})\right\rangle_{A}\right\},
\nonumber
\end{eqnarray}
where
\begin{equation}
\langle T j_{\mu }(x)j_{\nu }(y)\rangle_{A}\equiv 
   \frac {\int {\cal D}\psi {\cal D}\bar{\psi }e^{iS(\psi , A)} 
j_{\mu }(x)j_{\nu  }(y)}{\int {\cal D}\psi {\cal D}\bar{\psi }
e^{iS(\psi , A)}}.   
\label{3.2.3}
\end{equation}
Here $S(A)$ and $S(\psi ,A)$ are the gluon and quark-gluon parts of
the QCD action respectively, and  ${\rm det}(i\nabla )$ is the determinant of
Dirac operator in the external gluon field.

 The Regge limit $s\rightarrow\infty$ with $p_A^2$ and $p_B^2$ fixed 
corresponds 
to the following rescaling of the virtual photon 
momentum:  
\begin{equation}
p_A=\lambda p_1^{(0)}+\frac {p_A^2}{2\lambda p^{(0)}_1\cdot 
p_2}p_2,
\label{3.2.4}
\end{equation}
with $p_B$ fixed.  This is equivalent to
\begin{equation}
   p_{1}=\lambda p_{1}^{(0)}, \ \ \ p_{2}=p_{2}^{(0)} ,
\label{3.2.5}
\end{equation}
where $p^{(0)}_1$ and $p_2^{(0)}$ are fixed light-like vectors so that
$\lambda$ is a large parameter associated with the 
center-of-mass energy
($s=2\lambda p_1^{(0)}\cdot p_2^{(0)}$).
Let us study the asymptotics of high-energy $\gamma^*\gamma^*$ scattering 
from the fixed external field 
\begin{equation}
\int \! dx\int \! dz \delta (z_{\bullet})e^{-ip_{A}x-i(r,z)_{\perp}}
\langle T\{j_{\mu }(x+z)j_{\nu }(z)\}\rangle_{A}. 
\label{3.2.6}
\end{equation}
Instead of rescaling of the 
incoming
photon's momentum (\ref{3.2.4}), it is convenient to boost the 
external field instead:
\begin{eqnarray}
&&\int \! dx dz \delta (z_{\bullet})e^{-ip_{A}x-i(r,z)_{\perp}}
\langle T\{j_{\mu }(x+z)j_{\nu }(z)\}\rangle_{A}\nonumber\\
&=&\int \! dx dz \delta (z_{\circ})e^{-ip^{(0)}_{A}x-i(r,z)_{\perp}}
\langle T\{j_{\mu }(x+z)j_{\nu }(z)\}\rangle_{B} ,
\label{3.2.7}
\end{eqnarray}
where $p^{(0)}_{A}=p^{(0)}_{1}+{p_{A}^{2}\over s_{0}}p_{2}$ 
and the boosted
field
$B_{\mu }$ has the form
\begin{eqnarray}
B_{\circ}(x_{\circ},x_{\ast},x_{\perp})&=&\lambda A_{\circ}
({x_{\circ}\over \lambda},x_{\ast}\lambda,x_{\perp}) ,
\nonumber\\
B_{\ast}(x_{\circ},x_{\ast},x_{\perp})&=&{1\over\lambda}A_{\ast}
({x_{\circ}\over
\lambda},x_{\ast}\lambda,x_{\perp}) , \nonumber\\
B_{\perp}(x_{\circ},x_{\ast},x_{\perp})&=&A_{\perp}
({x_{\circ}\over \lambda},x_{\ast}\lambda,x_{\perp}) ,
\label{3.2.8}
\end{eqnarray}
where we used the notations $x_{\circ}\equiv x^{\mu }p_{1\mu }^{(0)},
~x_{\ast}\equiv
x^{\mu }p_{2\mu }$. 
The field
\begin{equation}
A_{\mu }(x_{\circ},x_{\ast},x_{\perp})=
A_{\mu }({2\over s_{0}}x_{\circ} p_{1}^{(0)}+{2\over
s_{0}}x_{\ast}
p_{2}+x_{\perp})
\label{3.2.9}
\end{equation}
is the original external field in the coordinates independent of 
$\lambda$, therefore we
may assume that the scales of $x_{\circ},x_{\ast}$ (and $x_{\perp}$) 
in the function
(\ref{3.2.9}) are $O(1)$. First, it is easy to see that at large 
$\lambda$ the field
$B_{\mu }(x)$ does not depend on $x_{\circ}$. Moreover, in the limit 
of very large
$\lambda$ the field $B_{\mu }$ has a
form of the shock wave. It is especially clear if one writes down the field
strength tensor $G_{\mu \nu }$ for the boosted field. If we assume that
the field strength
$F_{\mu \nu }$ for the external field $A_{\mu }$ vanishes at the infinity we
get
\begin{eqnarray}
G_{\circ  i}(x_{\circ },x_{\ast},x_{\perp})&=&\lambda F_{\circ
i}({x_{\circ}\over\lambda},x_{\ast}\lambda,x_{\perp}) 
\rightarrow \delta (x_{\ast})G_{i}(x_{\perp}) ,
\nonumber\\
G_{_{\ast} i}(x_{\circ},x_{\ast},x_{\perp})&=&
{1\over\lambda}F_{_{\ast} i}({x_{\circ}\over
\lambda},x_{\ast}\lambda,x_{\perp})\rightarrow 0 ,\nonumber\\
G_{\circ\ast}(x_{\circ},x_{\ast},x_{\perp})&=&
F_{\circ\ast}({x_{\circ}\over \lambda},x_{\ast}\lambda,x_{\perp})
\rightarrow 0 ,\nonumber\\
G_{ik}(x_{\circ},x_{\ast},x_{\perp})&=&
F_{ik}({x_{\circ}\over \lambda},x_{\ast}\lambda,x_{\perp})\rightarrow 0 ,
\label{3.2.10}
\end{eqnarray}
so the only component which survives the infinite boost is $F_{\circ\perp}$ 
and it
exists only within the thin ``wall" near $x_{\ast}=0$. In the rest of the 
space the
field $B_{\mu }$ is a pure gauge. Let us denote by $\Omega $ the corresponding
gauge matrix and by $B^{\Omega }$ the rotated gauge field which vanishes
everywhere except the thin wall:
\begin{equation}
B^{\Omega }_{\circ}=\lim_{\lambda\rightarrow\infty}
\frac {\partial ^{i}}{\vec{\partial}^2_{\perp}}
G^{\Omega}_{i\circ}(0,\lambda x_{\ast},x_{\perp})\rightarrow
\delta (x_{\ast})
\frac {\partial ^{i}}{\vec{\partial}^2_{\perp}}G^{\Omega}_{i}
(x_{\perp}), 
~B^{\Omega}_{\ast}=B_{\perp}=0 .
\label{3.2.11}
\end{equation}

\begin{figure}[htb]
\centerline{
\epsfysize=5cm
\epsffile{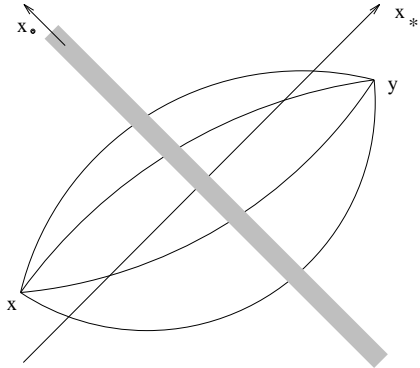}}
\caption{ Propagator in the shock-wave field.} \label{ofig13}
\end{figure}

To illustrate the method, consider 
at first the propagator of the scalar particle (say, the 
Faddeev-Popov ghost) in the shock-wave background. 
In Schwinger's notations we write down formally the propagator in the
external gluon field $A_{\mu }(x)$ as
\begin{equation}
G(x,y) =\hbox{\bf\Big($\!\!$\Big(}  x \Big| \frac {1}{P^2+i\epsilon} 
\Big| y\hbox{\bf\Big)$\!\!$\Big)} =
        \hbox{\bf\Big($\!\!$\Big(} x \Big| \frac {1}{(p + gA)^2+i\epsilon}
         \Big| y\hbox{\bf\Big)$\!\!$\Big)} ,
\label{3.2.12}
\end{equation}
where $\hbox{\bf($\!$(} x|y \hbox{\bf)$\!$)} =\delta ^{(4)}(x-y)$,
\begin{equation}
 \hbox{\bf($\!$(} x | p_{\mu } | y \hbox{\bf)$\!$)} =
   -i\frac {\partial }{\partial y^{\mu }}\delta ^{(4)}(x-y),~~~~
  \hbox{\bf($\!$(} x | A_{\mu } | y \hbox{\bf)$\!$)}
   =A_{\mu }(x)\delta ^{(4)}(x-y) .
\label{3.2.13}
\end{equation}
Here $|x\hbox{\bf)$\!$)}$ are the eigenstates of the coordinate operator
${\cal X}|x\hbox{\bf)$\!$)} =x|x\hbox{\bf)$\!$)}$ 
(normalized according to the second
line in the above
equation). From Eq.~(\ref{3.2.13}) it is also easy
to see that the eigenstates of
the free momentum operator $p$ are the plane waves
$|p\hbox{\bf)$\!$)}=\int d^4x\,e^{-ip\cdot x}|x\hbox{\bf)$\!$)}$.  
The path-integral 
representation of a Green function of scalar particle in the external
field has the form:
\begin{eqnarray}
&&{\hbox{\bf\Big($\!\!$\Big(}} x\Big|\frac {1}{{\cal P}^2}
\Big| y{\hbox{\bf\Big)$\!\!$\Big)}}=-i\int _{0}^{\infty}\! d\tau 
{\hbox{\bf\Big($\!\!$\Big(}} x\Big|e^{i\tau{\cal P}^2}
\Big| y{\hbox{\bf\Big)$\!\!$\Big)}}\label{3.2.14}\\
&=&
-i\int _{0}^{\infty}\! d\tau {\cal N}^{-1}
\int _{x(0)=y}^{x(\tau )=x}\!{\cal D}x(t)
e^{-i\int _{0}^{\tau }\!
dt{\dot{x}^{2}\over 4}}Pexp\{ig\int _{0}^{\tau }\! dt(B^{\Omega}_{\mu }(x(t))
\dot{x}^{\mu }(t)\} ,
\nonumber
\end{eqnarray}
where $\tau$ is Schwinger's proper time. It is clear 
that all the interaction 
with the external field $B^{\Omega}_{\mu}$ occurs 
at the point of the intersection 
of the path of the particle with the shock wave 
(see Fig.~\ref{ofig13}).
Therefore, it is convenient to rewrite at first the bare propagator
\begin{equation}
{\hbox{\bf\Big($\!\!$\Big(}} x\Big|\frac {1}{p^2}\Big| 
y{\hbox{\bf\Big)$\!\!$\Big)}}=
\frac{i}{4\pi^2(x-y)^2}=
-i\int _{0}^{\infty}\! d\tau {\cal N}^{-1}\int ^{x(\tau )=x}_{x(0)=y}\!
{\cal D}x(t)(\tau )e^{-i\int ^{\tau }_{0}\! dt{\dot{x}^{2}\over 4}}
\label{3.2.15}
\end{equation}
 marking 
the point of the intersection of integration path with the plane $z_{\ast}=0$. 
To
this end, consider the case $x_{\ast}>0,~y_{\ast}<0$ and insert 
\begin{equation}
1=\int \!d\tau '\dot{x}_{\ast}(\tau ')\delta (x_{\ast}(\tau '))
\label{3.2.16}
\end{equation}
in the path integral (\ref{3.2.15}). (Here $\tau'$ has the meaning of the 
time at 
which  the intersection with the plane $z_{\ast}=0$ takes place). We get
\begin{eqnarray}
&&{\hbox{\bf\Big($\!\!$\Big(}} x\Big|\frac {1}{p^2}
\Big| y{\hbox{\bf\Big)$\!\!$\Big)}}=-i\int _{0}^{\infty}\! 
d\tau 
{\hbox{\bf\Big($\!\!$\Big(}} x\Big|e^{i\tau{p^2}}
\Big| y{\hbox{\bf\Big)$\!\!$\Big)}}\label{3.2.17}\\
&=&
-i\int _{0}^{\infty}\! d\tau \int _{0}^{\tau}\! d\tau' {\cal N}^{-1}
\int _{x(0)=y}^{x(\tau )=x}\!{\cal D}x(t)\dot{x}_{\ast}(\tau ')
\delta (x_{\ast}(\tau '))
e^{-i\int _{0}^{\tau }\!dt{\dot{x}^{2}\over 4}}\nonumber\\
&=&-i\int _{0}^{\infty}\! d\tau 
\int _{0}^{\tau }\!d\tau'\!\int \! dz\delta(z_{*})
{\cal N}^{-1}\int _{x(\tau ')=z}^{x(\tau )=x}\!\!{\cal D}x(t)\!
e^{-i\int _{\tau '}^{\tau }\! dt{\dot{x}^{2}\over 4}}{\cal
N}^{-1}\nonumber\\
&\times&\int _{x(0)=y}^{x(\tau ')=z}\!\!{\cal D}x(t)
\dot{x}_{*}(\tau ')
e^{-i\int _{0}^{\tau '}\! dt{\dot{x}^{2}\over 4}} .
\nonumber
\end{eqnarray}
Making the shift of integration variable $\tau -\tau'\rightarrow\tau$,
we can rewrite the path integral (\ref{3.2.17}) in the form:
\begin{eqnarray}
&&-i\int _{0}^{\infty}\! d\tau \int _{0}^{\infty}\! d\tau '\int \! dz\delta
(z_{\ast})\label{3.2.18}\\
&\times&{\cal N}^{-1}\int _{x(0)=z}^{x(\tau )=x}\!{\cal D}x(t)\!
e^{-i\int _{0}^{\tau }\! dt{\dot{x}^{2}\over 4}}{\cal
N}^{-1}\int _{x(0)=y}^{x(\tau ')=z}\!{\cal D}x(t)\dot{x}_{\ast}
e^{-i\int _{0}^{\tau'}\! dt{\dot{x}^{2}\over 4}} .
\nonumber
\end{eqnarray}
Using Eq.~(\ref{3.2.15}) and similar formula
\begin{equation}
\int _{0}^{\infty}\! 
d\tau {\cal N}^{-1}\int ^{x(\tau )=x}_{x(0)=y}\!{\cal D}x(t)
\dot{x}_{\mu }(\tau )e^{-i\int ^{\tau }_{0}\! dt{\dot{x}^{2}\over 4}}=
\frac {i(x-y)_{\mu }}{\pi ^{2}(x-y)^{4}}~~,
\label{3.2.19}
\end{equation} 
we arrive at the following 
representation of the bare propagator 
(in the case of $x_{\ast}>0,~y_{\ast}<0$):
%
\begin{equation}
{\hbox{\bf\Big($\!\!$\Big(}} x\Big|\frac {1}{p^2+i\epsilon}\Big| 
y{\hbox{\bf\Big)$\!\!$\Big)}}=\int \!
dz\delta (z_{\ast})
\frac {1}{4\pi ^{2}(x-z)^{2}}
{y_{\ast}\over\pi ^{2}(z-y)^{4}}
\label{3.2.20}
\end{equation}
where $z$ is the point of the intersection of the path of the 
particle with the shock wave.

Now let us recall that our particle moves in the shock-wave external field 
and therefore each path in the functional integral (\ref{3.2.14}) is weighted 
with the additional gauge factor $Pe^{ig\int B_{\mu}dx_{\mu}}$. 
Since the  external field exists only within the infinitely thin wall at
$x_{\ast}=0$ we can replace the gauge factor along the actual path 
$x_{\mu }(t)$ by
the gauge factor along the straight-line path shown in Fig.~\ref{ofig13}.
It intersects the plane $z_{\ast}=0$ at the same point  
$(z_{\circ},z_{\perp})$ at which
the original path does. Since the shock-wave field outside the wall vanishes we
may formally extend the limits of this segment to infinity and write the
corresponding gauge factor as 
$U^{\Omega}(z_{\perp})=[\infty p_{1}+z_{\perp},-\infty p_{1}+z_{\perp}]$. The
error brought by replacement of the original path $inside$ the wall by the
segment of straight line parallel to $p_{1}$ is $\sqrt{m^2\over s}$. Indeed,
the
time of the transition of the particle through the wall is proportional to the
thickness of the wall which is $\sim{m^2\over s}$. It indicates that the 
particle can
deviate in the perpendicular directions inside the wall only to the distances
$\sqrt{m^2\over s}$. Thus, if
the particle intersects this wall at some point $(z_{\ast},z_{\perp})$
the gauge factor $Pe^{ig\int B^{\Omega}_{\mu}dx_{\mu}}$ reduces to 
$U^{\Omega}(z_{\perp})$.
One can now repeat for the path integral (\ref{3.2.14}) the steps which
lead us  from  path-integral representation of bare propagator (\ref{3.2.15})
 to the formula 
(\ref{3.2.20}); the only difference will be the factor 
$U^{\Omega}(z_{\perp})$ in the 
point of the intersection of the path with the plane $z_{\ast}=0$:
%
\begin{equation}
{\hbox{\bf\Big($\!\!$\Big(}} x\Big|
\frac {1}{{\cal P}^2}\Big| y{\hbox{\bf\Big)$\!\!$\Big)}}=\int \!
dz\delta (z_{\ast})
\frac {1}{4\pi ^{2}(x-z)^{2}}U^{\Omega}(z_{\perp})
\frac {y_{\ast}}{\pi ^{2}(z-y)^{4}}
\label{3.2.21}
\end{equation}
(in the region $x_{\ast}>0,y_{\ast}<0$). It is easy to see that 
the propagator in the region
$x_{\ast}<0,y_{\ast}>0$ differs from Eq.~(\ref{3.2.21}) by the replacement
$U^{\Omega}\leftrightarrow U^{\Omega \dagger }$. Also, the propagator 
outside the shock-wave wall (at $x_{\ast},y_{\ast}<0$ or $x_{\ast},y_{\ast}>0$)
coincides with the bare
propagator.  The final answer for the Green function of the scalar particle 
in the $B^{\Omega}$
background can be written down as:
\begin{eqnarray}
{\hbox{\bf\Big($\!\!$\Big(}} x\Big|
\frac {1}{{\cal P}^2}\Big| y{\hbox{\bf\Big)$\!\!$\Big)}}&=&
i\frac {1}{4\pi ^{2}(x-y)^{2}}\theta(x_{\ast} y_{\ast})+
\int \! dz\delta (z_{\ast})
\frac {1}{4\pi ^{2}(x-z)^{2}}\label{3.2.22}\\
&\times&\{U^{\Omega}(z_{\perp})\theta (x_{\ast})\theta
(-y_{\ast})-U^{\Omega \dagger }
(z_{\perp})\theta (y_{\ast})\theta (-x_{\ast})\}
\frac {y_{\ast}}{\pi
^{2}(z-y)^{4}}.
\nonumber
\end{eqnarray}
We see that the propagator in the shock-wave background is a convolution of the
free  propagation up to the plane $z_{\ast}=0$, instantaneous interaction with
the  shock wave described by the Wilson-line operator $U^{\Omega}$ 
($U^{\dagger\Omega}$), 
and another 
free propagation from $z$ to the final point (see Fig.~\ref{ofig13})
One can check that the Green function (\ref{3.2.22}) is 
continuous as $x_\ast\rightarrow 0$ (or $y_\ast\rightarrow 0$).

\begin{figure}[htb]
\centerline{
\epsfysize=6cm
\epsffile{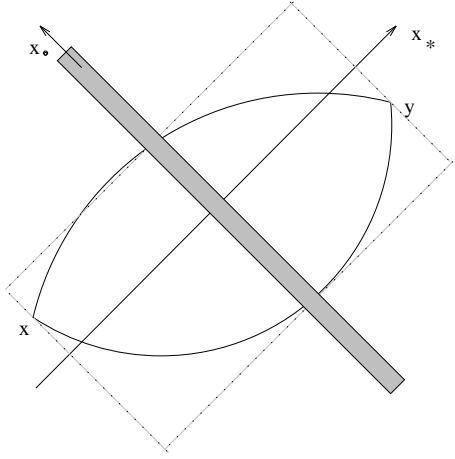}}
\caption{ Quark-antiquark propagation in the shock wave.} \label{ofig14}
\end{figure}

In order to get the propagator in the original field $B_{\mu}$ 
we must perform back the gauge rotation with the $\Omega$ matrix. 
It is convenient to represent the result in the following form:
\begin{eqnarray}
{\hbox{\bf\Big($\!\!$\Big(}} x\Big|
 {1\over{\cal P}^2}\Big| y{\hbox{\bf\Big)$\!\!$\Big)}}\!\!\!\!&=&
\!\!\!\! \frac {i}{4\pi ^{2}(x-y)^{2}}[x,y]\theta (x_{\ast} y_{\ast})
+\!\!\int \! dz\delta
(z_{\ast})\frac {1}{4\pi ^{2}(x-z)^{2}}\!\!\!\!\!
\label{3.2.23}\\
&\times& \!\!\!\!\!\!\!
\{U(z_{\perp};x,y)\theta (x_{\ast})\theta
(-y_{\ast})-U^{\dagger}(z_{\perp};x,y)\theta (y_{\ast})\theta (-x_{\ast})\}
\frac {y_\ast}{\pi^{2}(z-y)^{4}} ,\nonumber
\end{eqnarray}
where
\begin{eqnarray}
&&U(z_{\perp};x,y)=[x,z_{x}][z_{x},z_{y}][z_{y},y]~,\nonumber\\
&&{}z_{x}\equiv ({2\over s_{0}}z_{\circ} p^{(0)}_{1}+{2\over s_{0}}x_{\ast}
p_{2},z_{\perp}),~~~~~z_{y}=z_{x}(x_{\ast}\leftrightarrow y_{\ast})
\label{3.2.24}
\end{eqnarray}
is a gauge factor for the contour made from segments of straight lines as shown
in Fig.~\ref{ofig14}. Since the field $B_{\mu }$ outside the shock-wave
wall is a pure gauge, the precise form of the contour does not matter as long
as it starts at the point $x$, intersects the wall at the point $z$ in the
direction collinear to $p_{2}$, and ends at the point $y$. We have chosen this 
contour in such a way that the
gauge factor (\ref{3.2.24}) is the same for the field $B_{\mu }$ and for the
original field $A_{\mu }$ (see Eq.~(\ref{3.2.8})).

The quark propagator in a shock-wave background can be calculated 
in a similar way (see Appendix 7.2),
\begin{eqnarray}
&&\hspace{-5mm}{\hbox{\bf\Big($\!\!$\Big(}} x\Big|
\frac {1}{{\not\! {\cal P}}}\Big| y{\hbox{\bf\Big)$\!\!$\Big)}}~=~
-~\frac {{\not\! x}-{\not\! y}}{2\pi ^{2}(x-y)^{4}}[x,y]
\theta (x_{\ast} y_{\ast})~+~i\int \! dz\delta
(z_{\ast}) \frac {{\not\! x}-{\not\! z}}{2\pi ^{2}(x-z)^{4}}
\nonumber\\
&&\hspace{-5mm}\times~
\{U(z_{\perp};x,y)
\theta (x_{\ast})\theta
(-y_{\ast})-U^{\dagger}(z_{\perp};x,y)\theta (y_{\ast})
\theta (-x_{\ast})\}
\frac { {\not\! z}-{\not\! y}}{2\pi
^{2}(z-y)^{4}}.\label{3.2.25}
\end{eqnarray}
For the quark-antiquark amplitude in the shock-wave field 
(see Fig.~\ref{ofig14})
 we get
\begin{eqnarray}
&&\hspace{-5mm}{\rm Tr}\gamma_{\mu }{\hbox{\bf\Big($\!\!$\Big(}} x\Big|
\frac {1}{{\not\! {\cal P}}}\Big| y{\hbox{\bf\Big)$\!\!$\Big)}}
\gamma_{\nu }
{\hbox{\bf\Big($\!\!$\Big(}} y\Big|\frac {1}{{\not\! {\cal P}}}
\Big| x{\hbox{\bf\Big)$\!\!$\Big)}}\label{3.2.26}\\
&&\hspace{-5mm}=~
\frac {{\rm Tr}\gamma_{\mu }({\not\! x}-{\not\! y})\gamma_{\nu }({\not\! y}-
{\not\! x})}{4\pi ^{4}(x-y)^{8}}
\theta (x_{\ast} y_{\ast})
~-~\theta (-x_{\ast} y_{\ast})\int \! dzdz'\delta (z_{\ast})
\delta (z'_{\ast})\nonumber\\
&&\hspace{-5mm}\times~{\rm Tr}\gamma_{\mu }\frac {{\not\! x}-{\not\! z}}{2\pi
^{2}(x-z)^{4}}{\not\! p}_{2}\frac {{\not\! z}-
{\not\! y}}{2\pi ^{2}(z-y)^{4}}\gamma_{\nu }
\frac {{\not\! y}-{\not\! z}'}{2\pi ^{2}(y-z')^{4}}{\not\! p}_{2}
\frac { {\not\! z}'-{\not\! x}}{2\pi
^{2}(z'-x)^{4}}{\rm U}(z_{\perp};z'_{\perp}),\nonumber
\end{eqnarray}
where we can write down the gauge factor U$(z_{\perp};z'_{\perp})\equiv
U(z_{\perp};x,y)U^{\dagger}(z'_{\perp};y,x)$ as a product of two infinite 
Wilson-lines operators
connected by gauge segments at $\pm \infty$,
\begin{eqnarray}
&&\hspace{-8mm}{\rm U}(z_{\perp};z'_{\perp})\nonumber\\
&&\hspace{-8mm}=
\lim_{u\rightarrow\infty}\Big\{[up_{1}+z_{\perp},-up_{1}+z_{\perp}]
[-up_{1}+z_{\perp},-up_{1}+z'_{\perp}][-up_{1}+
z'_{\perp},up_{1}+z'_{\perp}]\nonumber\\
&&\hspace{-8mm}\times [up_{1}+z'_{\perp},up_{1}+z_{\perp}] \Big\}~=~
U_z[z_{\perp},z'_{\perp}]_{-}U^{\dagger}_{z'}[z'_{\perp},z_{\perp}]_{+} .
\label{3.2.27}
\end{eqnarray}
Here we use the notations 
\begin{equation}
[x_{\perp},y_{\perp}]_{+}\equiv  
[\infty p_{1}+z_{\perp},\infty p_{1}+z'_{\perp}],~~
[x_{\perp},y_{\perp}]_{-}\equiv  
[-\infty p_{1}+z_{\perp},-\infty p_{1}+z'_{\perp}].
\label{endfactor}
\end{equation}
As we mentioned
above, the precise form of the connecting contour at infinity 
does not matter as long as
it is outside the shock wave. We have chosen this contour in such a way that
the gauge factor (\ref{3.2.27}) is the same for the field $B_{\mu }$ and for
the original field $A_{\mu }$ (see Eq.~(\ref{3.2.8})). Now, substituting our
result  for
quark-antiquark propagation (\ref{3.2.26}) in the right-hand side 
of Eq.~(\ref{3.2.6}),
one obtains 
\begin{eqnarray}
\lefteqn{
   \int \! d^{4}x \int \! d^{4}z \, \delta (z_\bullet) e^{-i(r,z)_{\perp}} 
e^{-ip_{A}\cdot  x}
   \langle T\{j_{A }(x+z)j'_{A }(z)\}\rangle_{A}
}
\hspace*{0.5in}
\nonumber\\
   &=& \sum e_{i}^{2}\int \frac {d^{2}k_{\perp}}{4\pi ^{2}}
   I^A(k_{\perp},r_{\perp}) {\rm Tr}\{ U(k_{\perp})U^{\dagger}
   (r_{\perp}-k_{\perp}) \} ,
\label{3.2.28}
\end{eqnarray}
where the impact factor $I^A$ is given by Eq.~(\ref{fla10d}). For brevity,
we omit the end gauge factors (\ref{endfactor}).
 
 Formula (\ref{3.2.27}) describes a quark and antiquark moving fast
through an external gluon field. After integrating over gluon
fields in the functional integral we obtain the virtual photon
scattering amplitude (\ref{3.2.2}). It is convenient to rewrite it
in the factorized form:
\begin{equation}
   {\cal A}(p_{A},p_{B})=i{s\over 2}
   \sum e_{i}^{2}
   \int \frac {d^{2}k_{\perp}}{4\pi ^{2}}I^{A}(k_{\perp},r_{\perp})
\langle \langle{\rm Tr}\{\hat U(k_{\perp}) 
\hat U^{\dagger }(r_{\perp}-k_{\perp})\}
\rangle\rangle .
\label{3.2.29}
\end{equation}
where $I^A(p_{\perp})=e^A_{\mu}e^A_{\nu}I^A_{\mu\nu}(p_{\perp})$.
The gluon fields in $U$ and $U^{\dagger }$ have been promoted to
operators, a fact which we signal by replacing $U$ by $\hat U$, etc.
The reduced matrix
elements of the
operator ${\rm Tr}\{ \hat U(k_{\perp}) \hat U^{\dagger }(r_{\perp}-k_{\perp})\}$
between the ``virtual photon states" are defined as follows:
\begin{eqnarray}
   \langle \langle {\rm Tr}\{ \hat U(k_{\perp}) 
  \hat U^{\dagger }(r_{\perp}-k_{\perp})\}\rangle\rangle 
   &=&
   \int \! d^{2}x_{\perp} e^{-i(kx)_{\perp}}
 \langle \langle {\rm Tr}\{ \hat U(x_{\perp}) \hat U^{\dagger }(0) \} 
 \rangle\rangle 
\nonumber\\
   \langle \langle{\rm Tr} \{ \hat U(x_{\perp}) \hat U^{\dagger }(0)\}
    \rangle\rangle 
   &\equiv &-\int \! d^{4}z \delta (z_{\ast})e^{i(r,z)_{\perp}}
\int \! d^{4}y 
e^{-ip_{B}\cdot y}
\label{3.2.30}\\
&&\quad
   \langle 0| T\{ {\rm Tr} \{ \hat U(x_{\perp}) \hat U^{\dagger }(0) \}
         j_B(y+z)j'_B(z)\}
   |0\rangle .
\nonumber
\end{eqnarray}
This matrix element describes the propagation of the ``color dipole'' in the
 background of the shock wave created by the second virtual photon.

It is worth noting that for a real photon our definition of the
reduced matrix element can be rewritten as
\begin{equation}
   \langle \epsilon , p_{B} |
     {\rm Tr} \{ \hat U(x_{\perp}) \hat U^{\dagger }(x'_{\perp}) \}
   |\epsilon ', p_{B}+\beta p_{B}\rangle
   ~=~ 2\pi \delta (\beta ) \,
  \langle \langle {\rm Tr}\{ \hat U(x_{\perp}) 
\hat U^{\dagger }(x'_{\perp}) \}
  \rangle\rangle ,
\nonumber
\end{equation}
where $\epsilon$ and $\epsilon'$ represent the polarizations of the photon
states. The factor $2\pi \delta (\beta)$ reflects the fact that the forward
matrix element of the operator 
$\hat U(x_{\perp}) \hat U^{\dagger }(x'_{\perp})$
contains an unrestricted integration along $p_{1}$. Taking the
integral over $\beta$ one reobtains Eq.~(\ref{3.2.30}).

\subsection{Regularized Wilson-line operators}

\begin{figure}[htb]
\centerline{
\epsfysize=5cm
\epsffile{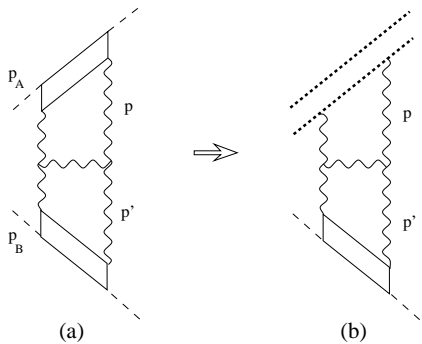}}
\caption{A typical Feynman diagram for the $\gamma^*\gamma^*$ scattering 
amplitude (a) and the corresponding two-Wilson-line operator 
(b).} \label{fig4}
\end{figure}

In the Regge limit (\ref{3.2.4}) we have formally obtained the operators 
$\hat{U}$
ordered  along the 
 light-like lines. Matrix elements of such operators contain divergent 
longitudinal integrations which reflect the fact that light-like gauge factor 
corresponds to a quark moving with speed of light (i.e., with infinite 
energy). 
This divergency can be already seen at the one-loop level
if one calculates the contribution to the matrix element of the
two-Wilson-line operator $\hat{U}(x_{\perp})\hat{U}^{\dagger}(y_{\perp})$ 
between the ``virtual photon states".
As I mentioned above, the reason for this divergence is that 
we have replaced the fast-quark propagators in the ``external 
field" represented by two gluons 
 coming from the bottom part of the diagram in Fig.~\ref{fig4}a by the
light-like Wilson lines in Fig.~\ref{fig4}b. 
The integration over rapidities of the gluon $\eta_p$ in the matrix element
of the light-like Wilson-line operator 
$\hat{U}(x_{\perp})\hat{U}^{\dagger}(y_{\perp})$
 is formally 
unbounded , consequently we need some 
regularization of the Wilson-line operator which 
 cuts off the fast gluons. 
As demonstrated in Ref.~35, 
it can be done
 by changing the slope of the
supporting line. If we wish the 
longitudinal integration stop at
$\eta=\eta_0$, we should order our gauge factors $U$ along a 
line parallel
to $p_\zeta=p_1+\zeta p_2$ where $\eta_0=\ln\zeta$.\footnote{The situation  
here is again quite similar to the usual OPE for
DIS. Recall that when  separating the Feynman integrals over loop momenta
$p$ into the coefficient  functions (with $p^{2}\gg\mu ^{2}$) and matrix
elements ($p^{2}\ll\mu ^{2}$) we
expand hard propagators in powers of soft external fields. As a result
of this expansion we formally obtain the expressions of the type
 $\bar{\psi }(\lambda e_{1})[\lambda e_{1},0]\psi (0)$ with external fields
lying
exactly on the light
 cone. In operator language it corresponds to the matrix element of the
same light-cone operator 
$\hat{\bar{\psi }}(\lambda e_{1})[\lambda
e_{1},0]\hat{\psi }(0)$  normalized at the point $\mu ^{2}$ in order to ensure
the
restriction that matrix elements of this operator do not contain 
virtualities larger
than $\mu ^{2}$. Moreover, in principle we can regularize these light-cone
operators
for DIS by changing the slope of the supporting line (say, take
$e=e_{1}+{\mu ^{2}\over Q^{2}}e_{2}$).  The only reason why we use the 
regularization by counterterms is that, unlike the regularization by
the slope, counterterms 
are governed by renormalization-group equations.} 
We define
 \begin{eqnarray}
\hat{U}^{\zeta} (x_{\perp} )&=& [\infty p^{\zeta}+x_{\perp} ,-\infty p^{\zeta}
+x_{\perp} ] , \nonumber\\
\hat{U}^{\dagger\zeta} (x_{\perp})&=&[-\infty
p^{\zeta}+x_{\perp},\infty p^{\zeta} +x_{\perp}] . \label{3.3.1}
 \end{eqnarray}
Matrix elements of these operators 
coincide with matrix elements of the operators
 $\hat{U}$ and $\hat{U}^{\dagger}$ calculated with the restriction
$\alpha<\sigma=\sqrt{p_A^2\over s\zeta}$ imposed in the internal loops 
(and external
tails).
Let us demonstrate this
using the simple example of the matrix element of the operator 
$\hat{U}^{\zeta}(k_{\perp})\hat{U}^{\dagger\zeta}(r_{\perp}-k_{\perp})$ 
coming from the diagram shown in
Fig.~\ref{fig4}. It has the form
\begin{eqnarray}
&& \!\!\!\!\! -{i\over 2}g^{6}\int \! \frac {d\alpha_{p}}{2\pi } 
\frac {d^{4}p'}{16\pi ^{4}}
\,  \frac {[(\alpha_{p}-2\alpha'_k)\beta'_ks-(\vec{k}+\vec{k}')^2_{\perp}]  
\Phi ^{B}(k')}{(\zeta \alpha_{p}^{2}s+\vec{k}^{2}_{\perp}-i\epsilon)^{2} \, 
(\alpha'_k\beta'_ks-
\vec{p'}^{2}_{\perp}+i\epsilon)^{2}} 
\nonumber\\
&\times&\frac{1}{[-(\alpha_{p}-\alpha')(\alpha_{p}\zeta +\beta'_k)s-
(\vec{k}-\vec{k}')^2_{\perp}+i\epsilon]} ,
\label{3.6}
\end{eqnarray}
where the numerator comes from the product of two three-gluon 
vertices (\ref{2.2.2})
\begin{equation}    
{4\over s^{2}}\Gamma_{\ast\bullet}{}^{\sigma}(k,-k')
\Gamma_{\ast\bullet\sigma}(k,-k') =  
=(\alpha_{k}-2\alpha'_{k})\beta'_{k}s-(\vec{k}+\vec{k}')^2_{\perp} .
\label{3.3}
\end{equation}
As we shall see below, the logarithmic contribution comes from the region
$\sqrt {m^{2}\over\zeta s} \gg \alpha_k \gg \alpha'_k \sim {m^2\over s}$,
$1 \gg \beta _{p}' \gg \beta _{p} = 
-\zeta \alpha_k \sim \sqrt {m^{2}\zeta \over s}$.
In this region one can perform the integration over $\beta'_k$  by taking
the residue at the pole
$\left[ -(\alpha_{p}-\alpha')(\alpha_{p}\zeta +\beta'_k)s-
(\vec{k}-\vec{k}')^2_{\perp}+
i\epsilon \right]^{-1}$.
The result is~\footnote{In the region we are investigating, we can neglect
   the $\beta'_k$ dependence  of the lower quark loop.}
\begin{eqnarray}
&&\!\!\!\!
   \frac {g^{6}}{s} \int  \frac {d\alpha_k}{2\pi } 
\frac {d\alpha'_k}{2\pi }
   \int \frac {d^{2}k'_{\perp}}{4\pi ^{2}}
   \left[ \Theta (\alpha_k>\alpha'_k>0 + \Theta (0>\alpha'_k>
\alpha_k) \right]
\label{3.7}\\
&\times&
\frac {\left( \vec{k}^{2}_{\perp}+ {p'}_{\perp}^{2}
-\alpha_k^{2}\zeta s/2 \right) \, 
\Phi ^{B}\left(\alpha_k' p_{1}-(\alpha_k\zeta +
{(\vec{k}-\vec{k}')^2_{\perp}\over \alpha_{p} s})p_{2}+
k'_{\perp} \right)}{|\alpha_{p}-\alpha'_k| \, 
(\zeta \alpha_{p}^{2}s+\vec{k}^{2}_{\perp}-i\epsilon)^{2} \, 
[\frac {\alpha'_k}{\alpha_k}(\vec{k}-\vec{k}')^2_{\perp}+
\vec{p'}^{2}_{\perp}+i\epsilon]^{2}} .
\nonumber
\end{eqnarray}
We see that the  integral over $\alpha_{p}$ is logarithmic
in the region 
$\sqrt {m^{2}\over\zeta s} \gg \alpha_{p} \gg \alpha'_k \sim {m^2\over s}$ 
(cf. Eq.~(18)). 
The lower limit of this logarithmical
integration is provided by the matrix
element itself ($\beta_{k}\sim 1$ in the lower quark bulb) while
the upper limit, at $\alpha_k^{2} \sim m^{2}/\zeta s$
is enforced by the non-zero $\zeta $ and the result has the form
\begin{equation}
   \langle \!\langle  {\rm Tr} {\hat{U}^{\zeta }}
(k_{\perp}) {\hat{U}^{\dagger\zeta }}(-k_{\perp})\rangle \!\rangle _{{\rm
Fig.~\ref{fig4}}}  ~=~
   {g^{6}\over 8\pi }
   \ln \left( {s\over m^{2}\zeta } \right)  \int \!\frac {d^{2}k'_{\perp}}{4\pi ^{2}} \frac {\vec{k}^{2}_{\perp} 
+ {p'}_{\perp}^{2}}{\vec{k}^{4}_{\perp} \vec{p'}^{4}_{\perp}}
   I^{B}(k'_{\perp}) .
\label{3.8}
\end{equation}

\begin{figure}[htb]
\centerline{
\epsfxsize=11cm
\epsffile{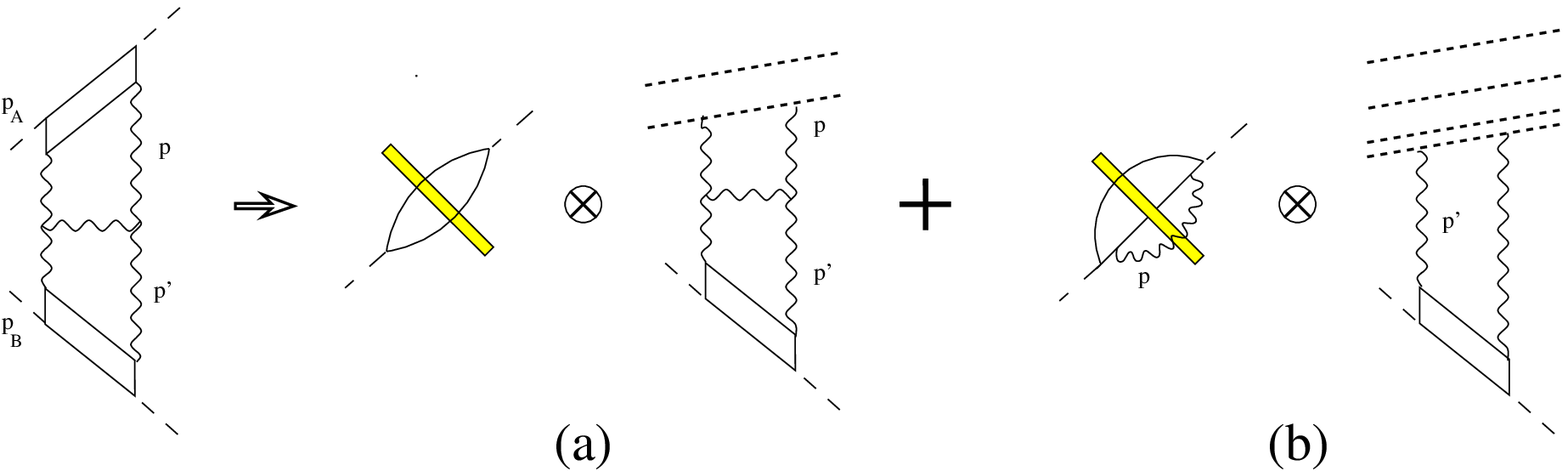}}
\caption{Decomposition into product of coefficient function
and matrix element of the two-Wilson-line operator for a typical Feynman
diagram. (Double Wilson line  corresponds to the fast-moving gluon.)}
\label{figxz}
\end{figure}

Similarly to the case of usual light-cone expansion, we expand the 
amplitude in a set of
``regularized'' Wilson-line operators $\hat{U}^\zeta$ (see Fig.~\ref{figxz}):
\begin{eqnarray}
A(p_A,p_B\Rightarrow p'_A,p'_B)&=&
\label{heope}
\sum\int d^2x_1...d^2x_nC(x_1,...x_n:\zeta) \\
&\times&\langle p_B|{\rm Tr}\{\hat{U}^\zeta(x_1)
\hat{U}^{\dagger\zeta}(x_2)...\hat{U}^\zeta(x_{n-1})
\hat{U}^{\dagger\zeta}(x_n)|p'_B\rangle .
\nonumber
\end{eqnarray}
The coefficient 
functions in front of Wilson-line operators (impact factors) 
will contain logarithms 
$\sim g^2\ln 1/\sigma$ and the matrix elements $\sim g^2\ln {s\sigma\over m^2}$.
 Similar to DIS, when we calculate the amplitude,
 we add the terms $\sim g^2\ln 1/\sigma$ coming from
the coefficient functions (see Fig.~\ref{figxz}b) to the terms $\sim g^2\ln
{\sigma\over m^2/s}$ coming from matrix elements (see Fig.~\ref{figxz}a) 
so
that the dependence on the ``rapidity divide" $\sigma$ cancels resulting 
in the usual high-energy factors  $g^2\ln {s\over m^2}$ which are 
responsible for the BFKL pomeron, cf.~(\ref{2.3.3}). 

In the LLA, the light-like operators $\hat{U}$ and $\hat{U}^{\dagger}$ in 
Eq.~(\ref{3.2.29})
should be replaced by the Wilson-line operators $\hat{U}^\zeta$ and
$\hat{U}^{\dagger\zeta}$ ordered along $n\parallel p_A$. Indeed, let us compare 
the matrix element (\ref{3.8}) shown in Fig.~6b to the corresponding
physical amplitude (\ref{2.2.1}) shown in Fig.~6a.  The
integral in Eq.~(\ref{2.2.1}) is similar to the one for the matrix element of
the operator (\ref{3.8}), except that there is now a factor of the 
upper quark bulb and
the integral over $p_{\perp }$. If we calculate only the contribution of 
the diagram
in Fig.~6a , we would get (cf. Eq.~(\ref{2.2.19}))
\begin{eqnarray}
&\sim&
   i{g^{6}\over 4\pi } \ln \left( {s\over m^{2}} \right)
   \int \! \frac {d^{2}k_{\perp}}{4\pi ^{2}} 
   \frac {d^{2}k'_{\perp}}{4\pi ^{2}}
   \frac {\vec{k}^{2}_{\perp}+\vec{p'}^{2}_{\perp}}{\vec{k}^{4}_{\perp} 
\vec{p'}^{4}_{\perp}} 
   I^{A}(k_{\perp}) I^{B}(k'_{\perp})
\label{3.11}
\end{eqnarray}
which agrees with the with estimate Eq.~(\ref{3.8}), if we set
$\zeta  = {p_{A}^{2}\over s}$.  This corresponds to making the line in the
path-ordered exponential collinear to the momentum of the
photon.

\subsection{One-loop evolution of Wilson-line operators.}

As we demonstrated in previous section, with the LLA accuracy, the
improved version of the factorization formula
Eq.~(\ref{3.2.28})  has the operators
$\hat{U}$ and $\hat{U}^{\dagger}$ ``regularized'' at 
$\zeta  \sim {p_{A}^{2}\over s}$:
\begin{eqnarray}
\lefteqn{
   \int  d^{4}x \int  d^{4}z \,
   \delta (z_{\bullet}) e^{-ip_{A}\cdot x-i(r,z)_{\perp}} 
T\{j_{A}(x+z)j'_{A }(z)\}}\qquad
\label{3.4.1}\\
   &=&
   \sum _{i} e_{i}^{2}  \int  \frac {d^{2}k_{\perp}}{4\pi ^{2}}
   I^{A}(k_{\perp},r_{\perp}){\rm Tr}\{\hat{U}^{\zeta ={m^{2}\over s}}(k)
   \hat{U}^{\dagger \zeta ={m^{2}\over s}}(r-k)\} .
\nonumber
\end{eqnarray}
In the next-to-leading order in $\alpha_s$ we
will have the corrections\\ 
$\sim \alpha_s{\rm Tr}\hat{U}(x_{\perp})\hat{U}^\dagger(y_{\perp})   
{\rm Tr}\hat{U}(y_{\perp})\hat{U}^\dagger(z_{\perp})$, see Fig.~\ref{figxz}.
 
 Next we derive the equation for the evolution of
these operators with respect to slope $\zeta$ (in the LLA). 
 In order to find the behavior of the matrix elements of
the operators 
$\hat{U}^{\zeta}(x_{\perp})\hat{U}^{\dagger \zeta}(y_{\perp})$ on the 
slope $\zeta$ we
must take the
matrix element of this operator ``normalized" at $\zeta_{1}$ and integrate 
over the momenta with 
$\sigma_{1}=\sqrt{m^2\over s\zeta_1}>\alpha>\sigma_{2}=\sqrt{m^2\over s\zeta_2}$ 
(similar to the case of ordinary Wilson OPE
where
in order to find the dependence of the light-cone operator on the normalization
point $\mu $ we integrate over the momenta with virtualities
$\mu _{1}^{2}>p^{2}>\mu _{2}^{2}$). The result will be the operators 
$\hat{U}$ and $\hat{U}^{\dagger}$
``normalized" at the slope $\zeta_{2}$ times the coefficient functions
determining
the kernel of the evolution equation.  
The calculation of the kernel is essentially identical to the
 calculation of the
impact factor with the only difference of having initial gluons instead of
quarks. 
Here we will present only the outline of the calculations; 
the details can be found
in Appendix C.

\begin{figure}[htb]
\centerline{
\epsfxsize=11cm
\epsffile{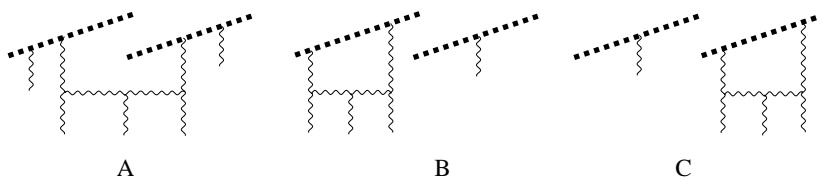}}
\caption{ One-loop diagrams for the evolution of the
   two-Wilson-line operator.} \label{fig:fig8}
\end{figure}

In the first
order in $\alpha _{s}$ there are two one-loop diagrams for the matrix 
element of
operator $\hat{U}(x_{\perp})\hat{U}^{\dagger}(y_{\perp})$ in external field 
(see Fig.~\ref{fig:fig8}). 
This external field is made from slow gluons with $\alpha<\zeta_2$.
Like the case of the fast quark propagator considered above, 
it is convenient to go 
to the rest frame of ``fast'' gluons, as a consequence
the ``slow'' gluons will form a thin pancake. 

 Let us start with the diagram shown in Fig.~\ref{fig:fig8}a. 
We will calculate the
one-loop evolution of the operator  
$\hat{U}(x_{\perp})\otimes\hat{U}^{\dagger}(y_{\perp})\equiv 
\{\hat{U}(x_{\perp})\}^{i}_{j}
\{\hat{U}^{\dagger}(y_{\perp})\}^{k}_{l}$  
with the non-convoluted color indices. In the LLA, the slope 
$p^\zeta$ 
of the operators $U$ can be replaced by $p_1$. 
Using the expression for the axial-gauge gluon propagator in the 
external field (\ref{c7})~\footnote{It can be
demonstrated that further terms in
expansion in
powers of gluon propagator (\ref{c5}) beyond those given 
in Eq.~(\ref{c6}) 
do not contribute in
the LLA.}
 we obtain:
\begin{eqnarray}
&&\hspace{-5mm}\hat{U}(x_{\perp})
\otimes\hat{U}^{\dagger}(y_{\perp})\rangle_A\label{3.4.3}\\
&&\hspace{-3mm}=-ig^2\int du [\infty p_1,up_1]_xt^{a}[up_1,
-\infty p_1]_x\int dv [-\infty p_1,vp_1]_yt^{b}
[vp_1,\infty p_1]_y\nonumber\\
&&\hspace{-5mm}\times~
\hbox{\bf\Big($\!\!$\Big(} up_1+x_{\perp}\Big|(p_{1\xi }-{\cal
P}_{\bullet}\frac {p_{2\xi }}{p\cdot p_{2}})
{\cal O}^{\xi\eta}(p_{1\eta }
-\frac {p_{2\eta }}{p\cdot p_{2}}{\cal
P}_{\bullet})\Big| vp_1+y_{\perp}\hbox{\bf\Big)$\!\!$\Big)}_{ab}.
\nonumber
\end{eqnarray}
Hereafter we use the space-saving notation
\begin{equation}
[un,vn]_x\equiv [un+x_{\perp},vn+x_{\perp}].
\label{savespace}
\end{equation}

We may drop the
terms
proportional to ${\cal P}_{\bullet}$ in the parenthesis since they 
lead to the terms
proportional to the integrals of total derivatives, namely
\begin{eqnarray}
&&\int du [\infty p_1,up_1]t^{a}
[up_1,-\infty p_1]p_{1\mu }(D^{\mu }\Phi
(up_1,...))_{ab}\nonumber\\
&=&\int du\frac {d}{du}\{[\infty p_1,up_1]t^{a}
[up_1,-\infty p_1]
(\Phi (up_1,...))_{ab}\}=0
\label{3.4.4}
\end{eqnarray}
and similar for the total derivative with respect to $v$. Now, we can
rewrite Eq.~(\ref{3.4.3}) in the form
\begin{eqnarray}
&&\hspace{-0.5cm}\langle\hat{U}(x_{\perp})
\otimes\hat{U}^{\dagger}(y_{\perp})\rangle_A~=~
-ig^2\int du [\infty
p_1,up_1]_x t^{a}[up_1,-\infty p_1]_x\label{3.4.5}\\
&\otimes&
\int dv [-\infty p_1,vp_1]_yt^{b}
[vp_1,\infty p_1]_y
\hbox{\bf\Big($\!\!$\Big(}
up_1+x_{\perp} \Big|{\cal O}_{\bullet\bullet}\Big|
vp_1+y_{\perp}\hbox{\bf\Big)$\!\!$\Big)}_{ab} . \nonumber
\end{eqnarray}

\begin{figure}[htb]
\centerline{
\epsfysize=3.7cm
\epsffile{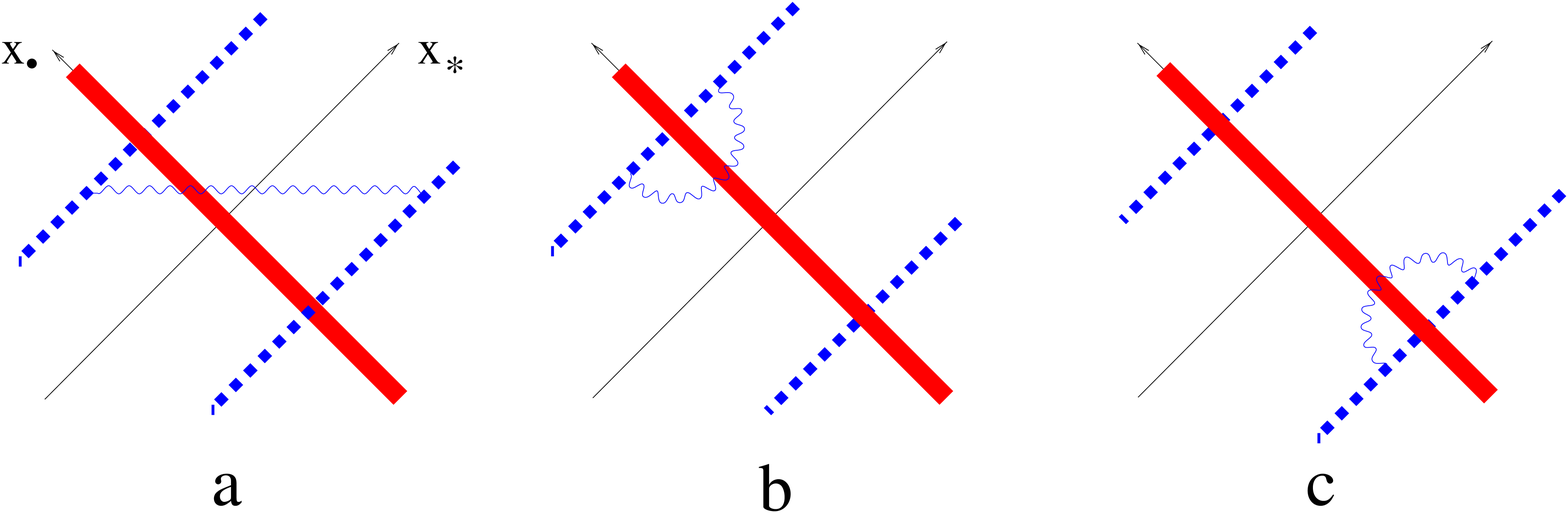}}
\caption{Path integrals describing one-loop diagrams for 
Wilson-line operators in
the shock-wave field background.}\label{ofig18}
\end{figure}

As in the calculation of the quark propagator,  it is convenient to go 
to the rest frame of ``fast'' gluons. In this frame
the ``slow'' gluons will form a thin pancake shown in 
 Fig.~\ref{ofig18}. At first, we consider the case $x_{\ast}>0,~y_{\ast}<0$.
It is clear from the picture that we can rewrite Eq.~(\ref{3.4.5}) 
as follows:
\begin{eqnarray}
\langle\hat{U}(x_{\perp})\hat{U}^{\dagger}(y_{\perp})\rangle_{A}&=&
\label{3.4.6}
-ig^{2}t^{a}U(x_{\perp})\otimes
t^{b}U^{\dagger}(y_{\perp})\\
&\times&\int _{0}^{\infty}\! du\!\int _{-\infty}^{0}\! dv
\hbox{\bf\Big($\!\!$\Big(} up^{(0)}_A+
x_{\perp}\Big|{\cal O}_{\bullet\bullet}\Big| vp^{(0)}_{A}+
y_{\perp}\hbox{\bf\Big)$\!\!$\Big)}_{ab}
\nonumber
\end{eqnarray}
(we shall calculate only the contribution $\sim U$ which comes from the
region $x_{\ast}>0, y_{\ast}<0$ - the term $\sim U^{\dagger}$ coming from 
$x_{\ast}>0, y_{\ast}<0$ is
similar). Technically it is convenient to find at first
the derivative of the integral of gluon propagator in the right-hand side  
of Eq.~(\ref{3.4.6})
with respect to $x_{\perp}$. Using the thin-wall approximation we obtain 
\begin{eqnarray}
\hbox{\bf\Big($\!\!$\Big(} x\Big|
{\cal O}_{\bullet\bullet}\Big| y\hbox{\bf\Big)$\!\!$\Big)}&=&\label{3.4.7}
{s^{2}\over 2}\int \! dz\delta (z_{\ast})
\frac {\ln(x-z)^{2}}{16\pi ^{2}x_{\ast}}\\
&\times& \{2[FF](z_{\perp})-i[DF](z_{\perp})\}
\frac {1}{4\pi ^{2}(z-y)^{2}},\nonumber
\end{eqnarray}
where 
\begin{eqnarray} 
\ [DF](x_{\perp})&\stackrel{\rm def}{\equiv}&\int du
[\infty p_{1}, up_{1}]_x
D^{\alpha}F_{\alpha\bullet }(up_{1}+x_{\perp})[up_{1},-\infty p_{1}]_x ,
\nonumber\\ 
\ [ FF](x_{\perp})&\stackrel{\rm def}{\equiv}&
\int \! du\int \! dv \Theta (u-v) [\infty p_{1}, up_{1}]_x
F^{\xi }_{\bullet }(up_{1}+x_{\perp})\nonumber\\
&\times&[up_{1},vp_{1}]_x
F_{\xi \bullet }(vp_{1}+x_{\perp})[vp_{1}, -\infty p_{1}]_x.
\label{3.4.8}
\end{eqnarray}
It is easy to see that the operators in braces are in
fact the total derivatives of $U$ and $U^{\dagger}$ 
with respect to translations in the
perpendicular directions,
\begin{eqnarray}
\vec{\partial}_{\perp}^{2}U(x_{\perp})&\equiv &
\frac {\partial ^{2}}{\partial x_{i}\partial
x_{i}}U(x_{\perp})=-i[DF](x_{\perp})+2[FF](x_{\perp}),\nonumber\\
\vec{\partial}_{\perp} ^{2}U^{\dagger}(x_{\perp})&\equiv &
\frac {\partial ^{2}}{\partial x_{i}\partial
x_{i}}U^{\dagger}(x_{\perp})=i[DF](x_{\perp})+2[FF](x_{\perp}),
\label{3.4.9}
\end{eqnarray}
(note that $\vec{\partial}_{\perp}^{2}U=-\partial^{2}U$).

For the derivative of the gluon propagator
$(x|p_i{\cal O}|y)$ we obtain:
\begin{eqnarray}
&&\!\!\!\!-ig^{2}\int \! du\!\int \! dv \hbox{\bf\Big($\!\!$\Big(}
up^{(0)}_A+x_{\perp}\Big| p_{i}{\cal O}_{\bullet\bullet}\Big|vp^{(0)}_{A}+
y_{\perp}\hbox{\bf\Big)$\!\!$\Big)}_{ab}\label{3.4.10}\\
&=&{g^{2}\over 16\pi ^{4}}\int \! dz_{\perp}\!\int _{0}^{\infty} 
\frac {du}{u}dv\int \!
dz_{\bullet}\nonumber\\
&\times&\frac {(x_{\perp}-z_{\perp})_{i}
[\vec{\partial}^{2}_{\perp} U(z_{\perp})]_{ab}}{[u(u\zeta
s-2z_{\bullet})-(\vec{x}-\vec{z})^{2}_{\perp}-i\epsilon]
[v(v\zeta s+2z_{\bullet})-
(\vec{y}-\vec{z})^{2}_{\perp}-i\epsilon]} .
\nonumber
\end{eqnarray}
The integration over $z_{\bullet}$ can be performed by taking the residue;
 the result is
\begin{equation}
-i{g^{2}\over 16\pi ^{3}}\int \! dz_{\perp}\int _{0}^{\infty} \frac {du}{u}
dv\frac {(x_{\perp}-z_{\perp})_{i}[\vec{\partial}^{2}_{\perp}
U(z_{\perp})]_{ab}}{[(\vec{x}-\vec{z})^{2}_{\perp}v+(\vec{y}-
\vec{z})^{2}_{\perp}v-uv(u+v)\zeta s+i\epsilon]}.
\label{3.4.11}
\end{equation}
This integral diverges logarithmically when $u\rightarrow 0$ --- in other words
when the emission of quantum gluon occurs in the vicinity of the shock wave.
(Note that if we had done integration by parts, the divergence would be at
$v\rightarrow 0$, therefore there is no asymmetry between $u$ and $v$). 
The size of the
shock wave $z_{\ast}\sim m^{-1}{\sigma_2\over \sigma_1}$ 
(where $1/m$ is the characteristic transverse
size) serves as the lower cutoff for this integration and we obtain
\begin{eqnarray}
&&\!\!\!\!-i{g^{2}\over 16\pi ^{3}}\ln{\sigma_1\over\sigma_2}\int \! dz_{\perp}
\int _{0}^{1} \frac {d\alpha}{\alpha}
\frac {(x_{\perp}-z_{\perp})_{i}[\vec{\partial}^{2}_{\perp}
U(z_{\perp})]_{ab}}{[(\vec{x}-\vec{z})^{2}_{\perp}\bar{\alpha}+
(\vec{y}-\vec{z})^{2}_{\perp}\alpha]}
\nonumber\\
&=&-{g^{2}\over 16\pi ^{3}}\ln{\sigma_1\over\sigma_2}
\hbox{\bf\Big($\!\!$\Big(} x_{\perp}\Big|{p_{i}\over \vec{p}_{\perp}^2}
(\vec{\partial}_{\perp}^{2}U){1\over
\vec{p}_{\perp}^2}\Big| y_{\perp}\hbox{\bf\Big)$\!\!$\Big)}_{ab} ,
\label{3.4.12}
\end{eqnarray}
(recall that $\bar{\alpha}\equiv 1-\alpha$). Thus, 
the contribution of the diagram in 
 Fig.~\ref{ofig18}a  in the
LLA takes the form 
%
\begin{eqnarray}
&&\hspace{-5mm}\langle\hat{U}(x_{\perp})\hat{U}^{\dagger}(y_{\perp})\rangle_{A}
~ = ~ -\left({g^{2}\over 2\pi }\ln{\sigma_1\over\sigma_2}\right) 
\Bigg\{t^{a}U(x_{\perp})\otimes t^bU^{\dagger}(y_{\perp})
\hbox{\bf\Big($\!\!$\Big(} x_{\perp}
\Big|{1\over \vec{p}_{\perp}^2}(\vec{\partial}_{\perp}^{2}U)\nonumber\\
&&\hspace{-5mm}\times
{1\over \vec{p}_{\perp}^2}
\Big| y_{\perp}\hbox{\bf\Big)$\!\!$\Big)}_{ab}+
 U(x_{\perp})t^a\otimes U^{\dagger}(y_{\perp})t^b
\hbox{\bf\Big($\!\!$\Big(} x_{\perp}
\Big|{1\over \vec{p}_{\perp}^2}(\vec{\partial}_{\perp}^{2}U^{\dagger})
{1\over \vec{p}_{\perp}^2}
\Big| y_{\perp}\hbox{\bf\Big)$\!\!$\Big)}_{ab}\Bigg\}.
\label{3.4.13}
\end{eqnarray}
where we have added the term coming from $x_\ast<0, y_\ast>0$.
A corresponding result for the diagram shown in Fig.~\ref{ofig18}b
 can be obtained by comparing 
the space-time picture Fig.~\ref{ofig18}b for this process 
with Fig.~\ref{ofig18}a,
%
\begin{eqnarray}
\langle\hat{U^{\zeta }}(x_{\perp})
\otimes\hat{U^{\dagger\zeta}}(y_{\perp})\rangle_{A}
&=&
\left({g^{2}\over 2\pi }\ln {\sigma_1\over \sigma_2}\right)
U(x_{\perp})\otimes t^aU^{\dagger}(y_{\perp})t^b\nonumber\\
&\times&\hbox{\bf\Big($\!\!$\Big(} y_{\perp}\Big|{1\over \vec{p}_{\perp}^2}
(\vec{\partial}^{2}U){1\over \vec{p}_{\perp}^2}
\Big| y_{\perp}\hbox{\bf\Big)$\!\!$\Big)}_{ab} .
\label{3.4.14}
\end{eqnarray}
Likewise, the diagram in  Fig.~\ref{ofig18}c yields
%
\begin{eqnarray}
\langle\hat{U^{\zeta }}(x_{\perp})
\otimes\hat{U^{\dagger\zeta}}(y_{\perp})\rangle_{A}
&=&
\left({g^{2}\over 2\pi }\ln {\sigma_1\over \sigma_2}\right)
t^{a}U(x_{\perp})t^{b} \otimes U^{\dagger}(y_{\perp})\nonumber\\
&\times&\hbox{\bf\Big($\!\!$\Big(} x_{\perp}\Big|{1\over \vec{p}_{\perp}^2}
(\vec{\partial}^{2}U)
{1\over \vec{p}_{\perp}^2}\Big|
x_{\perp}\hbox{\bf\Big)$\!\!$\Big)}_{ab} . 
\label{3.4.14a}
\end{eqnarray}

The total result for the one-loop evolution of two-Wilson-line
operator  is the sum of Eqs.~(\ref{3.4.13}),  (\ref{3.4.14}),  
and (\ref{3.4.14a}), 
\begin{eqnarray}
&&\langle\{\hat{U}^{\zeta_1}(x_{\perp})\}^{i}_{j}
\{\hat{U}^{\dagger\zeta_1}(y_{\perp})\}^{k}_{l}\rangle_A
={g^2\over 8\pi^3}\ln{\sigma_1\over\sigma_2}\int dz_{\perp}\label{3.4.15}\\
&&\times~
 \left\{ -\left[\{\hat{U}^{\dagger\zeta_2}(z_{\perp})
\hat{U}^{\zeta_2}(x_{\perp})\}^{k}_{j}
\{\hat{U}^{\zeta_2}(z_{\perp})\hat{U}^{\dagger\zeta_2}(y_{\perp})\}^{i}_{l}
\right.\right. \nonumber\\
&&+~ 
\{\hat{U}^{\zeta_2}(x_{\perp})\hat{U}^{\dagger\zeta_2}
(z_{\perp})\}^{i}_{l}\{\hat{U}^{\dagger\zeta_2}
(y_{\perp}) \hat{U}^{\zeta_2}(z_{\perp})\}^{k}_{j} 
\nonumber\\
&&-~\left. 
\delta ^{k}_{j}
\{\hat{U}^{\zeta_2}(x_{\perp})\hat{U}^{\dagger\zeta_2}
(y_{\perp})\}^{i}_{l}-\delta ^{i}_{l}
\{\hat{U}^{\dagger\zeta_2}(y_{\perp})\hat{U}^{\zeta_2}
(x_{\perp})\}^{k}_{j}\right]
\frac {(\vec{x}-\vec{z},\vec{y}-\vec{z})_{\perp}}
{(\vec{x}-\vec{z})^{2}_{\perp}(\vec{y}-\vec{z})^{2}_{\perp}}
\nonumber\\
&&+~ 
\left[\{\hat{U}^{\zeta_2}(z_{\perp})\}^{i}_{j}
{\rm Tr}\{\hat{U}^{\zeta_2}(x_{\perp})\hat{U}^{\dagger\zeta_2}(z_{\perp})\}-
N_{c}\{\hat{U}^{\zeta_2}(x_{\perp})\}^{i}_{j}\right]
\{\hat{U}^{\dagger\zeta_2}(y_{\perp})^{k}_{l}\}
\nonumber\\
&&\times~\frac {1}{(\vec{x}-\vec{z})^{2}_{\perp}}~+~ 
\{\hat{U}^{\zeta_2}(x_{\perp})\}^{i}_{j} \left[\{\hat{U}^{\dagger\zeta_2}
(z_{\perp})\}^{k}_{l}
{\rm Tr}\{\hat{U}^{\zeta_2}(z_{\perp})\hat{U}^{\dagger\zeta_2}(y_{\perp})\}
\right. 
\nonumber\\
&&-~\left.\left.
N_{c}\{\hat{U}^{\dagger\zeta_2}(y_{\perp})\}^{k}_{l}\right]
\frac {1}{(\vec{y}-\vec{z})^{2}_{\perp}}\right\}.
\nonumber
\end{eqnarray}
The evolution of a general $n$-Wilson-line operator is presented in  
Appendix 7.3.\footnote{
A more careful analysis performed in  Appendix shows that the 
Wilson lines $U$ and $U^{\dagger}$ are connected by gauge links at infinity,
see Eq.~(\ref{b18}).
}

\subsection{BFKL pomeron from the evolution of the Wilson-line operators}

As we demonstrated in  Sec.~3.2, with the LLA accuracy the
improved version of the factorization formula
Eq.~(\ref{3.2.28})  has the operators
$U$ and $U^{\dagger}$ ``regularized'' at $\zeta  \sim {p_{A}^{2}\over s}$:
\begin{eqnarray}
\hspace{-2mm}&&\!\!\!\!   \int  d^{4}x \int  d^{4}z \,
   \delta (z_{\bullet}) e^{-ip_{A}\cdot x-i(r,z)_{\perp}} 
T\{j_{A}(x+z)j'_{A }(z)\}\qquad
\label{3.12}\\
\hspace{-2mm}&=&
   \sum _{i} e_{i}^{2}  \int  \frac {d^{2}k_{\perp}}{4\pi ^{2}}
   I^{A}(k_{\perp},r_{\perp}){\rm Tr}\{U^{\zeta ={m^{2}\over s}}(k)
   U^{\dagger \zeta ={m^{2}\over s}}(r-k)\}+O(g^2) .
\nonumber
\end{eqnarray}
In the next-to-leading order in $\alpha_s$ we
will have the corrections \\
$\sim \alpha_s {\rm Tr}U(x_{\perp})U^\dagger(y_{\perp})
{\rm Tr}U(y_{\perp})U^\dagger(z_{\perp})$, see Fig.~\ref{figxz}.
The matrix element of this
operator 
$\langle\!\langle U^{\zeta }(x_{\perp})U^{\dagger \zeta }(y_{\perp})
\rangle\!\rangle$ 
(see Eq.~(\ref{3.2.30}) for the
definition) describes the gluon-photon scattering at large 
energies $\sim s$. (Hereafter we will wipe the label $(\hat{})$ from
the notation of the operators).
The behavior of this matrix element with  energy is determined by the 
dependence on the
``normalization point" $\zeta $.
From the one-loop results for the
evolution of the operators $U$ and $U^{\dagger}$ (\ref{3.4.15})
 it is easy to obtain the following evolution
equation:\cite{ing,yura}\footnote
{The similar non-linear equation describing the
multiplication of pomerons was
suggested in Ref.~38  
and proved in Ref.~39 
in the double-log approximation}
%
\begin{eqnarray}
\zeta 
\frac {\partial }{\partial \zeta } {\cal U}(x_{\perp} ,y_{\perp} )
&=&
-{\alpha_sN_{c}\over 4\pi^{2}}
\!\int \! dz_{\perp} \Big\{{\cal U}(x_{\perp} ,z_{\perp} )+
{\cal U}(z_{\perp} ,y_{\perp} )-{\cal U}(x_{\perp},y_{\perp})
\nonumber\\
&+&
{\cal U}(x,z){\cal U}(z,y)\Big\}
\frac {(\vec{x}-\vec{y})_{\perp}^2}
{(\vec{x}_{\perp}-\vec{z}_{\perp})^{2}
(\vec{z}_{\perp}-\vec{y}_{\perp} )^{2}} ,
\label{master}
\end{eqnarray}
where
\begin{equation}
{\cal U}(x_{\perp},y_{\perp})\equiv
\frac {1}{N_{c}}({\rm Tr}\{U(x_{\perp})[x_{\perp},y_{\perp}]_{-}
U^{\dagger}(y_{\perp})[y_{\perp},x_{\perp}]_{+}\}-N_{c}) 
\label{V}
\end{equation}
(cf. Eq.~(\ref{3.2.27})). Note that right-hand side of this equation is 
both infrared (IR) and 
ultraviolet (UV) finite.\footnote{The IR 
finiteness is due to the fact that ${\rm Tr}{UU^{\dagger}}$ 
corresponds
to the colorless state in t-channel, as a consequence the IR
 divergent parts coming 
from the diagrams in Figs.~\ref{ofig18}a, \ref{ofig18}b, and \ref{ofig18}c 
cancel out. If we had the exchange by color state in t-channel, the 
result will be IR divergent (cf. Eq.~(\ref{2.5.2})).
} 
We see
that as a result of the evolution, the two-line operator
${\rm Tr}\{UU^{\dagger}\}$ is the same operator (times the kernel) plus the
 four-line
operator ${\rm Tr}\{UU^{\dagger}\}{\rm Tr}\{UU^{\dagger}\}$. The result of 
the evolution of the four-line
operator will be the same operator times some kernel plus the six-line operator
of the type 
${\rm Tr}\{UU^{\dagger}\}{\rm Tr}\{UU^{\dagger}\}{\rm Tr}\{UU^{\dagger}\}+
{\rm Tr}\{UU^{\dagger} UU^{\dagger}\}{\rm Tr}\{UU^{\dagger}\}$ and
so on. Therefore it is instructive to consider at first the linearization of  
the
Eq.~(\ref{master}) with the number of operators $U$  conserved
during the evolution.

The linear evolution
of the  two-line operator ${\cal U}(x_{\perp},y_{\perp})$ is governed by the
BFKL equation~\footnote{If $F(k_{\perp},r_{\perp})$ satisfies 
the BFKL Eq.~(\ref{2.4.9}) then 
\begin{eqnarray}
{\cal U}(x_{\perp},y_{\perp})&=&\label{svyaz}
\int dk_{\perp} dr_{\perp} 
e^{i(\vec{k},\vec{x})_{\perp}+i(\vec{r}-\vec{k},\vec{y})_{\perp}}\\
&\times&\Big( 
{F(k_{\perp},r_{\perp})\over \vec{k}_{\perp}^2
(\vec{r}-\vec{k})_{\perp}^2}-
{1\over 2}[\delta(k_{\perp})+\delta(r_{\perp}-k_{\perp})]
\int dk'_{\perp} 
{F(k'_{\perp},r_{\perp})\over
(\vec{k}')_{\perp}^2(\vec{r}-\vec{k}')_{\perp}^2}\Big). \nonumber
\end{eqnarray}
}
\begin{eqnarray}
&&\!\!\!\!\zeta \frac {\partial }{\partial \zeta } {\cal U}(x_{\perp},y_{\perp})
\label{5.4}\\
&&\!\!\!\!=
-{\alpha_{s}\over 4\pi
^{2}}
N_{c}\int \!
 dz_{\perp}\left\{{\cal U}(x_{\perp},z_{\perp})+
{\cal U}(z_{\perp},y_{\perp})-
{\cal U}(x_{\perp},y_{\perp})\right\}
{(\vec{x}-\vec{y})_{\perp}^2
\over (\vec{x}-\vec{z})_{\perp}^2(\vec{z}-\vec{y})_{\perp}^2}.
\nonumber
\end{eqnarray}
Let us start from the simplest case of forward matrix elements 
(which describes, for example, the small-x DIS from the virtual photon).  
Then  the equation (\ref{5.4}) takes the form
%
\begin{equation}
\zeta \frac {\partial }{\partial \zeta } 
\langle\!\langle {\cal U}(x_{\perp})\rangle\!\rangle =
-{\alpha_{s}\over 4\pi
^{2}}
N_{c}\int \!
 dz_{\perp}[{\cal U}(x-z_{\perp})+
{\cal U}(z_{\perp})-{\cal U}(x_{\perp})]
{\vec{x}_{\perp}^2\over (\vec{x}-\vec{z})_{\perp}^2\vec{z}_{\perp}^2},
\label{5.4a}
\end{equation}
where $\langle\!\langle {\cal U}(x_{\perp})\rangle\!\rangle\equiv 
\langle\!\langle {\cal U}(x_{\perp},0)\rangle\!\rangle$ 
(see Eq.~(\ref{3.2.30})).
The eigenfunctions of this
equation
are powers $(x_{\perp}^{2})^{-{1\over 2}+i\nu }$ and the eigenvalues are
$-{\alpha _{s}\over\pi }N_{c}\chi (\nu )$, where
$\chi (\nu )=-{\rm Re}\psi ({1\over 2}+i\nu )-C$. Therefore, the 
evolution of the
operator ${\cal U}$ takes the form:
%
\begin{eqnarray}
\langle\!\langle {\cal U}^{\zeta _{1}}(x_{\perp})\rangle\!\rangle&=&
\int \! \frac {d\nu }{2\pi ^{2}}
(\vec{x}_{\perp}^{2})^{{1\over 2}+i\nu }
\left(\frac {\zeta _{1}}{\zeta _{2}}\right)
^{-{\alpha _{s}\over\pi }N_{c}\chi
(\nu )}\nonumber\\
&\times&\int \! dz_{\perp}(\vec{z}_{\perp}^{2})^{-{3\over 2}-i\nu }
\langle\!\langle {\cal U}^{\zeta _{2}}(z)\rangle\!\rangle
\label{5.5}
\end{eqnarray} 

We may proceed with this evolution as long as the upper limit of our
logarithmic integrals over $\alpha$, $\sqrt {p_{A}^{2}\over\zeta s}$, is
much
larger than the lower limit ${p_B^2\over s}$ determined by the 
lower quark bulb,
see the discussion in Sec.~3.3.  It is
convenient to
stop evolution at a certain point $\zeta _{0}$ such as
\begin{equation}
\zeta _{0}=\sigma^{2}{s\over m^{2}},~~~\sigma\ll 1,~~~~
g^{2}\ln\sigma\ll 1,
\label{5.6}
\end{equation}
then the relative energy between the Wilson-line operator 
${\cal U}^{\zeta _{0}}$ and
lower virtual photon will be $s_{0}=m^{2}\sigma^{2}$ which is big 
enough to apply
our
usual high-energy approximations (such as pure gluon exchange 
and substitution
$g_{\mu \nu }\rightarrow {2\over s_{0}}p_{2\mu }p_{1\nu }$) 
but small in a sense that
one does not need take into account the difference between 
$g^{2}\ln{s\over m^{2}}$
and $g^{2}\ln{s\over m^{2}\sigma^{2}}$. Finally, the evolution 
(\ref{5.4})
takes the
form:
%
\begin{equation}
\langle\!\langle {\cal U}^{\zeta ={m^2\over s}}(x_{\perp})\rangle\!\rangle=
\! \int \! \frac {d\nu }{2\pi ^{2}}
(x_{\perp}^{2})^{{1\over 2}+i\nu }
(\frac {s}{m^{2}})^{{2\alpha _{s}\over\pi }N_{c}\chi (\nu )}
\! \int \! dz_{\perp}(z_{\perp}^{2})^{-{3\over 2}-i\nu }
\langle\!\langle {\cal U}^{\zeta _{0}}(z_{\perp})\rangle\!\rangle .
\label{5.7}
\end{equation}
Now let us rewrite this evolution in terms of original operators 
$UU^{\dagger}$ in the
momentum representation. One obtains:
%
\begin{eqnarray}
&&\langle\!\langle {\rm Tr}\{U^{\zeta ={m^2\over s}}(p_{\perp})
U^{\dagger \zeta ={m^2\over s}}(-p_{\perp})\}\rangle\!\rangle
~=~\int \!
\frac {d\nu }{2\pi ^{2}}
(\vec{p}_{\perp}^2)^{-{3\over 2}-i\nu }\label{5.8}\\
&&~~~~~~~~~~\times\Big(\frac {s}{m^{2}}\Big)^
{{2\alpha _{s}\over\pi }N_{c}\chi (\nu )}
\int \! dp'_{\perp}(\vec{p'}_{\perp}^2)^{{1\over 2}+i\nu }
\langle\!\langle {\rm Tr}\{U^{\zeta _{0}}(p'_{\perp})
U^{\dagger \zeta _{0}}(-p'_{\perp})\}\rangle\!\rangle
\nonumber
\end{eqnarray}
where we omit the gauge links at infinity (\ref{endfactor}) for brevity. 
Since we
neglect the logarithmic corrections $\sim g^{2}\ln\sigma$ the matrix element
of our operator $U^{\zeta _{0}}U^{\dagger\zeta_{0}}$ coincides with impact
factor $I^{B}$ up to $O(g^{2})$ corrections:
%
\begin{eqnarray}
\lefteqn{\langle\!\langle {\rm Tr}\{U^{\zeta _{0}}(p_{\perp})
U^{\dagger \zeta _{0}}(-p_{\perp})\}\rangle\!\rangle}\label{5.9}\\
&=&g^{4}{N_{c}^{2}-1\over
2}\sum e_{i}^{2}\int \frac {d\alpha}{\pi s}
\frac {\Phi ^{B}(\alpha_{p}p_{1}-\zeta
_{0}\alpha_{p}p_{2}+p_{\perp})}{ (\zeta _{0}\alpha_{p}^{2}+
\vec{p}_{\perp}^2)^{2}}\nonumber\\
&=&g^{4}{N_{c}^{2}-1\over 2}\sum e_{i}^{2}\left({1\over \vec{p}_{\perp}^4}
I^{B}(p_{\perp})
-\delta(p_{\perp})\int dp'_{\perp}{1\over \vec{p'}_{\perp}^4}
I^B(p'_{\perp})\right).
\nonumber
\end{eqnarray}
Combining Eqs.~(\ref{3.2.29}), (\ref{5.8}), and (\ref{5.9}) we reproduce 
the leading logarithmic result for virtual 
$\gamma \gamma $ scattering (\ref{2.3.12}).

In the case of small-x DIS from the nucleon the matrix element of 
the operator $UU^{\dagger}$ 
describes the propagation of 
the ``color dipole''\cite{kop} in the nucleon. The evolution of
the
matrix element $\mbox{$\langle N|$} {\cal U}\mbox{$|N\rangle $}$ is the same as
Eq.~(\ref{5.8}) with the only difference that the lower impact factor $I^{B}$
should be substituted by the nucleon impact factor $I^{N}$ determined by the
matrix element of the operator $UU^{\dagger}$ between the nucleon 
states:\footnote{This is called ``hard pomeron" contribution 
to the structure functions
of DIS since the transverse momenta in our loop integrals are large ($\sim
Q^{2}$), at least in the lowest orders in perturbation theory. 
However, due to the 
diffusion in transverse momenta the characteristic size of the 
$\vec{p}_{\perp}^2$ 
in the
middle of gluon ladder is $Q^2e^{-\sqrt {g^{2}\ln s}}$ 
(see the discussion in Sec.~2.4), so at
very
small $x$ the region $p_{\perp}\sim \Lambda _{QCD}$ may 
become important. It
corresponds to the contribution of the  ``soft"  pomeron 
which
is constructed from non-perturbative gluons in our language 
and must be added
to the hard-pomeron result.}
\begin{equation}
\langle N,p_{B}|{\rm
Tr}\{U^{\zeta_0}(x_{\perp})U^{\dagger\zeta_0}(0)\}|N,p_{B}+\beta p_{2}
\rangle=2\pi \delta (\beta)\int \!\frac {dp_{\perp}}{4\pi^2}
e^{i(px)_{\perp}}{1\over \vec{p}_{\perp}^4}I^{N}(p_{\perp}) 
\label{5.12}
\end{equation}
where $2\pi \delta (\beta)$ reflects the fact that matrix 
element of the 
operator
$UU^{\dagger}$ contains
unrestricted integration along $p^{\zeta _{0}}$, 
(cf. Eq.~(\ref{3.2.30})). The
nucleon impact factor $I^{B}(p_{\perp})$ defined in (\ref{5.12}) is a
phenomenological
low-energy characteristic of the nucleon. 
In the BFKL evolution it plays a
role similar to that of a nucleon structure function at 
low normalization point
for DGLAP evolution. In principle, it can be estimated 
using QCD sum rules or
phenomenological models of nucleon.

In conclusion, let us present the results for the linear 
evolution for the
non-forward case. Due to the conformal invariance 
of the tree-level QCD 
the eigenfunctions of the equation (\ref{5.4}) are 
powers~\cite{lip86}
\begin{equation}
\left({(\vec{x}-\vec{y})_{\perp}^2\over 
(\vec{x}-\vec{x}_0)_{\perp}^2(\vec{y}-\vec{x}_0)_{\perp}^2}\right)^
{{1\over 2}+i\nu}
\label{xyuz}
\end{equation}
where $x_0$ is arbitrary. The eigenvalues are the same as for 
the forward case, $-{\alpha_s\over\pi}N_c\chi(\nu)$. 
The corresponding formula for
the result of the evolution of the two-Wilson-line operator has 
the form:
\begin{eqnarray}
{\cal U}^{\zeta_1}(x_{\perp},y_{\perp})
&=&\int d\nu d^2x_0{\nu^2\over\pi^4}
\left({(\vec{x}-\vec{y})_{\perp}^2\over
(\vec{x}-\vec{x}_0)_{\perp}^2
(\vec{y}-\vec{x}_0)_{\perp}^2}\right)^{{1\over
2}-i\nu} \nonumber\\
&\times&\left({\zeta_1\over\zeta_2}\right)^{-
{\alpha_s\over\pi} N_c\chi(\nu))}{\cal U}^{\zeta_2}(x_0,\nu)
\label{xzxz}
\end{eqnarray}
where
\begin{equation}
{\cal U}^{\zeta}(x,\nu)~\equiv~\int dx'\int dy'
{1\over (\vec{x}'-\vec{y}')_{\perp}^4}
\left({(x'-y')^2\over (\vec{x}'-\vec{x})_{\perp}^2
(\vec{y}'-\vec{x})_{\perp}^2}\right)^{{1\over 2}+i\nu}
{\cal U}^{\zeta}(x'_{\perp},y'_{\perp})
\label{unfwd}
\end{equation}

It is worth noting that at large momentum transfers 
$-t=\vec{r}_{\perp}^2\gg m_N^2$ the nucleon impact factor is determined by 
the well-studied  electric and magnetic form factors of the nucleon
\begin{equation}
I_N(k_{\perp}, r_{\perp})\stackrel{\vec{k}_{\perp}^2\gg m^2}{=}
\delta_{\lambda\lambda'}F_1^{p+n}(t)+
{1\over 2ms}\bar{u}(p',\lambda')\not\! p_1\!\not\! r_{\perp} 
u(p,\lambda)F_2^{p+n}(t)
\label{dvcs28},
\end{equation}
which gives an opportunity to calculate the amplitude of 
deeply virtual Compton scattering from the nucleon at small $x$
without any model assumptions.\cite{bkucha}

\subsection{Non-linear evolution of Wilson lines}

\begin{figure}[htb]
\centerline{
\epsfysize=5cm
\epsffile{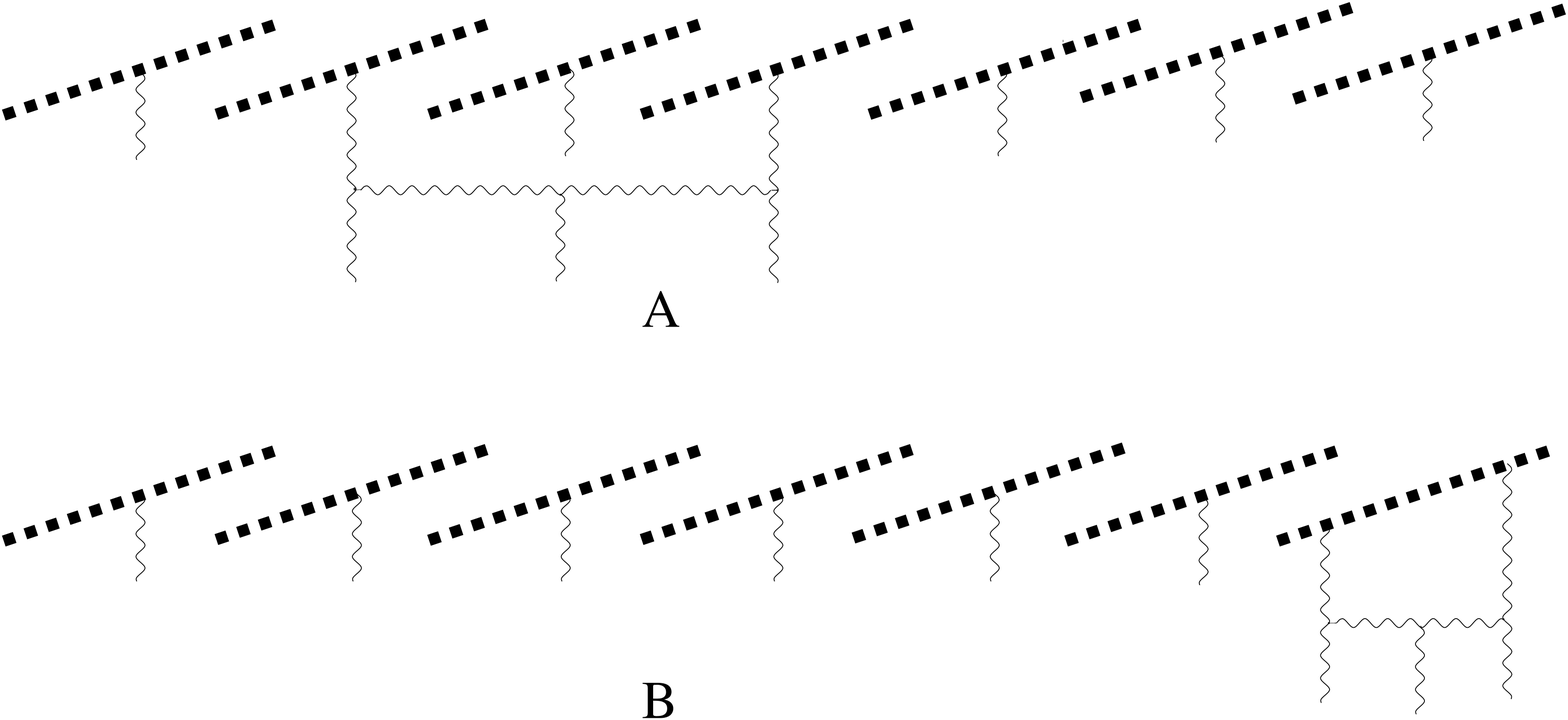}}
\caption{Typical diagrams for the one-loop 
evolution of the $n$-line operator.} \label{ofig19}
\end{figure}

Unlike the linear evolution, the general picture  is
very
complicated: not only the number of operators $U$ and $U^{\dagger }$ 
increase after each
evolution but they form increasingly complicated structures like 
those displayed in Eq.~(\ref{5.23}) below.
In the leading log approximation the evolution of the $2n$-line
operators such as
${\rm Tr}\{UU^{\dagger }\}{\rm Tr}\{UU^{\dagger }\}...
{\rm Tr}\{UU^{\dagger }\}$ comes from either self-interaction
diagrams  or from the pair-interactions ones (see Fig.~\ref{ofig19})
The one-loop evolution equations for these operators can be constructed using
the pair-wise kernels calculated in the Appendix C. 
For instance, the evolution equation for the
four-line operator appearing in the right-hand side 
of Eq.~(\ref{master}) has the form:
\begin{eqnarray}
&&\zeta \frac {\partial }{\partial \zeta }
{\rm Tr}\{U_{x}[x,z]_{-}U^{\dagger}_{z}[z,x]_{+}\}
{\rm Tr}\{U_{z}[z,y]_{-}
U^{\dagger}_{y}[y,z]_{+}\}\label{5.22}\\
&=&-{g^{2}\over 16\pi ^{3}}
\int \! dt_{\perp} \Bigg\{\Big[{\rm Tr}\{U_{x}[x,t]_{-}
U^{\dagger}_{t}[t,x]_{+}\} 
{\rm Tr}\{U_{t}[t,z]_{-}U^{\dagger}_{z}[z,t]_{+}\}\nonumber\\ 
&-&
N_{c}{\rm Tr}\{U_{x}[x,z]_{-}
U^{\dagger}_{z}[z,x]_{+}\}\Big]{\rm Tr}\{U_{z}[z,y]_{-}
U^{\dagger}_{y}[y,z]_{+}\}
\frac {(\vec{x}-\vec{z})_{\perp}^2}{(\vec{x}-\vec{t})_{\perp}^2
(\vec{z}-\vec{t})_{\perp}^2}\nonumber\\
&+&{\rm Tr}\{U_{x}[x,z]_{-}
U^{\dagger}_{z}[z,x]_{+}\}
\frac {(\vec{y}-\vec{z})_{\perp}^2}{(y-t)_{\perp}^2
(\vec{z}-\vec{t})_{\perp}^2}\nonumber\\
&\times&\Big[{\rm Tr}\{U_{z}[z,t]_{-}U^{\dagger}_{t}[t,z]_{+}\}
{\rm Tr}\{U_{t}[t,y]_{-}U^{\dagger}_{y}[y,t]_{+}\} \nonumber\\
&-&
N_{c}{\rm Tr}\{U_{z}[z,y]_{-}U^{\dagger}_{y}[y,z]_{+}\}\Big]
\nonumber\\
&+&
\Big[{\rm Tr}\{U_{x}[x,z]_{-}U^{\dagger}_{z}[z,t]_{+}U_{t}[t,y]_{-}
U^{\dagger}_{y}[y,z]_{+}
U_{z}[z,y]_{-}U^{\dagger}_{t}[t,x]_{+}\} \nonumber\\
&+&
{\rm Tr}\{U_{x}[x,t]_{-}U^{\dagger}_{t}[t,z]_{+}U_{z}[z,y]_{-}
U^{\dagger}_{y}[y,t]_{+}
U_{t}[t,z]_{-}U^{\dagger}_{z}[z,x]_{+}\}\nonumber\\
&-&
2{\rm Tr}\{U_{x}[x,y]_{-}U^{\dagger}_{y}\}[y,x]_{+}\Big]\nonumber\\
&\times&
\left[-\frac {(\vec{x}-\vec{t},\vec{y}-\vec{t})_{\perp}}
{(\vec{x}-\vec{t})_{\perp}^2(y-t)_{\perp}^2}-
\frac {1}{(\vec{z}-\vec{t})_{\perp}^2}\right.\nonumber\\
&+&\left.
\frac {(\vec{x}-\vec{t},\vec{z}-\vec{t})_{\perp}}
{(\vec{x}-\vec{t})_{\perp}^2(\vec{z}-\vec{t})_{\perp}^2}+
\frac {(\vec{z}-\vec{t},\vec{y}-\vec{t})_{\perp}}
{(\vec{z}-\vec{t})_{\perp}^2(\vec{y}-\vec{t})_{\perp}^2}\right]\Bigg\} ,
\nonumber
\end{eqnarray}
where we have displayed the end gauge links (\ref{b18}) explicitly. Note
that each of the separate contributions (\ref{5.20}) and (\ref{5.21})
corresponding to the diagrams in Fig.~\ref{ofig35}a and
\ref{ofig35}b diverges at large $t$ while the
total result (\ref{5.22}) is convergent. This is the 
usual cancellation of the IR divergent contributions between the
emission of the real
(Fig.~\ref{ofig35}a) and virtual
(Fig.~\ref{ofig35}b) gluons from the colorless
object (corresponding to the l.h.s. of Eq.~(\ref{5.22}))
(cf Eq.~(\ref{master})).

Thus, the result of the evolution of the operator 
in the right-hand side  of Eq.~(\ref{3.4.1})
has a generic form:
\begin{eqnarray}
&&{\rm Tr}\{U^{\zeta}_{x}[x,y]_{-}
U^{\dagger\zeta}_{y}[y,x]_{+}\}
\Rightarrow \sum
_{n=0}^{\infty}(\alpha_s\ln\frac {  \zeta }{\zeta _{0}})^{n}
\int dz^{1}dz^{2}...dz^{n}\nonumber\\
&\times&
\Big[A_{n}(x,z^{1},z^{2},...z^{n},y)
{\rm Tr}\{U^{\zeta _{0}}_{x}[x,1]_{-}
U^{\dagger \zeta _{0}}_{1}[1,x]_{+}\}\nonumber\\
&\times&{\rm Tr}\{U^{\zeta _{0}}_{1}[1,2]_{-}
U^{\dagger \zeta _{0}}_{2}[2,1]_{+}\} \dots
{\rm Tr}\{U^{\zeta _{0}}_{n}[n,y]_{-}U^{\dagger \zeta _{0}}_{y}
[y,n]_{+}\}\nonumber\\
&+&B_{n}(x,z^{1},z^{2},...z^{n},y)\nonumber\\
&\times&{\rm Tr}\{U^{\zeta _{0}}_{x}[x,1]_{-}
U^{\dagger \zeta _{0}}_{1}[1,2]_{+}U^{\zeta _{0}}_{2}[2,3]_{-}
U^{\dagger \zeta _{0}}_{3}[3,1]_{+}U^{\zeta _{0}}_{1}[1,2]_{-}
U^{\dagger \zeta _{0}}_{2}[2,x]_{+}\}\nonumber\\
&\times&{\rm Tr}\{U^{\zeta _{0}}_{3}[3,4]_{-}
U^{\dagger \zeta _{0}}_{4}
[4,3]_{+}\}...
{\rm Tr}\{U^{\zeta _{0}}_{n}[n,y]_{-}
U^{\dagger \zeta _{0}}_{y}[y,n]_{+}\}+...
\nonumber\\
&+&
N_{c}^{n} C_{n}(x,z^{1},z^{2},...z^{n},y;)
{\rm Tr}\{U^{\zeta _{0}}_{x}
[x,y]_{-}
U^{\dagger \zeta _{0}}_{y}[y,x]_{+}\}\Big] ,
\label{5.23}
\end{eqnarray}
where $U^{(\dagger )}_{n}\equiv U^{(\dagger )}(z^{n}_{\perp})$,
$[i,j]\equiv [x_{i},x_{j}]$ and\\ 
$A_{n}(x,z^{1},z^{2},\ldots,z^{n},y)$, 
$B_{n}(x,z^{1},z^{2},\ldots,z^{n},y)$, 
$\ldots$, $C_{n}(x,z^{1},z^{2},\ldots,z^{n},y)$
are the meromorphic functions that can be obtained by using the
Eqs.(\ref{5.20},\ref{5.21}) $n$ times which give us a sort of 
Feynman rules for
calculation of these coefficient functions. If we now evolve 
our operators from
$\zeta \sim {p_A^2\over s}$ to $\zeta _{0}$ given by Eq.~(\ref{5.6}) we 
shall obtain a
series
(\ref{5.23}) of matrix elements of the 
operators $(U)^{n}(U^{\dagger })^{n}$ normalized at
$\zeta _{0}$.
These matrix elements correspond to small energy $\sim m^{2}$ 
and they can be
calculated either perturbatively (in the case the ``virtual
photon" matrix element ) or
using some model calculations such as QCD sum rules in the case 
of nucleon
matrix element corresponding to small-$x$ $\gamma ^{\ast}p$ DIS . 
It should be
mentioned that in the case of virtual photon scattering considered above 
we can
calculate the matrix elements of operators $UU^{\dagger }...UU^{\dagger }$ 
perturbatively. Because $U=1+ig\int \!A_{\mu }dx_{\mu }+...$, in the leading
order in $\alpha_{s}$ we can replace by 1 all but two $U(U^{\dagger})$'s, 
so we return to the BFKL picture
describing the evolution of the two operators $UU^{\dagger}$. 
The non-linear equation (\ref{master}) enters the game in the
situation like small-x DIS from a nucleon or nucleus when the matrix 
elements of the
operators $UU^{\dagger }...UU^{\dagger }$ are non-perturbative, consequently 
there is no reason to expect that extra $U$ and $U^{\dagger }$ will 
lead to extra smallness.
In this case, at the low ``normalization point" $\zeta _{0}$ one 
must take into
account the whole series of the operators in the right-hand side 
of Eq.~(\ref{5.23}),
indicating the need for all the coefficients $a_{n},b_{n}...c_{n}$.
Recently, these coefficients were calculated by Y. Kovchegov~\cite{yura}
for the case of
DIS from the large nuclei in the McLerran-Venugopalan model, and the 
results indicate that the non-linear equation (\ref{master}) leads to 
unitarization of the pomeron in this case.\cite{yura}

The zoo of different Wilson-line operators (\ref{5.23})
may be reduceded by using the
dipole picture.\cite{mu94,nnn}
Technically, it arises when in each order in
$\alpha _{s}\ln(\frac {\zeta }{\zeta _{0}})$ we keep only the term
${\rm Tr}\{U^{\zeta _{0}}_{x}U^{\dagger \zeta _{0}}_{1}\}
{\rm Tr}\{U^{\zeta _{0}}_{1}U^{\dagger \zeta _{0}}_{2}\} ... 
{\rm Tr}\{U^{\zeta
_{0}}_{n}U^{\dagger \zeta _{0}}_{y}\}$-subtractions~\footnote{By
``subtractions" we mean this operator with some of
the ${\rm Tr}\{U_{k}U^{\dagger}_{k+1}\}$ substituted by $N_{c}$.}  
in right-hand side of Eq.~(\ref{5.23});
for example, in Eq.~(\ref{5.22}) we keep the two first terms and disregard the
third one. In other words, we take into account
only those diagrams in Fig.~\ref{ofig35}
which connect the Wilson lines belonging to the same 
${\rm Tr}\{U_{k}U^{\dagger}_{k+1}\}$. (This corresponds to the virtual photon 
wave function in the large-$N_c$ approximation). The diagrams 
of the corresponding effective theory are obtained by
multiple
iteration of Eq.~(\ref{master}) and give a picture  where each ``dipole"
${\rm Tr}\{U_{k}U^{\dagger}_{k+1}\}$ can create two dipoles 
according to Eq.~(\ref{master}). The
motivation of this approximation is given in Refs.~24, 25,  
and the discussion of unitarization of the BFKL pomeron in the dipole picture
is presented in Ref.~41. 

\subsection{Operator expansion for diffractive high-energy scattering}

The nonlinear term in the equation (\ref{master})
describes the triple vertex of hard pomerons in QCD.
In order to see that, it is convenient to consider some process
which is dominated by the three-pomeron vertex  --- the best
example is the diffractive dissociation of the virtual photon.

The relevant operator expansion for diffractive scattering is obtained by
direct generalization  of our approach to the 
diffractive processes.\cite{difope} The
total cross section for  diffractive scattering has the form:
\begin{equation}
\sigma^{\rm diff}_{\rm tot}=\int dx e^{iqx}\int {d^3p'\over(2\pi)^3}\sum_X
\langle p|j_{\mu}(x)|p'+X\rangle\langle p'+X|j_{\nu}(0)|p\rangle ,
\label{1.12}
\end{equation}
$p$ and $p'$ are the nucleon momenta and $\sum_X$ means the summation over
all the intermediate states. We can formally write down this cross section 
as a ``diffractive matrix element" (cf. Ref.~43): 
\begin{eqnarray}
\sigma^{\rm diff}_{\rm tot}&=&W^{\rm diff}_{\mu\nu}
\stackrel{\rm def}{\equiv}\int dxe^{iqx}
\langle p|T\{j^-_{\mu}(x)j^+_{\nu}(0)\}|p\rangle ,
\label{1.13}
\end{eqnarray}
where~\footnote{The difference between 
Eq.~(\ref{1.12}) and the last line in
Eq.~(\ref{1.14}) is that $j$'s are Heisenberg operators in (\ref{1.12})
while in Eq.~(\ref{1.14}) the operators stand in the 
interaction representation}
\begin{eqnarray}
&&\!\!\!\! \langle p|T\{j^-_{\mu}j^+_{\nu}
e^{i\int dz({\cal L}^+(z)-{\cal L}^-(z))}\}|p\rangle \label{1.14}\\
&\stackrel{\rm def}{\equiv}&\int {d^3p'\over(2\pi)^3}\sum_X
\langle p|\tilde{T}\{j_{\mu}(x)e^{-i\int dz{\cal L}(z)}\}|p'+X\rangle 
\nonumber\\
&\times&\langle p'+X|T\{j_{\nu}(0)e^{i\int dz{\cal L}(z)}\}|p\rangle .
\nonumber
\end{eqnarray}
The superscript ``--'' marks the fields to the left of the cut and $+$ to the
right. The definition of the T-product of the 
fields with $\pm$ labels is as follows: the $+$ fields are time-ordered, 
the $-$ fields stand 
in inverse time order (since 
they correspond to the complex conjugate amplitude), and $-$
fields stand always to the left of the $+$ ones. Therefore, the
diagram technique with the double set of fields is 
the following: contraction of two $+$ fields is the usual 
Feynman propagator ${\not\! p\over p^2+i\epsilon}$ (for the quark field), 
contraction of two $-$ 
fields is the complex
conjugated propagator ${\not\! p\over p^2-i\epsilon}$, and the contraction of 
the $-$ field with the $+$ one is the ``cut propagator" 
$2\pi\delta(p^2)\theta(p_0)\not\! p$.\footnote{We will use 
the $-+$ perturbative propagator only for hard momenta, 
hence the additional emitted nucleon with momentum p' 
(constructed from soft quarks) can be  factorized
\begin{eqnarray}
&&\!\!\!\! \sum_X\mbox{$\langle 0|$}\psi(x)|p'+X\rangle\langle p'+X|
\bar{\psi}(0)\mbox{$|0\rangle $}\label{1.15}\\
&\simeq&\sum_X\mbox{$\langle 0|$}\psi(x)|X\rangle\langle X|
\bar{\psi}(0)\mbox{$|0\rangle $}
\otimes|p'\rangle\langle p'|
=\int{d^4p\over (2\pi)^4i}
\not\! p 2\pi\delta(p^2)\theta(p_0)\otimes|p'\rangle\langle p'| .
\nonumber
\end{eqnarray}} 
This diagram technique for calculating T-products of double set of 
fields exactly reproduces the Cutkosky rules for calculation 
of cross sections. The light-cone expansion of the diffractive matrix
element (\ref{1.13}) gives operator definition of the diffractive parton
distributions.\cite{diffdist}

Let us discuss the high-energy operator expansion for the
diffractive amplitude $W^{\rm diff}_{\mu\nu}$. Similarly to the case
of usual amplitude (\ref{3.2.29}), we get in the lowest order in $\alpha_s$
\begin{eqnarray}
W^{\rm diff}&=&
\sum_{{\rm flavors}} \! e_{i}^{2} \int {d^{2}k_{\perp}\over 4\pi^2}
   I^A(k_{\perp},0)\nonumber\\
&\times&\langle N|{\rm Tr}\{W^{\zeta=m^2/s}(k_{\perp})
W^{\dagger,\zeta=m^2/s}(-k_{\perp})\} |N\rangle,
\label{1.16}
\end{eqnarray}
where $W(k_{\perp})$ is a Fourier transform of
\begin{equation}
W(x_{\perp})=V^{\dagger}(x_{\perp})U(x_{\perp}),~~~W^{\dagger}(x_{\perp})=
U^{\dagger}(x_{\perp})V(x_{\perp}) .
\label{1.17}
\end{equation}
Here $U(x_{\perp})$ denotes the  Wilson-line operator 
constructed from $+$ fields and $V(x_{\perp})$ denotes the 
same operator 
constructed from $-$ fields:
\begin{equation}
U^{\zeta} (x_{\perp} )=[\infty p_1+x_{\perp},-\infty p_1
+x_{\perp}]^+,~~~V(x_{\perp})= 
[\infty e+x_{\perp} ,-\infty e +x_{\perp} ]^-.
\label{1.18}
\end{equation}
After integration over fast quarks, the slope of the Wilson lines 
is $\zeta=x_B\equiv{Q^2\over s}$, see the discussion in Sec.~3.3. 

\begin{figure}[htb]
\centerline{
\epsfysize=8cm
\epsffile{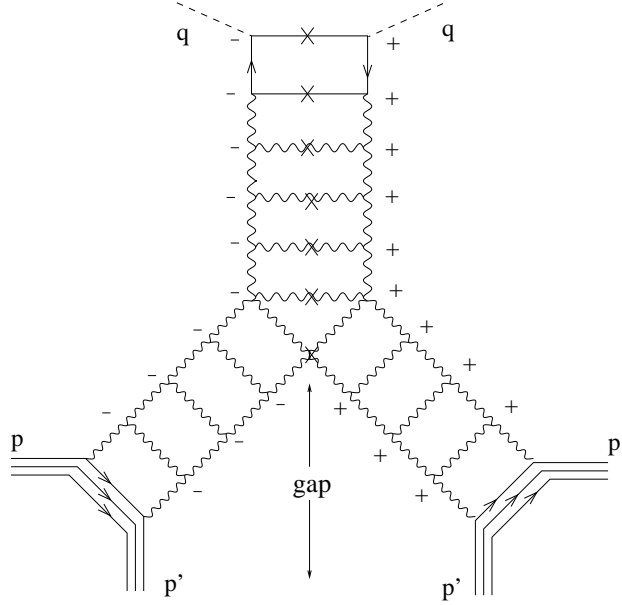}}
\caption{Amplitude of diffractive scattering in the LLA-$N_c$
approximation.} \label{ofig20}
\end{figure}

The evolution equation (with respect to the slope of the 
supporting line)
turns out to have the same form as Eq.~(\ref{master}) for non-diffractive
amplitudes: 
\begin{eqnarray}
&&\zeta \frac{d}{d\zeta} {\cal W}(x_{\perp},y_{\perp})=
\label{dmaster}
-{\alpha_s N_c \over 4\pi^2}\int dz_{\perp}
{(\vec{x}-\vec{y})_{\perp}^2\over (\vec{x}-\vec{z})_{\perp}^2
(\vec{z}-\vec{y})_{\perp}^2}\\
&\times&\Big\{{\cal W}(x_{\perp},z_{\perp})
+{\cal W}(x_{\perp},z_{\perp})-
{\cal W}(x_{\perp},y_{\perp})+
{\cal W}(x_{\perp},z_{\perp}){\cal W}(z_{\perp},y_{\perp})\Big\} ,
\nonumber
\end{eqnarray}
where
\begin{equation}
{\cal W}(x_{\perp},y_{\perp})\stackrel{\rm def}{\equiv}
{1\over N_c}{\rm Tr}\{W(x_{\perp} )W^{\dagger} (y_{\perp} )\}~-~1 ,
\label{1.20}
\end{equation}
(cf. Eq.~(\ref{V}). 
Similarly, the linear evolution is:
\begin{eqnarray}
\langle N|{\cal W}^{\zeta_1}(x_{\perp},0)|N\rangle&=&\label{1.21}
\int \! \frac {d\nu }{2\pi ^{2}}
(\vec{x}_{\perp}^{2})^{{1\over 2}+i\nu }
\left(\frac {\zeta _{1}}{\zeta _{2}}
\right)^{-{3\over 2}\omega(\nu )} \\
&\times&\int \! dz_{\perp}(\vec{z}_{\perp}^{2})^{-{1\over 2}-i\nu }
\langle N|{\cal W}^{\zeta_2}(z_{\perp},0)|N\rangle ,
\nonumber
\end{eqnarray}
where $\omega(\nu)=2{\alpha_s\over\pi}N_c\chi(\nu)$, see Eq.~(\ref{2.3.9}).
Let us now describe the diffractive amplitude in 
LLA and in leading order in $N_c$. In this approximation 
we must take into account the non-linearity in the Eq.~(\ref{dmaster}) 
only once, the rest of the evolution is linear. The result is
(roughly speaking) the three two-gluon BFKL ladders 
which couple in a certain point, see Fig.~\ref{ofig20}.
For the case of diffractive DIS, 
this evolution has the form (cf. Ref.~45): 
\begin{eqnarray}
&&\!\!\!\!\langle N|\int dy_{\perp} {\cal W}^{\zeta=x_B}
(x_{\perp}+y_{\perp},y_{\perp})|N\rangle\label{1.22}\\
&=&{\alpha_s N_c\over 8\pi^3}\int d\nu d\nu_1 dx_1 d\nu_2 dx_2
(\vec{x}_{\perp}^2)^{{1\over 2}+i\nu} 
\left((\vec{x}_1-\vec{x}_2)_{\perp}^2\right)^{-{1\over 2}+i(\nu_1+\nu_2-\nu)}
\nonumber\\
&\times& 
{\nu_1^2\nu_2^2\over \pi^8} 
\Theta(\nu;\nu_1,\nu_2)\int{d^2p'_{\perp}\over 4\pi^2}
\int^s_{Q^2}{dM^2\over M^2}
\left({s\over M^2}\right)^{\omega(\nu)}
\left({M^2\over Q^2}\right)^{\omega(\nu_1)+\omega(\nu_2)} \nonumber\\
&\times&\langle p|{\cal U}^{\zeta_0}(x_1,\nu_1)|p'\rangle\langle p'|
{\cal U}^{\zeta_0}(x_2,\nu_2)|p\rangle ,
\nonumber
\end{eqnarray}
where $M^2$ is the invariant mass of the produced particles, and
\begin{eqnarray}
\Theta(\nu;\nu_1,\nu_2)&=& 
{\Gamma({1\over 2}-i(\nu+\nu_1-\nu_2))\Gamma({1\over 2}-i(\nu-\nu_1+\nu_2))
\over
\Gamma({1\over 2}+i(\nu+\nu_1-\nu_2))\Gamma({1\over 2}+i(\nu-\nu_1+\nu_2))}
\nonumber\\
&\times&
{\Gamma^2({1\over 2}+i\nu)\over
\Gamma^2({1\over 2}+i\nu)} 
\Omega({1\over 2}+i\nu,{1\over 2}-i\nu_1,{1\over 2}-i\nu_2)
\end{eqnarray}
is a certain numerical function of
three conformal weights (the explicit form was found in Ref.~46 )  
which has a maximum 
$\Theta(0,0,0)=2\pi^7{}_4F_3({1\over
2}){}_6F_5({1\over 2})\simeq 7766.679$ . 
The value of $M^2$ determines the
rapidity gap: from $\eta=\ln{s\over Q^2}$ to  $\eta=\ln{M^4\over Q^2s}$  we
have a production of particles  described by  the cut part of the ladder in
Fig. \ref{ofig20} which brings in the factor $\left({s/M^2}\right)^{\omega(\nu)}$ 
while from $\eta=\ln{M^4\over Q^2s}$ to $\eta=\ln x_B$ we 
have a rapidity gap so there are two 
independent BFKL ladders which bring in the factors 
$\left({M^2/Q^2}\right)^{\omega(\nu_1)}$ and 
$\left({M^2/Q^2}\right)^{\omega(\nu_2)}$.
Since the intercept of the BFKL pomeron $\omega_0>0$, 
this cross section increases with the growth of the rapidity gap. 

The coupling of BFKL ladder with non-zero momentum transfer to a 
nucleon is described by the matrix element $\langle p'|{\cal U}(x,\nu)|p\rangle$.
As we discussed in the previous section, at large momentum transfer it can be 
approximated by the electromagnetic form factor of the nucleon, 
\begin{eqnarray}
{\cal U}(x,\nu)&=& \int dx' dy'
\left({(\vec{x}'-\vec{y}')_{\perp}^2\over 
(\vec{x}'-\vec{x})_{\perp}^2(\vec{y}'-\vec{x})_{\perp}^2}
\right)^{{1\over 2}+i\nu}
{1\over(\vec{x}'-\vec{y}')_{\perp}^4}\label{1.23}\\
&\times&\int {dk_{\perp}\over 4\pi^2}
{dr_{\perp}\over 4\pi^2}
e^{i(\vec{k},\vec{x})_{\perp}+i(\vec{r}-\vec{k},\vec{y})_{\perp}}
\Big(\delta_{\lambda\lambda'}F_1^{p+n}
(-{\vec r}_{\perp}^2)\nonumber\\
&+&
{1\over 2ms}\bar{u}(p',\lambda')\not\! p_1\!\not\! r_{\perp} 
u(p,\lambda)F_2^{p+n}(-{\vec r}_{\perp}^2)\Big) . \nonumber
\end{eqnarray}
If one interpolates the form factors by the dipole formulas, 
the diffractive amplitude in the LLA-$N_c$ approximation 
(\ref{1.22}) can be calculated numerically.

The non-linar equation (\ref{dmaster}) can be applied to  the 
diffractive DIS from the nuclei. In this case there is an additional
large parameter, the atomic number $A$,
and therefore one should take into account 
the multitude of the non-linear vertices rather than one vertex as
in Fig.~\ref{ofig20}. These ``fan'' diagrams were summed up in 
Ref.~47 
resulting in a cross section which has a maximum at 
a certain rapidity gap (unlike the LLA-$N_c$ model for the nucleon
where the cross section increases with the rapidity).

\section{Factorization and effective action for high-energy scattering}

\subsection{Factorization formula for high-energy scattering}

Unlike usual factorization, the coefficient functions and 
matrix elements enter
the expansion  (\ref{fla1.1}) on equal footing. 
We could have integrated first
over  slow fields (having the rapidities close to that of $p_B$)
and the expansion would have the form:
\begin{equation}
A(s,t)=\sum\int d^2x_1...d^2x_nD^{i_1...i_n}(x_1,...x_n)
\langle p_A|{\rm Tr}\{U_{i_1}(x_1)...U_{i_n}(x_n)\}|p'_A\rangle .
\label{fla1.2}
\end{equation}
In this case, the coefficient functions $D$ are the results of integration 
over slow fields ant the matrix elements of the $U$ operators contain only the
large rapidities $\eta >\eta_0$. The symmetry between 
Eqs. (1) and (2)
calls for a factorization formula which would have this symmetry between 
slow and fast fields in explicit form. 

I will demonstrate that one can combine the operator expansions 
(\ref{fla1.1}) and (\ref{fla1.2}) in the following way:\cite{prl}
\begin{eqnarray}
A(s,t)&=&\sum{i^n\over n!}\int d^2x_1...d^2x_n\label{fla1.3d}\\
&\times&\langle p_A|U^{a_1i_1}(x_1)...U^{a_ni_n}(x_n)|p'_A\rangle 
\langle p_B|U^{a_1}_{i_1}(x_1)...U^{a_n}_{i_n}(x_n)|p'_B\rangle ,
\nonumber
\end{eqnarray}
where $U^a_i\equiv {\mathop{\rm Tr}}(\lambda^aU_i)$ ($\lambda^a$ 
are the Gell-Mann matrices). It is possible 
to rewrite this factorization 
formula in a more visual form if we agree that operators 
$U$ act only on states 
$B$ and $B'$ and introduce the notation $V_i$ for the same operator as 
$U_i$ only acting on the $A$ and $A'$ states:
\begin{eqnarray}
&A(s,t)=\langle p_A|\langle p_B|
\exp\left(i\!\int\! d^2xV^{ai}(x)
U^a_i(x)\right)|p'_A\rangle|p'_B\rangle .
\label{fla1.3}
\end{eqnarray}
\begin{figure}[htb]
\centerline{
\epsfysize=6cm
\epsffile{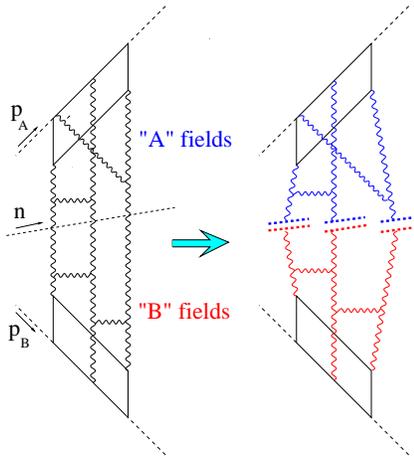}}
\caption{Structure of the factorization formula. 
The vector $n$ gives the 
direction of
the ``rapidity divide" between fast and slow fields.}
\label{kxdfig1}
\end{figure}
The supporting lines of both $U$ and $V$ operators are collinear to 
the vector $n$ corresponding to the ``rapidity divide'' $\eta_0$. 
The explicit form of this vector is
$n=\sigma p_1+\tilde{\sigma}p_2$,  where
$\tilde{\sigma}={m^2\over s\sigma}$ and $\ln \sigma/\tilde{\sigma}=\eta$.
In a
sense, formula (\ref{fla1.3}) amounts to writing the coefficient functions in 
Eq.~(\ref{fla1.1}) (or Eq.~(\ref{fla1.2})) 
as matrix elements of 
Wilson-line operators. 
Eq.~(\ref{fla1.3}) illustrated in Fig.~\ref{kxdfig1} is our main tool for
factorizing in rapidity space. 

In order to understand how this expansion can be generated by the 
factorization
formula of Eq.~(\ref{fla1.3}) type we have to rederive the 
operator expansion in axial gauge $A_{\bullet}=0$ with an additional condition 
$\left.A_{*}\right|_{x_*=-\infty}=0$ (the existence of such a gauge was 
illustrated in Ref.~49 
by an explicit construction). It is important to 
note that with
with power accuracy (up to corrections $\sim \sigma$) our gauge condition may
be replaced by 
$n^{\mu}A_{\mu}=0$. In this gauge the coefficient
functions are given by Feynman diagrams in the external field
\begin{equation}
B_i(x)=U_i(x_{\perp})\Theta(x_*),\qquad\qquad  B_{\bullet}=B_{*}=0,
\label{fla10}
\end{equation} 
which is a gauge rotation of our shock wave (it is easy to see that the 
only nonzero component of the field strength tensor 
$F_{\bullet i}(x)=U_i(x_{\perp})\delta(x_*)$ corresponds to shock wave). 
The Green functions in external field (\ref{fla10}) can be obtained
from a generating functional with a source responsible for this external field. 
Normally, the source for given external field $\bar{{\cal A}}_{\mu}$ is just 
$J_{\nu}=\bar{D}^{\mu} \bar{F}_{\mu\nu}$, so in our case the only non-vanishing
 contribution  is $J_{*}(B)=\bar{D}^i\bar{F}_{i*}$. However, 
we have a problem because the field
which we try to create by this source does not decrease at infinity. To 
illustrate the problem, suppose that we use another light-like gauge
${\cal A}_{*}=0$ for a calculation of the propagators in the external field 
(\ref{fla10}). In this case, the only would-be nonzero contribution
to the source term in the functional integral 
$\bar{D}^i\bar{F}_{i_{\bullet}}{\cal A}_{*}$ vanishes,
 and it looks like 
we do not need a source at all to generate the field $B_{\mu}$!
(This is of course wrong since $B_{\mu}$ is not a classical solution).
What it really means is that the source in this case lies entirely at the 
infinity. Indeed, when we are trying to make an external field 
$\bar{{\cal A}}$ 
in the
functional integral by the source $J_{\mu}$ we need to make a shift
${\cal A}_{\mu}\rightarrow {\cal A}_{\mu}+\bar{{\cal A}}_{\mu}$ 
in the functional integral
\begin{eqnarray}
&\int{\cal D}{\cal A} \exp\left\{iS({\cal A})-i\!\int\! d^4x 
J^a_{\mu}(x){\cal A}^{a\mu}(x)\right\} ,
\label{fla11d}
\end{eqnarray}
after which the linear term $\bar{D}^{\mu} \bar{F}_{\mu\nu}{\cal A}^{\nu}$ 
cancels out
with our source term $J_{\mu}{\cal A}^{\mu}$ and the quadratic terms
lead to the Green functions in the external field $\bar {\cal A}$.
(Note that the classical action $S(\bar{{\cal A}})$ for our external 
field $\bar{{\cal A}}=B$ (\ref{fla10}) vanishes). 
However, in order to reduce the linear
term $\int d^4x\bar{F}^{\mu\nu}\bar{D}_{\mu}{\cal A}_{\nu}$ in the functional 
integral to the form 
$\int d^4x\bar{D}^{\mu} \bar{F}_{\mu\nu}{\cal A}^{\nu}(x)$ we need to perform
an integration by parts, and if the external field does not decrease 
there will be 
additional surface terms at infinity. In our case we are trying to make the
external field $\bar{{\cal A}}=B$, consequently the linear term which need to
be canceled by the source is
\begin{eqnarray}
&{2\over s}\int\! dx_{\bullet}dx_{*}d^2x_{\perp} \bar{F}_{i\bullet}
\bar{D}_{*}{\cal A}^{i}=
\left.\int\! dx_{*}d^2x_{\perp} \bar{F}_{i\bullet}
{\cal A}^{i}\right|^{x_{\bullet}=\infty}_{x_{\bullet}=-\infty}.
\label{fla12}
\end{eqnarray}
This contribution comes entirely from the boundaries of integration. If we
recall that in our case 
$\bar{F}_{\bullet i}(x)=U_i(x_{\perp})\delta(x_*)$ we can finally rewrite 
the linear term as
\begin{eqnarray}
&\int\! d^2x_{\perp} U_i(x_{\perp})
\{{\cal A}^{i}(-\infty p_2+x_{\perp})-{\cal A}^{i}(\infty p_2+x_{\perp})\} .
\label{fla13}
\end{eqnarray}
The source term which we must add to the exponent in the functional 
integral to cancel the linear term after the shift is given by Eq.~(\ref{fla13})
with the minus sign. Thus, Feynman diagrams in the external
field (\ref{fla10}) in the light-like gauge ${\cal A}_{*}=0$ are generated
 by the functional integral
\begin{equation}
\!\int\!{\cal D}{\cal A} \exp\Big\{iS({\cal A})+
i\!\int\! d^2x_{\perp} 
U^{ai}(x_{\perp})[{\cal A}^a_i(\infty p_2+x_{\perp})
-{\cal A}^{ai}(-\infty p_2+x_{\perp})]\Big\}.
\label{5.1.15a}
\end{equation}
In an arbitrary gauge the source term in the exponent in Eq.~(\ref{5.1.15a}) 
can be rewritten in the form
\begin{equation}
2i\int d^2x_{\perp}{\mathop{\rm Tr}} \{U^i(x_{\perp})\int^{\infty}_{-\infty} 
dv[-\infty p_2,vp_2]_{x_{\perp}}
F_{*i}(vp_2+x_{\perp})[vp_2,-\infty p_2]_{x_{\perp}}\} .
\label{fla15}
\end{equation} 
Therefore, we have found the generating functional for our Feynman diagrams in
the  external field (\ref{fla10}). 

\begin{figure}[htb]
\centerline{
\epsfysize=7cm
\epsffile{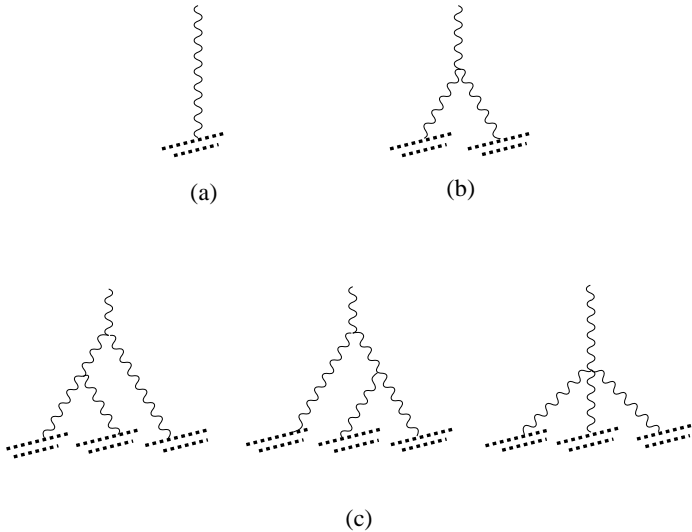}}
\caption{Perturbative diagrams for the classical field  
(\protect\ref{fla10}).} 
 \label{fig6}
\end{figure}

It is instructive to see how the source (\ref{fla15}) creates the field
(\ref{fla10}) in perturbation theory. To this end, we
must calculate the field 
\begin{eqnarray}
\bar{\cal A}_{\mu}(x)&=&\!\int\!{\cal D}{\cal A} {\cal A}_{\mu}(x)
\exp\Big\{iS({\cal A})+
2i\!\int d^2x_{\perp}{\mathop{\rm Tr}} \{U^i(x_{\perp})\nonumber\\
&\times&\int^{\infty}_{-\infty} 
dv[-\infty p_2,vp_2]_{x_{\perp}}
F_{*i}(vp_2+x_{\perp})[vp_2,-\infty p_2]_{x_{\perp}}\}\Big\}
\label{fla15a}
\end{eqnarray}
by expansion of both 
$S({\cal A})$ and gauge links in the source term (\ref{fla15}) in powers
of $g$ (see Fig.~\ref{fig6}). 
In the first order one gets
%
\begin{equation}
\bar{\cal A}^{(0)}_{\mu}(x)=i\int^{\infty}_{-\infty}dv 
\int dz_{\perp}U^{ia}(z_{\perp})\langle{\cal A}_{\mu}(x)
F^a_{*i}(vp_2+z_{\perp})\rangle,
\end{equation}
where $\langle{\cal O}\rangle\equiv\int\! {\cal D}{\cal A}e^{iS_0}{\cal O}$.
  Now we must choose a proper gauge for our calculation. We are trying
to create a field (\ref{fla10}) perturbatively and therefore the 
gauge for our perturbative calculation must be compatible with the form
(\ref{fla10}), otherwise,  we will end up with the gauge rotation of 
the field $B(x)$. 
(For example, in Feynman gauge we
will get the field $\bar{\cal A}_{\mu}$ of the form of the shock wave 
$\bar{\cal A}_i=\bar{\cal A}_{\ast}=0,~
\bar{\cal A}_{\bullet}\sim \delta(x_{\ast})$). It is convenient
to choose the temporal gauge ${\cal A}_0=0$~\footnote{The 
gauge ${\cal A}_{\ast}=0$ 
which we used above is too singular for the perturbative calculation.
In this gauge one must first regulate the external field (\ref{fla10}) 
by the replacement 
$U_i\theta(x_{\ast})\rightarrow U_i\theta(x_{\ast})e^{-\epsilon x_{\bullet}}$
and let $\epsilon\rightarrow 0$ only in the final results.}
with the boundary condition $\left.{\cal A}\right|_{t=-\infty}=0$ where
\begin{equation}
{\cal A}_{\mu}(t, \vec{x})=\int^{t}_{-\infty}dt'F_{0\mu}(t',\vec{x}) .
\label{gau}
\end{equation}
In this gauge we obtain
%
\begin{eqnarray}
&&\hspace{-3mm}\bar{\cal A}^{(0)}_{\mu}(x)~=~\int\!{dp\over (2\pi)^3}
\left(g_{\mu\nu}-2
{p_{\mu}(p_1+p_2)_{\nu}+(\mu\leftrightarrow\nu)\over 
s(\alpha+\beta+i\epsilon)}+
{4p_{\mu}p_{\nu}\over s(\alpha+\beta+i\epsilon)^2}\right)
\nonumber\\
&&\hspace{-3mm}\times~
{1\over \alpha\beta s-\vec{p}_{\perp}^2+i\epsilon}\int dz_{\perp}
e^{-i\alpha x_{\bullet}-
i\beta x_{\ast}+i\vec{p}_{\perp}(\vec{x}-\vec{z})_{\perp}}p_{2\nu}
\delta(\alpha{s\over 2})
\partial_jU^{ja}(z_{\perp})
\end{eqnarray}
where $\delta(\alpha{s\over 2})$ comes from the 
$\int dve^{iv\alpha{s\over 2 }}$.  
(Note that the form of the singularity ${1\over(p_0+i\epsilon)}$ which
follows from Eq.~(\ref{gau}) differs from the conventional prescription
$V.p.{1\over p_0}$). Recalling that in terms of  Sudakov variables 
$dp={s\over 2}d\alpha d\beta dp_{\perp}$ one easily gets 
$\bar{\cal A}^{(0)}_{\ast}=\bar{\cal A}^{(0)}_{\bullet}=0$ and
%
\begin{equation}
\bar{\cal A}^{(0)}_{i}(x)=\theta(x_{\ast})\int\!{dp\over (2\pi)^2}
{1\over \vec{p}_{\perp}^2}\int dz_{\perp
}e^{i\vec{p}_{\perp}(\vec{x}-\vec{z})_{\perp}}
\partial_i
\partial_jU^{ja}(z_{\perp}) ,
\end{equation}
or more formally, 
%
\begin{eqnarray}
\bar{\cal A}^{(0)}_{i}(x)&=&-\theta(x_{\ast})
{1\over\vec{\partial}_{\perp}^2}\partial_i\partial_jU^{j}(x_{\perp})
\nonumber\\
&=&
U_i(x_{\perp})\theta(x_{\ast})-\theta(x_{\ast}){1\over\vec{\partial}_{\perp}^2}
(\vec{\partial}_{\perp}^2g_{ij}+\partial_i\partial_j)U^{j}(x_{\perp}) ,
\end{eqnarray}
(in our notations $\vec{\partial}_{\perp}^2\equiv-\partial_i\partial^i$).
Now,
since $U_i(x)$ is a pure gauge field (with respect to transverse
coordinates) we have $\partial_i U_j-\partial_j U_i=i[U_i,U_j]$, so
%
\begin{equation}
\bar{\cal A}^{(0)}_{i}(x)=
U_i(x_{\perp})\theta(x_{\ast})-\theta(x_{\ast})
ig{\partial^j\over\vec{\partial}_{\perp}^2}
[U_i,U_j])(x_{\perp}).
\label{pr7}      
\end{equation}
Consequently, we have reproduced the field (\ref{fla10}) up to the correction
of of $g$.   We will demonstrate now that this $O(g)$ correction is canceled
by the next-to-leading term in the expansion of the  exponent of the source
term in Eq.~(\ref{fla15a}). In the next-to-leading order one gets 
(see Fig.~\ref{fig6}b) 
%
\begin{eqnarray}
\bar{\cal A}^{(1)}_{\mu}(x)&=&
g\int\!dy\int\!dz_{\perp}dz'_{\perp}U^{ja}(z_{\perp})U^{kb}(z'_{\perp})
\label{nlo}\\
&\times&\Big\langle{\cal A}_{\mu}(x)
2{\rm Tr}\{\partial^{\alpha}{\cal A}^{\beta}(y)
[{\cal A}_{\alpha}(y),{\cal A}_{\beta}(y)]\} \nonumber\\
&\times&\int\! dv F^a_{*j}(vp_2+z_{\perp})
\int\! dv'F^b_{*k}(vp_2+z'_{\perp})\Big\rangle . 
\nonumber
\end{eqnarray}
It is easy to see that 
$\bar{\cal A}^{(1)}_{\ast}=\bar{\cal A}^{(1)}_{\bullet}=0$
and
\begin{eqnarray}
\bar{\cal A}^{(1)}_{i}(x)&=&
-g\int\!dy\int{dp\over (2\pi)^4i}e^{-ip(x-y)}{1\over p^2}\\
&\times&\left(\partial^k[{\cal A}^{(0)}_{i}(y),{\cal A}^{(0)}_{k}(y)]+
[{\cal A}^{(0)k}(y),\partial_i{\cal A}^{(0)}_{k}(y)-
(i\leftrightarrow k)]\right). 
\nonumber 
\end{eqnarray}
Since ${\cal A}^{(0)}_{k}$ is given by Eq.~(\ref{pr7}), this reduces to
%
\begin{eqnarray}
\bar{\cal A}^{(1)}_{i}(x)=
-g\theta(x_{\ast})\!\int\!dy_{\perp}{dp_{\perp}\over
(2\pi)^2} {e^{-ip_{\perp}(x-y)_{\perp}}\over \vec{p}_{\perp}^2}
i\partial^k([U_{i}(y),U_{k}(y)])+O(g^2).
\end{eqnarray}
The right-hand side of this expressions cancels the second term in 
Eq.~(\ref{pr7}) and we obtain 
\begin{equation}
\bar{\cal A}_{i}(x)=
U_i(x_{\perp})\theta(x_{\ast})+O(g^2).
\label{pr8}      
\end{equation}
Likewise, one can check that the contributions $\sim g^2$ coming the diagrams 
in Fig.~\ref{fig6}c cancel the $g^2$ term in the Eq.~(\ref{pr8}). 
Taking into 
account arbitrary number of the tree-gluon vertex iterations,
one gets the expression
$U_i(x_{\perp})\theta(x_{\ast})$ without any corrections. 

\begin{figure}[htb]
\centerline{
\epsfysize=5cm
\epsffile{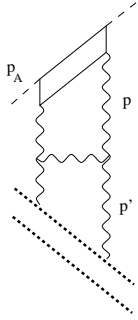}}
\caption{A typical loop diagram in the external field 
created by the Wilson-line 
source (\protect\ref{fla15}).} \label{fig7}
\end{figure}

We have found the generating functional 
for the diagrams in the external field (\ref{fla10}) which give the coefficient
functions in front of our Wilson-line operators $U_i$.
Note that formally we obtained the source term with the gauge link 
ordered along the light-like line, a potentially dangerous situation. 
Indeed, it it is easy to see that already the first loop diagram shown
in Fig.~\ref{fig7} is divergent. 
The reason is that the longitudinal integrals
over  $\alpha_p$ are unrestricted from below (if the Wilson 
line is light-like).
However,  this is not what we want for the
coefficient functions because they should include only the integration over
the  region $\alpha_p>\sigma$ (the region $\alpha_p<\sigma$ 
belongs to matrix elements, see the discussion in Sec.~3). Therefore,
we must impose somehow this condition $\alpha_p>\sigma$ in our Feynman diagrams
created by the source (\ref{fla15}). Fortunately we already faced similar
problem --- how to impose a condition $\alpha_p<\sigma$  on the matrix
elements of operators $U$ (see Fig.~\ref{fig4}) --- 
and we solved that problem by changing the slope of the supporting
 line. We demonstrated that in order to cut the integration over 
 large $\alpha>\sigma$ 
from matrix elements of Wilson-line operators $U_i$  we need 
to change the slope of these Wilson-line operators
to  $n=\sigma p_1+\tilde{\sigma} p_2$. Similarly,  
if we want to cut the integration over
small $\alpha_p<\sigma$ from the
coefficient functions we need to order the gauge 
factors in  Eq.~(\ref{fla15}) along (the same) vector 
$n=\sigma p_1+\tilde{\sigma} p_2$.\footnote{Note that the 
diagram in Fig.~\ref{fig7} is the diagram in Fig.~\ref{fig4}b turned
upside down.  In the Fig.~\ref{fig4}b diagram we have a restriction
$\alpha<\sigma$. It is easy to see that 
this implies a restriction $\beta>\tilde{\sigma}$ 
if one chooses to write down the rapidity 
integrals in terms of $\beta$'s rather than $\alpha$'s. Turning
the diagram upside down amounts to interchange of $p_A$ and $p_B$, 
leading to ({\bf i}) replacement of the slope of the Wilson line by 
$\tilde{\sigma}p_1+\sigma p_2$ and ({\bf ii})
replacement $\alpha\leftrightarrow \beta$ in the integrals. 
Thus, the restriction
$\beta>\tilde{\sigma}$ imposed by the line collinear to 
$\sigma p_1+\tilde{\sigma} p_2$ in diagram in Fig.~\ref{fig4}b means
the restriction $\alpha>\tilde{\sigma}$ by the line 
$\parallel~\tilde{\sigma}p_1+\sigma p_2$ in the Fig.~\ref{fig7} diagram. 
After renaming $\sigma$ by $\bar{\sigma}$ we obtain 
 the desired result.}
 
Therefore, the final form of
the generating functional for the Feynman diagrams (with  $\alpha>\sigma$
cutoff) in the external field  (\ref{fla10}) is 
\begin{eqnarray} \int\!{\cal
D}{\cal A} {\cal D}\Psi {\rm exp}\left\{iS({\cal A},\Psi)+ i\int d^2x_{\perp}
U^{ai}(x_{\perp})V^a_i(x_{\perp})\right\} , \label{fla16} 
\end{eqnarray} 
where
\begin{equation} 
V_i(x_{\perp})=\label{fla17}
\int^{\infty}_{-\infty} dv[-\infty n,vn]_x n^{\mu}F_{\mu
i}(vn+x_{\perp}) [vn,-\infty n]_x,
\nonumber
\end{equation}
and $V^a_i\equiv {\rm Tr}(\lambda^aV_i)$ as usual. For completeness, 
we have added 
integration over quark fields so $S({\cal A},\Psi)$ is the full QCD action.
 
 Now we can assemble the different parts of the factorization 
formula (\ref{fla1.3}). We have written down the generating functional integral 
for the diagrams with $\alpha>\sigma$ in the external fields with 
$\alpha<\sigma$; what remains now is to write down the integral over 
these ``external'' fields. 
Since this
integral is completely independent of (\ref{fla16}) we will use a different
notation ${\cal B}$ and $\chi$ for the $\alpha<\sigma$ fields. We have 
\begin{eqnarray}
&&\!\!\!\!\int\!\! {\cal D}A{\cal D}\bar{\Psi}{\cal D}\Psi e^{iS(A,\Psi)} 
j(p_{A})j(p'_{A})j(p_{B})j(p'_{B})
\label{fla18}\\ 
&=&\int\!\! {\cal D}{\cal A}{\cal D}\bar{\psi}{\cal D}
\psi e^{iS({\cal A},\psi)} j(p_{A})j(p'_{A})
\int\!\! {\cal D}{\cal B}{\cal D}\bar{\chi}{\cal D}\chi \nonumber\\
&\times&
j(p_{B})j(p'_{B})
e^{iS({\cal B},\chi)} \exp\Big\{i\!\int\! d^2x_{\perp} 
U^{ai}(x_{\perp})V^{a}_i(x_{\perp})\Big\}.\nonumber
\end{eqnarray}
The operator $U_i$ in an arbitrary gauge is
given by the same formula (\ref{fla17}) as operator $V_i$
with the only difference that the gauge links and $F_{{\bullet} i}$ 
are constructed from the fields 
${\cal B}_{\mu}$. This is our factorization formula (\ref{fla1.3}) 
in the functional integral representation.

The functional integrals over ${\cal A}$ fields give logarithms of the
type $g^2\ln{1/\sigma}$ while the integrals over slow ${\cal B}$ fields give
powers of $g^2\ln (\sigma s/m^2)$. With logarithmic accuracy, 
they add up to
$g^2\ln s/m^2$. However, there will be
additional terms $\sim g^2$ due to mismatch coming from the region 
of integration near the dividing point $\alpha\sim\sigma$, where the
details of the cutoff in the matrix elements of the operators 
$U$ and $V$ 
become important. Therefore, one should expect the corrections 
of order of 
$g^2$ to the effective action $\int dx_{\perp} U^iV_i$
of the type 
\begin{eqnarray}
&&\!\!\!\!\exp\Big\{i\int d^2x_{\perp}U_i(x_{\perp})V_i(x_{\perp})+
\label{flad1}
i\int dx_{\perp}
dy_{\perp}dz_{\perp}\\
&\times&U_i(x_{\perp})U_i(y_{\perp})
V_i(z_{\perp}) V_i(t_{\perp}) 
K(x_{\perp}-t_{\perp},y_{\perp}-t_{\perp},z_{\perp}-t_{\perp})\Big\}
\nonumber
\end{eqnarray}
where $K$ is a calculable kernel. In general,
the fact that the fast quark moves along the straight line has nothing
to do with perturbation theory (cf. Ref.~50), 
therefore it is 
natural to expect the
non-perturbative generalization of the factorization formula 
constructed from the same Wilson-line operators $U_i$ and $V_i$.

\subsection{Effective action for given interval of rapidities}

\begin{figure}[htb]
\centerline{
\epsfysize=6cm
\epsffile{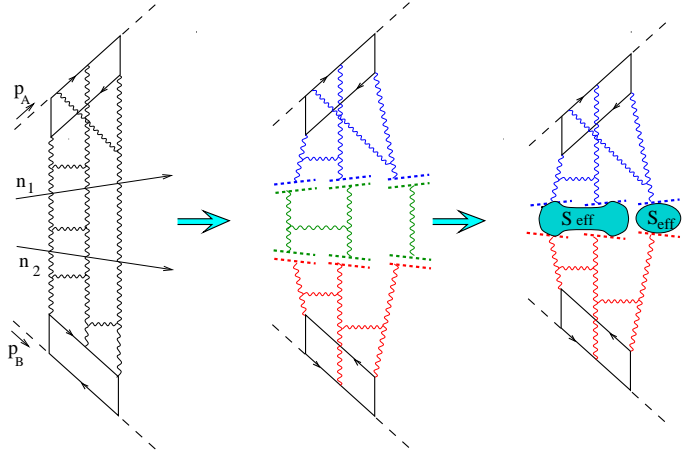}}
\caption{The effective action for the interval of rapidities 
$\eta_0>\eta>\eta'_0$. The two  vectors
 $n$ and $n'$ correspond to ``rapidity divides" $\eta_0$ and $\eta'_0$ 
bordering our chosen region of rapidities.} 
\label{ofig24}
\end{figure}

The factorization formula gives us a starting point for a new approach
to the analysis of the high-energy effective action. 
Consider another rapidity $\eta'_0$ in the region between $\eta_0$ and 
$\eta_B=\ln m^2/s$. If we use the factorization formula 
(\ref{fla18}) once more, this time dividing between the rapidities 
greater and smaller than $\eta'_0$, we get the expression 
for the amplitude (\ref{2.1.1}) in the form 
(see Fig.~\ref{ofig24}):\footnote{Strictly speaking, the l.h.s. 
of Eq.~(\ref{fla19}) contains an extra $16\pi^4\delta(p_A+p'_A-p_B-p'_B)$ in 
comparison to the amplitude (\ref{2.1.1}).}
\begin{eqnarray}
iA(s,t)&=&\int\!\! {\cal D}Ae^{iS(A)} 
j(p_{A})j(p'_{A})j(p_{B})j(p'_{B})
\label{fla19}\\ 
&=&\int\!\! {\cal D}{\cal A} e^{iS({\cal A})} j(p_{A})j(p'_{A})
\int\!\! {\cal D}{\cal B}
e^{iS({\cal B})}j(p_{B})j(p'_{B}) \nonumber\\
&\times&\int\! {\cal D}{\cal C}e^{iS({\cal C})}
e^{i\!\int\! d^2x_{\perp} 
V^{ai}(x_{\perp})Y^a_i(x_{\perp})+i\!\int\! d^2x_{\perp} 
W^{ai}(x_{\perp})U^a_i(x_{\perp})}.
\nonumber
\end{eqnarray}
(For brevity, 
we do not display the quark fields.) In this formula the operators $V_i$ (made from ${\cal A}$ fields) 
are given by Eq.~(\ref{fla17}), the operators $Y_i$ are also given by 
Eq.~(\ref{fla17}) but constructed from the 
${\cal C}$ fields instead, and the operators $W_i$ (made from 
${\cal C}$ fields) and $U_i$ 
(made from ${\cal B}$ fields) are aligned along the 
direction $n'=\sigma'p_1+\tilde{\sigma}'p_2$ 
corresponding to the rapidity $\eta'$ (as usual, 
$\ln \sigma'/\tilde{\sigma}'=\eta'$ 
where
$\tilde{\sigma}'=m^2/s\sigma'$),
\begin{eqnarray}
V_i({\cal A})_{x_{\perp}}&=&\int^{\infty}_{-\infty} 
dv[-\infty n,vn]_x
n^{\mu}F_{\mu i}(vn+x_{\perp})
[vn,-\infty n]_x,\nonumber\\
Y_i({\cal C})_{x_{\perp}}&=&\int^{\infty}_{-\infty} 
dv[-\infty n,vn]_x
n^{\mu}F_{\mu i}(vn+x_{\perp})
[vn,-\infty n]_x,\nonumber\\
W_i({\cal C})_x &=&\int^{\infty}_{-\infty} 
dv[-\infty n',vn']_x
n^{'\mu}F_{\mu i}(vn'+x_{\perp})
[vn',-\infty n']_x,\nonumber\\
U_i({\cal B})_{x_{\perp}}&=&\int^{\infty}_{-\infty} 
dv[-\infty n',vn']_x
n^{'\mu}F_{\mu i}(vn'+x_{\perp})
[vn',-\infty n']_x.
\nonumber
\end{eqnarray}
In conclusion, we have factorized the functional integral 
over ``old'' ${\cal B}$ fields 
into the product of two integrals over ${\cal C}$ and ``new" ${\cal B}$
fields.

Now, let us integrate over the ${\cal C}$ fields  
and write down the result in terms of an effective action. 
Formally, one obtains:
\begin{equation}
iA(s,t)=
\int\!\! {\cal D}{\cal A} e^{iS({\cal A})} j(p_{A})j(p'_{A})
\int\!\! {\cal D}{\cal B}
e^{iS({\cal B})}j(p_{B})j(p'_{B})
e^{iS_{\rm eff}(V,U;{\sigma\over\sigma'})} ,
\label{fla21}
\end{equation}
where the effective action for the rapidity interval between $\eta$ and 
$\eta'$ is defined as 
\begin{equation}
e^{iS_{\rm eff}(V,U;{\sigma\over\sigma'})}=
\int\! {\cal D}{\cal C}e^{iS({\cal C})}
e^{i\!\int\! d^2x_{\perp} 
V^{ai}(x_{\perp})Y^a_i(x_{\perp})+i\!\int\! d^2x_{\perp} 
W^{ai}(x_{\perp})U^a_i(x_{\perp})} ,
\label{fla22}
\end{equation}
($U_i\equiv U^{\dagger}{i\over g}\partial_iU$ and 
$V_i\equiv V^{\dagger}{i\over g}\partial_iV$ as usually). 
This formula gives a
rigorous definition for the effective action for a  given interval 
in rapidity.

\begin{figure}[htb]
\centerline{
\epsfysize=3cm
\epsffile{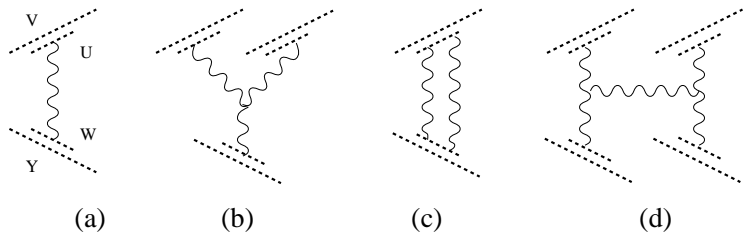}}
\caption{Lowest order terms in the perturbative expansion 
of the effective action.} \label{fig8}
\end{figure}

\begin{figure}[htb]
\centerline{
\epsfysize=4cm
\epsffile{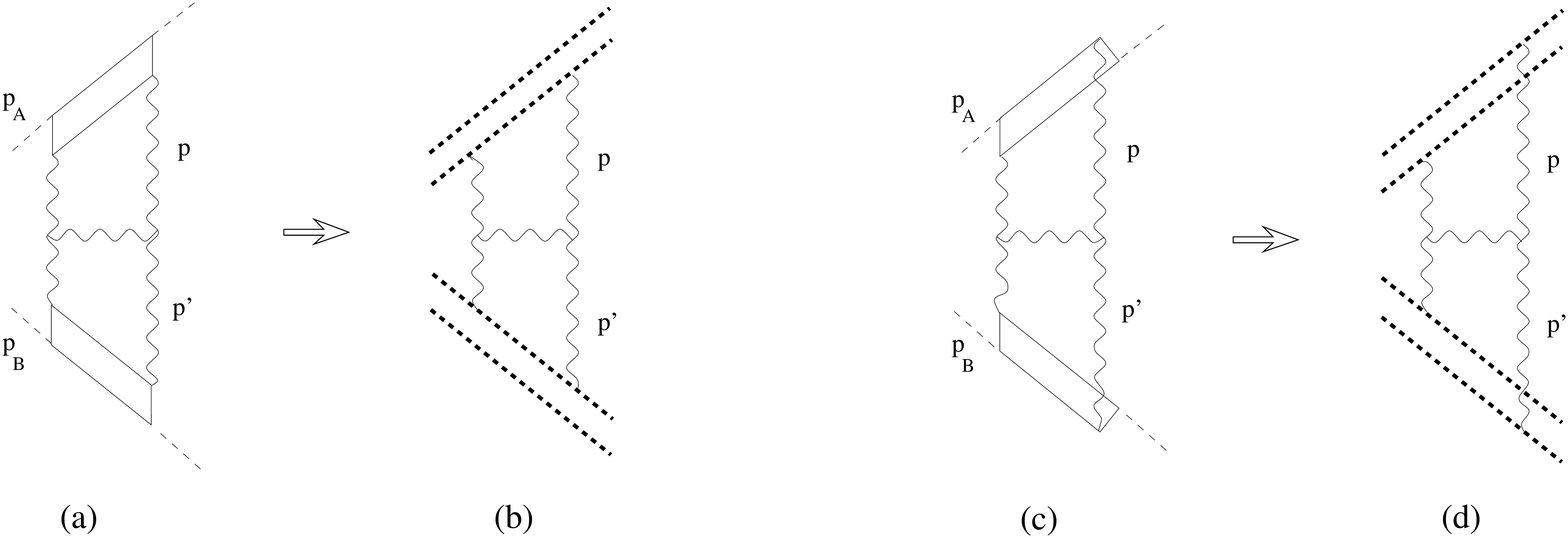}}
\caption{Counting of loops for Feynman diagrams (a),(c) and 
the corresponding
Wilson-line operators (b),(d).}
\label{fig4a}
\end{figure}

Next step would be to perform explicitly the integrations over the 
longitudinal momenta in the right-hand side 
of Eq.~(\ref{fla22}) and obtain the answer
for the integration over
our rapidity region (from $\eta_0$ to $\eta'_0$) in terms of two-dimensional 
theory in the transverse coordinate space,\footnote{Historically, 
the idea how to reduce
QCD at high energies to the two-dimensional effective theory was first
suggested in Ref.~51 
where the leading term in Eq.~(\ref{fla23})
was obtained. However, careful analysis of the assumptions made in this paper
shows that the authors considered the fixed-angle limit of the  theory
($s,t\rightarrow\infty$) rather than the Regge limit (where $\rightarrow\infty$
but $t$ is fixed). It turns out that the first term in Eq.~(\ref{fla23}) 
is the same for both limits, but the subsequent terms differ.
}
hopefully resulting in
the unitarization of the BFKL pomeron. At present, the  
known how to do this. One can obtain, however, a first few terms in the 
expansion of effective action in powers of $V_i$ and $U_i$. The easiest way 
to do this is to expand gauge factors $Y_i$ and $W_i$ in right-hand side of 
Eq.~(\ref{fla22}) in powers of ${\cal C}$ fields and calculate the relevant 
perturbative diagrams (see Fig.~\ref{fig8}).
The first few terms in the effective action at the one-log 
level~\footnote{This ``one-log" level corresponds to 
one-loop level for usual Feynman diagrams.
Superficially, the diagram in Fig.~\ref{fig8}d looks like tree diagram in
comparison to diagram in Fig.~\ref{fig8}c which has one loop. 
However, both of the diagrams in Fig.~\ref{fig8}c and d  contain 
integration over longitudinal momenta
(and thus the factor $\ln{\sigma\over\sigma'}$) so in the longituduinal
space the diagram in Fig.~\ref{fig8}d is also a loop diagram. 
This happens because for diagrams with 
Wilson-line operators the 
counting of number of loops literally corresponds to the counting of the 
number of loop integrals only for the transverse momenta. For
the longitudinal variables, the diagrams which look like 
trees may contain logarithmical loop integrations. This property is 
illustrated in Fig.~\ref{fig4a}: the Wilson-line diagram shown in 
Fig.~\ref{fig4a}b has two loops and the diagram shown in Fig.~\ref{fig4a}d  
is a tree but both of them originated from  Feynman diagrams 
shown in Fig.~\ref{fig4a}a and c with equal number of loops. 
To avoid confusion, we will use the term ``one-log 
level" instead of ``one-loop level."}
have the form:\cite{fizrev,many} 
\begin{eqnarray}
&&S_{\rm eff}~=~\int d^2x 
V^{ai}(x)U^a_i(x)\label{fla23}\\
&&-~{g^2\over 64\pi^3}\ln{\sigma\over\sigma'}
\Bigg(N_c\int d^2x d^2y V^a_{i,i}(x)\ln^2(x-y)^2U^a_{j,j}(y)
\nonumber\\
&&+~{f_{abc}f_{mnc}\over 4\pi^2}\!\int\! d^2x d^2y d^2x' d^2y'd^2z
V^a_{i,i}(x)V^m_{j,j}(y)
U^b_{k,k}(x')U^n_{l,l}(y')\nonumber\\
&&
\ln{(x-z)^2\over(x-x')^2}\ln{(y-z)^2\over(y-y')^2}
\left({\partial\over\partial z_i}\right)^2\ln{(x'-z)^2\over(x-x')^2}
\ln{(y'-z)^2\over(y-y')^2}\Bigg) +\ldots,
\nonumber
\end{eqnarray}
where we we use the notation 
$V^a_{i,j}(x)\equiv {\partial\over \partial x_j}V^a_i(x)$ etc. 
The first term (see Fig.~\ref{fig8}a) looks like the 
corresponding term in the 
factorization 
formula (\ref{fla18}), only the directions of the supporting lines are 
now strongly different.\footnote{Strictly speaking, 
the contribution coming from the diagram shown in Fig.~\ref{fig8}a
has the form 
$\int d^2x V^{ai}(x){\partial_i\partial_j\over\partial^2}U^{aj}(x)$
which differs from the first term in the right-hand side  
of Eq.~(\ref{fla23}) by
$\int d^2x V^{ai}(x){1\over\partial^2}
(\partial^2g_{ij}-\partial_i\partial_j)U^{aj}(x)$. Yet, 
 it may be demonstrated that this discrepancy 
(which is actually
$\sim O(g)$ for a a pure gauge field $U_i$) is canceled by the 
contribution from the diagram with the three-gluon vertex shown in 
Fig.~\ref{fig8}b 
just
as in the case of perturbative calculation of ${\cal A}_i$ discussed in
Sec.~3.}  The second
term shown in Fig.~\ref{fig8}c is the first-order expression for the 
reggeization of the gluon (\ref{2.5.3})
and the third term (see Fig.~\ref{fig8}d) is the gluon emission term in the
BFKL kernel (\ref{2.2.19a}) in the impact parameter representation.

\begin{figure}[htb]
\centerline{
\epsfysize=3.5cm
\epsffile{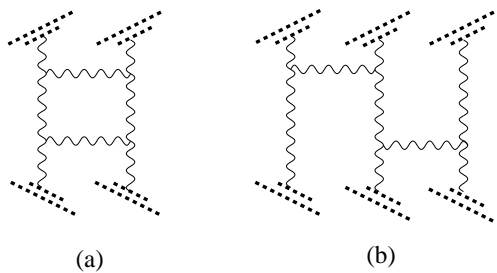}}
\caption{Typical perturbative diagrams in the next
$\left(\ln{\sigma\over\sigma'}\right)^2$ order.}
\label{fig9}
\end{figure}

Let us discuss subsequent terms in the perturbative expansion (\ref{fla23}).
There can be two types of the logarithmical contributions. First is
the ``true" loop contribution coming from the diagrams of the Fig.~\ref{fig9}a 
type. This diagram is an iteration of the Lipatov's Hamiltonian. In addition, 
in the same $(\ln{\sigma\over\sigma'})^2$ order there is another 
contribution
coming from the diagram shown in Fig.~\ref{fig9}b. 
In perturbation theory, these two contributions are of the same order
of magnitude. 

The situation is different for the case of scattering of two heavy nuclei. 
Assuming that the effective coupling constant is still small 
due to the high
density,\cite{lrmodel} we see that $g\ll 1$, yet the sources are strong
($\sim {1\over g}$)  so $gU_i\sim gY_i\sim 1$.
In this case, the diagram in Fig.~\ref{fig9}a
has the order 
$g^4U_i^2V_i^2\left(\ln{\sigma\over\sigma'}\right)^2
\sim \left(\ln{\sigma\over\sigma'}\right)^2$ while the ``tree" 
Fig.~\ref{fig8}b diagram is 
\begin{equation}
\sim g^4U_i^3V_i^3\left(\ln{\sigma\over\sigma'}\right)^2
\sim {1\over g^2}\left(\ln{\sigma\over\sigma'}\right)^2 .
\label{dupelog}
\end{equation}
In this approximation, first we shall sum up the tree diagrams. As usual, the
best way do this is to use the semiclassical
method which will be discussed in Sec.~5. In the next paragraph
we will consider the intermediate situation with one weak source and 
one strong source.

\subsection{Effective action for one weak and one strong source}

 Consider again the DIS from a nucleon or nucleus where the  high-energy 
behavior is governed by the non-linear 
evolution equation (\ref{master}). In this section we will translate the 
evolution results (\ref{5.23}) into the effective action language (see also
Refs.~52, 53). 
In the case DIS one of the
sources (corresponding to quark-antiquark pair) is weak while the other
(describing the nucleon or nuclei) is strong.

\begin{figure}[htb]
\centerline{
\epsfysize=3cm
\epsffile{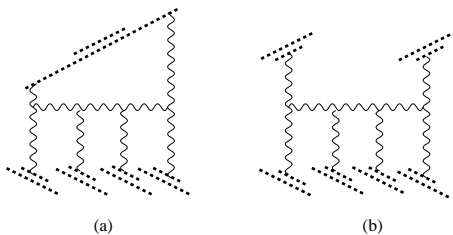}}
\caption{Perturbative diagrams for the effective action 
in the case of one weak source and one strong one.}
\label{fig10}
\end{figure}

 For example, if the source $V_i$ is weak 
 (and hence $gV_i$ is a valid small parameter) but the source $U_i$ is 
 not weak (so that $gV_i\sim 1$ is {\it not} a small parameter), one must 
 take
 into account the diagrams shown in Fig.~\ref{fig10}a and b.
The multiple rescatterings in Fig.~\ref{fig10}a,b  describe the motion of the 
gluon emitted by the 
weak source $V_i$ in the strong external field $A_i=U_i\theta(x_{\ast})$
created by the source $U_i$. The result of the calculation of the 
diagram in
Fig.~\ref{fig10}a presented in a form of the evolution of the 
Wilson-line operators $U_i$ can be easily obtained using the evolution 
equations (\ref{5.21})
%
\begin{eqnarray}
U^a_i(x_{\perp})\!\!&\rightarrow& 
\!\! U^a_i(x_{\perp})\label{dob1}\\
&-&\!\! 
{g^2\over 8\pi^3}\ln{\sigma\over\sigma'}
\int dy_{\perp}{1\over(\vec{x}-\vec{y})_{\perp}^2}
\Big(f^{abc}\big(U^{\dagger}_x\partial_iU_y\big)^{bc}+
N_cU^a_i(x_{\perp})\Big)+\ldots,
\nonumber
\end{eqnarray}
where dots stand for the terms with higher powers of 
$g^2\ln{\sigma\over\sigma'}$. This evolution equation 
means that if we integrate over the rapidities $\eta_0>\eta>\eta'_0$  
in the matrix elements of the 
operator $Y_i$ we will get the expression (\ref{dob1}) constructed 
from the operators $U_i$ with rapidities up to $\eta'_0$ times 
 factors proportional to  
 $g^2(\eta_0-\eta'_0)\equiv g^2\ln{\sigma\over\sigma'}$.
Therefore, the corresponding contribution to the effective action 
at the one-log level takes the form
%
\begin{eqnarray}
&&\hspace{-5mm}\int dx_{\perp}V^a_i(x_{\perp}) U^{ai}(x_{\perp})~~
\rightarrow
~~ \int dx_{\perp} V^a_i(x_{\perp})U^{ai}(x_{\perp})\label{dob2}\\
&&\hspace{-5mm}+~
{g^2\over 8\pi^3}\ln{\sigma\over\sigma'}
\int\! dx_{\perp} dy_{\perp}
{1\over(\vec{x}-\vec{y})_{\perp}^2}\Big(i\big(V^{i}(x_{\perp})
U^{\dagger}_x\partial_iU_y\big)^{aa}-
N_cV^{ai}(x_{\perp})U^a_i(x_{\perp})\Big)
\nonumber
\end{eqnarray}
where the first term is the lowest-order 
effective action ($\equiv$ the first term in Eq.~(\ref{fla23})) and 
the second term contains new information. 
To check the second term, we may expand it in
powers of the source $U_i$, then it is easy to see that 
the first nontrivial term in this expansion coincides with the 
gluon-reggeization term in Eq.~(\ref{fla23}). 

\begin{figure}[htb]
\centerline{
\epsfysize=3cm
\epsffile{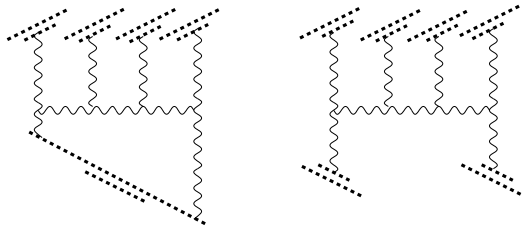}}
\caption{Effective action for the strong source $V$ and the weak source $U$.}
\label{dobavka.fig}
\end{figure}

Apart from the (\ref{dob2}) term, 
there is another contribution to the one-loop evolution equations 
coming from the diagrams in Fig.~\ref{fig10}b. It can be easily 
obtained using formulas (\ref{5.20}) from the Appendix,
\begin{eqnarray}
&&\!\!\!\! U_i^a(x_{\perp})U_j^b(y_{\perp})\rightarrow\label{dob3}
-{g^2\over 4\pi^3}\ln{\sigma\over\sigma'}\\
&\times&\Big(\nabla^x_i\Big[\int dz_{\perp}
{(\vec{x}-\vec{z},\vec{y}-\vec{z})_{\perp}
\over(\vec{x}-\vec{z})_{\perp}^2
(\vec{y}-\vec{z})_{\perp}^2}
(U^{\dagger}_xU_y+1-U^{\dagger}_xU_z- U^{\dagger}_zU_y)\Big]
\stackrel{\leftarrow}{\nabla}^y_j\Big)^{ab} ,
\nonumber
\end{eqnarray}
where
\begin{eqnarray}
\nabla^x_i{\cal O}(x_{\perp})&\equiv& 
{\partial\over\partial x^i}{\cal O}(x_{\perp})-
iU_i(x_{\perp}){\cal O}(x_{\perp}),\nonumber\\
{\cal O}(y_{\perp})\stackrel{\leftarrow}{\nabla}^y_i&\equiv& 
-{\partial\over\partial y^i}{\cal O}(y_{\perp})-
i{\cal O}(y_{\perp})U_i(y_{\perp}),
\label{dob4}
\end{eqnarray}
are the ``covariant derivatives" (in the adjoint representation). 
The corresponding term in effective action is
\begin{eqnarray}
&&\!\!\!\! {ig^2\over 8\pi^3}\ln{\sigma\over\sigma'}
\int dx_{\perp}dy_{\perp} \left(\nabla^x_iV^a_i\right)(x_{\perp})
\int dz_{\perp}
{(\vec{x}-\vec{z},\vec{y}-\vec{z})_{\perp}
\over(\vec{x}-\vec{z})_{\perp}^2(\vec
{y}-\vec{z})_{\perp}^2} \nonumber\\
&\times&
\big(U^{\dagger}_xU_y+1-U^{\dagger}_xU_z- U^{\dagger}_zU_y\big)^{ab}
\left(\nabla^y_jV^b_j\right)(y_{\perp}).
\label{dob5}
\end{eqnarray}
The final form of the one-log effective action for this case 
is the sum of the expressions 
(\ref{dob2}) and (\ref{dob5}),
\begin{eqnarray}
&&\!\!\!\! S^{(I)}_{\rm eff}(V_i,U_j) = \int d^2x 
V^{ai}(x)U^a_i(x)
+{g^2\over 8\pi^3}\ln{\sigma\over\sigma'}\int\! dx_{\perp} dy_{\perp}
{1\over(\vec{x}-\vec{y})_{\perp}^2}\nonumber\\
&\times&\Big(i\big(V^{i}(x_{\perp})U^{\dagger}_x\partial_iU_y\big)^{aa}-
N_cV^{ai}(x_{\perp})U^a_i(x_{\perp})\Big)
\nonumber\\
&+&{ig^2\over 8\pi^3}\ln{\sigma\over\sigma'}\int dx_{\perp}dy_{\perp}
\nabla^x_iV^{ai}(x_{\perp})\int dz_{\perp}
{(\vec{x}-\vec{z})_{\perp}\cdot (\vec{y}-\vec{z})_{\perp}
\over(\vec{x}-\vec{z})_{\perp}^2(\vec{y}-\vec{z})_{\perp}^2}
\nonumber\\
&\times&\big(U^{\dagger}_xU_y+1-U^{\dagger}_xU_z- U^{\dagger}_zU_y\big)^{ab}
\nabla^y_jV^{bj}(y_{\perp}),
\label{dob6}
\end{eqnarray}
where $V_i$ is a weak source and $U_i$ is
a strong one. It is clear that if 
the source $V_i$ is strong and $U_i$ is weak
diagrams the effective action $S^{(II)}_{\rm eff}(V_i,U_j)$
will have the similar form with the replacement $V\leftrightarrow U$
coming from the diagram shown in Fig.~\ref{dobavka.fig}. 

\begin{figure}[htb]
\centerline{
\epsfysize=3cm
\epsffile{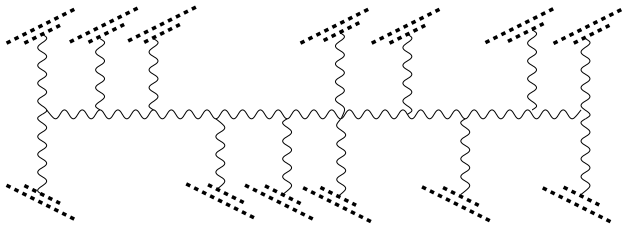}}
\caption{ Cluster expansion of the effective action.} 
\label{cluster}
\end{figure}

As we mentioned above, in the case of two strong sources the 
$\left(\ln{\sigma\over\sigma'}\right)^2$ terms start from the diagram 
shown in Fig.~\ref{fig9}b (see Eq.~(\ref{dupelog})), hence  
Fig.~\ref{fig9} and Fig.~\ref{dobavka.fig} complete the list of diagrams which
contribute to the effective action at the one-log level. 
Higher-order diagrams start from 
higher powers of $\ln{\sigma\over\sigma'}$. The analog of LLA here is a  
cluster expansion with the parameter 
$(U-1)(V-1)\ln{\sigma\over\sigma'}$ shown in Fig.~\ref{cluster}. 
Of course, the diagrams of Fig.~\ref{cluster} give the terms
$\sim\ln{\sigma\over\sigma'}$  too, but in the leading order 
the kernel of
the corresponding evolution equation is determined by 
Fig.~\ref{fig9} and Fig.~\ref{fig10}. Thus, the one-log answer for 
two strong
sources can be guessed by comparison of the answers for $S_{\rm eff}(V_i,U_j)$
with  $V_i\sim 1,~U_i\sim {1\over g}$ and with $U_i\sim 1,~V_i\sim {1\over g}$.
Instead of doing that, we will obtain
the one-log result for two strong sources using the semiclassical method and
check that it agrees with (\ref{dob6}).

It means 
that the one-log answer in the general case can be guessed by
comparison of the answers for $S_{\rm eff}(V_i,U_j)$ with 
$V_i\sim 1,~U_i\sim {1\over g}$ and with $U_i\sim 1,~V_i\sim {1\over g}$
Instead of doing that, we will obtain
the one-log result for two strong sources  using the semiclassical method and
check that it agrees with (\ref{dob6}).

\section{High-energy effective action in sQCD} 

\subsection{Effective action and collision of two shock waves}

\begin{figure}[htb]
\centerline{
\epsfysize=3cm
\epsffile{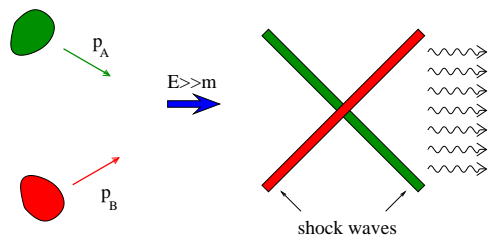}}
\caption{Scattering of two shock waves. } 
\label{shockwaves}
\end{figure}

The functional integral (\ref{fla22})
which defines the effective action is the usual QCD functional integral 
with two sources corresponding to the two colliding shock waves,
see Fig.~\ref{shockwaves}.\cite{tok} 
Instead of calculation of perturbative diagrams we can use the
semiclassical approach which is relevant when the coupling constant
is relatively small but the  characteristic fields are large 
-- in other words, when
$g^2\ll 1$  but $gV_i\sim gU_i \sim 1$. As was discussed in Ref.~4, 
this situation is realized in the heavy-ion 
collisions where the coupling constant is defined by the parton saturation
scale $Q_s$, which is estimated to be $\sim 1$ GeV at 
RHIC and $\sim 2-3$ GeV at LHC.\cite{finn,mu00} Even if we
consider the $\gamma^\ast\gamma^\ast$ scattering, the number of gluons in the
middle of the rapidity region  may become very large leading to the saturation
at high energies so in the  middle of the rapidity region we will se the
scattering of two  strong shock
waves.

If both sources are strong,  one can  calculate the functional integral
(\ref{fla22}) by expansion around the new stationary point corresponding to the
classical wave created by the collision of the shock waves.
With leading log accuracy, we can replace the vector $n$ by $p_1$ and the
vector $n'$ by $p_2$. Then the functional integral (\ref{fla22}) 
takes the  form 
\begin{equation}
e^{iS_{\rm eff}(V,U;{\sigma\over\sigma'})}=
\int\! {\cal D}Ae^{iS_{QCD}(A)}
e^{i\!\int\! d^2x_{\perp} 
V^{ai}(x_{\perp})Y^{a}_i(x_{\perp})
+i\!\int\! d^2x_{\perp}W^{ai}
U^a_i(x_{\perp})},
\label{5.1.1}
\end{equation}
where now
\begin{equation}
Y^{a}_i(x_{\perp})=\int^{\infty}_{-\infty} dv
\hat{F}_{\bullet i}(vp_1+x_{\perp}),\qquad
W^{a}_i=\int^{\infty}_{-\infty} dv
\tilde{F}_{\ast i}(vp_2+x_{\perp}).
\label{5.1.2}
\end{equation}
Hereafter we use the notations 
\begin{eqnarray}
\hat{\cal O}(x)&=&
[-\infty p_1+x,x]{\cal O}(x)[x,-\infty p_1+x], \nonumber\\
\tilde{\cal O}(x)&=&
[-\infty p_2+x,x]{\cal O}(x)[x,-\infty p_2+x].
\label{5.1.3}
\end{eqnarray}
Note that we changed the name for the gluon fields in the integrand 
from ${\cal C}$ back to $A$. 

As usual, the classical equation for the saddle point $\bar{A}$ in the 
functional integral (\ref{5.1.1}) is
\begin{equation}
\left.{\delta \over \delta A}\left(S_{QCD}+\!\int\! d^2x_{\perp} 
V^{ai}(x_{\perp})Y^{a}_i(x_{\perp})
+\!\int\! d^2x_{\perp}W^{ai}U^a_i(x_{\perp}\right)\right|_{A=\bar{A}}=0.
\label{5.1.4}
\end{equation}
To write them down explicitly we need the first variational 
derivatives of the source terms with respect to gauge field. 
We have:
\begin{eqnarray}
&&\hspace{-1.5cm} \delta Y_i=\delta \hat{A}_i(\infty p_1+x_{\perp})-
\delta {A}_i
(-\infty p_1+x_{\perp})-
\int^{\infty}_{-\infty}du
\hat{\nabla}_i\delta\hat{A}_i(u p_1+x_{\perp}),\nonumber\\
&&\hspace{-1.5cm} \delta W_i=\delta \tilde{A}_i(\infty p_2+x_{\perp})-
\delta {A}_i
(-\infty p_2+x_{\perp})-
\int^{\infty}_{-\infty}du
\tilde{\nabla}_i\delta\tilde{A}_i(u p_2+x_{\perp}),
\label{5.1.5}
\end{eqnarray}
where 
\begin{eqnarray}
\hat{\nabla}_i{\cal O}(x)&\equiv&\partial_i{\cal O}(x)
-i[Y_i(x_{\perp})+A_i(-\infty p_1+x_{\perp}),{\cal O}(x)] ,
\nonumber\\
\tilde{\nabla}_i{\cal O}(x)&\equiv&\partial_i{\cal O}(x)
-i[W_i(x_{\perp})+A_i(-\infty p_2+x_{\perp}),{\cal O}(x)] . 
\label{5.1.6}
\end{eqnarray}
Therefore the explicit form of the classical equations (\ref{5.1.4}) 
for the wave 
created by the collision is
\begin{eqnarray}
D^{\mu}{\bar F}_{\mu i}&=&0,\label{5.1.7}\\
D^{\mu}{\bar F}_{\ast\mu}&=&
\delta({2\over s}x_{\bullet})[{2\over s}x_{\ast}p_1, 
-\infty p_1]_{x_{\perp}}\hat{\nabla}_i V^i(x_{\perp})
[-\infty p_1,{2\over s}x_{\ast}p_1]_{x_{\perp}},\nonumber\\
D^{\mu}{\bar F}_{\bullet\mu}&=&
\delta({2\over s}x_{\ast})[{2\over s}x_{\bullet}p_2, 
-\infty p_2]_{x_{\perp}}\tilde{\nabla}_i U^i(x_{\perp})
[-\infty p_2,{2\over s}x_{\bullet}p_2]_{x_{\perp}}.
\nonumber
\end{eqnarray}

These equations define the classical field created by the collision of two 
shock waves.\footnote{They are essentially equivalent to the classical 
equations describing the collision of two heavy nuclei in Ref.~55. 
However, we
do not impose the additional boundary conditions at $x_{\parallel}^2=0$.}
Unfortunately, it is not clear how to solve these equations.\footnote{In 
Ref.~56, 
the numerical solution was suggested.}
One can  start with the trial field
which is a superposition of the two shock waves (\ref{fla10}), 
and improve
it by taking into account the interaction between the shock waves order by
order.\cite{fizrev} The
parameter of this expansion is the  commutator $g^2[U_i,V_k]$. Actually, there
are two independent commutators,
 \begin{eqnarray}
&&L_1=L_1^at^a, \qquad L^a_1=if^{abc}U^a_jV^{bj},\nonumber\\
&&L_2=L_2^at^a, \qquad L^a_2=i\epsilon_{ik}f_{abc}U^{bi}V^{ck}, 
\label{5.1.8}
\end{eqnarray}
where $\epsilon_{ik}$ is the totally antisymmetric tensor in two transverse
dimensions ($\epsilon_{12}=1$). In these notations $[U_i,V^i]=L_1$ and
$[U_i,V_k]-(i\leftrightarrow k)=\epsilon_{ik}L_2$. It can be demonstrated that 
each extra commutator brings a factor $\ln{\sigma\over\sigma'}$ (each 
commutator means higher term in the cluster expansion in Fig.~\ref{cluster}),
 thus this approach is a kind of LLA. It is
convenient to choose the trial field in the form~\footnote{In 
the paper of Ref.~3, 
I used a slightly different trial configuration 
${\bar A}^{(0)}_{\ast}={\bar A}^{(0)}_{\bullet}=0,
~{\bar A}^{(0)}_i=\theta(x_{\bullet})V_i+\theta(x_{\ast})U_i$. The 
difference $\Delta_i$ is corrected by the  ${\bar A}^{(1)}$ term, so 
the results for the total field 
${\bar A}^{(0)}+{\bar A}^{(1)}$ are the same.
}
\begin{equation}
{\bar A}^{(0)}_{\ast}={\bar A}^{(0)}_{\bullet}=0,\qquad
{\bar A}^{(0)}_i=\theta(x_{\bullet})V_i+\theta(x_{\ast})U_i
+\theta(x_{\bullet})\theta(x_{\ast})\Delta_i 
\label{5.1.9}
\end{equation}
where $\Lambda_i(x_{\perp})=U_i(x_{\perp})+V_i(x_{\perp})+\Delta_i(x_{\perp})$ 
is 
a pure gauge field satisfying the gauge
condition $\partial_i\Delta_i-i[\Lambda_i,\Delta_i]=0$. 
The explicit form of $\Delta_i$ is
\begin{eqnarray}
\Delta^i(x_{\perp})&=&
ig\epsilon^{ik}
\big(U^{\dagger}{\partial_k\over\vec{\partial}_{\perp}^2}U
+V^{\dagger}{\partial_k\over\vec{\partial}_{\perp}^2}V
-{\partial_k\over\vec{\partial}_{\perp}^2}\big)
L_2+O(L^2)\label{5.1.10}\\
&=&-ig\int
dz_{\perp}{\epsilon^{ik}(x-z)_k\over
2\pi(\vec{x}-\vec{z})_{\perp}^2}(U_xU^{\dagger}_z+
V_xV^{\dagger}_z-1)L_2(z_{\perp})
dz_{\perp} +O(L^2) .
\nonumber
\end{eqnarray}
In the first nontrivial order one gets:
\begin{eqnarray} 
{\bar A}^{(1)}_i&=&-{i\over 2\pi^2}
\int dz_{\perp}{1\over
-x_{\parallel}^2+(\vec{x}-\vec{z})_{\perp}^2+
i\epsilon}\Delta_i(z_{\perp})\nonumber\\
&=&-{g\over
4\pi^2}\int dz_{\perp}
{\epsilon_{ik}(x-z)^k\over(\vec{x}-\vec{z})_{\perp}^2}
\ln\left(1-{(\vec{x}-\vec{z})_{\perp}^2\over
x_{\parallel}^2+i\epsilon}\right)L_2(z_{\perp}), \nonumber\\ 
{\bar A}^{(1)}_{\bullet}&=&{gs\over 16\pi^2}\int dz_{\perp} {1\over
x_{\ast}+i\epsilon}\ln(-x_{\parallel}^2+(\vec{x}-\vec{z})_{\perp}^2+i\epsilon)
L_1(z_{\perp}),\nonumber\\ 
{\bar A}^{(1)}_{\ast}&=&-{gs\over 16\pi^2}\int dz_{\perp}
{1\over x_{\bullet}+i\epsilon}\ln(-x_{\parallel}^2+
(\vec{x}-\vec{z})_{\perp}^2+i\epsilon)
L_1(z_{\perp}),\label{5.1.11}
\end{eqnarray}
where $x_{\parallel}^2\equiv{4\over s}x_{\ast}x_{\bullet}$ is a 
longitudinal part of $x^2$. These fields are
obtained in the background-Feynman gauge. The corresponding expressions for
field strength have the form 
\begin{eqnarray}
{\bar F}^{(1)}_{\bullet\ast}&=&{gs\over 4\pi^2}
\int dz_{\perp}{1\over -x_{\parallel}^2+
(\vec{x}-\vec{z})_{\perp}^2+i\epsilon}L_1(z_{\perp})
\label{5.1.12},\\
{\bar F}^{(1)}_{ik}&=&{g\over 2\pi^2}\epsilon_{ik}
\int dz_{\perp}{1\over
-x_{\parallel}^2+(\vec{x}-\vec{z})_{\perp}^2+i\epsilon}L_2(z_{\perp}) , 
\nonumber\\
{\bar F}^{(1)}_{\bullet i}&=&
{gs\over 8\pi^2}
\!\int\! dz_{\perp}{(x-z)^k\over -x_{\parallel}^2+
(\vec{x}-\vec{z})_{\perp}^2+i\epsilon}
\left({g_{ik}L_1(z_{\perp})\over x_{\ast}-i\epsilon} +
{\epsilon_{ik}L_2(z_{\perp})\over
x_{\ast}+i\epsilon}\right)\nonumber\\
&-&i[{\bar A}^{(1)}_{\bullet},{\bar A}^{(0)}_i] ,
\nonumber\\
{\bar F}^{(1)}_{\ast i}&=&-
{gs\over 8\pi^2}
\!\int\! dz_{\perp}{(x-z)^k\over -x_{\parallel}^2+
(\vec{x}-\vec{z})_{\perp}^2+i\epsilon}
\left({g_{ik}L_1(z_{\perp})\over x_{\bullet}-i\epsilon}-
{\epsilon_{ik}L_2(z_{\perp})\over
x_{\bullet}+i\epsilon}\right)\nonumber\\
&-&i[{\bar A}^{(1)}_{\ast},{\bar A}^{(0)}_i].
\nonumber 
\end{eqnarray}

\begin{figure}[htb]
\centerline{
\epsfysize=3cm
\epsffile{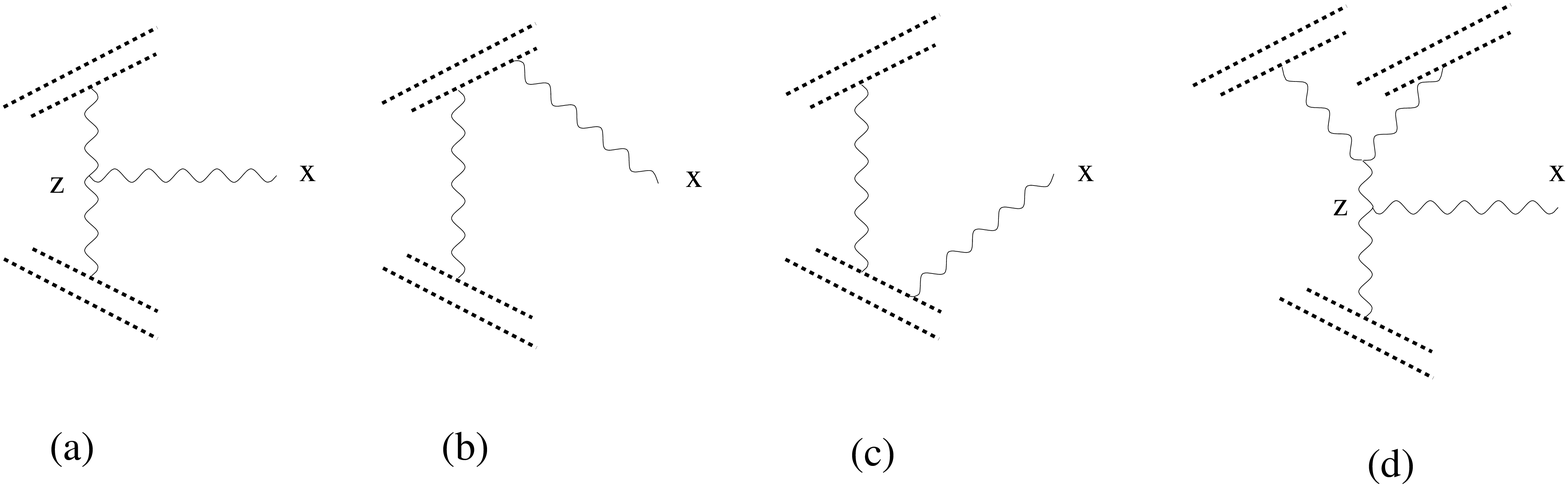}}
\caption{Perturbative Feynman diagrams for the field strength
(\protect\ref{5.1.12}).}  
\label{fig11}
\end{figure}

In terms of usual Feynman diagrams (when we expand in powers of source 
just like in Sec.~4.2) these expressions come from the diagrams
shown in Fig.~\ref{fig11}.
When we sum up the three contributions
from the diagrams in Figs.~\ref{fig11}a, \ref{fig11}b, and \ref{fig11}c the
three-gluon vertex in Fig.~\ref{fig11}a is replaced by the 
effective Lipatov's
vertex (\ref{2.2.14})
and we get (\ref{5.1.12}) up to the terms 
${1\over \partial^2}\partial_i\partial_kU^k$ and 
${1\over \partial^2}\partial_j\partial_kV^k$ standing in place of $U_i$ 
and $V_j$.
However, as we have discussed in Sec.~3, the difference 
$U_i-{1\over \partial^2}\partial_i\partial_kU^k=
g{\partial_k\over \partial^2}[U_i,U_k]$ (which has an additional power of g) 
will be canceled by the next-order perturbative diagrams 
of the Fig.~\ref{fig11}d type.

 Let us now find the effective action
\begin{equation}
\bar{S}_{\rm eff}=S_{QCD}(\bar{A})+
\!\int\! d^2x_{\perp} 
V^{ai}(x_{\perp})\bar{Y}^{a}_i(x_{\perp})
+\!\int\! d^2x_{\perp}\bar{W}^{ai}
U^a_i(x_{\perp})
\label{5.1.13}
\end{equation}
in the semiclassical approximation. In the trivial order the only non-zero 
field strength components are 
${\bar F}^{(0)}_{\bullet i}=\delta({2\over s}x_{\ast})U_i(x_{\perp})$ and
${\bar F}^{(0)}_{\ast i}=\delta({2\over s}x_{\bullet})V_i(x_{\perp})$,
hence we get the familiar  expression 
$S^{(0)}=\int d^2x_{\perp}V^{ai}U^{a}_i$. In the next order one has
\begin{eqnarray}
&&\hspace{-1cm}S^{(1)}~=~\int d^4x\left(-{2\over s}{\bar F}_{\ast}^{(1)ai}{\bar
F}^{(1)a}_{\bullet i}- 
{1\over 4}{\bar F}^{(1)a}_{ik}{\bar F}^{(1)aik}+{2\over
s^2} {\bar F}^{(1)a}_{\ast\bullet}
{\bar F}^{(1)a}_{\ast\bullet}\right) \nonumber\\
&&\hspace{-5mm} +~2\!\int\! d^2x_{\perp}\int\! du\Big({\rm Tr}V^{i}
\left([-\infty p_1,up_1]_x
{\bar F}_{\bullet i}(up_1+x_{\perp})
[up_1,-\infty p_1]_x\right)^{(1)}\nonumber\\
&&\hspace{-5mm}+~{\rm Tr}U^{i} \left([-\infty p_2,up_2]_x
{\bar F}_{\ast i}(up_2+x_{\perp})[up_2,\infty
p_2]_x\right)^{(1)}\Big). 
\label{5.1.14}
\end{eqnarray} 
Above, we have seen that the effective action
contains $\ln{\sigma\over\sigma'}$ (see Eq.~(\ref{fla23})). 
With logarithmic accuracy, the right-hand side  
of Eq.~(\ref{5.1.14}) reduces to
\begin{eqnarray}
S^{(1)}&=&-{2\over s}\int d^4x {\bar F}^{(1)ai}_{\ast}(x)
{\bar F}_{\bullet i}^{(1)a}(x).
\label{5.1.15}\\
&+&\int d^2x_{\perp}2{\rm Tr}L_1(x_{\perp})
\left([x_{\perp},-\infty p_2+x_{\perp}]^{(1)}-
[x_{\perp},-\infty p_1+x_{\perp}]^{(1)}\right).
\nonumber
\end{eqnarray}
The first term contains the integral over
$d^4x={2\over s}dx_{\bullet}dx_{\ast}d^2x_{\perp}$. In order to separate the 
longitudinal divergencies 
from the infrared divergencies in the transverse space we will
work in the $d=2+2\epsilon$ transverse dimensions. 
It is convenient first to perform the 
integral over $x_{\ast}$ determined by a residue in the
point $x_{\ast}=0$. The integration over remaining 
light-cone variable 
$x_{\bullet}$ then factorizes in the form 
$\int_{0}^{\infty}dx_{\bullet}/x_{\bullet}$ or
$\int_{-\infty}^{0}dx_{\bullet}/x_{\bullet}$.
This integral reflects our usual longitudinal logarithmic divergencies,
which arise from the replacement of vectors $n$ and $n'$ in (\ref{fla22}) 
by the light-like vectors $p_1$ and $p_2$. 
In the momentum space this logarithmical divergency has the form 
$\int d\alpha/\alpha$. 
It is clear that when $\alpha$ is close to $\sigma$ (or $\sigma'$) we 
can no longer approximate $n$ by $p_1$ (or $n'$ by $p_2$). Therefore, 
in the leading log approximation this divergency should be replaced by 
$\ln{\sigma\over\sigma'}$,
\begin{eqnarray}
\int_{0}^{\infty}dx_{\bullet}{1\over x_{\bullet}}=
\int_{0}^{\infty}d\alpha {1\over \alpha}\rightarrow 
\int_{\sigma}^{\sigma'}d\alpha {1\over \alpha}~=~\ln{\sigma\over\sigma'} .
\label{5.1.16}
\end{eqnarray}
The (first-order) gauge links in the second term in the right-hand side  
of Eq.~(\ref{5.1.15}) have the logarithmic divergence of the same origin,
\begin{eqnarray} 
\! [x_{\perp},-\infty p_1+x_{\perp}]^{(1)}&=&-{i\over 8\pi^2}
\int^{0}_{-\infty}\!{dx_{\ast}\over x_{\ast}}\int d^2z_{\perp}
{\Gamma(\epsilon)\over(\vec{x}-\vec{z})_{\perp}^{2\epsilon}}
L_1(z_{\perp}),\nonumber\\ 
\! [x_{\perp},-\infty p_2+x_{\perp}]^{(1)}&=&{i\over 8\pi^2}
\int^{0}_{-\infty}\!{dx_{\bullet}\over x_{\bullet}}\int d^2z_{\perp}
{\Gamma(\epsilon)\over(\vec{x}-\vec{z})_{\perp}^{2\epsilon}}
L_1(z_{\perp}),
\label{5.1.17}
\end{eqnarray}
which should also be replaced by $\ln{\sigma\over\sigma'}$.\footnote{The 
fields ${\bar A}_{\bullet}$ and ${\bar A}_{\ast}$ in Eq.~(\ref{5.1.11}) look 
like they
satisfy the condition $x_{\ast}A_{\bullet}+x_{\bullet}A_{\ast}=0$ implying
the fact that $P\exp ig\int du e^{\mu}A_{\mu}(un+x_{\perp})=0$ for any vector
$e=\varsigma p_1+\tilde{\varsigma}p_2$. One may suspect that the proper limit
at $e^2\rightarrow 0$ is to set $[x_{\perp},-\infty p_1+x_{\perp}]$ and
$[x_{\perp},-\infty p_2+x_{\perp}]$ to 0. However, careful analysis with the
slope of the $Y$ operators $n=\sigma p_1+\tilde{\sigma} p_2$ instead of $p_1$
and the slope of $W$ operators $n'=\sigma'p_1+\tilde{\sigma}'p_2$ instead of
$p_2$ shows that 
\begin{eqnarray}
[x_{\perp},-\infty e+x_{\perp}]&=&\label{dobav1}
{i\over 16\pi^2}\int d^2z_{\perp}
{\Gamma(\epsilon)\over(\vec{x}-\vec{z})_{\perp}^{2\epsilon}}
L_1(z_{\perp})\\
&\times&\Big({\sigma'/\tilde{\sigma}'+\varsigma/\tilde{\varsigma}
\over \sigma'/\tilde{\sigma}'-\varsigma/\tilde{\varsigma}} 
\ln{\tilde{\varsigma}\over\varsigma}
{\sigma'\over\tilde{\sigma}'}-
{\varsigma/\tilde{\varsigma}+\sigma/\tilde{\sigma}
\over \varsigma/\tilde{\varsigma}-\sigma/\tilde{\sigma}} 
\ln{\tilde{\varsigma}\over\varsigma}{\sigma\over\tilde{\sigma}}\Big)
\nonumber
\end{eqnarray}
leading to (\ref{5.1.17}) if $\varsigma\rightarrow \sigma$ or
$\varsigma\rightarrow \sigma'$.
}
Performing the
remaining integration over $x_{\perp}$ in the first term in right-hand side of 
Eq.~(\ref{5.1.15}) we obtain the the first-order classical action in the form
\begin{eqnarray}
S^{(1)}&=&\label{5.1.18} 
-{ig^2\over 8\pi^2}\ln{\sigma\over\sigma'}\\
&\times&\int d^2x_{\perp}
d^2y_{\perp}\big(L^a_1(x_{\perp})L^a_1(y_{\perp})+
L^a_2(x_{\perp})L^a_2(y_{\perp})\big)
{\Gamma(\epsilon)\over(\vec{x}-\vec{y})_{\perp}^{2\epsilon}}
\nonumber
\end{eqnarray}
or
\begin{equation}
S^{(1)}={ig^2\over 2\pi}\ln{\sigma\over\sigma'}
\int d^2x_{\perp} \left(L_1^a{1\over \vec{\partial}_{\perp}^2}L_1^a+
L_2^a{1\over \vec{\partial}_{\perp}^2}L_2^a\right).
\label{5.1.19}
\end{equation}
Note that in the trivial order the three terms in Eq.~(\ref{5.1.13}) are
equal up to the different sign of the $S(\bar{A})$ term. 
It can be demonstrated that this is true in the first order, too:
\begin{equation}
\int d^2x_{\perp}2{\rm Tr}V^i\bar{Y}_i^{(0+1)}= \int d^2x_{\perp}
2{\rm Tr}\bar{W}_i^{(0+1)}U_i= -S(\bar{A})^{(0+1)} .
\label{5.1.20}
\end{equation}
 A more accurate version of Eq.~(\ref{5.1.19}) has the form (see Appendix 7.5) 
%
\begin{eqnarray}
S^{(1)}&=&\label{5.1.21}
{ig^2\over 2\pi}\ln{\sigma\over\sigma'}\int d^2x_{\perp}\\
&\times&
\Bigg(L_1^a{1\over \vec{\partial}_{\perp}^2}L_1^a+
L_2^a\big(U^{\dagger}{1\over \vec{\partial}_{\perp}^2}U+
V^{\dagger}{1\over \vec{\partial}_{\perp}^2}V-
{1\over \vec{\partial}_{\perp}^2}\big)^{ab}L_2^b
\nonumber\\
&+&
L_1^a\big({\partial_i\over\vec{\partial}_{\perp}^2}
U^{\dagger}{\partial_k\over\vec{\partial}_{\perp}^2}U
-U\leftrightarrow V\big)L^b_2\epsilon^{ik} \nonumber\\
&-&L_2^a\epsilon^{ik}
\big(U^{\dagger}{\partial_i\over\vec{\partial}_{\perp}^2}
U{\partial_k\over\vec{\partial}_{\perp}^2}
-U\leftrightarrow V\big)^{ab}L^b_1\Bigg)+O([U,V]^3).
\nonumber
\end{eqnarray}

 It is easy to see that in the case of one weak and one strong  
 source this expressions coincides with (\ref{dob5}) (up to the terms of higher 
 order in weak source which we neglect anyway).
 
\begin{figure}[htb]
\centerline{
\epsfysize=4cm
\epsffile{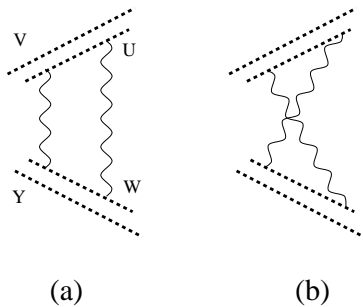}}
\caption{Lowest-order diagrams for gluon reggeization.} 
\label{figreg}
\end{figure}

At $d=2$ we have an infrared pole in $S^{(1)}$ which must be canceled by the 
corresponding divergency in the trajectory of the reggeized gluon. 
The gluon reggeization is not a classical effect in our approach, rather 
it is a quantum correction coming from the loop corresponding to the determinant
of the operator of second derivative of the action
\begin{equation}
\left.{\delta \over \delta A_{\mu}}{\delta \over \delta
A_{\nu}}\left(S_{QCD}+\!\int\! d^2x_{\perp} 
V^{ai}(x_{\perp})Y^{a}_i(x_{\perp}) +\!\int\!
d^2x_{\perp}W^{ai}U^a_i(x_{\perp}\right)\right|_{A=\bar{A}}. 
\label{5.1.22}
\end{equation} 
The lowest-order diagrams are shown in 
Fig.~\ref{figreg}
and the explicit form of the second derivative 
of the Wilson-line operator is 
\begin{eqnarray}
\delta Y_i&=&i
\int^{\infty}_{-\infty}du\int^{u}_{-\infty}dv
[\delta\hat{A}_i(u p_1+x_{\perp}),
\hat{\nabla}_i\delta\hat{A}_i(v p_1+x_{\perp})],\nonumber\\
\delta W_i&=&i
\int^{\infty}_{-\infty}du\int^{u}_{-\infty}dv
[\tilde{A}_i(u p_2+x_{\perp}),
\tilde{\nabla}_i\delta\tilde{A}_i(u p_2+x_{\perp})] .
\label{5.1.23}
\end{eqnarray}
Now one easily gets the contribution of the 
Fig.~\ref{figreg} diagrams in the form 
\begin{eqnarray}
S_{\rm r}&=&\label{5.1.24}
{g^2N_c\over 8\pi^3}\ln{\sigma\over\sigma'}
\int d^2x_{\perp} d^2y_{\perp} \\
&\times&\Big(V_i^a(x_{\perp})U^{ai}(y_{\perp})-
V_i^a(x_{\perp})U^{ai}(x_{\perp})\Big)
{\Gamma^2(1+\epsilon)\over
((\vec{x}-\vec{y})_{\perp}^2)^{(1+2\epsilon)}} .
\nonumber
\end{eqnarray}
A more accurate form
of this equation reads:
\begin{eqnarray}
S_{\rm r}&=&\label{5.1.25}
{g^2N_c\over 8\pi^3}\ln{\sigma\over\sigma'}
\int d^2x_{\perp} d^2y_{\perp} 
{\Gamma^2(1+\epsilon)\over
((\vec{x}-\vec{y})_{\perp}^2)^{(1+2\epsilon)}} \\
&\times& \Big\{-V_i^a(x_{\perp})U^{ai}(x_{\perp})
+{1\over N_c}\Big(
V^{i}(x_{\perp})\{U(x_{\perp})U^{\dagger}(y_{\perp})\nonumber\\
&+&
V(x_{\perp})V^{\dagger}(y_{\perp})-1\}
U^{i}(y_{\perp})\Big)^{aa}\Big\}+O([U,V]),\nonumber
\end{eqnarray}
where ${\cal O}^{aa}\equiv$Tr$O$ in the gluonic representation. In 
the case of one strong and one weak source it coincides 
with (\ref{dob2}) (up
to the higher powers of weak source).
 
The complete first-order ($\equiv$ one-log) expression 
for the effective action is
the sum of $S^{(0)}$, $S^{(1)}$, and $S_{\rm r}$,
\begin{eqnarray}
S_{\rm eff}\!\!\! &=&\!\!\!
\int d^2xV^{ai}(x)U^{a}_i(x)~+~ 
{ig^2\over 8\pi^2}\ln{\sigma\over\sigma'}\!\int\! d^2x
d^2y
\Bigg\{-{\Gamma(\epsilon)\over(\vec{x}-\vec{z})_{\perp}^{2\epsilon}} 
\label{5.1.26}\\
&\times&\!\!\! \Big(L^a_1(x)L^{a}_1(y)+L^a_2(x)L^{b}_2(y)
\big(U^{\dagger}_xU_y +V^{\dagger}_xV_y-1\big)^{ab}\Big)
\nonumber\\ 
&+&\!\!\! \int\! d^2z 
{\epsilon^{ij}
(x-z)_i(z-y)_j\over \pi
(\vec{x}-\vec{z})_{\perp}^2(\vec{z}-\vec{y})_{\perp}^2}\nonumber\\
&\times&\!\!\! 
\Big(L^a_1(x)\big(U^{\dagger}_zU_y -U\leftrightarrow V\big)^{ab}
L^{b}_2(y)-
L^a_2(x)\big(U^{\dagger}_xU_z -U\leftrightarrow V\big)^{ab}
L^{b}_1(y)\Big)
\Bigg\}\nonumber\\ 
&+&\!\!\! {g^2N_c\over 8\pi^3}\ln{\sigma\over\sigma'}
\int d^2x_{\perp} d^2y_{\perp} 
{\Gamma^2(1+\epsilon)
\over((\vec{x}-\vec{y})_{\perp}^2)^{(1+2\epsilon)}}
\Bigg\{-V_i^a(x_{\perp})U^{ai}(x_{\perp})\nonumber\\
&+&\!\!\! 
{1\over N_c}\Big( V^{i}(x_{\perp})\{U(x_{\perp})
U^{\dagger}(y_{\perp})+
V(x_{\perp})V^{\dagger}(y_{\perp})-1\}
U^{i}(y_{\perp})\Big)^{aa}\Bigg\}.\nonumber
\end{eqnarray}
In the case of one weak and one strong source this expression coincides with
(\ref{dob6}) up to the higher powers of weak source. (As we discussed in 
Sec.~4.3, the new nontrivial terms in the case of two strong sources start from
$[Y,V]^3\ln^2{\sigma\over\sigma'}$).  

As usual, in the case of scattering of white objects the logarithmic
infrared divergence $\sim {1\over\epsilon}$ cancels. 
For example, for the case of one-pomeron exchange the relevant term
in the expansion of $e^{iS_{\rm eff}}$ has the form 
\begin{eqnarray}
&&-{g^2\over 16\pi^2}\ln{\sigma\over\sigma'}\int d^2x_{\perp} d^2y_{\perp} 
f^{dam}(V_j^aU^{mj}g_{ik}+V^a_iU^m_k-V^a_kU^m_i)(x_{\perp})\nonumber\\
&\times&{\Gamma(\epsilon)\over(\vec{x}-\vec{y})_{\perp}^{2\epsilon}}
f^{dbn}(V_l^bU^{nl}g^{ik}+V^{bi}U^{mk}-V^{bk}U^{mi})(y_{\perp})\nonumber\\
&+&
{g^2N_c\over 16\pi^3}\ln{\sigma\over\sigma'}
\int d^2x_{\perp}V^a_i(x_{\perp})U^{ai}(x_{\perp})\int d^2y_{\perp} 
d^2y'_{\perp} (V_j^b(y_{\perp})-V_j^b(y'_{\perp}))
\nonumber\\
&\times&{\Gamma^2(1+\epsilon)\over((\vec{y}-\vec{y}')_{\perp}^2)^{(1+2\epsilon)}}
(U^{bj}(y_{\perp})-U^{bj}(y'_{\perp})).
\label{5.1.27}
\end{eqnarray}
It is easy to see that the terms $\sim {1\over\epsilon}$  cancel if 
we project Eq.~(\ref{5.1.27}) onto 
colorless state in t-channel (that is, replace $V^{ai}V_j^b$ by 
${\delta_{ab}\over N_c^2-1}V^{ci}V_j^c$). It is worth noting that in the 
two-gluon approximation the right-hand side  
of the Eq.~(\ref{5.1.27}) gives the BFKL kernel 
(\ref{2.2.30}).

\begin{figure}[htb]
\centerline{
\epsfysize=3cm
\epsffile{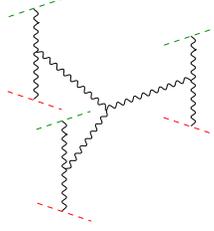}}
\caption{Typical next-to-leading order contribution to $S_{\rm
eff}$.}
\label{figa16} 
\end{figure}

As an illustration, let us present the next-to-leading contribution
to the effective action $\simeq[U,V]^3\ln{\sigma\over\sigma'}$ coming from the
diagrams of Fig.~\ref{figa16} type. 
\begin{eqnarray}
S_{\rm eff}\!\!&=&\!\!g^3f_{abc}
\ln{\sigma\over\sigma'}\int dx_{\perp} dy_{\perp} d_{\perp}z
\big[K_1(x_{\perp},y_{\perp},z_{\perp})L^a_1(x_{\perp}) 
L_1^b(y_{\perp}) 
L_2^c(z_{\perp})\nonumber\\
&+&\!\!K_2(x_{\perp},y_{\perp},z_{\perp})L_2^2(x_{\perp}) 
L_2^b(y_{\perp}) L_2^c(z_{\perp})\big] ,
\label{5.1.28}
\end{eqnarray}
where
\begin{eqnarray}
&&\hspace{-8mm}K_i(x,y,z)=\int {d^2p_1\over 4\pi^2}{d^2p_2\over
4\pi^2}K_i(p_1,p_2,-p_1-p_2)e^{ip_1\cdot(x-z)+ip_2\cdot(y-z)},
\nonumber\\ 
&&\hspace{-8mm}K_1(p_1,p_2,p_3)={i\over 2\pi^2}
{\epsilon_{ik}p_1^ip_2^k\over p_1^2p_2^2p_3^2}
\left(\ln p_3^2-{p_1^2\over p_1^2-p_2^2}\ln p_1^2-
{p_2^2\over p_2^2-p_1^2}
\ln p_2^2 \right)  ,
\nonumber\\ 
&&\hspace{-8mm}K_2(p_1,p_2,p_3)=-{i\over 4\pi^2}
{\epsilon_{ik}p_1^ip_2^k\over p_1^2p_2^2}
\left({1\over p_1^2-p_3^2}\ln {p_1^2\over p_3^2}+ 
{1\over p_2^2-p_3^2}\ln
{p_2^2\over p_3^2}\right) . 
\label{5.1.29}
\end{eqnarray}

\subsection{Effective action as integral over Wilson lines} 

\begin{figure}[htb]
\centerline{
\epsfysize=9cm
\epsffile{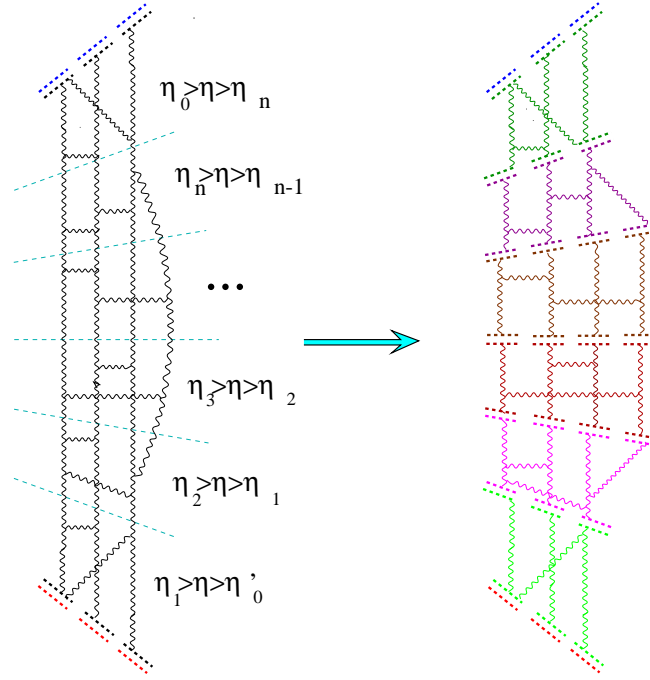}}
\caption{Effective action factorized in $n$ functional integrals.}
\label{figentimes} 
\end{figure}

In this section we will rewrite the functional integral for the 
effective
action (\ref{fla22}) in terms of Wilson-line variables. To this end, 
let us use
the factorization formula (\ref{fla18}) $n$ times as shown in 
Fig.~\ref{figentimes}.
The effective action factorizes then into a product 
of $n$ independent functional integrals over the gluon fields labeled
by index $k$:
%
\begin{eqnarray}
e^{iS_{\rm eff}(U,V;\eta)}&=&\int\! DA_1\ldots DA_{n+1} 
\ \exp i\Big\{V_iY_{n+1}^i +S(A_{n+1})\label{nov1}\\
&+&
W_{n+1,i}Y_n^i+S(A_n)+\ldots +W_{2i}Y^i_1+S(A_1)+W_1^iU_i\Big\} ,
\nonumber
\end{eqnarray}
where the integrals over $x_{\perp}$ and summation over the color indices are
implied.  As usual, $Y^i_k={i\over g}Y^{\dagger}_k\partial^iY_k$ and 
$W^i_k={i\over g}W^{\dagger}_k\partial^iW_k$
where
\begin{eqnarray}
Y_k(x_{\perp})&=&P\ \exp ig\int_{-\infty}^{\infty}du \ n_k^{\mu}
A_{k,\mu}(un^k+x_{\perp}) ,\nonumber\\
W_k(x_{\perp})&=&P\ \exp ig\int_{\infty}^{\infty}du \ n_{k-1}^{\mu}
A_{k,\mu}(un^{k-1}+x_{\perp}),
\label{nov2}
\end{eqnarray}
and the vectors $n_k$ are ordered in rapidity: 
$\eta_0>\eta_n>\eta_{n-1}\ldots\eta_2>\eta_1>\eta'_0$.
To disentangle integrations over different $A^k$ we use the formula 
%
\begin{eqnarray}
e^{i\int dx_{\perp}W_iY^i}&=&
\det (\partial_i-igW_i)(\partial^i-igY^i)\label{nov3}\\
&\times&
\int DV(x_{\perp})DU(x_{\perp})e^{i\int dx_{\perp}W_iU^i+
i\int dx_{\perp}V_iY^i-i\int dx_{\perp}V_iU^i}.
\nonumber
\end{eqnarray}
The determinant gives the perturbative non-logarithmic corrections 
of the same order as the corrections to the factorization 
formula (\ref{flad1}). In the LLA they can be ignored, 
consequently, we obtain
%
\begin{eqnarray}
e^{iS_{\rm eff}(U,V)}&=&
\int DA_1\ldots DA_{n+1} DU_1DV_1\ldots DU_nDV_n 
\nonumber\\
&\times&
\exp i\Big\{V_iY_{n+1}^i+S(A_{n+1})+
W_{n+1}^iU_{n,i}-V_{n,i}Y^i_n+\ldots
\nonumber\\
&+&W_{3i}U^i_2-V_2^iU_{2i}
+V_{2,i}Y^i_2+S(A_2)+
W_{2,i}U_1^i-V_{1,i}U_1^i\nonumber\\
&+&V_{1,i}Y^i_1+S(A_1)+
W_1^iU_i\Big\} .
\label{nov4}
\end{eqnarray}
Now we can integrate over the gluon fields $A_k$, 
\begin{equation}
\int DA_k e^{V_{k,i}Y_k^i+S(A_k)+W_{k,i}U^{i}_{k-1}}=
e^{iS_{\rm eff}(V^k,U^{k-1};\Delta\eta)}~;
\label{nov5}
\end{equation}
at sufficiently small $\Delta\eta$
\begin{equation}
S_{\rm eff}(V^k,U^{k-1};\Delta\eta)=V_{k,i}U_{k-1}^i-
i\Delta\eta K(V_k,U_{k-1})
+O(\Delta\eta^2),
\label{nov6}
\end{equation}
where K is the kernel calculated in the previous section,
%
\begin{eqnarray}
K(V,U)&=&-\alpha_s\int d^2x_{\perp} \nonumber\\
&\times&
\Bigg\{L_1^a{1\over \vec{\partial}_{\perp}^2}L_1^a+
L_2^a\big(U^{\dagger}{1\over \vec{\partial}_{\perp}^2}U+
V^{\dagger}{1\over \vec{\partial}_{\perp}^2}V-
{1\over \vec{\partial}_{\perp}^2}\big)^{ab}L_2^b
\nonumber\\
&+&
L_1^a\big({\partial_i\over\vec{\partial}_{\perp}^2}U^{\dagger}
{\partial_k\over\vec{\partial}_{\perp}^2}U
-U\leftrightarrow V\big)L^b_2\epsilon^{ik} \nonumber\\
&-&L_2^a\epsilon^{ik}
\big(U^{\dagger}{\partial_i\over\vec{\partial}_{\perp}^2}
U{\partial_k\over\vec{\partial}_{\perp}^2}
-U\leftrightarrow V\big)^{ab}L^b_1\nonumber\\ 
&+&{i\over 4\pi}\Big(V_i\big(U^{\dagger}
(\ln\vec{\partial}_{\perp}^2)U+
V^{\dagger}(\ln\vec{\partial}_{\perp}^2)V-
(\ln\vec{\partial}_{\perp}^2)\big)U^i\Big)^{aa}\Bigg\}.
\label{flakern}
\end{eqnarray}

Performing the integrations over $A^k$ we get
\begin{eqnarray}
e^{iS_{\rm eff}(U,V)}&=&\int DV_1DU_1\ldots DV_nDU_n 
\ \exp\Big\{iV_iU^i_{n}+K(V,U_{n})\Delta\eta \label{nov7}\\
&-&iV_{n,i}U^i_{n}+iV_{n,i}U_{n-1}^i+
K(V_{n},U_{n-1})\Delta\eta
+\ldots -iV_{2i}U^{i}_2 \nonumber\\
&-&iV_2^iU_{1i}+K(V_2,U_1)\Delta\eta-iV_1^iU_{1i}+iV_1^iU_i+
K(V_1,U)\Delta\eta\Big\} .
\nonumber
\end{eqnarray}
In the limit $n\rightarrow\infty$ we obtain the following functional
integral for the effective action
\begin{eqnarray}
e^{iS_{\rm eff}(U,V)}&=&\left.\int DV(\eta)DU(\eta)\right|_{U(\eta_0')=U}
~\exp\Big\{iV^a_iU^{ai}(\eta) \label{nov8}\\
&+& \int_{\eta'_0}^{\eta_0}d\eta\Big(-iV^{ai}(\eta)\dot{U}^a_i(\eta)
+ K(V(\eta),U(\eta)\Big)\Big\} .
\nonumber
\end{eqnarray}
where we displayed the color indices explicitly.
This looks like the functional integral over the canonical coordinates $U$ and 
canonical momenta $V$ with the (non-local) Hamiltonian $K(V,U)$.
The rapidity $\eta$ serves as the time variable for
this system. Let us demonstrate that perturbative expansion for the
functional integral (\ref{nov8}) determines the effective field theory for
reggeized gluons. To get the perturbative series for the functional integral
(\ref{nov8}), we write down $U(\eta)$ and $V(\eta)$ as 
\begin{eqnarray}
U(x_{\perp},\eta)=e^{-ig\phi(x_{\perp},\eta)},\qquad\qquad
V(x_{\perp},\eta)=e^{-ig\pi(x_{\perp},\eta)},
\label{nov9} 
\end{eqnarray}
($\phi^a(x_{\perp},\eta)$ and $\pi^a(x_{\perp},\eta)$ are scalar fields)
and expand in powers of $g$. In the leading order in $g$ we obtain 
%
\begin{eqnarray}
e^{iS_{\rm eff}(\phi,\pi)}&=&\left.\int
D\pi(\eta)D\phi(\eta)\right|_{\phi(\eta'_0)=\phi}~
\exp\Bigg\{-i\partial_i\pi^a\partial_i\phi^a(\eta_0) \nonumber\\
&+&
2{\rm Tr}\int_{\eta'_0}^{\eta_0}d\eta\Big(i\partial_i\pi(\eta)
\left({\partial\over\partial\eta}+{\alpha_s\over 4\pi}N_c
\ln\vec{\partial}_{\perp}^2\right)\partial_i\phi(\eta)
\nonumber\\
&-&\alpha_s
[\bar{\partial}\pi(\eta),\tilde{\partial}\phi(\eta)]
{1\over\vec{\partial}_{\perp}^2}
[\tilde{\partial}\pi(\eta),\bar{\partial}\phi(\eta)]\Big)\Bigg\} ,
\label{nov10}
\end{eqnarray}
where 
$\tilde{\partial}\equiv\partial_1+i\partial_2$,
$\bar{\partial}\equiv\partial_1- i\partial_2$.  
The bare propagator for
these fields is (cf. Ref.~2) 
%
\begin{eqnarray}
&&\langle\phi(x_{\perp},\eta)~\phi(y_{\perp},\eta')\rangle=0,
\qquad\qquad
\langle\pi(x_{\perp},\eta)~\pi(y_{\perp}, \eta')\rangle=0 ,
\nonumber\\
&&\langle\phi(x_{\perp},\eta)~\pi(y_{\perp},\eta')\rangle=
\theta(\eta-\eta')
\hbox{\bf\Big($\!\!$\Big(}x_{\perp}
\Big|{i\over \vec{p}_{\perp}^2}\Big|
y_{\perp}\hbox{\bf\Big)$\!\!$\Big)} . 
\label{nov11}
\end{eqnarray}
The $\theta$ function in this formula satisfies the condition 
$\theta(0)=0$ as 
can be easily seen from the limiting formula (\ref{nov7}). 
It is convenient to include the 
$g^2\pi\vec{\partial}_{\perp}^2\ln\vec{\partial}_{\perp}^2\phi$ 
in the
kinetic term rather than in the interaction Hamiltonian. 
Since this expression
is IR divergent one should at first consider the regularized 
$S_{\rm eff}$
\begin{eqnarray}
e^{iS_{\rm eff}(\phi,\pi)}&=&\int D\phi D\pi
\exp\Bigg\{2{\rm Tr}\int_{\eta'_0}^{\eta_0}d\eta\Big\{i\partial_i\pi
\left({\partial\over\partial\eta}+{\alpha_s\over 4\pi}N_c
\ln{\vec{\partial}_{\perp}^2\over\mu^2}\right)\partial_i\phi
\nonumber
\\
&-&\alpha_s
[\partial\phi,\bar{\partial}\pi]{1\over\vec{\partial}_{\perp}^2+
\mu^2}
[\partial\phi,\bar{\partial}\pi]\Big\}
-i\partial_i\pi^a\partial_i\phi^a(\eta_0)\Bigg\} 
\label{nov12}
\end{eqnarray}
and then take the limit $\mu^2\rightarrow 0$. (Alternatively, one can use the
regularization $d=2+\epsilon$ for the number of transverse
dimensions as it was done in Sec.~5.1.) The propagator takes the form
\begin{eqnarray} 
&&\hspace{-1.5cm}\langle\phi(x_{\perp},\eta)~\phi(y_{\perp},\eta')
\rangle=0,
\qquad\qquad 
\langle\pi(x_{\perp},\eta)~\pi(y_{\perp},\eta')\rangle=0,
\label{nov13} \\ 
&&\hspace{-1.5cm}\int
{dp_{\perp}\over 4\pi^2}
e^{ip(x-y)_{\perp}}\langle\phi(x_{\perp},\eta)\pi(y_{\perp},
\eta')\rangle=\theta(\eta-\eta') {i\over
\vec{p}_{\perp}^2}
e^{-{\alpha_s\over 4\pi}N_c(\eta-\eta')\ln{p^2\over \mu^2}} ,
\nonumber
\end{eqnarray} 
which coincides with the propagator of the reggeized gluon (\ref{2.5.2}).

Since the only non-vanishing Green functions are
$$
\langle\phi(x_1,\eta)\ldots \phi(x_m,\eta)\pi(y_1,\eta')\ldots 
\pi(y_n,\eta')\rangle
$$
with $m=n$, the number of reggeized gluons is conserved. 
It is easy to
see that the Feynman rules for the Green function 
$$
\langle \phi(x_1,\eta)\ldots \phi(x_n,\eta)\pi(y_1,\eta')\ldots 
\pi(y_n,\eta')\rangle
$$ 
reproduce the diagrams for the quantum mechanics of $n$
particles with Lipatov's Hamiltonian (\ref{2.5.4}) (see Fig.~10). 

\begin{figure}[htb]
\centerline{
\epsfysize=4cm
\epsffile{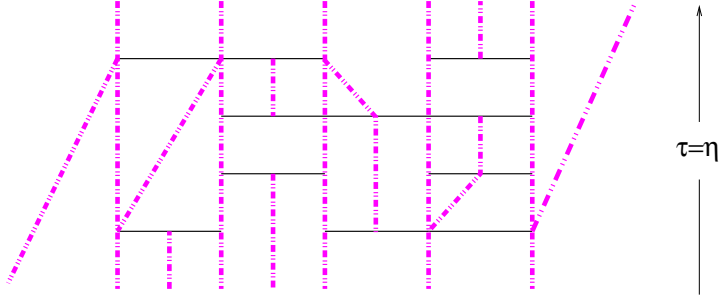}}
\caption{Feynman diagrams for the field theory of reggeized gluons.}
\label{figa19.fig} 
\end{figure}

In the next order in the expansion (\ref{nov9}) we get 
\begin{eqnarray}
e^{iS_{\rm eff}(\phi,\pi)}&=&
\left.\int D\pi(\eta)D\phi(\eta)\right|_{\phi(\eta'_0)=\phi}~\exp\Bigg\{ 
-i\partial_i\pi^a\partial_i\phi^a(\eta_0)\label{nov14}\\
&+& 2{\rm Tr}\int_{\eta'_0}^{\eta_0}
\Big\{i\partial_i\pi(\eta)
\left({\partial\over\partial\eta}+{\alpha_s\over 4\pi}N_c
\ln\vec{\partial}_{\perp}^2\right)\partial_i\phi(\eta) \nonumber\\
&-&\alpha_s
[\partial\phi,\bar{\partial}\pi]{1\over\vec{\partial}_{\perp}^2}
[\partial\phi,\bar{\partial}\pi]\Big\}+
i{g^3\over 4\pi}K_{(3)}(\phi,\pi)+{g^4\over 4\pi}K_{(4)}(\phi,\pi)
\Bigg\} ,
\nonumber
\end{eqnarray}
where
%
\begin{eqnarray}
K_{(3)}(\phi,\pi)&=&\hspace{-2mm}
\Bigg\{\big[[\partial_i\phi,\phi],\partial_i\pi\big]
{1\over\vec{\partial}_{\perp}^2} [\partial_j\phi,\partial_j\pi]
\label{nov15} \\
&+&\hspace{-2mm}
\Big(\big[[\partial_i\phi,\phi],\partial_j\pi
\big]+2\big[\phi,[\partial_i\phi,\partial_j\pi\big]\Big)
{1\over\vec{\partial}_{\perp}^2}
\big([\partial_i\phi,\partial_j\pi]-(i\leftrightarrow j)\big)
\nonumber\\
&-&\hspace{-2mm}
2[\partial_j\phi,\partial_j\pi]{\partial_i\over\vec{\partial}_{\perp}^2}
\phi^a{\partial_k\over\vec{\partial}_{\perp}^2}
\big(\big[t^a,[\partial_i\phi,\partial_k\pi]\big]-
(i\leftrightarrow k)\big)\Bigg\}
+\Big\{\pi\leftrightarrow\phi\Big\} 
\nonumber
\end{eqnarray}
and
\begin{eqnarray}
K_{(4)}(\phi,\pi)&=&\big[[\partial_i\phi,\phi],[\partial_i\pi,\pi]\big]
{1\over\vec{\partial}_{\perp}^2}
[\partial\phi,\bar{\partial}\pi]
\label{nov16} \\
&+&
\big[[\partial\phi,\phi],\partial_i\pi\big]{1\over\vec{\partial}_{\perp}^2}
\big[[\partial_i\phi,\phi],
\partial_i\pi \big]+\ldots . 
\nonumber
\end{eqnarray}
The number of reggeized gluons is no longer conserved, hence we get the field
theory of reggeized gluons with Feynman diagrams shown in 
Fig.~\ref{figa19.fig}.  
In higher orders we will get more complicated $\pi^m\phi^n$ vertices. 

It is intersting  to compare (\ref{nov4}) with Lipatov's effective action 
for reggeized gluons.\cite{many,lobzor} In these papers the reggeon 
is defined as a scalar field depending on both transverse and longitudinal 
coordinates. The integration of Lipatov's effective action over longitudinal
coordinates of the reggeons in the LLA
reproduces the first two (BFKL and three-pomeron) terms in the expansion
(\ref{nov4}). Hopefully, the integration of the Lipatov's action in the 
NLO LLA, NNLO LLA etc. will reproduce the expansion (\ref{nov4}) order by 
order in perturbation theory.

\subsection{Semiclassical approach to Wilson-line functional integral for
the effective action} 

Perturbation expansion (\ref{nov9}) is relevant when the characteristic 
$U_i$ and $V_i$ inside the functional integral (\ref{nov8}) are $\sim O(1)$.
However, we shall see below that at high energies
the characteristic fields in this functional integral seem to be large,
consequently the expansion (\ref{nov9}) may be useless. In this case, we
can try to calculate the functional integral (\ref{nov8}) semiclassically. 
The classical equations for the functional integral (\ref{nov8}) are
\begin{eqnarray}
(i\partial_i+g[V_i)\dot{U}^i&=&-{\delta \over V^{\dagger}\delta V}K(U,V) ,
\nonumber\\ 
(i\partial_i+g[U_i)\dot{V}^i&=&{\delta \over U^{\dagger}\delta U} K(U,V),
\label{nov17}
\end{eqnarray}
with the initial conditions
\begin{equation}
U(\eta)=U ~{\rm at}~\eta=\eta'_0,\qquad\qquad V(\eta)=V ~{\rm at}~\eta=\eta_0 .
\label{nov18}
\end{equation}
Let us denote the solution of these equation by $\bar{U}(x_{\perp},\eta)$ and
$\bar{V}(x_{\perp},\eta)$. In the LLA the semiclassical calculation of the 
Wilson-line integral (\ref{nov8}) 
is equivalent to the semiclassical calculation of the original functional
integral (\ref{fla22}). I will make a conjecture that the saddle point
of the original functional
integral (\ref{5.1.1}), satisfying the classical equations (\ref{5.1.7}),
corresponds to
the classical solution (\ref{fla18})  of the Wilson-line integral (\ref{nov8})
even beyond the LLA:
\begin{eqnarray}
&&\exp\left\{iV_i\bar{Y}^i(\eta_0)+i\bar{W}_i(\eta'_0)U^i+iS(\bar{A})\right\}
\label{nov19}\\
&=&\exp\Bigg\{iV_i\bar{U}_i(\eta_0)
+\int_{\eta'_0}^{\eta_0}d\eta\Big(-i\bar{V}^i(\eta)\dot{\bar U}_i(\eta)
+ K(\bar{V}(\eta),\bar{U}(\eta)\Big)\Bigg\},
\nonumber
\end{eqnarray}
where $\bar{A}$ is the classical solution of the equations (\ref{5.1.7}). 
(As in previous section, we do not display the integrals over the transverse
coordinates). Being a quantum correction, the gluon reggeization (\ref{5.1.25})
 exceeds the accuracy of the semiclassical approximation, hence we can drop the
last (reggeization) term in the kernel (\ref{flakern}).

Talking the variational derivative of
both sides of Eq.~(\ref{nov19})  with respect to $V$, we obtain
\begin{equation}
\bar{Y}_i(\eta)=\bar{U}_i(\eta).
\label{nov20}
\end{equation}
If we now take the derivative of both sides with respect to $\eta_0$, we get 
the equation 
\begin{equation}
iV_i\dot{\bar{Y}}_i(\eta_0)=iV_i\dot{\bar{U}}_i(\eta_0)=K(V,\bar{U}(\eta_0)),
\label{nov21}
\end{equation}
which may be used for the calculation of $K$. Correspondingly, one can 
differentiate with respect to $\eta'_0$ resulting in
\begin{equation}
-i\dot{\bar{V}}_i(\eta'_0)U_i=K(\bar{V}(\eta'_0),U).
\label{nov22}
\end{equation}
Since $V$ in Eq.~(\ref{nov21}) and $U$ in Eq.~(\ref{nov22}) are arbitrary, 
we may substitute $\bar{V}(\eta)$ and  $\bar{U}(\eta)$ instead :
\begin{equation}
i\bar{V}_i(\eta)\dot{\bar{U}}_i(\eta)=-i\dot{\bar{V}}_i(\eta)\bar{U}_i(\eta) =
K(\bar{V}(\eta),\bar{U}(\eta)). 
\label{nov23}
\end{equation}
The  exponential of the Wilson-line functional integral vanishes except
for the non-integral term $V_i\bar{U}^i(\eta_0)=V_i\bar{Y}^i(\eta_0)$, 
so 
\begin{eqnarray}
\exp\left\{iV_i\bar{Y}^i(\eta'_0)+i\bar{W}_i(\eta_0)U^i+iS(\bar{A})
\right\}&=&
\exp\big\{iV_i\bar{U}_i(\eta_0)\big\}\nonumber\\
&=&
\exp\big\{i\bar{V}_i(\eta'_0)U_i\big\}.
\label{nov24}
\end{eqnarray} 
Thus, in a semiclassical approximation 
(and with the assumption mentioned above)
we obtain 
\begin{equation}
S_{\rm eff}=V_i\bar{U}_i(\eta_0)=\bar{V}_i(\eta'_0)U_i =
-S(\bar{A}), 
\label{nov25}
\end{equation}
so that all the three terms in left-hand side of 
Eq.~(\ref{nov19}) contribute equally
up to a different sign for $S(\bar{A})$. We have checked it in LLA and
it is crucial to check it in the next-to-leading order.
From Eqs.~(\ref{nov23}) and (\ref{nov25}) we see that the effective action
in the semiclassical approximation can be written down also as
\begin{equation}
S_{\rm eff}=\bar{V}_i(\eta)\bar{U}_i(\eta)
\label{nov26}
\end{equation}
for arbitrary $\eta$. 

Instead of taking variational derivatives of the kernel $K(V,U)$, it
is possible to calculate $\dot{\bar{U}}\equiv\dot{\bar{Y}}$ directly. 
One obtains (cf. Eq.~(\ref{5.1.17}))
\begin{eqnarray} 
\! [x_{\perp},-\infty p_1+x_{\perp}]^{(1)}&=&{ig^2\over 2\pi}
\ln{\sigma\over\sigma'} \int d^2z_{\perp}
\label{nov27}\\
&\times&
\hbox{\bf\Big($\!\!$\Big(}x_{\perp}\Big|{1\over \vec{p}_{\perp}^2}\Big| 
z\hbox{\bf\Big)$\!\!$\Big)}
\Big(L_1(z_{\perp})+2[U_i(z_{\perp}),\Delta^i(z_{\perp})]\Big) ,
\nonumber\\ 
\! [\infty p_1+x_{\perp}, x_{\perp}]^{(1)}&=&-{ig^2\over 2\pi}
\ln{\sigma\over\sigma'}t^a\int d^2z_{\perp}
\nonumber\\
&\times&
\hbox{\bf\Big($\!\!$\Big(}x_{\perp}\Big|U^{\dagger}{1\over \vec{p}_{\perp}^2}U+
U^{\dagger}{1\over \vec{p}_{\perp}^2}(\vec{\partial}^2_{\perp}U)
{1\over \vec{p}_{\perp}^2}\Big| z\hbox{\bf\Big)$\!\!$\Big)}^{ab}
L^b_1(z_{\perp}) ,
\nonumber
\end{eqnarray}
and, therefore,
\begin{eqnarray}
&&\hspace{-8mm}[\infty p_1+x_{\perp},-\infty p_1+x_{\perp}]^{(1)}~=~
{ig^2\over \pi}
\ln{\sigma\over\sigma'}\int d^2z_{\perp}\nonumber\\
&&\hspace{-8mm}\times~\Bigg\{\hbox{\bf\Big($\!\!$\Big(}x_{\perp}\Big|
{1\over \vec{p}_{\perp}^2}\Big| z_{\perp}
\hbox{\bf\Big)$\!\!$\Big)}[U_i(z_{\perp}),\Delta^i(z_{\perp})]-t^a
\hbox{\bf\Big($\!\!$\Big(}x_{\perp}\Big|
U^{\dagger}{p^k\over \vec{p}_{\perp}^2}i(\partial_kU){1\over \vec{p}_{\perp}^2}
\Big| z_{\perp}\hbox{\bf\Big)$\!\!$\Big)}^{ab}L^b_1(z_{\perp})
\Bigg\}\nonumber\\
&&\hspace{-8mm}=~{ig^2\over \pi}t^a\ln{\sigma\over\sigma'}\int
d^2z_{\perp}\Bigg\{ \hbox{\bf\Big($\!\!$\Big(}x_{\perp}\Big|
-U^{\dagger}{p^k\over \vec{p}_{\perp}^2}i(\partial_kU)
{1\over \vec{p}_{\perp}^2}
\Big| z_{\perp}\hbox{\bf\Big)$\!\!$\Big)}^{ab}L^b_1(z_{\perp})\nonumber\\
&&\hspace{3cm}
+~\hbox{\bf\Big($\!\!$\Big(}x_{\perp}\Big|
{p_i\over \vec{p}_{\perp}^2}U^{\dagger}{p_k\over \vec{p}_{\perp}^2}U\Big| 
z\hbox{\bf\Big)$\!\!$\Big)}^{ab}\epsilon^{ik}
L^b_2(z_{\perp})\Bigg\}.
\label{nov28}
\end{eqnarray}
The derivative $\bar{U}^{\dagger}\dot{\bar U}$ is half of the coefficient in
front  of $\ln{\sigma\over\sigma'}$ in this formula so we obtain
\begin{equation}
\bar{U}^{\dagger}\dot{\bar U}={ig^2\over 2\pi}
\left(\bar{U}^{\dagger}{\partial^k\over\vec{\partial}_{\perp}^2}
(\partial_k\bar{U})\right)^{ab}{1\over \vec{\partial}_{\perp}^2}
\bar{L}^b_1-{ig^2\over 2\pi}
{\partial_i\over \vec{\partial}_{\perp}^2}\left(\bar{U}^{\dagger}
{\partial_k\over
\vec{\partial}_{\perp}^2}\bar{U}\right)^{ab}
\epsilon^{ik}\bar{L}^b_2 , \label{nov29}
\end{equation}
and, similarly, 
\begin{equation}
\bar{V}^{\dagger}\dot{\bar V}=-{ig^2\over 2\pi}
\left(\bar{V}^{\dagger}{\partial^k\over\vec{\partial}_{\perp}^2}
(\partial_k\bar{V})\right)^{ab}{1\over \vec{\partial}_{\perp}^2}
\bar{L}^b_1-{ig^2\over 2\pi}
{\partial_i\over \vec{\partial}_{\perp}^2}
\left(\bar{V}^{\dagger}{\partial_k\over
\vec{\partial}_{\perp}^2}\bar{V}\right)^{ab}
\epsilon^{ik}\bar{L}^b_2 , \label{nov30}
\end{equation}
where $\bar{U}\equiv{\bar U}(\eta),~\bar{V}\equiv{\bar V}(\eta)$. 

For illustration, let us present a first few terms in the
semiclassical expansion of the effective action, 
\begin{eqnarray}
\bar{S}_{\rm eff}&=&\int d^2x_{\perp}V_iU^i\label{nov31}\\
&+&\hspace{-1mm}{ig^2\over 2\pi}\ln{\sigma\over\sigma'}
\int d^2x_{\perp}
\Big(L_1^a{1\over \vec{\partial}_{\perp}^2}L_1^a
-{1\over g^2}\Delta^a_i\Delta^{a,i}+
2L_1^a{1\over \vec{\partial}_{\perp}^2}\big(U_i-V_i\big)^{ab}\Delta^{b,i}
\nonumber\\
&+&\hspace{-1mm}{1\over 2}\left({g^2\over 2\pi}
\ln{\sigma\over\sigma'}\right)^2
\Big\{L_1^a\Big({1\over\vec{\partial}_{\perp}^2}(\partial^kU^{\dagger})
{\partial_k\over\vec{\partial}_{\perp}^2}
U\Big)^{ab}-\Delta^{ak}U_k^{ab}{1\over\vec{\partial}_{\perp}^2}\Big\}\nonumber\\
&\times&\hspace{-1mm}
\Big((\partial^i-igU^i)(\partial_i-igV_i)\Big)^{bc}
\Big\{\Big(V^{\dagger}{\partial_j\over\vec{\partial}_{\perp}^2}(\partial^jV)
{1\over\vec{\partial}_{\perp}^2}
\Big)^{cd}L_1^d +{1\over\vec{\partial}_{\perp}^2}V_j^{cd}\Delta^{dj}\Big\} .
\nonumber
\end{eqnarray}

Once we know the solution of the Wilson-line classical equations 
(\ref{nov29})-(\ref{nov30}),
it is possible to restore $\bar{A}$. Suppose we want to find 
$\bar{A}(\eta_x,\tau,x_{\perp})$ where $\tau=x_{\parallel}^2$ and 
$\eta_x=\ln{x_*\over x_{\bullet}}$. Let us insert two factorization formulas
at $\eta_x+\delta\eta$ and  $\eta_x-\delta\eta$ and integrate over the
fields in the regions $\eta_0>\eta_x +\delta\eta$ and 
$\eta_x-\delta\eta>\eta>\eta'_0$ semiclassically. The final integration over
the region of rapidities  $\eta +\delta\eta>\eta>\eta_x-\delta\eta$ 
takes the form
\begin{equation}
\int\!\!\! D\!A \ \exp\left\{i\bar{V}^i(\eta_x+\Delta\eta)
Y_i(\eta_x+\Delta\eta)+
iW^i(\eta_x-\Delta\eta)\bar{U}_i(\eta_x-\Delta\eta)+iS(A)\right\}.
\label{nov32}
\end{equation}
(Here $\eta_x+\Delta\eta$ denotes the argument for the classical solution
$\bar{V}^i$ and the direction of the Wilson line for $Y_i$). Comparing this to
Eq.~(\ref{5.1.1}), we find that the field $\bar{A}(\eta_x,\tau,x_{\perp})$ is
given by expressions (\ref{5.1.11}) with $U\rightarrow \bar{U}(\eta)$, 
$V\rightarrow \bar{V}(\eta)$. Unfortunately, the accuracy is again up to
$[\bar{U}(\eta),\bar{V}(\eta)]^2$. Still, we see that the fields contain
logarithms of $\eta_x$ coming from and $\bar{U}(\eta)$  $\bar{V}(\eta)$ so our
assumption about large characteristic fields  in the functional integral
(\ref{fla22}) is justified.  Note that for the infinite Wilson line in $\eta_x$
direction we can get an (almost)  explicit expression in terms of $U\rightarrow
\bar{U}(\eta)$ and   $V\rightarrow \bar{V}(\eta)$ without the restriction 
$[\bar{U}(\eta),\bar{V}(\eta)]\ll 1$. It is easy to see that 
\begin{equation} 
[x_{\perp}-\infty n_{\eta},x_{\perp}+\infty n_{\eta}]
(i\partial_i+\bar{A}_i(x_{\perp}+\infty n_{\eta}) )
[x_{\perp}+\infty n_{\eta},x_{\perp}-\infty n_{\eta}]=
\Lambda_i(x_{\perp},\eta) ,
\label{nov33}
\end{equation}
where
$\Lambda_i(x_{\perp},\eta)=
\bar{U}(x_{\perp},\eta)+\bar{V}(x_{\perp},\eta)+\bar{\Delta}(x_{\perp},\eta)$  
is pure gauge field  satisfying the equation
\begin{equation}
\left(i\partial_i+[\bar{U}_i+\bar{V}_i,\right)\Delta_i=0 ,
\label{nov34}
\end{equation}
(see Eq.~(\ref{5.1.10}). Indeed, let us try to calculate 
the l.h.s. of the Eq.~(\ref{nov33}).
At small $\delta\eta$ all the contributions coming from 
$[x_{\perp}+\infty n_{\eta},x_{\perp}-\infty n_{\eta}]$ contain 
$\delta\eta$ (see Eq.~(\ref{dobav1})), hence they are small.
The only non-vanishing contribution comes from 
$\bar{A}_i(x_{\perp}+\infty n_{\eta})$ which coincide with
$\Lambda_i(x_{\perp},\eta)$ in the background-Feynman gauge (\ref{nov34}).

\section{Conclusions and outlook} 
First I would like to discuss the relation of this method to other
approaches to the
high-energy QCD discussed in the literature. 

By far, the most popular approach to high-energy pQCD is the 
direct summation of
Feynman diagrams (and related methods based on unitarity relations in 
$s$ and $t$  channels).
Although the majority of the results in pQCD, including the NLO BFKL kernel,
were obtained by this method, I think that even in pQCD, the Wilson-line
language, combined with the calculation of the propagators in the shock-wave 
background,
is technically more powerful. (Perhaps the comparison of the diagrammatic
calculation of the three-pomeron vertex in Ref.~45 
to the computation of the gluon propagator in the shock-wave background 
in Sec.~7.3 demonstrates this most clearly). 

The dipole picture~\cite{kop} has an advantage of visual interpretation 
of the 
high-energy scattering, especially in the case of DIS at 
small $x$.\cite{mu94,nnn} 
The dipole language is a light-cone version of the
Wilson-line approach combined with large-$N_c$ approximation for the wave
functions at small $x$.  However, it is hard to think about the effective
action in terms  of the dipoles, since in order to study the energy evolution
of the  effective action we must take into account not only the creation 
of the new dipoles, but their multiple creation and recombination, 
which is difficult to define in the framework of the dipole model.

The most close in spirit to our semiclassical method is the
renormalization-group approach to the 
high-energy scattering from the large nuclei advocated in the papers of 
L. McLerran and collaborators (see e.g. Refs.~4, 52, 58).
In this approach, the small-x
evolution of one strong shock wave (created by a source $\rho(x_{\perp})$) is
studied in the light-like gauge.  With such a choice of gauge, the second shock
wave can be treated perturbatively at the very end of the evolution process.
 In our terms, this amounts to the solution of classical
Eqs.~(\ref{5.1.7}) using the trial configuration $A_i=U_i\theta(x_{\ast})$
(instead of starting point $A_i=U_i\theta(x_{\ast})+V_i\theta(x_{\bullet})
+\Delta_i$
taken in this paper). Unfortunately, due to different gauges adopted in our
paper and Refs.~52, 58,  
the treatment of the boundary terms
 in the functional integral is
different, leading to the different sources for the shock waves and making
 hard to compare the intermediate formulas. However, since the first-order
(BFKL) results coincide I think these effective actions are essentially 
the same. 

In conclusion I would like to outline possible uses of this approach.
The ultimate goal is
to obtain the explicit expression for the effective action in all orders in
$\ln{s\over m^2}$. One possible prospect is that due to the conformal 
invariance
of QCD at the tree level our  future result for the effective action can be
formalized in terms of conformal  two-dimensional theory in external
two-dimensional ``gauge fields''  $V_i$ and $U_i$. 
So far, I was not able to use the conformal invariance because
it is not obvious how to implement it in terms of Wilson-line operators.
We can, however, expand Wilson lines 
back to gluons. The conformal properties of (reggeized) gluon
amplitudes are now well studied. In the coordinate space the BFKL kernel
is invariant under Mobius group and therefore the eigenfunctions of 
BFKL kernel are simply powers of coordinates. It is not clear which part
of the conformal symmetry survives for the  full effective action, yet
there is every reason to believe that it will simplify 
the structure of the answer even after reassembling of Wilson lines.  

The semiclassical approach 
developed above for the small-x processes in perturbative QCD 
can be applied for studying the heavy-ion collisions. As advocated in 
Ref.~4, 
the coupling constant for the
heavy-ion collisions may be relatively small due to high density. 
An estimation of the corresponding ``parton saturation scale'' 
 $Q_s$  gives $\sim 1$~GeV for RHIC and $\sim 2-3$~GeV for LHC,\cite{mu00} so
$g(Q_s)$ is a valid perturbative parameter. On the other
hand, the fields produced by colliding ions are large, so that the product $gA$
is not small, showing that the Wilson-line gauge factors $V$ and $U$ are of
order of 1. Thus, we have a perfect situation to try sQCD methods.

 It should however be mentioned that in this paper 
we considered the special case 
of the collision of the two shock waves, namely without any particles 
in the final state. It follows from the usual boundary conditions for Feynman 
amplitude (\ref{3.2.1}) which we calculate: 
no outgoing waves at $t\rightarrow\infty$ and no incoming fields at 
$t\rightarrow-\infty$ (the latter condition
is satisfied automatically by the
$\left.A\right|_{t\rightarrow-\infty}=0$ choice of gauge). 
However, people are usually interested in the process of particle production  
during the collision (see e.g. Ref.~59) 
since it gives the experimental probe of quark-gluon plasma. 
In this case, our approach must be modified for the new boundary conditions
--- we must solve the  classical equations (\ref{5.1.7}) with 
Feynman boundary conditions only at $t\rightarrow-\infty$. 
The boundary condition
at $t\rightarrow\infty$ depends on the problem under investigation: in the 
case if we are interested in the 
the total cross section (cut diagrams) we must calculate the 
double functional integral 
corresponding to the integration over the ``+'' fields to the right 
and the ``--'' fields to the left of the cut (see Ref.~43). 
(This is actually a functional-integral 
formalization of Cutkosky rules).
In this case we may use the usual (Feynman and c.c. Feynman)
propagators for each type of the fields. 
The boundary condition 
 requires that two
types of the field --- the left-side ``--'' fields and the right-side ``+''
ones  --- coincide at $t\rightarrow\infty$.  (This boundary condition is
responsible for the $\delta(p^2)\theta(p_0)$  propagators on the cut). 
Finally, to find the total cross section of the 
shock-wave collision in the semiclassical 
approximation, we must solve 
the double set of classical equations for ``+'' and ``--'' fields with the 
boundary condition that these fields coincide at infinity 
(cf. Ref.~60). 
The study is in progress.

\vskip0.5cm
\section*{Acknowledgments}

The author is grateful to Y.~Kovchegov, E.M.~Levin, L.N.~Lipatov, 
L.~McLerran, and R.~Venugopalan for valuable discussions.
 
This work was supported by the US Department of Energy under contract 
DE-AC05-84ER40150.

\section{Appendix}
\subsection{Wilson lines from Feynman diagrams}

\begin{figure}[htb]
\centerline{
\epsfysize=4cm
\epsffile{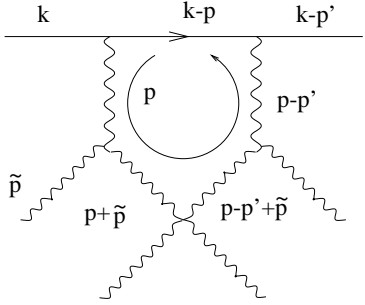}}
\caption{Typical diagram for the propagator of 
fast-moving quark.}
\label{fig:fig4}
\end{figure}

Let us demonstrate that the relevant operators are Wilson lines 
(\ref{flob3}).
The typical
contribution to the Green function of the fast-moving quark 
(with $\alpha_k\ll\sigma$) is shown in 
Fig.~\ref{fig:fig4}
where the gluons have $\alpha\ll\sigma$.
Consider the loop integral over $p$. Since we can neglect $\alpha_p$ 
as compared to $\alpha_k$, the quark propagator
with the momentum $k-p$ reduces to
\begin{equation}
\not\! p_2\frac{\not\! k-\not\! p}{(k-p)^2+i\epsilon}\not\! p_2\rightarrow 
\frac{\not\! p_2}{\beta_k-\beta_p-{(\vec{k}-\vec{p})_{\perp}^2\over\alpha_ks}+
i\epsilon\alpha_k}.
\label{2.3}
\end{equation}
Here we have used the fact that $g_{\mu\nu}$ in the numerator of the 
gluon propagator connecting the lines with very different rapidities 
($\equiv\alpha$'s) can be replaced by $\frac 2sp_{1\mu}p_{2\nu}$.

I will prove now that if I replace the propagator (\ref{2.3}) by
\begin{equation}
{\not\! p_2\over -\beta_p+i\epsilon\alpha_k},
\label{2.4}
\end{equation}
the value of the loop integral over $p$ remains unchanged.
Indeed, the integral over $p$ is the sum of the residue in the pole 
corresponding to the fast-quark propagator (\ref{2.3}) and/or the residues
in the slow-gluon propagators. Let us consider both residues in turn and
verify that the replacement (\ref{2.4}) does not affect the residues.

First, if I take the residue in the pole
\begin{equation}
\beta_p=\beta_k-\frac{(\vec{k}-\vec{p})_{\perp}^2}{\alpha_k s}
\label{2.5}
\end{equation}
corresponding tho the quark propagator,
the typical slow-gluon denominator takes the form 
\begin{eqnarray}
&&(\alpha_p+\tilde{\alpha_p})(\beta_p+\tilde{\beta_p})s - 
(p+\tilde{p})_{\perp}^2\label{2.6}\\
&=&(\alpha_p+\tilde{\alpha_p})\beta_p s - 
(p+\tilde{p})_{\perp}^2+(\alpha_p+\tilde{\alpha_p})\beta_k s-{\alpha_p+
\tilde{\alpha_p}\over\alpha_k}(\vec{k}-\vec{p})_{\perp}^2 .
\nonumber
\end{eqnarray}
The first two terms are or order of $m^2$ while the second 
two ones are $\sim {\alpha_p\over\alpha_k}m^2$ and hence 
they can be neglected, which corresponds to taking the residue at 
the pole $\beta_p=0$ in the propagator (\ref{2.4}). 
(Here we have used the fact that $\beta_k\sim{m^2\over \alpha_k s}$, see below). 

Second possibility corresponds to the residue taken at
\begin{equation}
\beta_p=
-\tilde{\beta}_p+{(p+\tilde{p})_{\perp}^2\over (\alpha_p+\tilde{\alpha_p})s}
\label{2.7}
\end{equation}
in one of the slow-gluon propagators. The quark propagator (\ref{2.3})
then takes the form 
\begin{equation}
\frac{\not\! p_2}{\tilde{\beta_p}+
{(p+\tilde{p})_{\perp}^2\over (\alpha_p+\tilde{\alpha_p})s}+
\beta_k-{(\vec{k}-\vec{p})_{\perp}^2\over\alpha_ks}+i\epsilon\alpha_k} .
\label{2.8}
\end{equation}
Again,  the first two terms in the denominator are 
$\sim {m^2\over\alpha_p s}$, while the second two ones are 
$\sim {m^2\over\alpha_p s}\ll {m^2\over\alpha_k s}$ and can be 
neglected which is exactly equivalent to replacing the Eq.~(\ref{2.3}) 
by Eq.~(\ref{2.4}).

Hence, we have proved that the propagator of the fast quark can be 
reduced to (\ref{2.4}) which is nothing but the eikonal 
gauge-factor in the momentum representation.

\subsection{Quark propagator in a shock-wave background.}

 Let us now find the quark propagator in the shock-wave background.
We start the path-integral representation of a quark Green function in the 
external field $B^{\Omega}$, 
\begin{eqnarray}
{\hbox{\bf\Big($\!\!$\Big(}} x\Big|\frac {1}{{\not\! {\cal P}}}\Big| 
y{\hbox{\bf\Big)$\!\!$\Big)}}&=&-i\int _{0}^{\infty}\! d\tau 
{\hbox{\bf\Big($\!\!$\Big(}} x\Big|{\cal P}e^{i\tau{\cal
P}^{2}}\Big| y{\hbox{\bf\Big)$\!\!$\Big)}}\nonumber\\
&=&
-i\int _{0}^{\infty}\! d\tau {\cal N}^{-1}
\int _{x(0)=y}^{x(\tau )=x}\!{\cal D}x(t)
\{{1\over 2}\not{\dot x}+\not\!{B^{\Omega}}(x(\tau ))\}
e^{-i\int _{0}^{\tau }\!
dt{\dot{x}^{2}\over 4}}\nonumber\\
&\times&
P\ \exp\{ig\int _{0}^{\tau }\! dt(B^{\Omega}_{\mu }(x(t))\dot{x}^{\mu }(t)+
{1\over 2}\sigma ^{\mu \nu }G^{\Omega}_{\mu \nu }(x(t))\} ,
\label{a6}
\end{eqnarray}
where $\sigma_{\mu \nu }\equiv {i\over2}
(\gamma _{\mu }\gamma _{\nu }-\gamma _{\nu }\gamma _{\mu })$.
First, it is easy to see that since in our external field (\ref{3.2.10})
the only nonzero components of the field tensor is $G^{\Omega}{\circ i}$ only
 the first two first term of the expansion of the exponent
$\exp\{\int \! dt{i\over 2}(\sigma G^{\Omega})\}$ in powers of $(\sigma G)$
survive. Indeed,
$\sigma^{\mu \nu }G^{\Omega}_{\mu \nu }=
{4i\over s_{0}}{\not\! p}^{0}_{2}\gamma ^{i}G^{\Omega}_{_{\circ} i}$
and therefore 
$(\sigma G^{\Omega})^{2}\sim ({\not\! p}_{2}\gamma ^{i})^{2}=0$ 
since ${\not\! p}_{2}$ commutes
with $\gamma ^{i}_{\perp}$. Consequently, the phase factor for the motion of
the particle  in the external field (\ref{3.2.10}) has the form
\begin{eqnarray}
&&Pe^{ig\int _{0}^{\tau }\! dt
B^{\Omega}_{\mu }(x(t))\dot{x}_{\mu }(t)}\label{a7}\\
&+&
{2\gamma ^{i}{\not\! p}_{2}\over s}\int _{0}^{\tau }\!dt'
Pe^{ig\int _{t'}^{\tau }\! dtB^{\Omega}_{\mu }(x(t))\dot{x}_{\mu }(t)}
gG^{\Omega}_{\circ i}(x(t'))Pe^{ig\int ^{t'}_{0}\! dtB^{\Omega}_{\mu }(x(t))
\dot{x}_{\mu }(t)} .
\nonumber
\end{eqnarray}
Let us consider the case $x_{\ast}>0,y_{\ast}<0$ as shown in 
Fig.~\ref{ofig13}.
Similarly to the case of scalar propagator, we can replace the gauge
factor along the actual path $x_{\mu }(t)$ by the gauge factor along the
straight-line path shown in Fig.~\ref{ofig14} which intersects the plane
$x_{\ast}=0$ at the same point  $(z_{\circ},z_{\perp})$ at which the original
path does.
The gauge factor (\ref{a8}) reduces to
\begin{equation}
U^{\Omega}(z_{\perp})+{\gamma ^{i}\!{\not\! p}_{2}\over\dot{x}_{\ast}(\tau ')}
i\partial _{i}U^{\Omega}(z_{\perp}) 
\label{a8}
\end{equation}
where the last term was obtained using the identity
\begin{eqnarray}
\frac {\partial }{\partial x_{i}}U(x_{\perp})&=&
-{2i\over s_{0}}\int \! dx_{\ast}
[\infty p^{(0)}_{1},{2\over s_{0}}x_{*}p^{(0)}_{1}
]_xG_{\circ i}({2\over
s_{0}}x_{*}p^{(0)}_{1}+
x_{\perp})\nonumber\\
&\times& 
[{2\over s_{0}}x_{*}p^{(0)}_{1}, 
-\infty p^{(0)}_{1}]_x ,
\label{a9}
\end{eqnarray}
and the factor $\dot{x}_{\ast}(\tau ')$ in Eq.~(\ref{a7}) comes from 
changing of
variable of integration from $t$ to $x_{\ast}(t)$. Similarly,
the phase factor for the term in the right-hand side of Eq.~(\ref{a6}) which
contains $\not\! B^{\Omega}(x(\tau ))={2\over s_{0}}{\not\! p}_{2}
B^{\Omega}_{\circ}(x(\tau ))$ in front
of
the gauge factor
Eq.~(\ref{a6}) can be reduced to
\begin{equation}
-{\not\! p}_{2}{\partial \over\partial x_{*}}[{2\over
s_{0}}x_{*}p^{(0)}_{1}+x_{\perp},-\infty+x_{\perp}]
=-{\not\! p}_{2}\delta (x_{*})[U(x_{\perp})-1] .
\label{a10}
\end{equation}
(The factor $\sim (\sigma G)$ is absent since it contains extra 
${\not\! p}_{2}$ and
${\not\! p}_{2}^{2}=0$). If we now insert the expression for the 
phase factors
(\ref{a7}), (\ref{a10})
into the path integral (\ref{a6}), we obtain (cf. Eq.~(\ref{3.2.17}))
\begin{eqnarray}
&-&{\not\! p}_{2}\delta (x_{*})[U^{\Omega}(x_{\perp})-1]\int _{0}^{\infty}\! 
d\tau {\cal
N}^{-1}\int ^{x(\tau )=x}_{x(0)=y}\!{\cal D}x(t)
e^{-i\int _{\tau }^{0}\! dt{\dot{x}^{2}\over 4}}
\label{a11}\\
&-&{i\over 2}\int _{0}^{\infty}\! d\tau \int _{0}^{\tau }\! 
d\tau '\int \! dz\delta
(z_{*})
{\cal N}^{-1}\int _{x(\tau ')=z}^{x(\tau )=x}\!{\cal D}x(t)\!
\not{\dot x}(\tau )e^{-i\int _{\tau '}^{\tau }\! dt{\dot{x}^{2}\over
4}}\nonumber\\
&\times&\{U^{\Omega}(z_{\perp})+{i\over \dot{x}_{*}(\tau ')}\not{\partial
}U^{\Omega}(z_{\perp}){\not\! p}_{2}\}{\cal N}^{-1}\int _{x(0)=y}^{x(\tau
')=z}\!{\cal D}x(t) \dot{x}_{*}(\tau ')
e^{-i\int _{\tau '}^{\tau }\! dt{\dot{x}^{2}\over 4}}.
\nonumber
\end{eqnarray}
Make a shift of time variable $\tau '$ and using Eqs.~(\ref{3.2.15})
 and (\ref{3.2.19}) to perform path integrals in the right-hand side of 
Eq.~(\ref{a11}),
it is easy to reduce the path-integral expression for the quark propagator in
the shock-wave field (\ref{3.2.11}) to
\begin{eqnarray}
{\hbox{\bf\Big($\!\!$\Big(}} x\Big|
\frac {1}{{\not\! {\cal P}}} \Big| y{\hbox{\bf\Big)$\!\!$\Big)}}&=&
\frac {{\not\! p}_{2}}{4\pi ^{2}(x-y)^{2}}\delta (x_{\ast})[U^{\Omega}-1]
(x_{\perp}) 
\label{a15}\\
&+& \int \! dz\delta (z_{\ast})
\frac {({\not\! x}-{\not\! z}){\not\! p}_{2}}{2\pi ^{2}(x-z)^{4}}
\{U^{\Omega}(z_{\perp})\frac {-2iy_{\ast}}{2\pi
^{2}(z-y)^{4}}\nonumber\\
&-&i\!\not{\partial }_{\perp} U^{\Omega}(z_{\perp})\frac {{\not\! p}_{2}}{4\pi 
^{2}(z-y)^{2}}\}\nonumber\\
&=&i\int \! dz\delta (z_{\ast})
\frac {({\not\! x}-{\not\! z}){\not\! p}_{2}}{2\pi ^{2}(x-z)^{4}}
U^{\Omega}(z_{\perp})
\frac {{\not\! z}-{\not\! y}}{2\pi ^{2}(z-y)^{4}}
\nonumber
\end{eqnarray} 
(in the region $x_{\ast}>0$, $y_{\ast}<0$). 
The propagator in the region
$x_{\ast}<0$, $y_{\ast}>0$ differs from Eq.~(\ref{a15}) by the replacement
$U^{\Omega}\leftrightarrow U^{\Omega \dagger }$. In addition, the propagator 
outside the
shock-wave wall (at $x_{\ast},y_{\ast}<0$ or $x_{\ast},y_{\ast}>0$) 
coincides with bare
propagator, so the final answer for the quark Green function in the 
$B^{\Omega}$
background can be written down as:
\begin{eqnarray}
{\hbox{\bf\Big($\!\!$\Big(}} x\Big|
\frac {1}{{\not\! {\cal P}}}\Big| y{\hbox{\bf\Big)$\!\!$\Big)}}&=&
-\frac {{\not\! x}-{\not\! y}}{2\pi ^{2}(x-y)^{4}}\nonumber\\
&+& i\int \! dz\delta (z_{\ast})
\frac {({\not\! x}-{\not\! z}){\not\! p}_2}{2\pi ^{2}(x-z)^{4}}
\{[U^{\Omega}-1](z_{\perp})\theta (x_{\ast})\theta
(-y_{\ast})\nonumber\\
&-&[U^{\Omega \dagger }-1](z_{\perp})\theta (y_{\ast})\theta (-x_{\ast})\}
\frac { {\not\! z}-{\not\! y}}{2\pi
^{2}(z-y)^{4}} ,
\label{a16}
\end{eqnarray}
where we have used the formula
\begin{equation}
i\int \! dz \delta (z_{\ast})
\frac {{\not\! x}-{\not\! z}}{2\pi ^{2}(x-z)^{4}}{\not\! p}_{2}
\frac {{\not\! z}-{\not\! y}}{2\pi ^{2}(z-y)^{4}}=
-\frac { {\not\! x}-{\not\! y}}{2\pi
^{2}(x-y)^{4}}(\theta (x_{\ast})-\theta (y_{\ast}))
\label{a17}
\end{equation}
to separate the bare propagator.

Now, one easily obtains the quark propagator (\ref{3.2.25}) in the original 
field $B_{\mu }$ Eq.~(\ref{3.2.8}) by making back the gauge rotation 
of the answer (\ref{a16}) with
matrix $\Omega ^{-1}$.

\subsection{One-loop evolution: Wilson lines in a shock-wave background.}
\label{app:B}

The convenient way to get the kernel of the evolution equation is to
calculate the derivative of the two-Wilson-line operator 
with respect to the slope of the supporting line. 
Formally one obtains:
\begin{eqnarray}
&&\zeta {\partial \over\partial \zeta }
{\rm Tr}\{\hat{U}(x_{\perp})\hat{U}^{\dagger}
(y_{\perp})\}\label{7.3.1}\\
&=&ig\zeta \int \! udu 
\Big({\rm Tr}\{[\infty,u]_xF_{\ast\bullet} (up^\zeta +x_{\perp})[u,-\infty]_x
\hat{U}^{\dagger}(y_{\perp})\}
\nonumber\\
&-&{\rm Tr}\{\hat{U}(x_{\perp})ig\zeta 
\int \! udu [-\infty,u]_yF_{\ast\bullet}
(up^\zeta+y_{\perp})[u,\infty]_y\}\Big) .
\nonumber
\end{eqnarray}
The kernel is the result of the calculation of the right-hand side 
of Eq.~(\ref{7.3.1}) in the shock-wave background.

\begin{figure}[htb]
\centerline{
\epsfysize=3.7cm
\epsffile{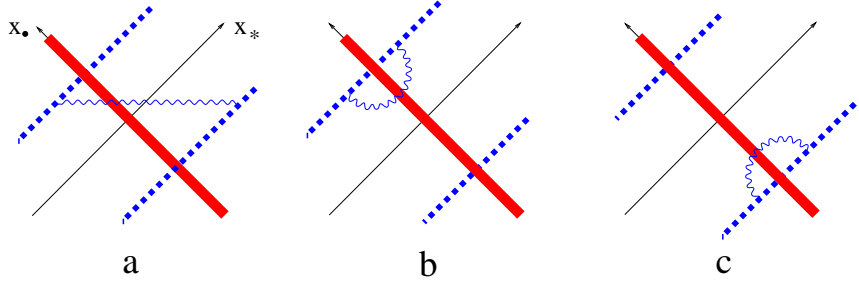}}
\caption{Path integrals describing one-loop diagrams for Wilson-line 
operators in the shock-wave field background.}
\label{fig:pint1}
\end{figure}

Consider the operators ${\hat{U}^{\zeta }}$ and 
${\hat{U}^{\dagger \zeta }}$ in the external
field formed by slow gluons with 
$\alpha\ll\sqrt{{m^2\over\ s\zeta}}$.
Making the rescaling (\ref{3.2.8}) we obtain:
\begin{eqnarray}
&&\langle[\infty p_{A},-\infty p_{A}]_x
[-\infty p_{A},\infty
p_{A}]_y\rangle_{A}\nonumber\\
&=&\langle[\infty p^{(0)}_{A},-\infty p^{(0)}_{A}]_x
[-\infty p^{(0)}_{A},\infty
p^{(0)}_{A}]_y\rangle_{B} ,
\label{b1}
\end{eqnarray}
where the shock-wave field is given by Eqs.~(\ref{3.2.8}) -- (\ref{3.2.10}). 
Equation~(\ref{7.3.1}) reduces to
\begin{eqnarray}
&&\zeta {\partial \over\partial \zeta}
\langle{\hat{U}}(x_{\perp}){\hat{U}}^{\dagger}(y_{\perp})\rangle_{A}
\label{b2}
\\
&=&ig{p_{A}^{2}\over
s_{0}}\int \! udu \langle[\infty p^{(0)}_{A}, up^{(0)}_{A}]_x
\hat{F}_{\ast\bullet}
(up^{(0)}_{A}+x_{\perp})[up^{(0)}_{A},-\infty p^{(0)}_{A}]_x
{\hat{U}}d(y_{\perp})\rangle_{B}\nonumber\\
&-&ig{p_{A}^{2}\over s_{0}}\int \! udu
\langle{\hat{U}}(x_{\perp})[-\infty p^{(0)}_{A}, up^{(0)}_{A}]_y
\hat{F}_{\ast\bullet} 
(up^{(0)}_{A}+y_{\perp})[up^{(0)}_{A},\infty p^{(0)}_{A}]_y\rangle_{B}.
\nonumber
\end{eqnarray}
Since the ($F_{\ast\circ}$) component of the field strength tensor 
(\ref{3.2.10})
vanishes for the
shock-wave field, the only nonzero contribution comes from the
diagrams with quantum gluons. In the lowest nontrivial order 
in $\alpha_{s}$ there
are three
diagrams shown in Fig.~\ref{fig:pint1}.
 
 Consider at first the diagram shown in Fig.~\ref{fig:pint1}a (which
corresponds to the case $x_{\ast}>0$, $y_{\ast}<0$). 
The relevant contribution to
the right-hand side of Eq.~(\ref{b2}) is:
\begin{eqnarray}
&-&g^2\int du [\infty p^{(0)}_A,
up^{(0)}_A]_xt^a[up^{(0)}_A,-\infty
p^{(0)}_A]_x 
\otimes \int dv [-\infty p^{(0)}_A,
vp^{(0)}_A]_yt^b\nonumber\\
&\times& [vp^{(0)}_A,\infty
p^{(0)}_A]_y{\hbox{\bf\Big($\!\!$\Big(}} up^{(0)}_A+x_{\perp}
\Big| up_{\ast}\{(p^{(0)}_{A\xi }-{\cal
P}_{\circ}\frac {p_{2\xi }}{p\cdot p_{2}})\nonumber\\
&\times& [\frac {1}{{\cal P}^{2}g_{\xi  \eta }+
2iG_{\xi \eta }}
-\frac {1}{{\cal P}^{2}g_{\xi \lambda }+2iG_{\xi \lambda
}}(D^{\alpha}G_{\alpha\lambda }
\frac {p_{2\rho }}{p\cdot p_{2}}
+\frac {p_{2\lambda }}{p\cdot p_{2}}
D^{\alpha}G_{\alpha\rho }\nonumber\\
&-&\frac {p_{2\lambda }}{p\cdot p_{2}}{\cal
P}^{\beta}D^{\alpha}G_{\alpha\beta}\frac {p_{2\rho }}{p\cdot p_{2}})
\frac {1}{{\cal P}^{2}g_{\rho \eta }+2iF_{\rho \eta }}+\ldots ]
\nonumber\\
&\times&(p^{(0)}_{A\eta }
-\frac {p_{2\eta }}{p\cdot p_{2}}{\cal
P}_{\circ})\}-v\left\{\ldots\right\}p_{\ast}\Big| vp^{(0)}_{A}+
y_{\perp}{\hbox{\bf\Big)$\!\!$\Big)}}_{ab} .
\label{b3}
\end{eqnarray}
As we discussed in Sec.~4,  terms in parentheses proportional to 
${\cal P}_{\circ}$
vanish after integration by parts (see. Eq.~(\ref{3.4.4})). Further, 
it is easy to
check that since the only nonzero component of field strength tensor for the
shock wave is $G_{\circ\perp}$ the expression in braces in Eq.~(\ref{b3}) 
can be reduced to ${\cal O}_{\circ\circ}$ where the
operator ${\cal
O}_{\mu \nu }$ is given
by Eq.~(\ref{c7}). 
 Starting from this point, it is convenient to perform the
calculation in the background of the rotated field $B^\Omega$ (\ref{3.2.11}) 
which is
$0$ everywhere except the shock-wave wall. 
(We shall make the rotation back to
field $B$ in the final answer). 
The gauge factors $[\infty,u]t^{a}[u,-\infty]$
and $[\infty,v]t^{b}[v,-\infty]$ in Eq.~(\ref{b3}) reduce 
to $t^{a}[\infty,-\infty]\otimes
t^{b}[-\infty,\infty]$ (at $x_{\ast}>0,y_{\ast}<0$) and we obtain:
\begin{equation}
-g^{2}t^{a}U^{\Omega}\otimes t^{b}U^{\dagger \Omega }\int \! du\int \! dv
(u-v){\hbox{\bf\Big($\!\!$\Big(}} up^{(0)}_A+
x_{\perp}\Big|p_{\ast}{\cal O}^{\Omega}_{\circ\circ}\Big|
vp^{(0)}_{A}+y_{\perp}{\hbox{\bf\Big)$\!\!$\Big)}}_{ab} \label{b4}
\end{equation}
where we have used the fact that the operator $p_{\ast}$ commutes with ${\cal
O}^{\Omega}$. Let us now derive the formula for the $(_{\circ\circ })$ 
component of the gluon
propagator $\hbox{\bf\Big($\!\!$\Big(}x\Big|{\cal O}^{\Omega}\Big| 
y\hbox{\bf\Big)$\!\!$\Big)}$ in the 
shock-wave background. The path-integral
representation of 
$\hbox{\bf\Big($\!\!$\Big(}x\Big|{\cal O}^{\Omega}_{\circ\circ}\Big| 
y\hbox{\bf\Big)$\!\!$\Big)}$ has the form 
\begin{eqnarray}
\!\!\!&&\!\!\!\!\! {\hbox{\bf\Big($\!\!$\Big(}} x\Big|4
\frac {1}{{\cal P}^{2}}G^{\xi
\Omega }_{~~\circ } \frac { 1}{{\cal
P}^{2}}G^{\Omega}_{\xi \circ }\frac {1}{{\cal P}^{2}}-\frac {1}{{\cal
P}^{2}}(D^{\alpha}G^{\Omega}_{\alpha\circ }\frac {s_{0}}{2p_{\ast}}
\label{b5}\\
&+&\!\!\!\!\! \frac {s_{0}}{2p_{\ast}}
D^{\alpha}G^{\Omega}_{\alpha\nu }-\frac {s_{0}}{2p_{\ast}}{\cal
P}^{\beta}D^{\alpha}G^{\Omega}_{\alpha\beta}\frac {s_{0}}{2p_{\ast}})
\frac {1}{{\cal
P}^{2}}\Big| y{\hbox{\bf\Big)$\!\!$\Big)}}\nonumber\\
&=&\!\!\! i\int _{0}^{\infty}\! d\tau \int _{0}^{\tau }\!
 d\tau '{\hbox{\bf\Big($\!\!$\Big(}} x\Big|e^{i(\tau 
-\tau '){\cal
P}^{2}}\Big\{G^{\alpha\Omega }_{\circ }\int _{0}^{\tau '}\! 
d\tau ''e^{i(\tau '-\tau
''){\cal P}^{2}}\nonumber\\
&\times&\!\!\!\!\! 
 G^{\Omega}_{\alpha\circ }e^{i\tau ''{\cal P}^{2}}-
{is_{0}\over 2p_{\ast}}D^{\alpha}
G^{\Omega}_{\alpha\circ}
e^{i\tau ''{\cal P}^{2}}\Big\}\Big| y{\hbox{\bf\Big)$\!\!$\Big)}}
\nonumber\\
&=&\!\!\!\!\! 
i\int _{0}^{\infty}\! d\tau {\cal N}^{-1}
\int _{x(0)=y}^{x(\tau )=x}\!{\cal D}x(t)
e^{-i\int _{0}^{\tau }\! dt{\dot{x}^{2}\over 4}}
\{4\int _{0}^{\tau }\!d\tau '\int _{0}^{\tau ' }\!d\tau ''\nonumber\\
&\times&\!\!\!\!\!
Pe^{ig\int _{\tau '}^{\tau }\! dtB^{\Omega}_{\mu }(x(t))\dot{x}_{\mu }(t)}
gG^{\Omega}_{\circ i}(x(\tau '))
Pe^{ig\int ^{\tau '}_{\tau ''}\! dtB^{\Omega}_{\mu }(x(t))\dot{x}_{\mu }(t)}
\int _{0}^{\tau '}\!d\tau ''\nonumber\\
&\times&\!\!\!\!\! 
Pe^{ig\int _{\tau ''}^{\tau '}\! dtB^{\Omega}_{\mu }(x(t))\dot{x}_{\mu }(t)}
gG^{\Omega}_{\circ i}(x(\tau ''))Pe^{ig\int ^{\tau ''}_{0}\! dt
B^{\Omega}_{\mu
}(x(t))\dot{x}_{\mu }(t)}\nonumber\\
&+&\!\!\!\!\! i\int _{0}^{\tau }\!d\tau '
Pe^{ig\int _{\tau '}^{\tau }\! dtB^{\Omega}_{\mu }(x(t))\dot{x}_{\mu }(t)}
{s_{0}\over \dot{x_{\ast}}(\tau ')}gD^{\alpha}
G^{\Omega}_{\alpha\circ }(x(\tau '))Pe^{ig\int
^{\tau '}_{0}\! dtB^{\Omega}_{\mu }(x(t))\dot{x}_{\mu }(t)}\} .
\nonumber
\end{eqnarray}
As we discussed above, the transition through the shock wave occurs in a 
short
time $\sim {1\over \lambda}$ so the gluon has no time to deviate in the 
transverse
directions and therefore the gauge factors in Eq.~(\ref{b5}) can be
approximated by segments of Wilson lines. One obtains then 
(cf. Eq.~(\ref{a6})):
\begin{eqnarray}
&&{\hbox{\bf\Big($\!\!$\Big(}} x\Big|
{\cal O}^{\Omega}_{\circ\circ}\Big| y{\hbox{\bf\Big)$\!\!$\Big)}}\label{b6}\\
&=&
{i\over 2}s_{0}^{2}\int _{0}^{\infty}\! d\tau 
\int _{0}^{\tau }\! d\tau '\int \!
dz\delta
(z_{*})
{\cal N}^{-1}\int _{x(\tau ')=z}^{x(\tau )=x}\!{\cal D}x(t)\!
e^{-i\int _{\tau '}^{\tau }\! dt{\dot{x}^{2}\over 4}}\nonumber\\
&\times&{1\over
\dot{x}_{\ast}(\tau ')}\{2[GG]^{\Omega}(z_{\perp})-
i[DG]^{\Omega}(z_{\perp})\}{\cal
N}^{-1}\int _{x(0)=y}^{x(\tau ')=z}\!{\cal D}x(t)
e^{-i\int _{\tau '}^{\tau }\! dt{\dot{x}^{2}\over 4}},
\nonumber
\end{eqnarray}
where $[GG]^{\Omega}$ and $[DG]^{\Omega}$ are the notations for the gauge 
factors
(\ref{3.4.9})
calculated for the background field $B^{\Omega}_{\mu }$,
\begin{eqnarray} 
\! [DG]^{\Omega}(x_{\perp})&=&
\int du[\infty p_{1}, up_{1}]_x
D^{\alpha}G^{\Omega}_{\alpha\circ }(up_{1}+x_{\perp})
[up_{1},-\infty p_{1}]_x , \nonumber\\  
\! [GG]^{\Omega}(x_{\perp})&=&\int \! du\int \! dv \Theta (u-v) 
[\infty p_{1}, up_{1}]_x
G^{\xi \Omega }_{\circ }(up_{1}+x_{\perp})\nonumber\\
&\times&[up_{1},vp_{1}]_x
G^{\Omega}_{\xi \circ }(vp_{1}+x_{\perp})[vp_{1},-\infty p_{1}]_x .
\label{b7}
\end{eqnarray}
As we noted in Sec.~4, the gauge factor $-i[DG]+2[GG]$ in braces
in Eq.~(\ref{b5}) is in
fact the total derivative of $U$ with respect to translations in the
perpendicular directions so we get 
\begin{eqnarray}
{\hbox{\bf\Big($\!\!$\Big(}} x\Big|
{\cal O}^{\Omega}_{\circ \circ }\Big| y{\hbox{\bf\Big)$\!\!$\Big)}}
&=&{i\over 2}s_{0}^{2}\int _{0}^{\infty}\! d\tau 
\int _{0}^{\tau }\! d\tau '\int \!
dz\delta(z_{*})\label{b8}\\
&\times&
{\cal N}^{-1}\int _{x(\tau ')=z}^{x(\tau )=x}\!{\cal D}x(t)\!
e^{-i\int _{\tau '}^{\tau }\! dt{\dot{x}^{2}\over 4}}{1\over
\dot{x}_{\ast}(\tau ')}\vec{\partial}_{\perp}^2U^{\Omega}(x_{\perp})
\nonumber\\
&\times&
{\cal N}^{-1}\int _{x(0)=y}^{x(\tau ')=z}\!{\cal D}x(t)
e^{-i\int _{\tau '}^{\tau }\! dt{\dot{x}^{2}\over 4}} .\nonumber
\end{eqnarray}
Using now the path-integral representation for bare propagator (\ref{3.2.15})
and the following formula
\begin{equation}
\int _{0}^{\infty}\! d\tau {\cal N}^{-1}
\int ^{x(\tau )=x}_{x(0)=y}\!{\cal D}x(t)
{1\over\dot{x}_{\ast}(0)}
e^{-i\int ^{\tau }_{0}\! dt{\dot{x}^{2}\over 4}}=i\frac {
\ln (x-y)^{2}}{16\pi ^{2}(x-y)_{\ast}}
\label{b9}
\end{equation}
we finally obtain the $(_{\circ \circ })$ component 
of the gluon propagator in the
shock-wave background in the form:
\begin{eqnarray}
&&{\hbox{\bf\Big($\!\!$\Big(}} x\Big|
{\cal O}^{\Omega}_{\circ\circ}\Big| y{\hbox{\bf\Big)$\!\!$\Big)}}=
{s_{0}^{2}\over 2}\int \! dz\delta (z_{\ast})
\frac {\ln(x-z)^{2}}{16\pi ^{2}x_{\ast}}\\
\label{b10}
&\times&[\vec{\partial}_{\perp}^{2}
U^{\Omega}(z_{\perp})\Theta (x_{\ast})
\Theta (-y_{\ast})-\vec{\partial}_{\perp}^{2}U^{\dagger \Omega }(z_{\perp})
\Theta (-x_{\ast})
\Theta (y_{\ast})]\frac {1}{4\pi ^{2}(z-y)^{2}} , \nonumber
\end{eqnarray}
where we have added the similar term corresponding to the case 
$x_{\ast}<0$, $y_{\ast}>0$.
We need also the ${\partial \over\partial x_{\circ}}$ derivative 
of this propagator
(see Eq.~(\ref{b4})) which is 
\begin{eqnarray}
&&{\hbox{\bf\Big($\!\!$\Big(}} x\Big|p_{\ast}
{\cal O}^{\Omega}_{\circ\circ}\Big| y{\hbox{\bf\Big)$\!\!$\Big)}}=
{is_{0}^{2}\over 64\pi ^{4}}\int \!
dz\frac {\delta (z_{\ast})}{(x-y)^{2}}\label{b11}\\
&\times&
[\vec{\partial}_{\perp}^{2}U^{\Omega}(z_{\perp})\Theta (x_{\ast})
\Theta (-y_{\ast})-\vec{\partial}_{\perp}^{2}
U^{\dagger \Omega }(z_{\perp})\Theta (-x_{\ast})\Theta (y_{\ast})]
\frac {1}{(z-y)^{2}} .
\nonumber
\end{eqnarray}
Substituting now the Eq.~(\ref{b11}) into Eq.~(\ref{b4}) one obtains
\begin{eqnarray}
&&{g^{2}\over  4\pi }{\hbox{\bf\Big($\!\!$\Big(}} x_{\perp}\Big|
\frac {1}{\vec{p}_{\perp}^{2}}\vec{\partial}_{\perp}^2U^{\Omega}
\frac {1}{\vec{p}_{\perp}^{2}}
\Big| y_{\perp}{\hbox{\bf\Big)$\!\!$\Big)}}_{ab}t^{a}
U^{\Omega}(x_{\perp})\otimes t^{b}
U^{\dagger \Omega }(y_{\perp})\nonumber\\
&+&{g^{2}\over  4\pi }{\hbox{\bf\Big($\!\!$\Big(}} x_{\perp}\Big|
\frac {1}{\vec{p}_{\perp}^{2}}
\vec{\partial}_{\perp}^2U^{\dagger \Omega }
\frac {1}{\vec{p}_{\perp}^{2}}\Big| y_{\perp}
{\hbox{\bf\Big)$\!\!$\Big)}}_{ab}
U^{\Omega}(x_{\perp})t^{a}\otimes U^{\dagger \Omega }(y_{\perp})t^{b} .
\label{b13}
\end{eqnarray}
which agrees with Eq.~(\ref{3.4.13}).

Let us consider now the diagram shown in Fig.~\ref{fig:pint1}c. The calculation
is very similar to the case of Fig.~\ref{fig:pint1}a diagram considered above
so we shall only briefly outline the calculation. One starts with the
corresponding contribution to the right-hand side of Eq.~(\ref{b2}) which 
has the form (cf.~(\ref{b3}):
\begin{eqnarray}
&-&\!\! 
g^{2}\zeta \int \! du \int \! dv \Theta (u-v) [\infty p^{(0)}_{A}+x_{\perp},
up^{(0)}_{A}+x_{\perp}]t^{a}[up^{(0)}_{A}+x_{\perp},vp^{(0)}_{A}+x_{\perp}]
\nonumber\\
&\times&\!\! 
t^{b}[vp^{(0)}_{A}+x_{\perp},-\infty
p^{(0)}_{A}+x_{\perp}]\otimes U^{\dagger}(y_{\perp})\nonumber\\
&\times&\!\! 
{\hbox{\bf\Big($\!\!$\Big(}} up^{(0)}_A+x_{\perp}
\Big|up_{\ast}\{(p^{(0)}_{A\xi }-{\cal
P}_{\circ}\frac {p_{2\xi }}{p\cdot p_{2}})
[\frac {1}{{\cal P}^{2}g_{\xi  \eta }+
2iG_{\xi \eta }}
-\frac {1}{{\cal P}^{2}g_{\xi \lambda }+2iG^{\Omega}_{\xi \lambda }}
\nonumber\\
&\times&\!\! 
\Big[D^{\alpha}G^{\Omega}_{\alpha\lambda }
\frac {p_{2\rho }}{p\cdot p_{2}}
+\frac {p_{2\lambda }}{p\cdot p_{2}}
D^{\alpha}G^{\Omega}_{\alpha\rho }-\frac{p_{2\lambda }}{p\cdot p_{2}}{\cal
P}^{\beta}D^{\alpha}G_{\alpha\beta}\frac{p_{2\rho }}{p\cdot p_{2}}\Big]
\nonumber\\
&\times&\!\! 
\frac {1}{{\cal
P}^{2}g_{\rho \eta }+2iG_{\rho \eta }}+\ldots](p^{(0)}_{A\eta }
-\frac {p_{2\eta }}{p\cdot p_{2}}{\cal
P}_{\circ})\}-v\left\{\ldots\right\}p_{\ast}\Big| vp^{(0)}_{A}+
x_{\perp}{\hbox{\bf\Big)$\!\!$\Big)}}_{ab} .
\label{b14}
\end{eqnarray}
As we demonstrated in Sec.~4, the terms in parentheses proportional to ${\cal
P}_{\circ}$ vanish and after that the operator in braces reduce to 
${\cal O}_{\circ\circ}$.
Again, it is convenient to make a gauge transformation to the rotated field
(\ref{3.2.11}) which is 0 everywhere except the shock wave.  Then the 
gauge factor
$[\infty,u]t^{a}[u,v]t^{b}[v,-\infty]$ in Eq.~(\ref{b14}) simplifies to 
$t^{a}[\infty,-\infty]
t^{b}$ (at $x_{\ast}>0,y_{\ast}<0$) and we obtain
\begin{equation}
-g^{2}t^{a}U^{\Omega} t^{b}\otimes U^{\dagger \Omega }\int \! du\int \! dv
(u-v){\hbox{\bf\Big($\!\!$\Big(}} up^{(0)}_A+x_{\perp}
\Big|p_{\ast}{\cal O}^{\Omega}_{\circ\circ}\Big| vp^{(0)}_{A}+
x_{\perp}{\hbox{\bf\Big)$\!\!$\Big)}}_{ab} .
\label{b15}
\end{equation}
Using the expression (\ref{b11}) for the gluon propagator in the shock-wave
background we can reduce Eq.~(\ref{b15}) to
\begin{equation}
-{g^{2}\over 4\pi } t^{a}U^{\Omega}(x_{\perp})t^{b}\otimes 
U^{\dagger\Omega}(y_{\perp})
{\hbox{\bf\Big($\!\!$\Big(}} x_{\perp}\Big|
\frac {1}{\vec{p}_{\perp}^{2}}
(\vec{\partial}_{\perp}^{2}U^{\Omega})\frac {1}{\vec{p}_{\perp}^{2}}\Big|
x_{\perp} {\hbox{\bf\Big)$\!\!$\Big)}}_{ab}.
\label{b16}
\end{equation}
The contribution of the diagram in Fig.~\ref{fig:pint1}b differs from 
Eq.~(\ref{b16}) only in change 
$U\leftrightarrow U^{\dagger}, ~x\leftrightarrow y$. 
Combining these expressions, one obtains the 
answer in the
rotated field (\ref{3.2.11}) in the form
\begin{eqnarray}
&&\!\!\!\!\! {g^{2}\over 16\pi ^{3}}\int \!dz_{\perp}
\Bigg\{\Big[\{U^{\dagger\Omega}(z_{\perp}) 
U^{\Omega}(x_{\perp})\}^{k}_{j} \{
U^{\Omega}(z_{\perp})U^{\dagger\Omega}(y_{\perp})\}^{i}_{l} 
\label{b17}\\
&+&\!\!\!\!\!
\{U^{\Omega}(x_{\perp})U^{\dagger\Omega}(z_{\perp})\}^{i}_{l}
\{U^{\dagger\omega}(y_{\perp})U^{\Omega}(z_{\perp})\}^{k}_{j}
\nonumber\\
&-&\!\!\!\!\!
\delta ^{k}_{j}\{U^{\Omega}(x_{\perp})
U^{\dagger\Omega}(y_{\perp})\}^{i}_{l}-\delta ^{i}_{l}
\{U^{\dagger\Omega}(y_{\perp})U^{\Omega}(x_{\perp})\}^{k}_{j}\Big]
\frac {(\vec{x}-\vec{z},\vec{y}-\vec{z})_{\perp}}
{(\vec{x}-\vec{z})_{\perp}^2(\vec{y}-\vec{z})_{\perp}^2}
\nonumber\\
&-&\!\!\!\!\!
\Big[\{U^{\Omega}(z_{\perp})\}^{i}_{j}{\rm Tr}\{U^{\Omega}(x_{\perp})
U^{\dagger\Omega}
(z_{\perp})\}-N_{c}
\{U^{\Omega}(x_{\perp})\}^{i}_{j})U^{\dagger\Omega}(y_{\perp})^{k}_{l}
\frac {1}{(\vec{x}-\vec{z})_{\perp}^2} 
\nonumber\\
&-&\!\!\!\!\!
\{U^{\Omega}(x_{\perp})\}^{i}_{j}
[U^{\dagger\Omega}(z_{\perp})^{k}_{l}
{\rm
Tr}\{U^{\Omega}(z_{\perp})U^{\dagger\Omega}(y_{\perp})\}-
N_{c}\{U^{\dagger\Omega}(y_{\perp})\}^{k}_{l})\Big]
\frac { 1}{(\vec{y}-\vec{z})_{\perp}^2}\Bigg\} . \nonumber
\end{eqnarray}
Now we must perform the gauge rotation back to the ``original" field 
$B_{\mu }$.
The answer is especially simple if we consider the evolution of the
gauge-invariant operator such as
$Tr\{U(x_{\perp})[x_{\perp},y_{\perp}]_{-}
U^{\dagger}(y_{\perp})[y_{\perp},x_{\perp}]_{+}\}$
where the Wilson lines are connected by gauge segments at the infinity. 
We have then
\begin{eqnarray}
&&\zeta {\partial \over\partial \zeta }\langle
{\rm Tr}\{{\hat{U}^{\zeta }}(x_{\perp})[x_{\perp},y_{\perp}]_{-}
{\hat{U}^{\dagger \zeta }}
(y_{\perp})[y_{\perp},x_{\perp}]_{+}\}\rangle_{A}=\nonumber\\
&=&
-{\alpha _{s}\over 4\pi^2} \int \! dz_{\perp}
{(\vec{x}-\vec{y})_{\perp}^2\over(\vec{x}-\vec{z})_{\perp}^2
(\vec{z}-\vec{y})_{\perp}^2}\nonumber\\
&\times&\Bigg({\rm Tr}\{U(x_{\perp})[x_{\perp},z_{\perp}]_{-}
U^{\dagger}(z_{\perp} )[z_{\perp},x_{\perp}]_{+}\} \nonumber\\
&\times&
{\rm Tr}\{U(z_{\perp})[z_{\perp},y_{\perp}]_{-}U^{\dagger}(y_{\perp})
[y_{\perp},z_{\perp}]_{+}\} \nonumber\\ 
&-&N_{c}{\rm
Tr}\{U(x_{\perp})[x_{\perp},y_{\perp}]_{-}U^{\dagger}(y_{\perp})
[y_{\perp},x_{\perp}]_{+}\}\Bigg)  ,
\label{b18}
 \end{eqnarray}
 where we have replaced the end gauge factors like
$\Omega (\infty p_{1}+x_{\perp})\Omega ^{\dagger }(\infty p_{1}+y_{\perp})$ and
$\Omega (-\infty p_{1}+x_{\perp})\Omega ^{\dagger }(-\infty p_{1}+y_{\perp})$
by segments of gauge line $[x_{\perp},y_{\perp}]_{+}$
and $[x_{\perp},y_{\perp}]_{-}$, respectively.
Since the background field $B_{\mu }$ is a pure gauge outside the shock wave
the
specific form of the contour in Eq.~(\ref{b18}) does not matter as long as it
has the same initial and final points. Finally, note that the gauge factors in
the right-hand side 
of Eq.~(\ref{b18}) preserve their form  after rescaling back to the
field $A_{\mu }$ so we reproduce the Eq.~(\ref{master}).

\begin{figure}[htb]
\centerline{
\epsfysize=5cm
\epsffile{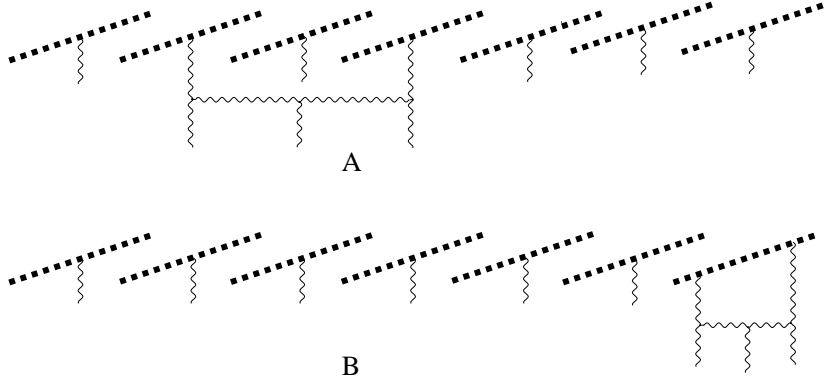}}
\caption{ Typical diagrams for the one-loop 
evolution of the $n$-line operator.}
\label{ofig35}
\end{figure}

In the general case, the evolution of the 2$n$-line
operators such as\\
${\rm Tr}\{UU^{\dagger }\}{\rm Tr}\{UU^{\dagger }\}...
{\rm Tr}\{UU^{\dagger }\}$ come from either self-interaction
diagrams  or from the pair-interactions 
ones (see Fig.~\ref{ofig35}).
These pair-wise kernels have the form ($U_{x}\equiv U(x_{\perp})$, etc.)
\begin{eqnarray}
&&\zeta \frac {\partial }{\partial \zeta }\{U_{x}\}^{i}_{j}
\{U^{\dagger}_{y}\}^{k}_{l}~=~{g^{2}\over 16\pi ^{3}}\int dz_{\perp}
\frac {(\vec{x}-\vec{z},\vec{y}-\vec{z})_{\perp}}
{(\vec{x}-\vec{z})_{\perp}^2(\vec{y}-\vec{z})_{\perp}^2}
\label{5.20}\\
&\times&
\Bigg(\{U^{\dagger}_{z}U_{x}\}^{k}_{j}
\{U_{z}U^{\dagger}_{y}\}^{i}_{l}+
\{U_{x}U^{\dagger}_{z}\}^{i}_{l}\{U^{\dagger}_{y}U_{z}\}^{k}_{j}-
\delta ^{k}_{j}
\{U_{x}U^{\dagger}_{y}\}^{i}_{l}-\delta ^{i}_{l}
\{U^{\dagger}_{y}U_{x}\}^{k}_{j}\Bigg) ,
\nonumber\\
&&\zeta \frac {\partial }{\partial \zeta }\{U_{x}\}^{i}_{j}
\{U_{y}\}^{k}_{l}~=~-{g^{2}\over 16\pi ^{3}}
\int dz_{\perp}
\frac {(\vec{x}-\vec{z},\vec{y}-\vec{z})_{\perp}}
{(\vec{x}-\vec{z})_{\perp}^2(\vec{y}-\vec{z})_{\perp}^2}
\nonumber\\
&\times&
\Bigg(\{U_{z}\}^{i}_{l}\{U_{y}U^{\dagger}_{z}U_{x}\}^{k}_{j}+
\{U_{x}U^{\dagger}_{z}U_{y}\}^{i}_{l}\{
U_{z}\}^{k}_{j}-\{U_{x}\}^{i}_{l}
\{U_{y}\}^{k}_{j}-\{U_{y}\}^{i}_{l}
\{U_{x}\}^{k}_{j}\Bigg) ,
\nonumber\\
&&\zeta \frac {\partial }{\partial \zeta }\{U^{\dagger}_{x}\}^{i}_{j}
\{U^{\dagger}_{y}\}^{k}_{l}~=~-{g^{2}\over 16\pi ^{3}}
\int dz_{\perp}
\frac {(\vec{x}-\vec{z},\vec{y}-\vec{z})_{\perp}}
{(\vec{x}-\vec{z})_{\perp}^2(\vec{y}-\vec{z})_{\perp}^2}
\nonumber\\
&\times&\Bigg(\{U^{\dagger}_{z}\}^{i}_{l}\{U^{\dagger}_{y}U_{z}
U^{\dagger}_{x}\}^{k}_{j}+\{U^{\dagger}_{x}U_{z}U^{\dagger}_{y}\}^{i}_{l}
\{U^{\dagger}_{z}\}^{k}_{j}-\{U^{\dagger}_{x}\}^{i}_{l}
\{U^{\dagger}_{y}\}^{k}_{j}-\{U^{\dagger}_{y}\}^{i}_{l}
\{U^{\dagger}_{x}\}^{k}_{j}\Bigg)
\nonumber
\end{eqnarray}
for the pair-interaction diagrams in Fig.~\ref{ofig35}a and
\begin{eqnarray}
\zeta \frac {\partial }{\partial \zeta }\{U_{x}\}^{i}_{j}&=&
-{g^{2}\over 16\pi
^{3}}
\int dz_{\perp}[U_{z}{\rm Tr}\{U_{x}U^{\dagger}_{z}\}-N_{c}U_{x}]
\frac {1}{(\vec{x}-\vec{z})_{\perp}^2},
\nonumber\\
\zeta \frac {\partial }{\partial \zeta }\{U^{\dagger}_{x}\}^{i}_{j}&=&
-{g^{2}\over 16\pi^{3}}
\int dz_{\perp}[U^{\dagger}_{z}{\rm Tr}\{U_{z}U^{\dagger}_{x}\}-
N_{c}U^{\dagger}_{x}]
\frac {1}{(\vec{x}-\vec{z})_{\perp}^2} ,
\label{5.21}
\end{eqnarray}
for the self-interaction diagrams of Fig.~\ref{ofig35}b type.

\subsection{Gluon propagator in the axial gauge.}
\label{app:GlueProp}

Our aim here is to derive the expression for the gluon propagator in the
external field in the
axial gauge. The propagator of the ``quantum" gauge field $A^{q}$ in the
external
``classical" field $A^{cl}$ in the axial gauge $e_{\mu }A_{\mu }=0$ can be
represented as the following functional integral:
\begin{eqnarray}
G_{\mu \nu }^{ab}(x,y)&=&\lim_{w\rightarrow 0}N^{-1}\int DA
A^{qa}_{\mu }(x)A^{qb}_{\nu }(y)\label{c1}\\
&\times& e^{i\int dz {\rm Tr}\{ A^{q}_{\alpha}(z)
(D^{2}g^{\alpha\beta}-D^{\alpha}D^{\beta}-2iF_{cl}^{\alpha\beta}-{1\over
w}e^{\alpha}e^{\beta})A^{q}_{\beta}(z)\}},
\nonumber
\end{eqnarray}
where $D_{\mu }=\partial _{\mu }-igA^{cl}_{\mu }$. Hereafter we shall omit the
label ``cl" from the external field. This propagator can be formally written
down as
\begin{equation}
iG^{ab}_{\mu \nu }(x,y)={\hbox{\bf\Big($\!\!$\Big(}} x\Big|
\frac {1}{\Box^{\mu \nu }-
{\cal P}^{\mu }{\cal
P}^{\nu }+\frac {1}{w}e^{\mu }e^{\nu }}
\Big| y{\hbox{\bf\Big)$\!\!$\Big)}}^{ab} ,
\label{c2}
\end{equation}
where $\Box^{\mu \nu }={\cal P}^{2}g^{\mu \nu }+2iF^{\mu \nu }$. It is easy to
check that the operator in right-hand side of 
Eq.~(\ref{c2}) satisfies the recursion formula
\begin{eqnarray}
 \frac {1}{\Box^{\mu \nu }-{\cal P}^{\mu }{\cal P}^{\nu }+
{e^{\mu}e^{\nu }\over w}}&=&(\delta _{\mu }^{\xi }
-{\cal P}_{\mu }\frac {e^{\xi }}{{\cal P}e})\frac {1}{\Box^{\xi \eta }}(\delta
_{\nu }^{\eta }-
\frac {e^{\eta }}{{\cal P}e}{\cal P}_{\nu })+{\cal P}_{\mu }
\frac {w}{({\cal P}e)^{2}}
{\cal P}_{\nu }\nonumber\\
&-& 
\frac {1}{\Box^{\mu \alpha}-{\cal P}^{\mu }{\cal P}^{\alpha} +
{e^{\mu}e^{\alpha}\over w}}(D_{\lambda}F^{\lambda\alpha}\frac {e^{\xi }}{{\cal
P}e} -{\cal P}^{\alpha}\frac {1}{{\cal P}^{2}}D_{\lambda}F^{\lambda\xi })
\nonumber\\
&\times&
\frac{1}{\Box^{\xi \eta }}(\delta _{\nu }^{\eta }-
\frac {e^{\eta }}{{\cal P}e}{\cal P}_{\nu })\label{c3}
\end{eqnarray}
which gives the propagator as an expansion in powers of the operator
$D_{\lambda}F^{a}_{\lambda\alpha}=-g\bar{\psi }t^{a}\gamma _{\alpha}\psi $. 
We shall see below that
in the leading logarithmic approximation we need the terms not higher than the
first nontrivial order in this operator. With this accuracy
\begin{eqnarray}
\frac {1}{\Box^{\mu \nu }-{\cal P}^{\mu }{\cal P}^{\nu }+
\frac {1}{w}e^{\mu}e^{\nu }}&=& 
(\delta _{\mu }^{\xi }
-{\cal P}_{\mu }\frac {e^{\xi }}{{\cal P}e})\frac {1}{\Box^{\xi \eta }}(\delta
_{\nu }^{\eta }-
\frac {e^{\eta }}{{\cal P}e}{\cal P}_{\nu })+{\cal P}_{\mu }
\frac {w}{({\cal P}e)^{2}}
{\cal P}_{\nu }\nonumber\\
&-&(\delta _{\mu }^{\xi }-{\cal P}_{\mu }\frac {e^{\xi }}{{\cal P}e})
\frac {1}{\Box^{\xi \eta}}\Big(D_{\lambda}F^{\lambda\eta }
\frac {e^{\rho }}{{\cal P}e}
+\frac {e^{\eta }}{{\cal P}e}D_{\lambda}F^{\lambda\rho }\nonumber\\
&-&
\frac {e^{\eta }}{{\cal P}e}{\cal P}^{\beta}D_{\alpha}F^{\alpha\beta}
\frac {e^{\rho }}{{\cal P}e}\Big)\frac {1}{\Box^{\rho \sigma}}
(\delta _{\nu }^{\sigma}-
\frac {e^{\sigma}}{{\cal P}e}{\cal P}_{\nu }) .
\label{c4}
\end{eqnarray}
We take now $w\rightarrow 0$, obtaining the propagator in external field in
axial gauge in the form 
\begin{eqnarray}
iG_{\mu \nu }^{ab}(x,y)&=&
(\delta _{\mu }^{\xi }
-{\cal P}_{\mu }\frac {e^{\xi }}{{\cal P}e})\frac {1}{\Box^{\xi \eta }}(\delta
_{\nu }^{\eta }-
\frac {e^{\eta }}{{\cal P}e}{\cal P}_{\nu })-(\delta _{\mu }^{\xi }
-{\cal P}_{\mu }\frac {e^{\xi }}{{\cal P}e})\frac {1}{\Box^{\xi \eta }}
\nonumber\\
&\times&
\Big(D_{\lambda}F^{\lambda\eta }
\frac {e^{\rho }}{{\cal P}e}
+\frac {e^{\eta }}{{\cal P}e}D_{\lambda}F^{\lambda\rho }-
\frac {e^{\eta }}{{\cal P}e}{\cal P}^{\beta}D_{\alpha}F^{\alpha\beta}
\frac {e^{\rho }}{{\cal P}e}\Big)\nonumber\\
&\times&\frac {1}{\Box^{\rho \sigma}}(\delta _{\nu }^{\sigma}-
\frac {e^{\sigma}}{{\cal P}e}{\cal P}_{\nu })+\ldots
\label{c5}
\end{eqnarray}
where the dots stand for the terms of second (and higher) order in
$D^{\lambda}F_{\lambda\rho }$. It can be demonstrated that for our purposes 
a first few terms of the expansion of operators ${1\over\Box}$ in 
powers of $F_{\xi\eta}$ are enough, namely
\begin{equation}
iG_{\mu \nu }^{ab}(x,y)=\label{c6}
(\delta _{\mu }^{\xi }
-{\cal P}_{\mu }\frac {e^{\xi }}{{\cal P}e})
\left[\frac {\delta_{\xi\eta}}{{\cal P}^2}-2i\frac {1}{{\cal P}^2}
F_{\xi\eta}\frac {1}{{\cal P}^2}+{\cal O}_{\xi\eta}\right]
(\delta _{\nu
}^{\eta}- \frac {e^{\eta}}{{\cal P}e}{\cal P}_{\nu })+\ldots
\end{equation}
where the operator ${\cal O}$ stands for
\begin{eqnarray}
{\cal O}_{\mu \nu }&=& 4\frac {1}{{\cal P}^{2}}F^{\xi}_{~~\mu }
\frac { 1}{{\cal P}^{2}}F_{\xi \nu }\frac {1}{{\cal P}^{2}}\label{c7}\\
&-&\frac {1}{{\cal P}^{2}}
(D^{\alpha}F_{\alpha\mu }\frac {p_{2\nu }}{p\cdot p_{2}}
+\frac {p_{2\mu }}{p\cdot p_{2}}
D^{\alpha}F_{\alpha\nu }-\frac {p_{2\mu }}{2p\cdot p_{2}}{\cal
P}^{\beta}D^{\alpha}F_{\alpha\beta}\frac {p_{2\nu }}{2p\cdot p_{2}})
\frac {1}{{\cal P}^{2}} .
\nonumber
\end{eqnarray}

\subsection{First-order effective action.}

As we discussed in Sec.~5, in order to calculate the effective action
semiclassically we can  start with the trial configuration (\ref{5.1.9}).
Making the shift $A\rightarrow A+{\bar A}^{(0)}$ in the functional integral
(\ref{5.1.1}), we obtain
%
\begin{eqnarray}
e^{iS_{\rm eff}}&=&
\int \! D \! A \ \exp i\Bigg\{\int
dx_{\perp}V^a_i(x_{\perp})U^{ai}(x_{\perp})+ 2\int
dx_{\perp}\Delta^a_i(x_{\perp})A^{ai}(0,x_{\perp})\nonumber\\
&+& 
2{\rm Tr}\int
dx_{\perp}\Big[-{1\over 2}[U^i,\Delta_i]W_1+\big(L_1+{1\over 2}
[U^i,\Delta_i]\big)W_2\nonumber\\
&-&
 {1\over 2}[V^i,\Delta_i]Y_1+\big(-L_1+{1\over 2}[V^i,\Delta_i]\big)Y_2\Big]
\nonumber\\
&+&
{1\over 2}\int d^4x
A^{a\mu}\left({\bar D}^2g_{\mu\nu}-2ig{\bar F}_{\mu\nu}+
g^2{\cal G}_{\mu\nu}\right)^{ab}A^{b\mu}+
O(A^3)\Bigg\}
\label{app1}
\end{eqnarray}
where
\begin{eqnarray}
Y_1(x_{\perp}) =[x_{\perp}+\infty p_1, x_{\perp}]^{(1)},&\quad&
Y_2(x_{\perp})=[x_{\perp},x_{\perp} -\infty p_1]^{(1)},\nonumber\\
W_1(x_{\perp}) =[x_{\perp}+\infty p_2, x_{\perp}]^{(1)},&\quad &
W_2(x_{\perp})=[x_{\perp},x_{\perp} -\infty p_2]^{(1)},
\label{app2}
\end{eqnarray}
and the operator ${\cal G}_{\mu\nu}$ is the second variational 
derivative of the source
term with respect to $A_{\mu}$, $A_{\nu}$. The non-zero 
components of
${\cal G}_{\mu\nu}$ are
%
\begin{equation}
{\cal G}_{\bullet\bullet}=\delta({2\over s}x_{\ast})
\big(\partial_i-i[V_i,)U^i{s/2\over i\partial_{\ast}},\qquad
{\cal G}_{\ast\ast}=\delta({2\over s}x_{\bullet})
\big(\partial_i-i[U_i,)V^i{s/2\over i\partial_{\bullet}} ,
\label{app3}
\end{equation}
while all other components vanish. 
In the first order in our cluster expansion we obtain
%
\begin{eqnarray}
S^{(1)}_{\rm eff}&=&-2\hbox{\bf\Big($\!\!$\Big(}0,\Delta_i^a\Big|
\left({1\over {\bar D}^2g_{ik}-2ig{\bar
F}_{ik}}\right)^{ab}|0,\Delta_k)\label{app4}\\ 
&&\hspace{-1.7cm}
+{2g^2\over s^2}\Bigg\{\hbox{\bf\Big($\!\!$\Big(}0,L_1\Big|
{p_2^{\mu}\over \alpha
+i\epsilon}+ \hbox{\bf\Big($\!\!$\Big(}0,[U_i,\Delta_i]\Big|
{p_2^{\mu}\over
\alpha}- \hbox{\bf\Big($\!\!$\Big(}0,L_1\Big|
{p_1^{\mu}\over \beta +i\epsilon}+
\hbox{\bf\Big($\!\!$\Big(}0,[V_i,\Delta_i]\Big|{p_1^{\mu}\over
\beta}\Bigg\}^a\nonumber\\ 
&&\hspace{-1.7cm}
\times\left({1\over {\bar D}^2g_{\mu\nu}-2ig{\bar
F}_{\mu\nu}+g^2{\cal G}_{\mu\nu}}\right)^{ab}\nonumber\\ 
&& \hspace{-1.7cm}
\times\Bigg\{{p_2^{\nu}\over \alpha -i\epsilon}
\Big| 0,L_1 \hbox{\bf\Big)$\!\!$\Big)}+
{p_2^{\nu}\over \alpha}\Big| 0,[U^i,\Delta_i] 
\hbox{\bf\Big)$\!\!$\Big)}
-{p_1^{\nu}\over \beta -i\epsilon}\Big| 0,L_1 
\hbox{\bf\Big)$\!\!$\Big)}+ 
{p_1^{\nu}\over\beta}\Big| 0,[V^i,\Delta_i] 
\hbox{\bf\Big)$\!\!$\Big)}\Bigg\} 
\nonumber
\end{eqnarray}
where ${1\over\alpha}\equiv{1\over 2}
({1\over\alpha-i\epsilon}+{1\over\alpha+i\epsilon})$ 
( similarly for ${1\over \beta}$) and 
$\Big| 0,\Delta_i \hbox{\bf\Big)$\!\!$\Big)}\equiv 
\int dz_{\perp}\Big| 0,z_{\perp} \hbox{\bf\Big)$\!\!$\Big)}
\Delta_i(z_{\perp})$
etc. We will now demonstrate  that with $O[U,V]^2$ 
accuracy one can reduce 
${1\over {\bar D}^2g_{\mu\nu}-2ig{\bar F}_{\mu\nu}+
g^2{\cal G}_{\mu\nu}}$ in
right-hand side of  Eq.~(\ref{app4}) to 
${g_{\mu\nu}\over {\bar D}^2}$. 
Indeed,
\begin{eqnarray}
&&{1\over \bar{D}^2g_{\mu\nu}-2ig\bar{F}_{\mu\nu}+
g^2{\cal G}_{\mu\nu}}
\label{app5}\\
&=&
{g_{\mu\nu}\over \bar{D}^2}+2ig{1\over
\bar{D}^2}\bar{F}_{\mu\nu}{1\over\bar{D}^2} -4g^2{1\over
\bar{D}^2}\bar{F}_{\mu\xi}{1\over \bar{D}^2}
\bar{F}_{\xi\nu}{1\over \bar{D}^2}-g^2
{1\over \bar{D}^2}{\cal G}_{\mu\nu}{1\over \bar{D}^2}+\ldots .
\nonumber
\end{eqnarray}
It is easy to note that the term 
$\sim {1\over\bar{D}^2}\bar{F}_{\mu\nu}{1\over\bar{D}^2}$ 
does not contribute to right-hand side of Eq.~(\ref{app4}) 
because the 
relevant components of ${\bar F}_{\mu\nu}$ vanish: 
${\bar F}_{ik}={\bar F}_{\ast\bullet}=0$. 
Let us prove that the last term in the right-hand side  
of Eq.~(\ref{app5}) leads to the
contributions $\sim [U,V]^3$.  Consider 
the first term in the right-hand side  of Eq.~(\ref{app3}).  
The corresponding
contribution is  ${1\over
{\bar D}^2}\bar{F}_{i\bullet}{1\over {\bar D}^2}
\bar{F}_{\ast k}{1\over
{\bar D}^2}+(\bullet\leftrightarrow\ast)$. Because  
$\bar{F}_{\ast
i}=U_i+O(\Delta_i),~~\bar{F}_{i\bullet}=V_i+O(\Delta_i)$ this 
term is actually
proportional to  
$\Delta_i{1\over {\bar D}^2}V_i{1\over {\bar D}^2}U_k{1\over
{\bar D}^2}\Delta_k\sim [U,V]^3$.  
Let us now turn our attention to the second term in the 
right-hand side  of 
Eq.~(\ref{app4}). The relevant contributions have
the structure 
$L_1\left({4\over {\bar D}^2}{\bar F}_{\bullet i}
{1\over {\bar D}^2}{\bar F}_{\bullet
i}{1\over {\bar D}^2} - 
{1\over {\bar D}^2}
{\cal G}_{\bullet\bullet}\right)L_1{1\over {\bar D}^2}$,  
$L_1{1\over {\bar D}^2}{\bar F}_{\bullet
i}{1\over {\bar D}^2} {\bar F}_{\ast i}{1\over {\bar D}^2}L_1$,  
$[V_i,\Delta_i]\left({1\over {\bar D}^2}{\bar F}_{\bullet i}
{1\over {\bar D}^2}{\bar F}_{\bullet
i}{1\over {\bar D}^2} + 
{1\over {\bar D}^2}{\cal G}_{\bullet\bullet}
{1\over {\bar D}^2}
\right)[V_i,\Delta_i]$, 
$[V_i,\Delta_i]{1\over {\bar D}^2}{\bar F}_{\bullet i}
{1\over {\bar D}^2}{\bar F}_{\ast i}{1\over
{\bar D}^2}[U_i,\Delta_i]$,\\
 and similar expressions with 
$U\leftrightarrow V$,
$\ast\leftrightarrow \bullet$. All of them are clearly 
$\sim[U,V]^3$ except the
first term which is
\begin{equation}
g^2\hbox{\bf\Big($\!\!$\Big(}0,L_1\Big|{1\over \beta +
i\epsilon}\left({4\over \bar{D}^2}{\bar F}_{\bullet i}{1\over
{\bar D}^2}{\bar F}_{\bullet i}{1\over {\bar D}^2} - 
{1\over {\bar D}^2}{\cal G}_{\bullet\bullet}{1\over {\bar D}^2}
\right){1\over \beta -
i\epsilon}\Big| 0,L_1 \hbox{\bf\Big)$\!\!$\Big)} .
\label{app6}
\end{equation}
If we neglect the $[U,V]^3$ terms in cluster expansion, the Green
function in braces in right-hand side  of Eq.~(\ref{app6}) 
should be taken in the
$U_i\theta(x_{\ast})$ background. This Green function has the form
%
\begin{eqnarray}
&&\hbox{\bf\Big($\!\!$\Big(}x\Big|
-4{1\over {\bar D}^2}{\bar F}_{\bullet i}{1\over
{\bar D}^2}{\bar F}_{\bullet i}{1\over {\bar D}^2} + 
{1\over {\bar D}^2}{\cal G}_{\bullet\bullet}{1\over
{\bar D}^2}\Big| y\hbox{\bf\Big)$\!\!$\Big)}~=~
\hbox{\bf\Big($\!\!$\Big(}x\Big|{\cal O}_{\bullet\bullet}
 \Big| y\hbox{\bf\Big)$\!\!$\Big)}\label{app7}\\
&=&-i\theta(x_\ast)\theta(-y_\ast)U^{\dagger}(x_{\perp})
\int dz \delta(z_\ast)
\hbox{\bf\Big($\!\!$\Big(}x\Big|{1\over
p^2\alpha}\Big| z\hbox{\bf\Big)$\!\!$\Big)}
\vec{\partial}_{\perp}^2U(z_{\perp})
\hbox{\bf\Big($\!\!$\Big(}z\Big|{1\over p^2}
\Big| y\hbox{\bf\Big)$\!\!$\Big)} , \nonumber
\end{eqnarray}
plus the similar term $\sim \theta(-x_\ast)\theta(y_\ast)$. 
It is easy to see
that the terms $\sim \theta(x_\ast)\theta(-y_\ast)$ or
$\sim\theta(-x_\ast)\theta(y_\ast)$ do not contribute 
to Eq.~(\ref{app6}) --- 
recall that this term comes from the contraction of 
$L_1W_2(x)$ and $L_1Y_2(y)$
where both $x_\ast,y_\ast<0$.

\begin{figure}[htb]
\centerline{
\epsfysize=5cm
\epsffile{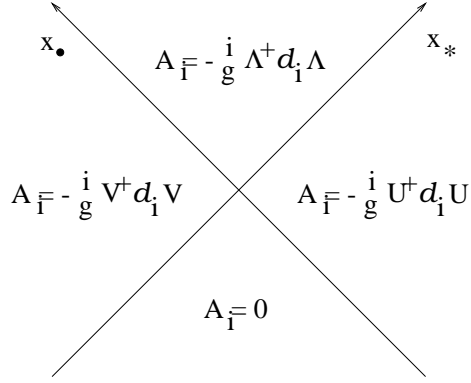}}
\caption{Trial field configuration.} 
\label{figapp}
\end{figure}

Thus, the $[U,V]^2$ term in cluster expansion of 
Eq.~(\ref{app4}) reduces to
%
\begin{eqnarray}
&&\hspace{-5mm}S^{(1)}_{\rm eff}~=~
\hbox{\bf\Big($\!\!$\Big(}0,\Delta_i\Big|{-2\over {\bar D}^2}\Big|
0,\Delta_i \hbox{\bf\Big)$\!\!$\Big)}\label{app8}\\ 
&&\hspace{-5mm}-~{g^2\over
s}\hbox{\bf\Big($\!\!$\Big(}0,L_1\Big|
\Big[{1\over \alpha+i\epsilon}{1\over
{\bar D}^2} {1\over \beta -i\epsilon}+
{1\over \beta+i\epsilon}{1\over
{\bar D}^2} {1\over \alpha -i\epsilon}\Big]
\Big| 0,L_1 \hbox{\bf\Big)$\!\!$\Big)}
\nonumber\\  
&&\hspace{-5mm}-~{g^2\over
s}\hbox{\bf\Big($\!\!$\Big(}0,L_1\Big|
{1\over \beta+i\epsilon} {1\over
{\bar D}^2}{1\over \alpha}\Big| 0,[U^i,\Delta_i] 
\hbox{\bf\Big)$\!\!$\Big)}
+
{g^2\over s}\hbox{\bf\Big($\!\!$\Big(}0,L_1\Big|
{1\over \alpha+i\epsilon}
{1\over
{\bar D}^2}{1\over \beta}\Big| 0,[V^i,\Delta_i] 
\hbox{\bf\Big)$\!\!$\Big)}
\nonumber \\
&&\hspace{-5mm}-~{g^2\over s}
\hbox{\bf\Big($\!\!$\Big(}0,[U^i,\Delta_i]\Big|{1\over \alpha} 
{1\over
{\bar D}^2}{1\over \beta-i\epsilon}\Big| 0,L_1
\hbox{\bf\Big)$\!\!$\Big)}
+
{g^2\over s}\hbox{\bf\Big($\!\!$\Big(}0,[V_i,\Delta^i]
\Big|{1\over \beta}
{1\over
{\bar D}^2}{1\over \alpha-i\epsilon}
\Big| 0,L_1 \hbox{\bf\Big)$\!\!$\Big)} . \nonumber
\end{eqnarray}
It is easy to see that the remaining Green function connect points 
belonging to
the different boundaries of the same sector in Fig.~\ref{figapp}.
It may be demonstrated that up to $[U,V]$ accuracy the only
effect  of the background field on the Green function 
with the arguments belonging to the same sector is 
the corresponding gauge factor: 
$\hbox{\bf\Big($\!\!$\Big(}x\Big|{1\over D^2}\Big| y\hbox{\bf\Big)$\!\!$\Big)}=
\Omega^{\dagger}(x_{\perp})
\hbox{\bf\Big($\!\!$\Big(}x\Big|{-1\over p^2}\Big| y\hbox{\bf\Big)$\!\!$\Big)}
\Omega^{\dagger}(y_{\perp})$,
where $\Omega$ is $U,~V$, or $\Lambda$.  We obtain
\begin{eqnarray} 
&&\hspace{-0.9cm}\hbox{\bf\Big($\!\!$\Big(}0,x_{\perp}
\Big|{1\over\alpha+i\epsilon}{1\over
D^2}{1\over\beta-i\epsilon}\Big| 0,y_{\perp}\hbox{\bf\Big)$\!\!$\Big)}= 
\hbox{\bf\Big($\!\!$\Big(}0,x_{\perp}\Big|{1\over\alpha+i\epsilon}{-1\over
p^2+i\epsilon}{1\over\beta-i\epsilon}\Big| 
0,y_{\perp}\hbox{\bf\Big)$\!\!$\Big)} ,
\label{app9}\\
&&\hspace{-0.9cm}
\hbox{\bf\Big($\!\!$\Big(}0,x_{\perp}\Big|{1\over\alpha+i\epsilon}
{1\over
D^2}{1\over\beta+i\epsilon}\Big| 0,y_{\perp}\hbox{\bf\Big)$\!\!$\Big)}= 
U^{\dagger}_x\hbox{\bf\Big($\!\!$\Big(}0,x_{\perp}\Big|{1\over\alpha+i\epsilon}
{-1\over
p^2+i\epsilon}{1\over\beta+i\epsilon}
\Big| 0,y_{\perp}\hbox{\bf\Big)$\!\!$\Big)}U_y ,
\nonumber\\
&&\hspace{-0.9cm}\hbox{\bf\Big($\!\!$\Big(}0,x_{\perp}\Big|
{1\over\alpha-i\epsilon}{1\over
D^2}{1\over\beta-i\epsilon}\Big| 0,y_{\perp}\hbox{\bf\Big)$\!\!$\Big)}= 
V^{\dagger}_x
\hbox{\bf\Big($\!\!$\Big(}0,x_{\perp}\Big|{1\over\alpha-i\epsilon}{-1\over
p^2+i\epsilon}{1\over\beta-i\epsilon}\Big| 0,y_{\perp}
\hbox{\bf\Big)$\!\!$\Big)}V_y .
\nonumber
\end{eqnarray}
In the leading log approximation~\footnote{This 
formula may obviously seem confusing since 
$\hbox{\bf\Big($\!\!$\Big(}0,x_{\perp}
\Big|{1\over p^2+i\epsilon}\Big| 0,y_{\perp}\hbox{\bf\Big)$\!\!$\Big)}=
{-i\over 4\pi^2(\vec{x}-\vec{y})_{\perp}^2}$, which does not have any
$\ln{\sigma\over\sigma'}$. However, careful analysis with the 
slope of the $Y$ operators $n=\sigma p_1+\tilde{\sigma} p_2$ instead of $p_1$
and the slope of $W$ operators $n'=\sigma'p_1+\tilde{\sigma}'p_2$ instead of
$p_2$, yields logarithmic contribution of the form
\begin{equation}
\hbox{\bf\Big($\!\!$\Big(}0,x_{\perp}\Big|
{\alpha\beta\over(\alpha+{\sigma\over\tilde{\sigma}}\beta-i\epsilon)
(\beta+{\tilde{\sigma}'\over\sigma'}\alpha+i\epsilon)}
{1\over p^2+i\epsilon}\Big| 0,y_{\perp}\hbox{\bf\Big)$\!\!$\Big)}=-{i\over
4\pi}\ln{\sigma\over\sigma'}\delta^2(x_{\perp}-y_{\perp}). 
\label{app10a}
\end{equation}
}
\begin{equation}
\hbox{\bf\Big($\!\!$\Big(}0,x_{\perp}\Big|
{-1\over p^2+i\epsilon}\Big| 0,y_{\perp}\hbox{\bf\Big)$\!\!$\Big)}={i\over
4\pi}\ln{\sigma\over\sigma'}\delta^2(x_{\perp}-y_{\perp}) ,
\label{app10}
\end{equation}
and
\begin{eqnarray}
\hbox{\bf\Big($\!\!$\Big(}0,x_{\perp}\Big|
{1\over\alpha\pm i\epsilon}{-2/s\over
p^2+i\epsilon}{1\over\beta-i\epsilon}\Big| 0,y_{\perp}
\hbox{\bf\Big)$\!\!$\Big)}\!\!\!\!&=&\!\!\!\!
\hbox{\bf\Big($\!\!$\Big(}0,x_{\perp}\Big|
{1\over\alpha\pm i\epsilon}{-2/s\over
p^2+i\epsilon}{1\over\beta+i\epsilon}\Big| 
0,y_{\perp}\hbox{\bf\Big)$\!\!$\Big)}\nonumber\\
&=&\!\!\!\!
{i\over 2\pi}\ln{\sigma\over\sigma'}
\hbox{\bf\Big($\!\!$\Big(}x_{\perp}\Big|{1\over \vec{p}_{\perp}^2}
\Big| y_{\perp}\hbox{\bf\Big)$\!\!$\Big)},
\label{app11}
\end{eqnarray}
so we get
%
\begin{eqnarray}
S^{(1)}_{\rm eff}&=&{-i\over 2\pi}\ln{\sigma\over\sigma'}
\Bigg(
\int dx_{\perp} \Delta^{ai}(x_{\perp}) \Delta^a_i(x_{\perp})~+~
g^2\int dx_{\perp}dy_{\perp} \nonumber\\
&\times&
\Bigg\{ L_1^a(x_{\perp})
\hbox{\bf\Big($\!\!$\Big(}x_{\perp}\Big|{1\over \vec{p}_{\perp}^2}
\Big|y_{\perp}\hbox{\bf\Big)$\!\!$\Big)}L^a_1(y_{\perp})\nonumber\\
&-&
L_1^a(x_{\perp})\hbox{\bf\Big($\!\!$\Big(}x_{\perp}\Big|{1\over 2}
(U^{\dagger}
{1\over \vec{p}_{\perp}^2}U+{1\over \vec{p}_{\perp}^2})
\Big|y_{\perp}\hbox{\bf\Big)$\!\!$\Big)}^{ab} [V_i,\Delta^i]^b(y_{\perp})
\nonumber\\
&+&
L_1^a(x_{\perp})
\hbox{\bf\Big($\!\!$\Big(}x_{\perp}\Big|{1\over 2}(V^{\dagger}
{1\over \vec{p}_{\perp}^2}V+{1\over \vec{p}_{\perp}^2})\Big| y_{\perp}
\hbox{\bf\Big)$\!\!$\Big)}^{ab}
[U_i,\Delta^i]^b(y_{\perp})\Bigg\}\Bigg) . 
\label{app12}
\end{eqnarray}
Finally, the effective actuion in the $[U,V]^2$ order in the cluster expansion
has the form
%
\begin{eqnarray}
&&\hspace{-13mm}S^{(1)}_{\rm eff}~=~-{ig^2\over 2\pi}
\ln{\sigma\over\sigma'}2{\rm
Tr} \Bigg\{
\int dx_{\perp} {1\over g^2}\Delta_i(x_{\perp}) \Delta^i(x_{\perp})+
\int dx_{\perp}dy_{\perp}\Big\{L_1(x_{\perp})\nonumber\\
&&\hspace{-13mm}\times
\hbox{\bf\Big($\!\!$\Big(}x_{\perp}
\Big|{1\over \vec{p}_{\perp}^2}
\Big| y_{\perp}\hbox{\bf\Big)$\!\!$\Big)}L_1(y_{\perp})+
2L_1(x_{\perp})\hbox{\bf\Big($\!\!$\Big(}x_{\perp}\Big|
{1\over \vec{p}_{\perp}^2}\Big| y_{\perp}
\hbox{\bf\Big)$\!\!$\Big)}[U_i-V_i,\Delta^i]( y_{\perp})
\Big\}\Bigg\}
\label{app13}
\end{eqnarray}
which coincides with Eq.~(\ref{5.1.21}).


\section*{References}
\begin{thebibliography}{99}


\bibitem{bfkl}
V.S. Fadin, E.A. Kuraev, and L.N. Lipatov,
{\it Phys. Lett.} {\bf B 60}, 50 (1975)
I.I. Balitsky and L.N. Lipatov,
{\it Sov. Journ. Nucl. Phys.} 
{\bf 28}, 822 (1978).

\bibitem{lobzor}
L.N. Lipatov, 
{\it Phys. Rept.} {\bf 286}, 131 (1997).

\bibitem{fizrev}
I. Balitsky, 
{\it Phys. Rev.} {\bf D60}, 014020 (1999).

\bibitem{lrmodel}
L. McLerran and R. Venugopalan,  
{\it Phys. Rev.} {\bf D49}, 2233 (1994);
{\it Phys. Rev.} {\bf D49}, 3352 (1994).

\bibitem{jklw}
J. Jalilian-Marian, A. Kovner, L. McLerran, and
H. Weigert, 
{\it Phys. Rev.} {\bf D55},5414 (1997).

\bibitem{finn}
K.J. Eskola, K. Kajantie, P.V. Ruuskanen, and K. Tuominen,
{\it Nucl. Phys.} {\bf B570}, 379 (2000).

\bibitem{mu00}
A.H. Mueller, 
{\it Nucl. Phys.}  {\bf B572}, 227 (2000);
R. Venugopalan, 
{\it Acta. Phys. Polon.}  {\bf B30}, 3731 (1999).

\bibitem{nacht}
O. Nachtmann,
{\it Annals Phys.}  {\bf 209}, 436 (1991).

\bibitem{collell}
J.C. Collins and R.K. Ellis 
{\it Nucl. Phys.}  {\bf B360}, 3 (1991).

\bibitem{drossbook}
J. R. Forshaw and D. A. Ross,
{\em 
Quantum Chromodynamics and the Pomeron},
Cambridge Lecture Notes in Physics, 9
(Cambridge Univ. Press, 1997).

\bibitem{mes}
I.I. Balitsky and L.N. Lipatov, 
{\it JETP Letters} {\bf 30}, 355 (1979).

\bibitem{lip86}
L.N. Lipatov, 
{\it Sov. Phys. JETP} {\bf 63}, 904 (1986). 

\bibitem{nlobfkl}
V.S. Fadin and L.N. Lipatov,
{\it Phys. Lett.} {\bf B 429}, 127 (1998);
G. Carnici and M. Ciafaloni,
{\it Phys. Lett.} {\bf B 430}, 349 (1998).

\bibitem{smallxren}
E. Levin, 
{\it Nucl. Phys.} {\bf B453}, 303 (1995).

\bibitem{levininst}
D.E. Kharzeev and E. Levin, 
{\it Nucl. Phys.} {\bf B578}, 351 (2000);\\
D.E. Kharzeev, Y. V. Kovchegov, and E. Levin,
Preprint BNL-NT-00-18, TAUP-2637-2000, Jul 2000,
[hep-ph/0007182].

\bibitem{mbraun1}
M.A. Braun,
{\it Eur.Phys.J.}{\bf C16}, 337 (2000), 
[hep-ph/0001268];
[hep-ph/0010041].

\bibitem{kovmuller}
Y.V. Kovchegov, A.H. Mueller, 
{\it Phys. Lett.} {\bf B 439}, 428 (1998).

\bibitem{white}
C. Coriano, A. R. White, and M. Wusthoff,
{\it Nucl. Phys.}  {\bf B493}, 397 (1997).

\bibitem{barmbra}
N. Armesto, J. Bartels, and M.A. Braun,
{\it Phys. Lett.} {\bf B 442}, 459 (1998).

\bibitem{dross}
J.R. Forshaw, D.A. Ross, A. Sabio-Vera,  
[hep-ph/0011047]. 

\bibitem{cico} M. Ciafaloni, D. Colferai, 
G.P. Salam, 
{\it Phys. Rev.} {\bf D60}, 114036 (1999).

\bibitem{many}
R. Kirschner, L.N. Lipatov, L. Szymanowski, 
{\it Nucl. Phys.}  {\bf B425}, 579 (1994);
{\it Nucl. Phys.}  {\bf B452}, 369 (1996).

\bibitem{glla}
J. Bartels,
{\it Nucl. Phys.}  {\bf B175}, 365 (1980);\\
J. Kwiecinski and M. Praszalowicz, 
{\it Phys. Lett.} {\bf B 94}, 413 (1980).

\bibitem{mu94}
A.H. Mueller, 
{\it Nucl. Phys.}  {\bf B415}, 373 (1994);
 A.H. Mueller and Bimal Patel, 
{\it Nucl. Phys.}  {\bf B425}, 471 (1994).

\bibitem{nnn}
N.N. Nikolaev and B.G. Zakharov,
{\it Phys. Lett.} {\bf B 332}, 184 (1994);
{\it Z. Phys.}  {\bf C64}, 631 (1994);
N.N. Nikolaev B.G. Zakharov, and V.R. Zoller,
{\it JETP Letters} {\bf 59}, 6 (1994).

\bibitem{regg2} 
V.S. Fadin, R. Fiore, M.I. Kotsky, 
{\it Phys. Lett.} {\bf B 359}, 181 (1995);
{\it Phys. Lett.} {\bf B 387}, 593 (1996).

\bibitem{grishi}
 I.A. Korchemskaya and G.P. Korchemsky,   
{\it Phys. Lett.} {\bf B 387}, 346 (1996).

\bibitem{lip9093} 
L.N. Lipatov, 
{\it Phys. Lett.} {\bf B 251}, 284 (1990);
{\it Phys. Lett.} {\bf B 309}, 394 (1993).

\bibitem{kortok}
G.P. Korchemsky, [hep-ph/9511370]. 

\bibitem{chengwubook}
 H. Cheng and T.T. Wu,
{\em Expanding Protons: Scattering at High Energies},
(MIT press, Cambridge, 1987).

\bibitem{lkf}
L.N. Lipatov, 
{\it JETP Letters} {\bf 59}, 571 (1994);\\
L.D. Faddeev and G.P. Korchemsky,
{\it Phys. Lett.} {\bf B 342}, 311 (1995).

\bibitem{janik}
R.A. Janik and J. Wosiek, 
{\it Phys. Rev. Lett.} {\bf 82}, 1092 (1999).

\bibitem{mbraun}
M.A. Braun, P. Gauron, and B. Nicolescu, 
{\it Nucl. Phys.}  {\bf B542}, 329 (1999).

\bibitem{lipodderon}
J. Bartels, L.N. Lipatov, and G.P. Vacca,
``A New Odderon Solution in Perturbative QCD'',
preprint DESY 99-115 (Dec 1999), 
[hep-ph/9912423]. 

\bibitem{ing}
I. Balitsky, 
{\it Nucl. Phys.}  {\bf B463}, 99 (1996).

\bibitem{eveq}
I. Balitsky and V.M. Braun, 
{\it Nucl. Phys.}  {\bf B311}, 541 (1989).

\bibitem{yura}
Yu.V. Kovchegov,  
{\it Phys. Rev.} {\bf D60}, 034008 (1999);
{\it Phys. Rev.} {\bf D61},074018 (2000).

\bibitem{GLR}
L.V. Gribov, E.M. Levin, and M.G. Ryskin, 
{\it Phys. Rept.} {\bf 100}, 1 (1983).

\bibitem{muchu}
A.H. Mueller and J.W. Qiu  
{\it Nucl. Phys.}  {\bf B268}, 427 (1986).

\bibitem{bkucha}
I. Balitsky and E. Kuchina
{\it Phys. Rev.} {\bf D62},074004  (2000).

\bibitem{mu95}
A.H. Mueller, 
{\it Nucl. Phys.}  {\bf B437}, 107 (1995).

\bibitem{difope}
I. Balitsky, 
[hep-ph/9706411]. 

\bibitem{keld}
I. Balitsky and V.M. Braun, 
{\it Phys. Lett.} {\bf B 222}, 121 (1989);
{\it Nucl. Phys.}  {\bf B361}, 93 (1991).

\bibitem{diffdist}
A. Berera and D.E. Soper,
{\it Phys. Rev.} {\bf D53}, 6162 (1996);
 M. Grazzini, L. Trentadue, and G. Veneziano,
{\it Nucl. Phys.}  {\bf B519}, 394 (1998);
J.C. Collins,
{\it Phys. Rev.} {\bf D57}, 3051 (1998).

\bibitem{bar3pom}
J. Bartels and M. Wusthoff,
{\it Z. Phys.}  {\bf C66}, 157 (1995).

\bibitem{tripom}
A. Bialas, H. Navelet, and R. Peschanski, 
{\it Phys. Rev.} {\bf D57}, 6585 (1998);
G.P. Korchemsky, 
{\it Nucl. Phys.}  {\bf B550}, 397 (1999).

\bibitem{kovlev}
Y. V. Kovchegov and E. Levin, 
{\it Nucl. Phys.}  {\bf B577}, 221 (2000).

\bibitem{prl}
  I. Balitsky, 
{\it Phys. Rev. Lett.} {\bf 81}, 2024 (1998).

\bibitem{wlup}
I.I. Balitsky, 
{\it Nucl. Phys.}  {\bf B254}, 166 (1985).

\bibitem{dosch}
  H.G. Dosch, E. Ferreira, and A. Kraemer,
{\it Phys. Rev.} {\bf D50}, 2015 (1994).

\bibitem{verlinde}
   H. Verlinde and E. Verlinde,
{\em  ``QCD at High Energies and Two-Dimensional Field Theory''},
 preprint PUPT-1319, 
 [hep-th/9302104].

\bibitem{larry1}
L. McLerran and R. Venugopalan, 
{\it Phys. Rev.} {\bf D50}, 2225 (1994);
A. Ayala, J. Jalilian-Marian, L. McLerran , and
R. Venugopalan, 
{\it Phys. Rev.} {\bf D52}, 2935 (1995).

\bibitem{kovner}
A. Kovner, L. McLerran and H. Weigert, 
{\it Phys. Rev.} {\bf D52}, 6231 (1995).

\bibitem{tok}
 I. Balitsky,
[hep-ph/9808215].


\bibitem{nucnuc}
A. Kovner, L. McLerran and H. Weigert, 
{\it Phys. Rev.} {\bf D52}, 3809 (1995);
M. Gyulassy and L. McLerran, 
{\it Phys. Rev.} {\bf C52}, 2219 (1997).

\bibitem{raju}
Alex Krasnitz, Raju Venugopalan,
{\it Nucl. Phys.}  {\bf B557}, 237 (1999);
{\it Phys. Rev. Lett.} {\bf 84}, 4309 (2000);
[hep-ph/0007108].

\bibitem{kop}
B.Z Kopeliovich, I.L. Lapidus, and Al.B. Zamolodchikov,
{\it JETP Letters} {\bf 33}, 612 (1981);\\
N.N. Nikolaev and B.G. Zakharov,
{\it Z. Phys.}  {\bf C53}, 331 (1992).

\bibitem{jamal}
J. Jalilian-Marian, A. Kovner, and
H. Weigert, 
{\it Phys. Rev.} {\bf D59},014015 (1999).

\bibitem{almuller}
Yu.V. Kovchegov, A.H. Mueller, 
{\it Nucl. Phys.}  {\bf B529}, 451 (1998).

\bibitem{mdolini}
I. Balitsky and V.M. Braun,
{\it Nucl. Phys.}  {\bf B380}, 51 (1992).
\end{thebibliography}
\end{document}